\documentclass[preprint]{aastex63}

\usepackage[english]{babel}
\usepackage[utf8x]{inputenc}
\usepackage{amsmath,amsfonts,amsthm}   
\usepackage{graphicx}          
\usepackage{latexsym}
\usepackage{placeins}					

\usepackage{hyperref}
\hypersetup{
		colorlinks=true,
		linktoc=all,
		linkcolor=blue,
}

\begin{document}
\title{Simulations for Planning Next-Generation Exoplanet Radial Velocity Surveys}
\date{\today}
\author{Patrick D Newman}
\email{pnewman8@gmu.edu}
\affiliation{Department of Physics and Astronomy, George Mason University, 4400 University Drive, MSN 3F3, Fairfax, VA 22030}
\author{Peter Plavchan}
\affiliation{Department of Physics and Astronomy, George Mason University, 4400 University Drive, MSN 3F3, Fairfax, VA 22030}
\author{Jennifer A. Burt}
\affiliation{Jet Propulsion Laboratory, California Institute of Technology, 4800 Oak Grove Dr., Pasadena, CA 91109, USA}
\author{Johanna Teske}
\affiliation{Carnegie Institution for Science, Earth and Planets Laboratory, 5241 Broad Branch Road, NW, Washington, DC 20015}
\author{Eric E. Mamajek}
\affiliation{Jet Propulsion Laboratory, California Institute of Technology, 4800 Oak Grove Dr., Pasadena, CA 91109, USA}

\author{Stephanie Leifer}
\affiliation{The Aerospace Corporation, 200 S Los Robles Ave \#150, Pasadena, CA 91101}
\author{B. Scott Gaudi}
\affiliation{Department of Astronomy, The Ohio State University, 140 W. 18th Ave., Columbus, OH 43210}

\author{Gary Blackwood}
\affiliation{Jet Propulsion Laboratory, California Institute of Technology, 4800 Oak Grove Dr., Pasadena, CA 91109, USA}

\author{Rhonda Morgan}
\affiliation{Jet Propulsion Laboratory, California Institute of Technology, 4800 Oak Grove Dr., Pasadena, CA 91109, USA}

\begin{abstract}
Future direct imaging missions such as HabEx and LUVOIR aim to catalog and characterize Earth-mass analogs around nearby stars. The exoplanet yield of these missions will be dependent on the frequency of Earth-like planets, and potentially the a priori knowledge of which stars specifically host suitable planetary systems. Ground or space based radial velocity surveys can potentially perform the pre-selection of targets and assist in the optimization of observation times, as opposed to an uninformed direct imaging survey. In this paper, we present our framework for simulating future radial velocity surveys of nearby stars in support of direct imaging missions. We generate lists of exposure times, observation time-series, and radial velocity time-series given a direct imaging target list. We generate simulated surveys for a proposed set of telescopes and precise radial velocity spectrographs spanning a set of plausible global-network architectures that may be considered for next generation extremely precise radial velocity surveys. We also develop figures of merit for observation frequency and planet detection sensitivity, and compare these across architectures. From these, we draw conclusions, given our stated assumptions and caveats, to optimize the yield of future radial velocity surveys in support of direct imaging missions. We find that all of our considered surveys obtain sufficient numbers of precise observations to meet the minimum theoretical white noise detection sensitivity for Earth-mass habitable zone planets, with margin to explore systematic effects due to stellar activity and correlated noise.
\end{abstract}

\keywords{methods: data analysis, techniques: radial velocities}

\section{Introduction} \label{sec:Introduction}

Future, ground-based, extreme precision radial velocity (EPRV) surveys aim to detect Earth-mass analogs orbiting nearby Sun-like stars, many of which overlap with the possible survey samples of future direct imaging mission concepts such as HabEx and LUVOIR \citep{2019arXiv191206219T,2020arXiv200106683G,2021JATIS...7b1220M,2020arXiv200613428P}. The National Academies Exoplanet Science Strategy (ESS) report in 2018 recommended that ``NASA and NSF should establish a strategic initiative in extremely precise radial velocities (EPRVs) to develop methods and facilities for measuring the masses of temperate terrestrial planets orbiting Sun-like stars'' \citep{NAP25187}. In response, NASA and the NSF established the EPRV Working Group in 2019 to ``recommend a ground-based program architecture and implementation plan accounting for the full scope of the ESS report findings.'' \citep{2021arXiv210714291C}. Herein, ``architecture'' refers to an observatory or set of observatories and spectrographs, target lists and survey strategies to meet the goal of the ESS recommendation, and we will use the term architecture throughout this paper in this context.

Consideration of different architectures necessitates an assessment through simulation of the potential yield of a future EPRV survey. Historically, RV survey yields have been based upon theoretical photon noise approximations from the number of targets, number of observations, survey duration, and single measurement precision \citep{2007ApJ...655..550G,2015arXiv150301770P,2007MNRAS.377.1610K,2016PASP..128f6001F}. However, cadence sampling biases, weather and other realistic observatory and spectrograph factors can impact actual survey yields \citep{2016PASP..128k4401H}. Additionally, with the increased sensitivity of EPRV instrumentation, stellar activity has impacted EPRV exoplanet detection sensitivity and. consequently, survey yields \citep{2015arXiv150301770P,2014Sci...345..440R,2015ApJ...805L..22R,2018ApJ...864L..28R,2015ApJ...801...79R,2015AnA...580A..31H,2015PhDT.......193H,2016AnA...585A.134D,2016ApJ...820L...5K,2016ApJ...821L..19N,2016MNRAS.459.3565V,2021arXiv210507005L}. There has been one community challenge to date taking HARPS cadence measurements for individual targets and testing the injection and recovery of planetary systems for modeling and accounting for stellar activity correlated noise \citep{2016AnA...593A...5D,2017AnA...598A.133D}.

More recently, \citet{2018MNRAS.479.2968H} carried out a EPRV survey simulation for the HARPS3 spectrometer. They considered three different cadences, four different possible planetary architectures, instrument white noise, and star spot correlated noise around a G6V star. They found that some common survey cadences would be insufficient to accurately find/characterize earth-mass habitable zone planets, though more aggressive ones often are. Similarly, \citet{2020AJ....159...23N} investigated the impact of realistic cadence aliases mixing with stellar activity in producing false-positive EPRV signals for the Kepler 20 and 131 systems.

Imperfectly modeled stellar activity can generate spurious signals of planets \citep{2016MNRAS.456L...6R}, and biases in sampling can also cause large discrepancies in the properties of a planet. \citep{2017MNRAS.471L.125R} However, current mitigation techniques should be able to mitigate stellar activity and detect planets at the 1-2 m/s level. \citet{2021AnA...649A..26L,2021MNRAS.tmp.1963R}

In this paper, we simulate possible dedicated EPRV surveys to search nearby stars to detect and characterize Earth-mass analogs as input targets prior to the launch of a direct imaging mission such as HabEx or LUVOIR \citep{2021arXiv210714291C}. The architectures considered herein are all potential next-generation EPRV global networks of observatories assumed to be operational prior to the launch of a direct imaging mission. Our goal herein is to to evaluate survey cadences for these potential architectures for their ability detect Earth-like planets around Sun-like stars. For this, we choose a characteristic figure of merit of a 10 cm/s semi-amplitude stellar reflex velocity measured at a minimum SNR of 10 (${\S}$\ref{sec:PlanetDetectionFigureofMerit}). This RV semi-amplitude is representative of an edge-on Earth-mass planet in Habitable Zone (HZ) of a Sun-like star. A great many hidden variables can change this value by approximately a factor of 2 -- e.g. inclination, eccentricity, HZ location, stellar mass, etc. Due to the approximate nature of this figure of merit, we explicitly look at distributions of our survey sensitivity and do not treat it as a simple cut-off. We do not consider correlated noise from stellar activity or otherwise, and this is the subject of a companion paper by \citet{Luhn2021}. We restrict the analysis herein to the consideration of sensitivity guided by uncorrelated measurement photon noise statistics \citep{2007ApJ...655..550G}. No injection and recovery tests with realistic exoplanet populations and stellar activity are performed on the resulting simulated data, but such efforts could be the subject of future work.

In ${\S}$\ref{sec:Simulation}, we describe the simulations, and the features therein that we consider and ignore for a set of plausible observatory and spectrograph architectures. In ${\S}$\ref{sec:Results}, we show the results for each architecture. In ${\S}$\ref{sec:ArchComp}, we compare the figures of merit across architectures. In ${\S}$\ref{sec:Discussion}, we present a discussion of the results and limitations and effects of survey simulations, and in ${\S}$\ref{sec:Conclusions} we present our conclusions. 

\subsection{Planet Detection Figure of Merit} \label{sec:PlanetDetectionFigureofMerit}
We use the theoretical best-case scenario for estimating the detection significance achieved for a 1 $M_\oplus$ planet in an edge-on, circular, 1 $\rm{au}$ orbit around a 1 $M_\odot$ Sun-like star with unknown orbital phase.\citep{2007ApJ...655..550G}

\begin{equation}
\label{eqn:SNR}
  SNR = \frac{K}{\sigma}\sqrt{\frac{N_{\mbox{obs}}}{2}} 
\end{equation}

\noindent Here $K$ is the stellar reflex velocity semi-amplitude due to the orbiting exoplanet, $\sigma$ is the typical single measurement precision, and $N_{\mbox{obs}}$ is the number of observations realized from our simulations over the survey duration, which is assumed to be greater than the exoplanet orbital periods of interest. 
We note that Equation $\ref{eqn:SNR}$ formally assumes uniform cadence in phase. For the large number of observations that we find are required for a robust detection, this is an excellent approximation for most cases, with the exception being planets with periods very close to one year.\citep{2018AJ....156..255B}
Both the instrument and photon noise are added in quadrature for $\sigma$. In cases of instruments operating at different precisions (Architecture VIII), it can be shown that the equation is expanded to:

\begin{equation}
\label{eqn:SNR2}
  SNR = K \sqrt{\frac{1}{2} \left(\frac{N_{\mbox{obs1}}}{2\sigma_1^2} + \frac{N_{\mbox{obs2}}}{2\sigma_2^2}\right)}
\end{equation}


Throughout this analysis we assume stellar activity can be perfectly modeled and subtracted from observed RVs.

\section{Survey Simulation Description \label{sec:Simulation}}
\begin{figure}[ht]
\includegraphics[width=0.99\textwidth]{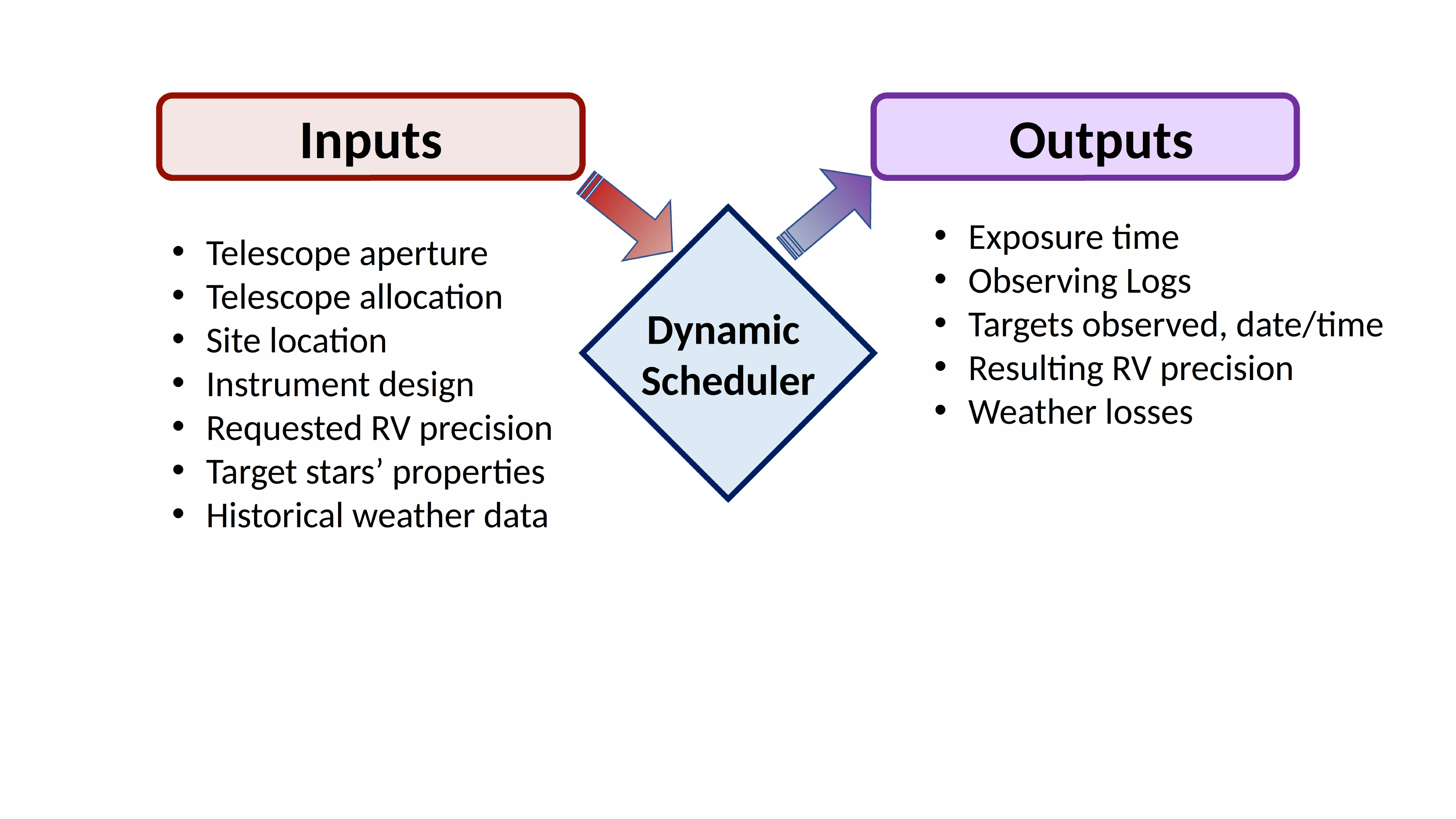}
\caption{Outline of the simulation pipeline, including the exposure time calculator (ETC) and dispatch scheduler.}
\label{fig:cartoon}
\end{figure}

In this section, we outline the survey simulations and the components\ref{fig:cartoon} and assumptions that went into these simulations. At the top level we simulate multiple radial velocity surveys to benchmark against one another. Each simulated radial velocity survey consists of a set of telescopes in a global network with different spectrographs and telescope specifications, which we call ``architectures'' (see ${\S}$\ref{sec:Architectures}), along with other additional miscellaneous simulation inputs (${\S}$\ref{sec:MiscInputs}). Then for each telescope site, we simulate the observations of a set of targets (see ${\S}$\ref{sec:TargetStarLists}) using a dispatch schedule that takes into account a variety of observational constraints and target prioritization ($\S$\ref{sec:DispatchScheduler}). The RV precision ($\S$\ref{sec:RVModel}) and exposure times are then simulated, and we conclude with describing the outputs from the simulations in ${\S}$\ref{sec:SimulationOutputs}.

\subsection{The Dispatch Scheduler} \label{sec:DispatchScheduler}
All survey simulations were carried out using a dispatch scheduler that selects which target to observe next ``on the fly'', instead of following a preset observing schedule. The dispatch scheduler simulates a survey in moderate detail to provide a realistic observation time series. Our version\footnote{Available at \url{https://github.com/pdn4kd/dispatch_scheduler}} is derived from the MINERVA scheduler \citep{2015JATIS...1b7002S,2015AAS...22525826N}. The scheduler takes a list of targets with name, right ascension, declination, and estimated observation times as inputs, and generates a time series of observations for each target (as well as rise and set times for the Sun and each of the potential targets). It also generates a nightly summary file containing the start and end times of the night, the simulated weather, the number of times each star was observed, the amount of time spent on each star which is impacted by the airmass at the time of observation, and whether the star was considered to be observable given our constraints. Constraints define the ordering and prioritization of targets to observe and can be divided into: weather, other natural constraints, observatory constraints, and prioritization weights. We discuss each of these in turn below. For simplicity, we do not take into account variable throughput from changing seeing conditions. However, the dispatch scheduler does account for airmass attenuation from Rayleigh-scattering.

\FloatBarrier

\subsubsection{Site Locations and Weather}
\begin{figure}[ht]
\includegraphics[width=0.99\textwidth]{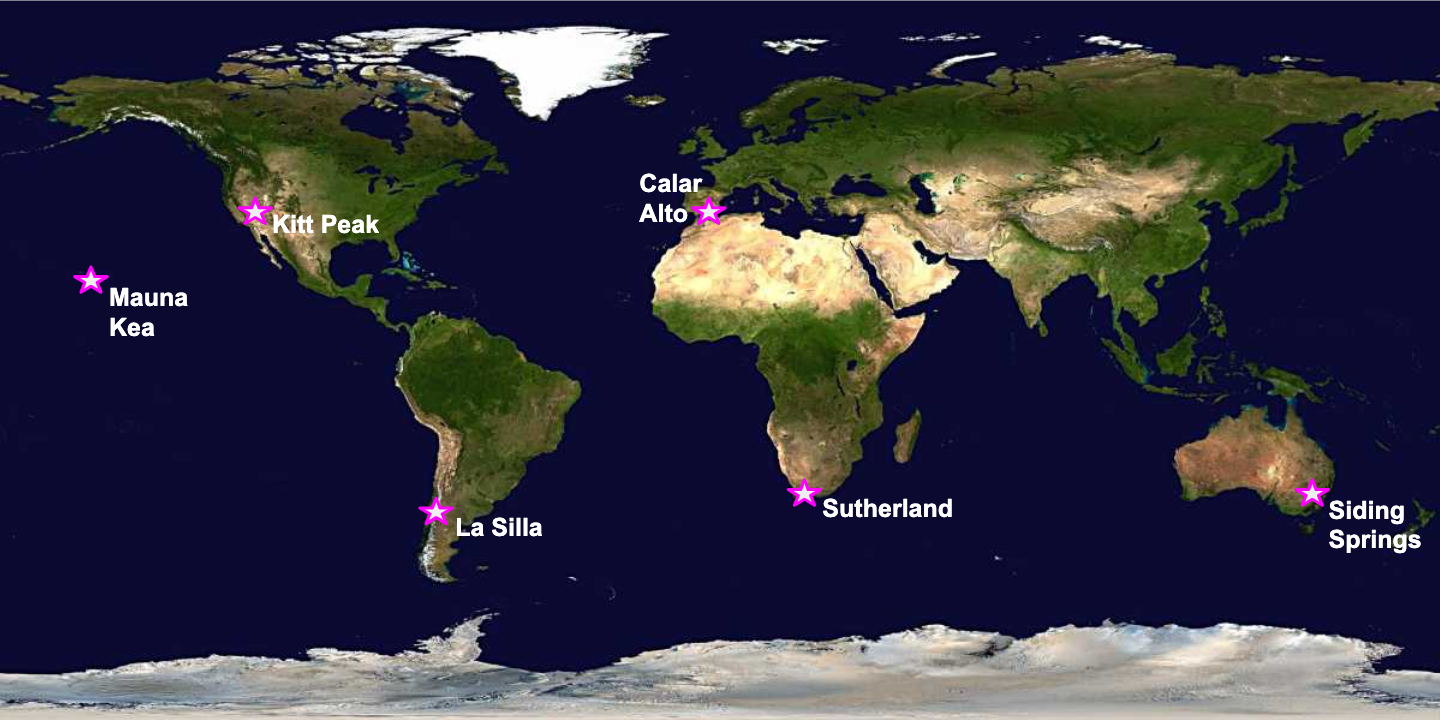}
\caption{Site locations}
\label{fig:locations}
\end{figure}

A representative set of six locations were chosen to achieve a global network of facilities. In the final round of simulations, these locations were: Mauna Kea, Kitt Peak, and Calar Alto in the North, along with Las Campanas, Sutherland, and Siding Spring in the South (figure \ref{fig:locations}). All of the architectures used all six sites. For the architectures with heterogeneous telescope compositions, the largest aperture telescopes were located at Mauna Kea and Las Campanas in our simulations. 

\begin{figure}[ht]
\label{fig:SiteWeather}
\includegraphics[width=0.98\textwidth]{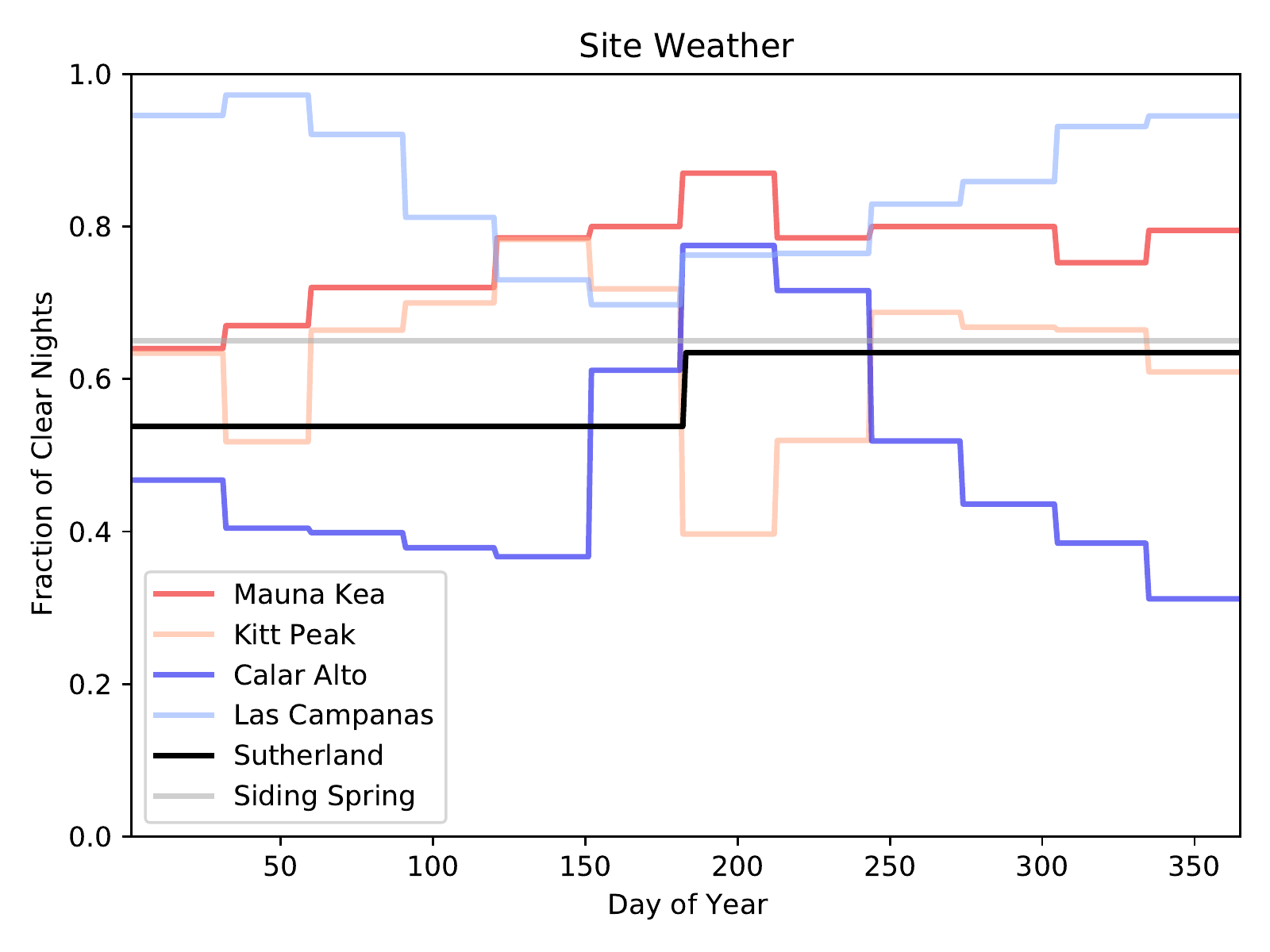}
\caption{Nightly weather probabilities for each site included in our simulations.}
\end{figure}

We use historical weather records for all observatory sites in order to determine the weather during our simulated survey. Specifically: Mauna Kea used a site survey \citep{1987PASP...99..560B}, Calar Alto and Sutherland used telescope operations reports \citet{2011hsa6.conf..637B,2016SPIE.9910E..0TV}, and Kitt Peak, Las Campanas, and Siding Spring used information from the observatory  websites.\footnote{\url{http://www-kpno.kpno.noao.edu/Images/wiynWeather_stats.jpeg} \url{https://www.eso.org/sci/facilities/lasilla/astclim/weather/tablemwr.html} \url{https://aat.anu.edu.au/about-us/AAT}} 
For each site, we compiled either monthly, or by semester in the case of Sutherland and yearly for Siding Spring, statistics for the fraction of clear nights (Figure \ref{fig:SiteWeather}). Finally, all surveys were assumed to span 10 years in duration (2020-01-01 to 2030-01-01).

\FloatBarrier

 We make the simplification that each night is assumed to be either entirely clear or entirely unusable, so there are no partial nights. Whether or not a night is lost due to weather is determined randomly at the start of the night, with the probabilities drawn from the historical monthly average. When monthly historical weather averages were not available, we instead used yearly averages for Siding Spring and semester averages for Sutherland. For simplification, we assume that nights lost due to weather do not exhibit night-to-night correlations, even though in reality it is typical to lose multiple nights in a row from weather patterns.

\subsubsection{Other Natural Constraints}
To further determine each target star's availability for observation we consider the telescope's latitude, longitude, and elevation, the time required for target acquisition, and the local horizon. Calculated sun-rise/set times are used to determine astronomical twilight which is when we allow observations to begin, and a minimum distance from the Moon (10 degrees for our surveys) that is required for a star to be targeted. Ephemerides for these are found via Astropy.

\subsubsection{Observatory Constraints}
For the observatory constraints, we consider telescope pointing limits (set to 30 degrees above the physical horizon), time for telescope slews between targets and for target acquisition (five minutes used for all observations), and total open shutter and detector readout times. 

\subsubsection{Target Prioritization}
Finally, as per the name, we use a dispatch scheduler prioritization scheme. The scheduler determines target choices from a user provided list of stars between each observation, rather than executing a preset observing order. Targets are ranked using a combination of hour angle (highest weight at zenith), and time since last observation. We assume a minimum separation between observations of two hours (e.g. within that time the target is at a minimum priority), with linearly increasing weight thereafter. The relative weighting of these two factors can be adjusted, and we fixed the relative weighting to produce cadences that spread the observations approximately evenly between the targets \citep{newman2}.

The dispatch scheduler assumes a single fixed observatory site. For handling up to the six telescope sites, a separate independent simulation was run. As such, we do not implement a mechanism for different telescope sites to coordinate prioritization of target observations with each other. In other words, we make the simplifying assumption that the telescope schedules are decoupled. On average, the different sites will randomly compensate for each others weather losses, but this is not optimized.

Despite the architectures in this survey having specified cadence goals/requirements, the dispatch scheduler does not attempt to reach any specific cadence. Rather, it observes every star in a given site's target list as frequently as possible, given the above constraints and weightings. As shown in the results, not all architectures can achieve their desired cadences. Reaching the cadences specified for each architecture below would require sculpting/optimizing the target list iteratively, which is beyond the scope and capability of these simulations.

\subsection{RV Model} \label{sec:RVModel}
RV precision and exposure times were calculated\footnote{Available at \url{https://github.com/pdn4kd/reimagined-palm-tree}.} using an implementation of the formalism presented in \cite{2015PASP..127.1240B}.
This RV model uses a set of BT Settl synthetic stellar spectra over the user specified instrument wavelength range, and a set of pre-calculated RV precision values for each wavelength and temperature bin in units of 1 photon per m/s in each 100 \AA\ wavelength bin. Correction factors for the RV content are used to account for variations in stellar surface gravity (log(g)), metallicity, and projected rotational velocity (v$\cdot$sin(i)). The finite size and throughput of of the spectrograph (in both resolution and pixels) is considered, though optical details beyond those are not. Instrument noise in the forms of dark current and readout noise are considered.
The important changes in our implementation are: allowing the spectrographs defined in Section \ref{sec:Architectures} to be in any wavelength range instead of just three specific ranges, and a simple atmospheric scattering model (Rayleigh plus a baseline). No telluric absorption lines are included, so our resulting RV precision estimates should be considered optimistic, especially in the near-infrared.
Additional details on the model and its calibration are in \citet{newman2}.

\subsection{Exposure Time Calculations} \label{sec:ExposureTimeCalculations}
Exposure times at each observing site are calculated with an exposure time calculator (ETC) in two steps. First, the RV and SNR are found for an observation that is just long enough to saturate the CCD at the star's peak brightness (approximated using Wein's Law) within the spectral grasp of the spectrograph as defined in Section \ref{sec:Architectures}. Second, this exposure time is scaled to the desired RV and SNR precision based upon an assumption that we are purely photon-noise limited (RV uncertainty $\sim$ SNR$^{-1}$ $\sim$ t$^{-1/2}$). This scaling dictates how many individual exposures (co-adds) are needed, and is combined with the instrumental readout time of co-add of 30 seconds to give a cumulative (``clock'') observation time. The SNR we use for the ETC is a per-pixel measurement unrelated to the SNR we calculate as a planet detection heuristic.

In order to average over short term stellar pulsation modes or p-mode oscillations, we implement a minimum clock observation time of 5 minutes and execute additional exposures as needed to achieve this minimum. For simplicity, we do not adjust this minimal time to account for the estimated p-mode oscillation time as a function of individual spectral type and surface gravity for each individual target (eg: \cite{2019AJ....157..163C}). This assumption is investigated (verses a 10 minute minimum), and we believe to to accurately describe the effect on observation rates in \ref{sec:pmode}. Finally, we assume that the observatory does not move onto the next target until the final co-add is read out; conversely, this can be thought of as an additional 30-second penalty on the assumed 5 minutes for target slews and acquisition overhead. 
 

\subsection{Architectures: Telescopes and Spectrographs} \label{sec:Architectures}

We consider a set of architectures of dedicated telescopes of different diameters equipped with high-resolution spectrometers to enable a next-generation global EPRV survey to search for and detect Earth-mass twins. We consider realistic telescope diameters that are representative of existing or proposed EPRV facilities and telescopes. All architectures we consider utilized the same notional ``baseline instrument'' design: a high resolution, optical band pass, extreme precision RV spectrograph loosely based on the design of NN-Explore’s NEID spectrograph located on the 3.5 m WIYN telescope \citep{2016SPIE.9908E..62R, 2016SPIE.9908E..7HS}. For all architectures, we also consider a ``defined instrument'' variation to these spectrograph parameters, motivated by the given telescope diameter or configuration. In the following sub-sections, we describe each architecture in detail.

\label{sec:ChampionDefinedArchitectures}
\begin{table}[ht]
{\tiny
\flushleft
	\begin{tabular}{| l | c | c | c | c |}
	\hline
	Architecture & I & IIa & IIb & V \\
\hline
Telescopes & 6x2.4 m & 2x6 m and 4x4 m & 6x4 m & 6x3 m \\
Collecting area by aperture & 2.4 m = 4.2 m$^2$ & 4 m = 9.5 m$^2$; 6 m = 27 m$^2$ & 4 m = 9.5 m$^2$ & 3 m = 6.3 m$^2$ \\
Time allocation & 100\% & 100\% & 100\% & 100\% \\
Wavelength coverage & 380-930 nm & 380 - 930 nm & 380 - 930 nm & 500-1700 nm \\
Spectral resolution & 180\,000 & 180\,000 & 180\,000 & 180\,000 \\
Total system efficiency & 6\% & 6\% & 6\% & 7\% \\
instrument noise floor & 10 cm/s & 5 cm/s & 5 cm/s & 10 cm/s \\
Required (peak) SNR/pix & 300 & 300 & 300 & 300 \\
Required RV precision & 10 cm/s & 10 cm/s & 10 cm/s & 10 cm/s \\
Observation cadence per star & 1 / night & 3 / night & 3 / night & 2 / telescope / night \\
\hline

	\end{tabular}	\\
	\begin{tabular}{| l | c | c | c |}
	\hline
	Architecture & VI & VIIIa & VIIIb \\
\hline
Telescopes & 6x arrays of 1 m & 2x10 m and 4x 3.5 m & 2x10 m and 6x2.4 m \\
Collecting area by aperture & 0.61m$^2$ each; array is 9.5 m$^2$ & 10 m = 75 m$^2$; 3.5 m = 9.5 m$^2$ & 10 m = 75 m$^2$; 2.4 m = 4.2 m$^2$ \\
Time allocation & 100\% & 25\% of 10 m; 100\% of 3.5m & 25\% of 10 m; 100\% of 2.4 m \\
Wavelength coverage & 500-800 nm & 380-930 nm & 380-930 nm \\
Spectral resolution & 150\,000 & 180\,000 & 180\,000 \\
Total system efficiency & 6\% & 6\% & 6\% \\
instrument noise floor & 10 cm/s & 5 cm/s &  5cm/s \\
Required (peak) SNR/pix & 300 & 1000 for the 10 m; 300 for 3.5 m & 1000 for the 10 m; 300 for 2.4 m \\
Required RV precision & 10 cm/s & 15 cm/s on 3.5 m; 5 cm/s on 10 m & 15 cm/s on 2.4 m; 5 cm/s on 10 m \\
Observation cadence per star & 1/night & 1/week on 10 m; 1/night on 3.5 m & 1/week on 10 m; 1/night on 2.4 m \\
\hline

	\end{tabular}	\\
}
\caption{The seven architectures' facility and instrumental properties. Architectures not listed (III, IV, VII) were dropped from direct consideration in earlier simulations. Architectures with a/b variants have different sizes/numbers of telescopes, but identical instruments for each variant.}
\label{table:ChampionDefinedArchitectures}
\end{table}
\FloatBarrier

\subsubsection{Architecture I}
This architecture utilizes a network of six identical, dedicated robotic 2.4-m telescopes, each of which is paired with one of the standardized EPRV spectrographs described above. The telescopes in this architecture spend 100\% of nightly observations on the EPRV survey (the survey targets are described in \ref{sec:TargetStarLists}). An example of such an observatory in existence today is the Automated Planet Finder telescope \citep{2014PASP..126..359V}; the APF is a fully automated, robotic facility that has been executing precise RV surveys since 2014, and is the largest aperture robotic RV telescope currently in operation. 

As with most architectures described herein, the telescopes are spread across both longitude and latitude - three in the northern hemisphere (Mauna Kea in Hawaii, Kitt Peak in Arizona, and Calar Alto in Spain) and three in the southern hemisphere (La Silla in Chile, Southerland in South Africa, and Siding Springs in Australia).
This architecture could potentially save costs from the identical telescope hardware. Alternatively, given the large number of existing (and sometimes underutilized) telescopes in this size class, a heterogeneous variant could re-purpose one or more.

\subsubsection{Architecture II}
Architecture II employs facilities at the same locations as Architecture I but uses larger aperture telescopes. It consists of two sub-architecture variations: IIa places 6-m telescopes at Mauna Kea and Las Campanas, and 4-m telescopes at the remaining locations, while IIb uses 4-m telescopes at all six sites. Like the telescopes in architecture I, all facilities in Architecture II are dedicated and spend the entirety of their observing time working on this particular survey. Given the number of existing 4-m and 6-m class telescopes around the globe, this architecture may more easily re-purpose existing observatories, with the caveat that only a handful of telescopes of their aperture are currently dedicated to obtaining PRVs.

\subsubsection{Architectures III and IV}
Architecture III consists of two 10-m class telescopes, one in the Northern hemisphere and one in the South, each of which has 50\% of its time assigned to carrying out this EPRV survey, matching the 10-m portion of Architecture VIII. Examples of analogous telescopes currently in operation include but are not limited to Keck and Gemini observatories. Architecture IV is a copy of Architecture VIII, with a small fraction of additional time on one or two 25-m class telescopes (eg: Thirty Meter Telescope or Giant Magellan Telescope) that is used exclusively for RV follow-up, and not for the primary RV survey being simulated herein. Thus, the simulated RV survey performances of Architectures III and IV can be evaluated from the simulated performance of Architecture VIII. Examples of planned 25-m class telescopes include the ELT, GMT and TMT; G-CLEF is a PRV spectrometer currently being built for the GMT \citep{2018SPIE10702E..1RS}.

\subsubsection{Architecture V}
The distinguishing feature of architecture V is the spectrograph, which would use adaptive optics to inject light into single mode fibers, and extends farther into the near-infrared than its counterparts. Examples of current or planned single-mode fiber PRV spectrometers include PARVI and iLocater \citep{2020JATIS...6a1002G,2016SPIE.9908E..19C}. Many of the unique advantages and challenges of this architecture are simulated at low fidelity herein, or are outside of the scope of these simulations. For example, we assume a spectral grasp that will extend into the visible with sufficient AO-corrections for efficient throughput coupling into the fibers than has currently been demonstrated on sky. For the simulations, we assume six 3-m telescopes are used.

\subsubsection{Architecture VI}
This architecture uses a fiber multi-plexed array of small (notionally 1-m) telescopes at each telescope site to get the light gathering power equivalent of a larger aperture at potentially lower cost. As simulated, each array is considered to be the same as a single 4-m class telescope (equivalent to architecture IIb). The spectrograph suffers from reduced spectral grasp and lower spectral resolution due to the penalty of multiple fibers for each telescope in the array, and drives the difference in the assumed instrument parameters. The heritage for this architecture is conceptually a scaled-up version of the MINERVA and MINERVA-Australis PRV telescope arrays \citep{2019PASP..131k5003A,2015JATIS...1b7002S}.

\subsubsection{Architecture VIII}
Architecture VIII is explicitly a hybrid architecture. It combines a pair of 10-m class telescopes operating at high precision and limited time allocation, one in each hemisphere, and a larger number (4-6) of smaller telescopes at lower precision and 100\% dedicated time allocation to the EPRV survey. Architecture VIII is the only architecture simulated with varying precision and/or varying amounts of allocated survey time available at each telescope site.
The VIIIa variant has four 3.5-m telescopes (comparable to architectures IIa/b and VI) located at four other locations distinct from the two 10-m telescopes. The VIIIb variant has six 2.4 m telescopes (similar to architecture I). We simulate two telescopes each (one 10-m and one 2.4-m) at both Mauna Kea and Las Campanas. In reality, these pairs of telescopes at Mauna Kea and Las Campanas would suffer identical weather losses. However, since we simulate each telescope independently, the telescopes at the same site have independently drawn weather losses. We reuse the 10-m simulations for all of Architectures VIIIa, VIIIb, and III.

\subsection{Target Star Lists} \label{sec:TargetStarLists}
Future exoplanet direct imaging NASA mission concept studies have produced prioritized target lists of nearby, bright stars. To identify a set of stars that would constitute a reasonable set of targets for a future EPRV survey of direct imaging targets, we cross-matched the HabEx and LUVOIR-A, LUVOIR-B, and Starshade Rendezvous\citep{2019BAAS...51g.106S} target lists \citep{2020arXiv200106683G,2019arXiv191206219T}, and gathered archival information on stellar properties relevant to each star's RV information content (i.e., spectral type, effective temperature, apparent magnitude, rotational velocity, metallicity and surface gravity). We first divide the stellar sample into three target prioritization levels listed below. All stars on the green and yellow lists must satisfy the criteria that 1) planets in the habitable zone would be visible to future direct imaging efforts, and 2) the host stars themselves are amenable to EPRV observations:

\begin{itemize}
  \item Green targets
  \begin{itemize}
    \item Spectral types F7-K9
    \item v$sin$i $<$ 5 km/s
    \item HabEx deep or nearest 50 or appearing on N $\geq$ 2 lists (including LUVOIR-A, LUVOIR-B, HabEx, Starshade Rendezvous)
    \item 106 stars
  \end{itemize}
  \item Yellow targets
  \begin{itemize}
    \item Spectral types F7-M
    \item v$sin$i $<$ 10 km/s
    \item Appears on at least one direct imaging mission concept list, but not a ``green'' target
    \item 123 stars
  \end{itemize}
  \item Red targets
  \begin{itemize}
    \item Spectral type hotter than F7
    \item v$sin$i $>$ 10 km/s
    \item 78 stars
  \end{itemize}
\end{itemize}

\begin{figure}[ht]
\includegraphics[width=0.99\textwidth]{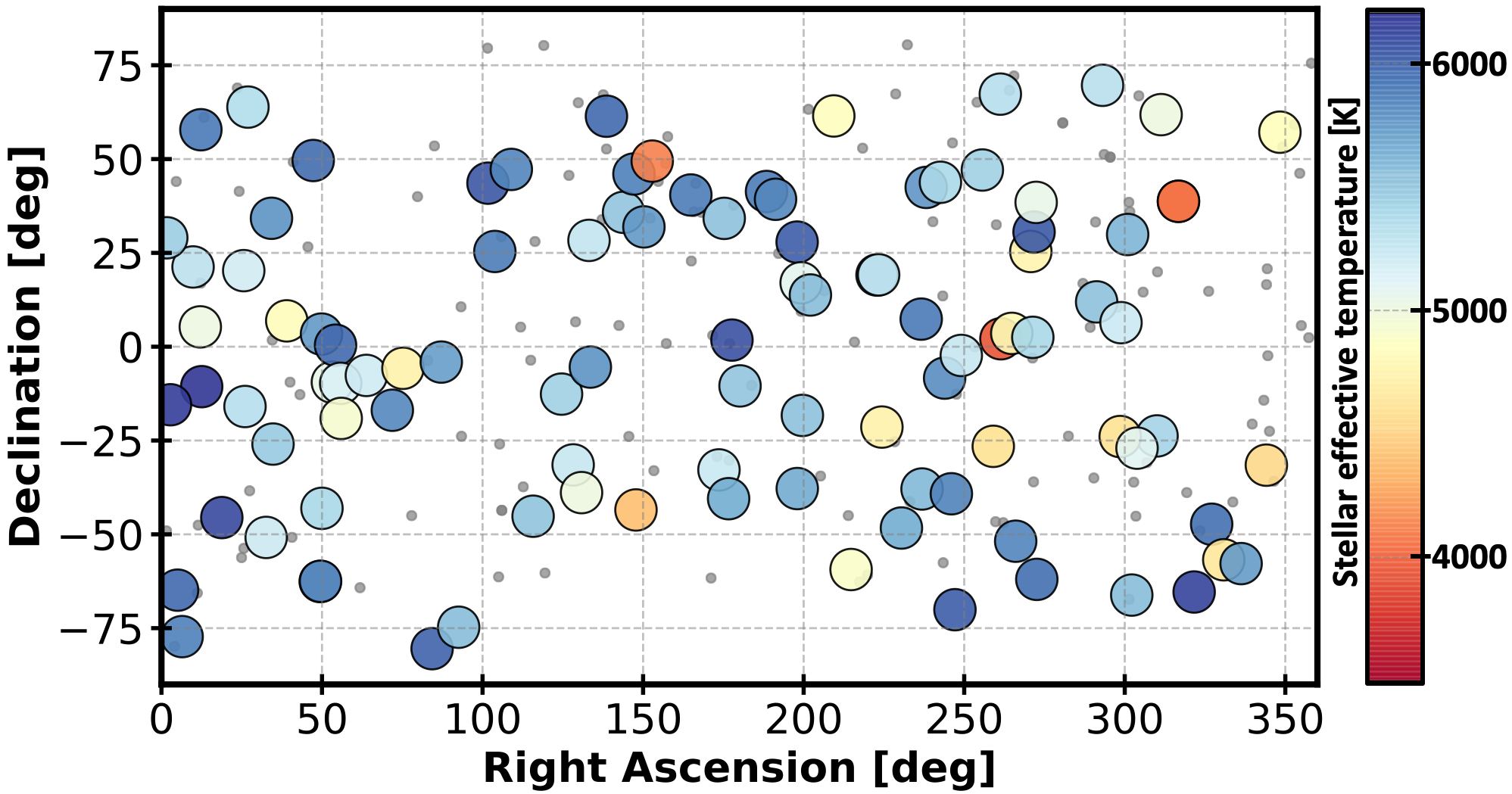}
\caption{The locations on the sky of all green and yellow list stars. Open circles represent green list targets, color coded by effective temperature, while the yellow list stars are depicted by the grey points. While the thin disk of nearby stars is effectively isotropic on the sky, the green prime list of 101 nearby bright stars shows some minor clustering on the sky which can impact RV survey efficiency.}
\label{fig:stars}
\end{figure}

An additional five stars were removed from the green target list to generate a final ``green prime'' list of 101 stars. These five stars possessed the longest (in the green list) estimated exposure times from ${\S}$1.4 to reach the desired RV precision, primarily due to a combination of spectral type and rotational velocity (e.g. these were corner cases). The resulting ``green prime'' star list of 101 stars and associated stellar properties are listed in Table 2, shown distributed on the sky in Figure \ref{fig:stars}, and constitute the sample used for our RV survey simulations. The yellow stars were both less favored by the direct imaging surveys, somewhat fainter, and in some cases had less RV information content due to rotational broadening, and thus were excluded from our simulations here. Similar the red targets were excluded due to their insufficient RV information content.

The green prime list of targets was in turn broken into a northern hemisphere sub-set ($\geq$ -5$\arcdeg$ dec, 51 stars), and a southern hemisphere sub-set ($\leq$ +5$\arcdeg$ dec, 58 stars) for observing with telescopes in their respective hemispheres, with 9 stars near the celestial equator appearing on both lists for cross-comparison. Since our simulations do not coordinate telescope target optimization, we did not dynamically optimize which stars were observed by which telescopes (e.g. a fluid declination cutoff that is optimized for each observing site and time of year). These particular declination cuts were chosen to have $\sim$10\% overlap between the target lists of both hemispheres.

\FloatBarrier

\startlongtable
\begin{deluxetable}{l | c | c | l | c | l | c | c | c | c | c |c}
\centerwidetable
\tabletypesize{\tiny}
\tablecaption{Stars in the ``green prime'' list. Name, spectral type, which hemispheres they were observed in, the properties used in the simulations, and their sources are included. Mass and radius for all stars are estimated from other properties.}
    \startdata
    Name & Right Ascension & Declination & Distance (pc) & v$sin$i & Spectral Type & T$\_{eff}$ & log(g) & Metallicity & Mass & Radius & Hemisphere \\
\hline
HIP 10138 & 02:10:24.00 & -50:49:31.1 & 10.787 $^{a}$ & 1.9 $^{b}$ & K1.5V $^{c}$ & 5217 $^{d}$ & 4.56 $^{d}$ & -0.23 $^{d}$ & 0.82 & 0.755 & South \\
HIP 101997 & 20:40:11.44 & -23:46:30.0 & 14.677 $^{a}$ & 1.8 $^{b}$ & G7.5IV-V $^{c}$ & 5414 $^{e}$ & 4.46 $^{e}$ & -0.31 $^{e}$ & 0.95 & 0.951 & South \\
HIP 102422 & 20:45:17.27 & +61:50:12.5 & 14.265 $^{f}$ & 1.7 $^{b}$ & K0IV $^{c}$ & 5002 $^{g}$ & 3.43 $^{g}$ & -0.09 $^{g}$ & 1.23 & 10.188 & North \\
HIP 104214 & 21:06:50.84 & +38:44:29.4 & 3.497 $^{a}$ & 1.8 $^{b}$ & K5V $^{c}$ & 4339 $^{h}$ & 4.43 $^{h}$ & -0.33 $^{h}$ & 0.68 & 1.019 & North \\
HIP 104217 & 21:06:52.19 & +38:44:03.9 & 3.495 $^{a}$ & 1.8 $^{b}$ & K7V $^{c}$ & 4045 $^{h}$ & 4.53 $^{h}$ & -0.38 $^{h}$ & 0.63 & 0.809 & North \\
HIP 105858 & 21:26:26.49 & -65:22:05.3 & 9.262 $^{f}$ & 3.4 $^{b}$ & F9V Fe-1.4 CH-0.7 $^{i}$ & 6150 $^{d}$ & 4.35 $^{d}$ & -0.66 $^{d}$ & 1.14 & 1.225 & South \\
HIP 10644 & 02:17:02.42 & +34:13:29.4 & 11.008 $^{a}$ & 2 $^{b}$ & G0.5V Fe-0.5 $^{c}$ & 5786 $^{d}$ & 4.29 $^{d}$ & -0.53 $^{d}$ & 1.08 & 1.406 & North \\
HIP 107649 & 21:48:15.61 & -47:18:10.4 & 15.561 $^{a}$ & 1.8 $^{b}$ & G0V Fe+0.4 $^{i}$ & 5946 $^{e}$ & 4.48 $^{e}$ & 0.01 $^{e}$ & 1.08 & 0.908 & South \\
HIP 10798 & 02:18:58.65 & -25:56:48.4 & 12.834 $^{a}$ & 2.7 $^{b}$ & G8V $^{i}$ & 5476 $^{d}$ & 4.61 $^{d}$ & -0.45 $^{d}$ & 0.94 & 0.673 & South \\
HIP 108870 & 22:03:17.44 & -56:46:47.3 & 3.639 $^{a}$ & 1.4 $^{b}$ & K4V(k) $^{i}$ & 4649 $^{d}$ & 4.63 $^{d}$ & -0.19 $^{d}$ & 0.72 & 0.643 & South \\
HIP 110649 & 22:24:56.19 & -57:47:47.8 & 20.454 $^{a}$ & 1.8 $^{b}$ & G2IV-V $^{i}$ & 5739 $^{j}$ & 4.15 $^{j}$ & 0.05 $^{j}$ & 1.02 & 1.941 & South \\
HIP 113283 & 22:56:23.83 & -31:33:54.6 & 7.608 $^{a}$ & 2.6 $^{b}$ & K4Ve $^{c}$ & 4555 $^{k}$ & 4.53 $^{k}$ & -0.01 $^{k}$ & 0.72 & 0.809 & South \\
HIP 114622 & 23:13:14.74 & +57:10:03.5 & 6.532 $^{a}$ & 1.8 $^{b}$ & K3V $^{c}$ & 4833 $^{d}$ & 4.59 $^{d}$ & 0 $^{d}$ & 0.75 & 0.705 & North \\
HIP 12114 & 02:36:03.83 & +06:53:00.1 & 7.235 $^{a}$ & 1.3 $^{b}$ & K3V $^{c}$ & 4829 $^{l}$ & 4.6 $^{l}$ & -0.16 $^{l}$ & 0.75 & 0.689 & North \\
HIP 14632 & 03:09:02.88 & +49:36:48.6 & 10.541 $^{f}$ & 3.6 $^{b}$ & G0V $^{c}$ & 5968 $^{d}$ & 4.19 $^{d}$ & 0.08 $^{d}$ & 1.08 & 1.770 & North \\
HIP 15330 & 03:17:44.47 & -62:34:36.8 & 12.039 $^{a}$ & 2.7 $^{b}$ & G2.5V Hdel1 $^{c}$ & 5712 $^{d}$ & 4.48 $^{d}$ & -0.24 $^{d}$ & 1.01 & 0.908 & South \\
HIP 15371 & 03:18:11.14 & -62:30:28.6 & 12.046 $^{a}$ & 2.7 $^{b}$ & G1V $^{c}$ & 5852 $^{d}$ & 4.43 $^{d}$ & -0.25 $^{d}$ & 1.07 & 1.019 & South \\
HIP 15457 & 03:19:21.54 & +03:22:11.9 & 9.14 $^{f}$ & 4.5 $^{b}$ & G5V $^{c}$ & 5749 $^{m}$ & 4.51 $^{m}$ & 0.08 $^{m}$ & 0.98 & 0.847 & Both \\
HIP 15510 & 03:19:53.22 & -43:04:17.6 & 6.043 $^{f}$ & 0.9 $^{b}$ & G6V $^{c}$ & 5398 $^{e}$ & 4.41 $^{e}$ & -0.41 $^{e}$ & 0.97 & 1.067 & South \\
HIP 1599 & 00:20:01.91 & -64:52:39.4 & 8.587 $^{f}$ & 4.9 $^{b}$ & F9.5V $^{i}$ & 5977 $^{n}$ & 4.51 $^{n}$ & -0.18 $^{n}$ & 1.11 & 0.847 & South \\
HIP 16537 & 03:32:56.42 & -09:27:29.9 & 3.216 $^{f}$ & 1.9 $^{b}$ & K2V $^{c}$ & 5050 $^{h}$ & 4.6 $^{h}$ & -0.09 $^{h}$ & 0.78 & 0.689 & South \\
HIP 16852 & 03:36:52.52 & +00:24:10.2 & 13.963 $^{f}$ & 3.7 $^{b}$ & F9IV-V $^{c}$ & 5971 $^{d}$ & 4.06 $^{d}$ & -0.09 $^{d}$ & 1.14 & 2.388 & Both \\
HIP 17378 & 03:43:14.96 & -09:45:54.7 & 9.041 $^{f}$ & 1 $^{b}$ & K0+IV $^{c}$ & 5144 $^{o}$ & 3.95 $^{o}$ & 0 $^{o}$ & 0.87 & 3.077 & South \\
HIP 17420 & 03:43:55.15 & -19:06:40.6 & 13.955 $^{a}$ & 3 $^{b}$ & K2.5V $^{i}$ & 4930 $^{p}$ & 4.41 $^{p}$ & -0.17 $^{p}$ & 0.76 & 1.067 & South \\
HIP 19849 & 04:15:17.64 & -07:38:40.4 & 4.985 $^{f}$ & 0.9 $^{b}$ & K0.5V $^{i}$ & 5202 $^{d}$ & 4.55 $^{d}$ & -0.28 $^{d}$ & 0.86 & 0.773 & South \\
HIP 2021 & 00:25:39.20 & -77:15:18.1 & 7.459 $^{f}$ & 3.4 $^{b}$ & G0V $^{i}$ & 5873 $^{h}$ & 3.98 $^{h}$ & -0.04 $^{h}$ & 1.08 & 2.871 & South \\
HIP 22263 & 04:47:36.21 & -16:56:05.5 & 13.241 $^{a}$ & 2.9 $^{b}$ & G1.5V CH-0.5 $^{i}$ & 5840 $^{q}$ & 4.5 $^{q}$ & 0.03 $^{q}$ & 1.04 & 0.867 & South \\
HIP 23311 & 05:00:48.68 & -05:45:03.5 & 8.848 $^{a}$ & 1.4 $^{b}$ & K3+V $^{r}$ & 4745 $^{d}$ & 4.57 $^{d}$ & 0.19 $^{d}$ & 0.75 & 0.738 & South \\
HIP 26394 & 05:37:08.79 & -80:28:18.0 & 18.28 $^{a}$ & 2.7 $^{b}$ & G0V $^{i}$ & 6003 $^{n}$ & 4.42 $^{n}$ & 0.09 $^{n}$ & 1.08 & 1.042 & South \\
HIP 27435 & 05:48:34.90 & -04:05:38.7 & 15.255 $^{a}$ & 2.7 $^{b}$ & G2V $^{r}$ & 5733 $^{n}$ & 4.51 $^{n}$ & -0.22 $^{n}$ & 1.02 & 0.847 & Both \\
HIP 29271 & 06:10:14.20 & -74:45:09.1 & 10.215 $^{a}$ & 2.3 $^{b}$ & G7V $^{i}$ & 5569 $^{d}$ & 4.43 $^{d}$ & 0.11 $^{d}$ & 0.96 & 1.019 & South \\
HIP 3093 & 00:39:22.09 & +21:15:04.9 & 11.137 $^{a}$ & 1.8 $^{b}$ & K0.5V $^{c}$ & 5303 $^{d}$ & 4.56 $^{d}$ & 0.18 $^{d}$ & 0.86 & 0.755 & North \\
HIP 32480 & 06:46:44.34 & +43:34:37.3 & 16.648 $^{a}$ & 3.6 $^{b}$ & F9V $^{r}$ & 6064 $^{j}$ & 4.33 $^{j}$ & 0.12 $^{j}$ & 1.14 & 1.283 & North \\
HIP 32984 & 06:52:18.37 & -05:10:25.3 & 8.749 $^{a}$ & 2.7 $^{b}$ & K3.5V $^{r}$ & 4758 $^{j}$ & 4.5 $^{s}$ & -0.07 $^{t}$ & 0.79 & 0.867 & South \\
HIP 33277 & 06:55:18.69 & +25:22:32.3 & 17.47 $^{a}$ & 2.9 $^{b}$ & G0V $^{r}$ & 5891 $^{j}$ & 4.36 $^{j}$ & -0.18 $^{j}$ & 1.08 & 1.197 & North \\
HIP 35136 & 07:15:50.11 & +47:14:25.5 & 16.867 $^{a}$ & 2.9 $^{b}$ & F9V $^{r}$ & 5849 $^{u}$ & 4.26 $^{u}$ & -0.33 $^{u}$ & 1.14 & 1.507 & North \\
HIP 37606 & 07:42:57.16 & -45:10:18.4 & 22.473 $^{a}$ & 4.3 $^{v}$ & G8IV-V $^{i}$ & 5526 $^{w}$ & 4.1 $^{w}$ & 0.19 $^{w}$ & 0.94 & 2.178 & South \\
HIP 3765 & 00:48:22.53 & +05:17:00.2 & 7.435 $^{a}$ & 1.8 $^{b}$ & K2V $^{r}$ & 5015 $^{d}$ & 4.6 $^{d}$ & -0.31 $^{d}$ & 0.78 & 0.689 & North \\
HIP 3821 & 00:49:05.10 & +57:48:59.6 & 5.953 $^{f}$ & 3.4 $^{b}$ & F9V $^{c}$ & 5904 $^{m}$ & 4.32 $^{m}$ & -0.25 $^{m}$ & 1.14 & 1.312 & North \\
HIP 3909 & 00:50:07.72 & -10:38:37.6 & 15.88 $^{a}$ & 3.9 $^{b}$ & F7V $^{r}$ & 6203 $^{x}$ & 4.27 $^{x}$ & -0.15 $^{x}$ & 1.21 & 1.473 & South \\
HIP 40693 & 08:18:23.78 & -12:37:47.2 & 12.564 $^{a}$ & 2.2 $^{b}$ & G8+V $^{i}$ & 5442 $^{d}$ & 4.53 $^{d}$ & -0.02 $^{d}$ & 0.93 & 0.809 & South \\
HIP 41926 & 08:32:52.26 & -31:30:09.7 & 12.182 $^{a}$ & 1.8 $^{b}$ & K1V $^{c}$ & 5243 $^{n}$ & 4.46 $^{n}$ & -0.41 $^{n}$ & 0.85 & 0.951 & South \\
HIP 42808 & 08:43:18.26 & -38:52:59.5 & 11.186 $^{a}$ & 2.7 $^{b}$ & K2.5V(k) $^{i}$ & 5005 $^{d}$ & 4.61 $^{d}$ & -0.01 $^{d}$ & 0.76 & 0.673 & South \\
HIP 43587 & 08:52:36.13 & +28:19:53.0 & 12.59 $^{a}$ & 2.2 $^{b}$ & K0IV-V $^{r}$ & 5270 $^{p}$ & 4.31 $^{p}$ & 0.31 $^{p}$ & 0.95 & 1.343 & North \\
HIP 43726 & 08:54:18.19 & -05:26:04.3 & 16.85 $^{a}$ & 2.4 $^{b}$ & G2V $^{c}$ & 5781 $^{e}$ & 4.44 $^{e}$ & 0.12 $^{e}$ & 1.02 & 0.996 & South \\
HIP 45333 & 09:14:20.55 & +61:25:24.2 & 19.659 $^{a}$ & 4.8 $^{b}$ & G0IV-V $^{r}$ & 5973 $^{d}$ & 4.13 $^{d}$ & 0.05 $^{d}$ & 1.08 & 2.033 & North \\
HIP 47080 & 09:35:40.03 & +35:48:38.8 & 11.203 $^{a}$ & 2.3 $^{b}$ & G8Va $^{c}$ & 5511 $^{d}$ & 4.46 $^{d}$ & 0.28 $^{d}$ & 0.94 & 0.951 & North \\
HIP 48113 & 09:48:35.18 & +46:01:16.4 & 18.904 $^{a}$ & 2.9 $^{b}$ & G0.5Va $^{c}$ & 5872 $^{d}$ & 4.1 $^{d}$ & 0.09 $^{d}$ & 1.08 & 2.178 & North \\
HIP 48331 & 09:51:06.68 & -43:30:05.9 & 11.285 $^{a}$ & 0.9 $^{b}$ & K6V(k) $^{i}$ & 4400 $^{e}$ & 4.36 $^{e}$ & -0.26 $^{e}$ & 0.65 & 1.197 & South \\
HIP 49081 & 10:01:01.02 & +31:55:29.0 & 14.926 $^{a}$ & 1.8 $^{b}$ & G3Va Hdel1 $^{c}$ & 5753 $^{m}$ & 4.3 $^{m}$ & 0.26 $^{m}$ & 1 & 1.374 & North \\
HIP 49908 & 10:11:23.36 & +49:27:19.7 & 4.869 $^{a}$ & 2.7 $^{b}$ & K7V $^{y}$ & 4131 $^{z}$ & 4.61 $^{aa}$ & 0.24 $^{z}$ & 0.63 & 0.673 & North \\
HIP 53721 & 10:59:28.22 & +40:25:48.4 & 13.802 $^{a}$ & 3.1 $^{b}$ & G1-V Fe-0.5 $^{c}$ & 5894 $^{d}$ & 4.3 $^{d}$ & 0.02 $^{d}$ & 1.07 & 1.374 & North \\
HIP 544 & 00:06:36.53 & +29:01:19.0 & 13.779 $^{a}$ & 3.6 $^{b}$ & G8V $^{r}$ & 5458 $^{d}$ & 4.52 $^{d}$ & 0.14 $^{d}$ & 0.94 & 0.828 & North \\
HIP 56452 & 11:34:29.95 & -32:50:00.0 & 9.544 $^{a}$ & 0.9 $^{b}$ & K0-V $^{i}$ & 5241 $^{d}$ & 4.59 $^{d}$ & -0.37 $^{d}$ & 0.88 & 0.705 & South \\
HIP 56997 & 11:41:03.03 & +34:12:09.2 & 9.579 $^{a}$ & 2.3 $^{b}$ & G8V $^{c}$ & 5528 $^{d}$ & 4.53 $^{d}$ & -0.05 $^{d}$ & 0.94 & 0.809 & North \\
HIP 57443 & 11:46:32.25 & -40:30:04.8 & 9.292 $^{a}$ & 2.7 $^{b}$ & G2V $^{i}$ & 5655 $^{d}$ & 4.44 $^{d}$ & -0.29 $^{d}$ & 1.02 & 0.996 & South \\
HIP 57757 & 11:50:41.29 & +01:45:55.4 & 10.929 $^{f}$ & 3.6 $^{b}$ & F9V $^{ab}$ & 6083 $^{h}$ & 4.08 $^{h}$ & 0.24 $^{h}$ & 1.14 & 2.281 & Both \\
HIP 58576 & 12:00:44.37 & -10:26:41.4 & 12.697 $^{a}$ & 1.8 $^{b}$ & G8IV $^{i}$ & 5510 $^{u}$ & 4.33 $^{u}$ & 0.25 $^{u}$ & 0.94 & 1.283 & South \\
HIP 5862 & 01:15:10.57 & -45:31:55.5 & 15.177 $^{a}$ & 4.7 $^{b}$ & F9V Fe+0.4 $^{i}$ & 6111 $^{ac}$ & 4.36 $^{ac}$ & 0.17 $^{ac}$ & 1.14 & 1.197 & South \\
HIP 61317 & 12:33:45.09 & +41:21:24.4 & 8.44 $^{f}$ & 2.8 $^{b}$ & G0V $^{c}$ & 5887 $^{m}$ & 4.34 $^{m}$ & -0.2 $^{m}$ & 1.08 & 1.253 & North \\
HIP 62207 & 12:44:59.68 & +39:16:42.9 & 17.565 $^{a}$ & 1.8 $^{b}$ & F9V Fe-0.3 $^{r}$ & 5842 $^{u}$ & 4.33 $^{u}$ & -0.5 $^{u}$ & 1.14 & 1.283 & North \\
HIP 64394 & 13:11:52.92 & +27:52:33.7 & 9.129 $^{f}$ & 4.5 $^{b}$ & F9.5V $^{c}$ & 6034 $^{m}$ & 4.44 $^{m}$ & 0.06 $^{m}$ & 1.11 & 0.996 & North \\
HIP 64408 & 13:12:03.47 & -37:48:11.3 & 20.295 $^{a}$ & 2.7 $^{b}$ & G4IV $^{i}$ & 5670 $^{o}$ & 3.9 $^{o}$ & 0.16 $^{o}$ & 0.99 & 3.452 & South \\
HIP 64797 & 13:16:50.67 & +17:01:04.1 & 10.985 $^{a}$ & 3.3 $^{b}$ & K2.5V(k) $^{r}$ & 5081 $^{d}$ & 4.62 $^{d}$ & -0.16 $^{d}$ & 0.76 & 0.658 & North \\
HIP 64924 & 13:18:24.97 & -18:18:31.0 & 8.555 $^{f}$ & 1.8 $^{b}$ & G6.5V $^{c}$ & 5537 $^{e}$ & 4.38 $^{e}$ & -0.03 $^{e}$ & 0.97 & 1.143 & South \\
HIP 65721 & 13:28:25.95 & +13:46:48.7 & 17.91 $^{a}$ & 3 $^{b}$ & G4V-IV $^{ad}$ & 5559 $^{o}$ & 4.05 $^{o}$ & -0.06 $^{o}$ & 0.99 & 2.444 & North \\
HIP 68184 & 13:57:32.10 & +61:29:32.4 & 10.078 $^{a}$ & 2 $^{ap}$ & K3V $^{ae}$ & 4851 $^{d}$ & 4.58 $^{d}$ & 0.11 $^{d}$ & 0.75 & 0.721 & North \\
HIP 69972 & 14:19:05.36 & -59:22:37.4 & 11.841 $^{a}$ & 0.9 $^{b}$ & K3IV $^{i}$ & 4903 $^{j}$ & 4.69 $^{j}$ & 0.32 $^{j}$ & 0.83 & 0.560 & South \\
HIP 72659 & 14:51:23.28 & +19:06:02.3 & 6.733 $^{a}$ & 3.5 $^{b}$ & G7V $^{r}$ & 5527 $^{u}$ & 4.6 $^{u}$ & -0.13 $^{u}$ & 0.96 & 0.689 & North \\
HIP 72848 & 14:53:24.04 & +19:09:08.2 & 11.51 $^{f}$ & 3.9 $^{b}$ & K0.5V $^{c}$ & 5291 $^{u}$ & 4.55 $^{u}$ & 0.08 $^{u}$ & 0.86 & 0.773 & North \\
HIP 73184 & 14:57:27.35 & -21:24:40.6 & 5.882 $^{a}$ & 3.5 $^{b}$ & K4V $^{c}$ & 4744 $^{j}$ & 4.76 $^{j}$ & 0.12 $^{j}$ & 0.72 & 0.477 & South \\
HIP 75181 & 15:21:49.57 & -48:19:01.1 & 14.688 $^{a}$ & 2.4 $^{b}$ & G2-V $^{i}$ & 5664 $^{n}$ & 4.39 $^{n}$ & -0.34 $^{n}$ & 1.02 & 1.117 & South \\
HIP 77257 & 15:46:26.75 & +07:21:11.7 & 11.819 $^{a}$ & 3.3 $^{b}$ & G0-V $^{c}$ & 5900 $^{d}$ & 4.17 $^{d}$ & -0.01 $^{d}$ & 1.09 & 1.854 & North \\
HIP 77358 & 15:47:29.41 & -37:54:56.9 & 15.26 $^{a}$ & 1.8 $^{b}$ & G7IV-V $^{i}$ & 5584 $^{e}$ & 4.4 $^{e}$ & 0.08 $^{e}$ & 0.96 & 1.092 & South \\
HIP 77760 & 15:52:40.19 & +42:27:00.0 & 15.832 $^{a}$ & 3.4 $^{b}$ & G0V Fe-0.8 CH-0.5 $^{r}$ & 5776 $^{u}$ & 3.83 $^{u}$ & -0.51 $^{u}$ & 1.08 & 4.056 & North \\
HIP 79248 & 16:10:24.21 & +43:49:06.1 & 17.942 $^{a}$ & 2.6 $^{b}$ & K0V $^{c}$ & 5388 $^{j}$ & 4.52 $^{j}$ & 0.46 $^{j}$ & 0.87 & 0.828 & North \\
HIP 79672 & 16:15:37.13 & -08:22:05.7 & 14.131 $^{a}$ & 2.7 $^{b}$ & G2Va $^{c}$ & 5814 $^{q}$ & 4.45 $^{q}$ & 0.06 $^{q}$ & 1.02 & 0.973 & South \\
HIP 7981 & 01:42:29.95 & +20:16:12.5 & 7.605 $^{a}$ & 0.1 $^{b}$ & K1V $^{c}$ & 5196 $^{u}$ & 4.5 $^{u}$ & -0.01 $^{u}$ & 0.85 & 0.867 & North \\
HIP 80337 & 16:24:01.24 & -39:11:34.8 & 12.908 $^{a}$ & 2.2 $^{b}$ & G1V CH-0.4 $^{i}$ & 5858 $^{n}$ & 4.5 $^{n}$ & 0.03 $^{n}$ & 1.07 & 0.867 & South \\
HIP 80686 & 16:28:27.80 & -70:05:04.8 & 12.177 $^{a}$ & 2.4 $^{b}$ & F9V $^{i}$ & 6030 $^{d}$ & 4.43 $^{d}$ & -0.08 $^{d}$ & 1.14 & 1.019 & South \\
HIP 8102 & 01:44:05.13 & -15:56:22.4 & 3.65 $^{f}$ & 0.9 $^{b}$ & G8V $^{c}$ & 5331 $^{h}$ & 4.44 $^{h}$ & -0.49 $^{h}$ & 0.94 & 0.996 & South \\
HIP 81300 & 16:36:21.18 & -02:19:25.8 & 9.92 $^{a}$ & 1.6 $^{b}$ & K0V(k) $^{i}$ & 5248 $^{d}$ & 4.55 $^{d}$ & 0.01 $^{d}$ & 0.87 & 0.773 & Both \\
HIP 83389 & 17:02:36.30 & +47:04:47.3 & 18.294 $^{a}$ & 1.8 $^{b}$ & G8V $^{af}$ & 5442 $^{ag}$ & 4.39 $^{ag}$ & -0.13 $^{ag}$ & 0.94 & 1.117 & North \\
HIP 8362 & 01:47:44.06 & +63:51:11.2 & 10.043 $^{a}$ & 0.9 $^{b}$ & G9V $^{r}$ & 5354 $^{d}$ & 4.53 $^{d}$ & 0.03 $^{d}$ & 0.9 & 0.809 & North \\
HIP 84478 & 17:16:13.68 & -26:32:36.3 & 5.95 $^{a}$ & 3.3 $^{b}$ & K5V(k) $^{i}$ & 4600 $^{d}$ & 4.7 $^{d}$ & -0.34 $^{d}$ & 0.68 & 0.547 & South \\
HIP 85235 & 17:25:00.90 & +67:18:24.1 & 12.793 $^{a}$ & 1.3 $^{b}$ & K0V $^{af}$ & 5327 $^{d}$ & 4.56 $^{d}$ & -0.42 $^{d}$ & 0.87 & 0.755 & North \\
HIP 85295 & 17:25:45.57 & +02:06:51.5 & 7.715 $^{a}$ & 3.5 $^{b}$ & K7V $^{ah}$ & 3941 $^{ai}$ & 4.68 $^{aj}$ & 0.19 $^{z}$ & 0.63 & 0.573 & Both \\
HIP 86400 & 17:39:17.02 & +03:33:19.7 & 11 $^{f}$ & 1.5 $^{b}$ & K3-V $^{c}$ & 4808 $^{d}$ & 4.56 $^{d}$ & -0.08 $^{d}$ & 0.75 & 0.755 & Both \\
HIP 86796 & 17:44:08.72 & -51:50:00.9 & 15.605 $^{a}$ & 3.8 $^{b}$ & G3IV-V $^{i}$ & 5845 $^{h}$ & 4.27 $^{h}$ & 0.35 $^{h}$ & 1 & 1.473 & South \\
HIP 88601 & 18:05:27.21 & +02:30:08.8 & 5.123 $^{a}$ & 3.7 $^{b}$ & K0-V $^{c}$ & 5394 $^{d}$ & 4.56 $^{d}$ & 0.07 $^{d}$ & 0.88 & 0.755 & Both \\
HIP 88745 & 18:07:01.61 & +30:33:42.7 & 15.739 $^{a}$ & 2.8 $^{b}$ & F9V mw $^{ak}$ & 6049 $^{d}$ & 4.18 $^{d}$ & -0.58 $^{d}$ & 1.14 & 1.812 & North \\
HIP 88972 & 18:09:37.65 & +38:27:32.1 & 11.096 $^{a}$ & 0.6 $^{b}$ & K2V $^{r}$ & 5048 $^{d}$ & 4.55 $^{d}$ & -0.2 $^{d}$ & 0.78 & 0.773 & North \\
HIP 89042 & 18:10:26.26 & -62:00:10.0 & 17.753 $^{a}$ & 4.2 $^{b}$ & G0V $^{i}$ & 5950 $^{al}$ & 4.31 $^{al}$ & 0.01 $^{al}$ & 1.08 & 1.343 & South \\
HIP 910 & 00:11:15.91 & -15:28:02.4 & 17.99 $^{a}$ & 4.8 $^{b}$ & F8V Fe-0.8 CH-0.5 $^{i}$ & 6169 $^{u}$ & 4.07 $^{u}$ & -0.34 $^{u}$ & 1.18 & 2.334 & South \\
HIP 95447 & 19:24:57.77 & +11:56:34.3 & 14.959 $^{a}$ & 1.9 $^{b}$ & G7IV Hdel1 $^{c}$ & 5530 $^{o}$ & 4.05 $^{o}$ & 0.34 $^{o}$ & 0.95 & 2.444 & North \\
HIP 96100 & 19:32:20.59 & +69:39:55.4 & 5.755 $^{f}$ & 1.8 $^{b}$ & K0V $^{c}$ & 5318 $^{u}$ & 4.59 $^{u}$ & -0.15 $^{u}$ & 0.87 & 0.705 & North \\
HIP 97944 & 19:54:17.82 & -23:56:24.3 & 14.107 $^{a}$ & 1.8 $^{b}$ & K2IV(k) $^{i}$ & 4600 $^{am}$ & 4.56 $^{am}$ & 0.25 $^{an}$ & 0.9 & 0.755 & South \\
HIP 98036 & 19:55:18.77 & +06:24:28.6 & 13.699 $^{f}$ & 2.7 $^{b}$ & G8IV $^{c}$ & 5223 $^{o}$ & 3.86 $^{o}$ & -0.17 $^{o}$ & 0.94 & 3.785 & North \\
HIP 98767 & 20:03:36.95 & +29:53:53.1 & 16.014 $^{a}$ & 0.8 $^{b}$ & G7IV-V $^{i}$ & 5606 $^{ao}$ & 4.44 $^{ao}$ & 0.25 $^{ao}$ & 0.96 & 0.996 & North \\
HIP 99240 & 20:08:41.86 & -66:10:45.6 & 6.108 $^{f}$ & 2 $^{b}$ & G8IV $^{i}$ & 5566 $^{e}$ & 4.24 $^{e}$ & 0.32 $^{e}$ & 0.94 & 1.578 & South \\
HIP 99825 & 20:15:16.58 & -27:01:57.1 & 8.799 $^{a}$ & 0.6 $^{b}$ & K2+V $^{i}$ & 5104 $^{d}$ & 4.54 $^{d}$ & 0.06 $^{d}$ & 0.77 & 0.791 & South \\

    \enddata
\end{deluxetable}
\let\thefootnote\relax\footnote{
$^{a }$\citet{GaiaDR2},
$^{b }$\citet{2005yCat.3244....0G},
$^{c }$\citet{1989ApJS...71..245K},
$^{d }$\citet{2013ApJ...764...78R},
$^{e }$\citet{2013AnA...555A.150T},
$^{f }$\citet{HIP2},
$^{g }$\citet{2015AnA...580A..24D},
$^{h }$\citet{2014AnA...564A.133J},
$^{i }$\citet{2006AJ....132..161G},
$^{j }$\citet{2005ApJS..159..141V},
$^{k }$\citet{2004AnA...415.1153S},
$^{l }$\citet{2012AnA...547A.106M},
$^{m }$\citet{2010MNRAS.403.1368G},
$^{n }$\citet{2008AnA...487..373S},
$^{o }$\citet{2015AnA...574A..50J},
$^{p }$\citet{2012AnA...547A..13T},
$^{q }$\citet{2014AnA...572A..48R},
$^{r }$\citet{2003AJ....126.2048G},
$^{s }$\citet{2003AJ....126.2015H},
$^{t }$\citet{2004AnA...415.1153S},
$^{u }$\citet{2005PASJ...57...27T},
$^{v }$\citet{2012AnA...542A.116A},
$^{w }$\citet{2006AnA...458..609D},
$^{x }$\citet{2008MNRAS.384..173F},
$^{y }$\citet{1991ApJS...77..417K},
$^{z }$\citet{2015ApJ...804...64M},
$^{aa }$\citet{2014MNRAS.444.3517M},
$^{ab }$\citet{1973ARAnA..11...29M},
$^{ac }$\citet{2014AnA...561A...7R},
$^{ad }$\citet{2002AJ....124.1670M},
$^{ae }$\citet{1984AJ.....89..702L},
$^{af }$\citet{1967AJ.....72.1334C},
$^{ag }$\citet{2015AnA...576A..94S},
$^{ah }$\citet{2002AJ....123.2002H},
$^{ai }$\citet{2005ApJ...626..446R},
$^{aj }$\citet{2007ApJS..168..297T},
$^{ak }$\citet{2001AJ....121.2148G},
$^{al }$\citet{2005AnA...437.1127S},
$^{am }$\citet{2004AnA...420..183A},
$^{an }$\citet{2016ApJ...826..171G},
$^{ao }$\citet{2013AnA...554A..84M},
$^{ap }$ Assumed
}

\FloatBarrier

\subsection{Other Simulation Inputs} \label{sec:MiscInputs}
We specify telescope size, available survey time (in fractions of a year), and the following instrument parameters: wavelength range, spectrograph resolution, overall efficiency, target RV precision photon noise, and target spectroscopic SNR per resolution element in Table \ref{table:ChampionDefinedArchitectures}. The remaining parameters, all related to the detectors such as read noise and dark current (see Table \ref{table:MiscInputs}), were kept the same across architectures. From these parameters, we calculated estimated exposure times for all stars for each telescope/site/instrument combination at an airmass of 1.0154 (10 degrees off zenith) as an intermediate step. These estimated exposure times were then scaled by the dispatch scheduler to calculate realized exposure times to achieve the desired precision. We calculated two sets of exposure times -- one that accounted for only the desired photon noise RV precision, and one that met both the desired photon noise RV precision and the spectroscopic SNR per resolution element (whichever was longer), to determine which requirement was driving the simulated exposure times and consequently the achieved survey cadence.

\begin{figure}[ht]
\includegraphics[width=0.99\textwidth]{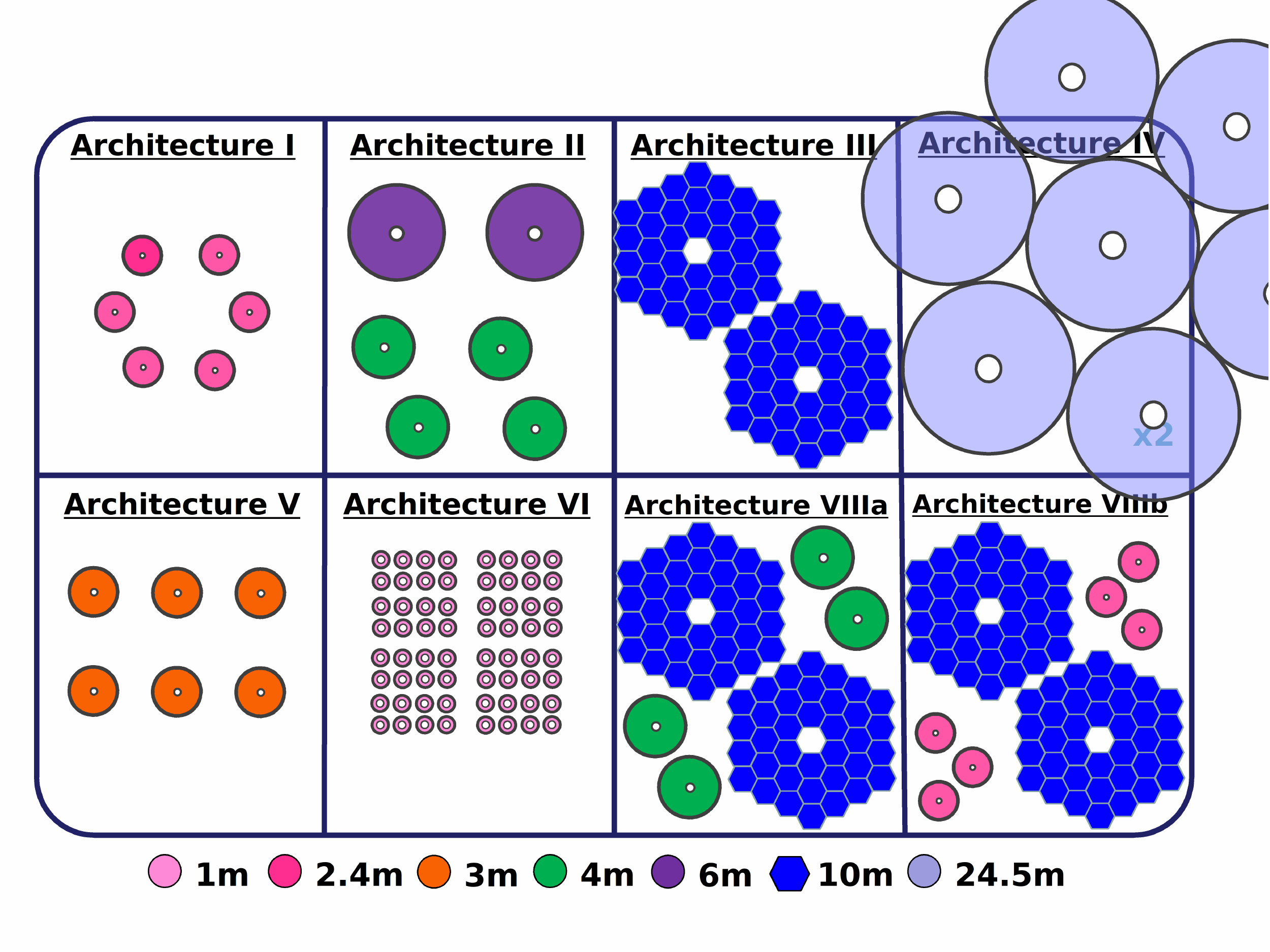}
\caption{Representative illustration of number and diameter of telescope in the architectures.}
\end{figure}
\FloatBarrier

\begin{table}[ht]
\centering
{\tiny
	\begin{tabular}{| c | c | c | c | c | c |}
	\hline
	Well Depth & Gain & Read Noise & Dark Current & Pixels per resolution element & Readout Time \\
	\hline
	90000 e-& 0.704225 ADU/e- & 4.5 e- & 3 e-/hour & 5 & 30 s \\
	\hline
	\end{tabular}
}
\caption{Notional detector properties derived from NEID's CCDs as a representative example. These values were used for all architecture simulation spectrograph assumptions, both the defined and baseline instruments.}
\label{table:MiscInputs}
\end{table}
\FloatBarrier

\subsection{Simulation Outputs} \label{sec:SimulationOutputs}
We generate outputs from both the exposure time calculator (ETC) and the dispatch scheduler. The ETC outputs are: Time observing with the shutter open, total time on a target including readouts, the number of exposures, and expected RV precision and SNR given its inputs and constraints. The dispatch scheduler outputs are: local rise/set times of the sun and target stars, site weather conditions (if a night was usable or not), and observation timeseries of the target stars.

\begin{figure}
  \centering
  \includegraphics[width=0.99\textwidth]{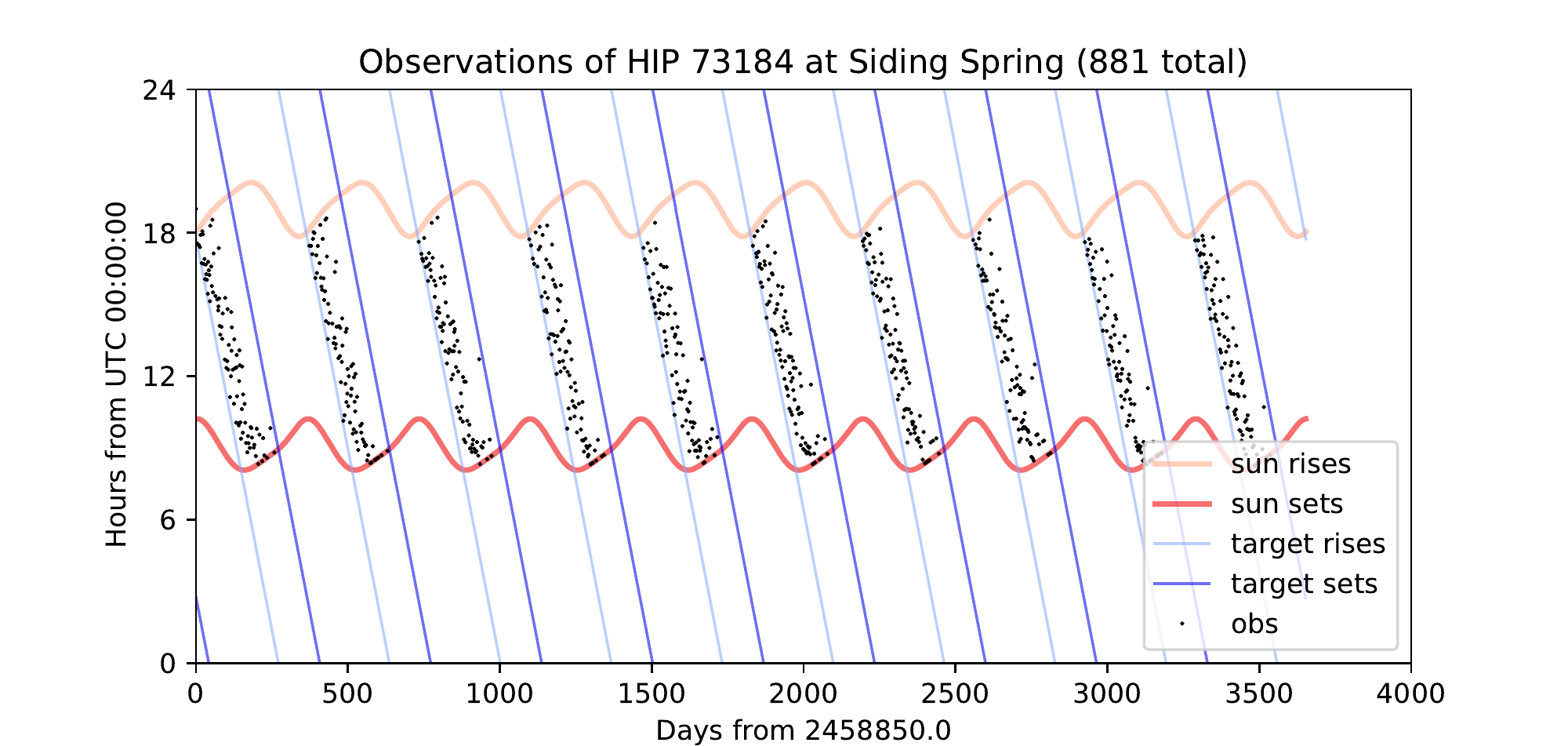}
  \caption{Observations of HIP 73184 (HD 131977) in the the architecture I simulation at Siding Spring Observatory. Local sunrise/set, star rise/set, and observation times for each day are shown.}
  \label{fig:HIP_74184a}
\end{figure}

\begin{figure}
  \centering
  \includegraphics[width=0.99\textwidth]{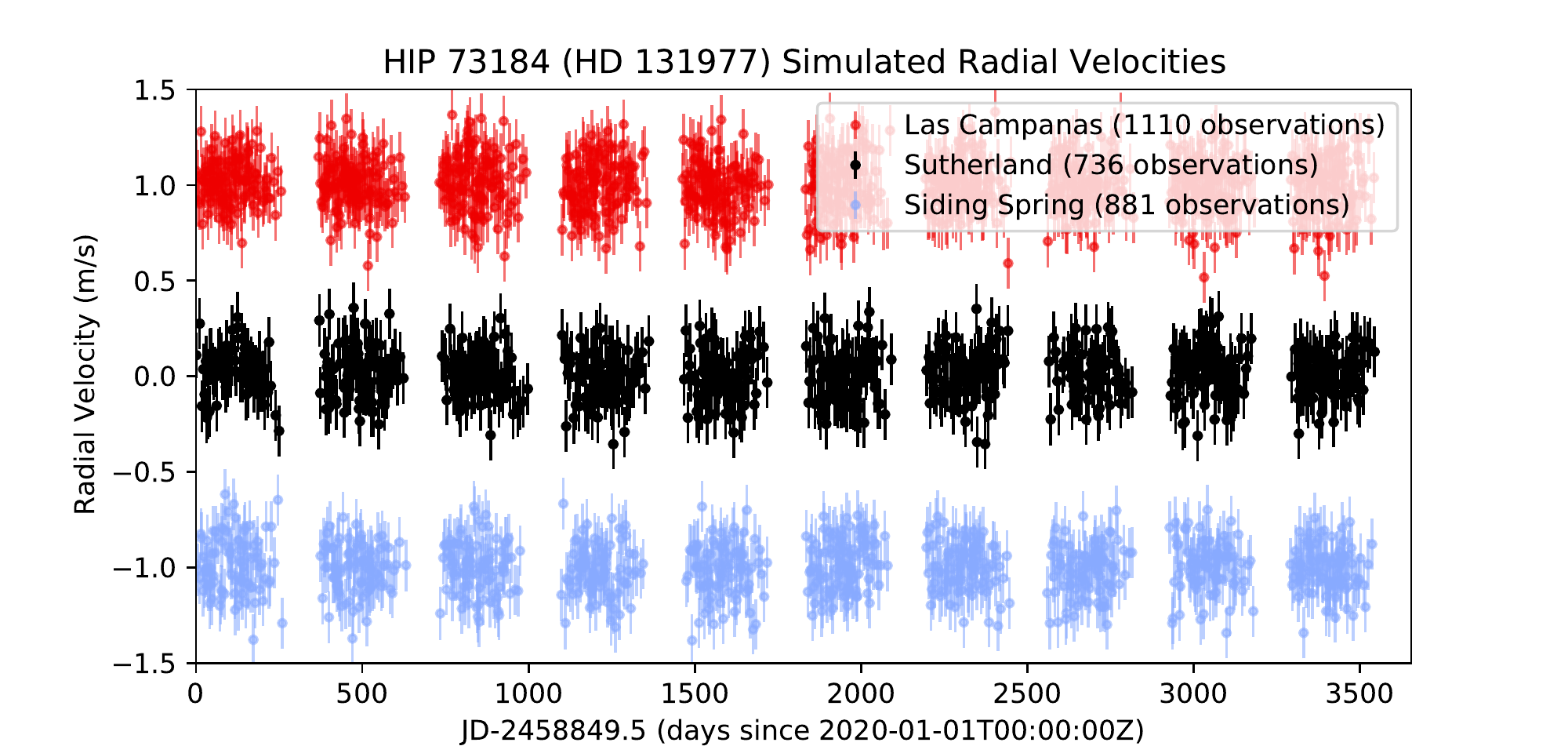}
  \caption{Simulated RV time-series for HIP 73184 (HD 131977) for Architecture I as observed by the three telescopes in its hemisphere (one each at Las Campanas, Sutherland, and Siding Spring). This star was chosen, as it has the median number of observations (2724) for the architecture. Note, only ``white'' noise is included in this simulation, despite the apparent correlated noise by eye during some seasons. The appearance of correlated noise could be partially due to the variable single measurement uncertainties from airmass extinction for a fixed exposure time over the course of a season.}
  \label{fig:HIP_74184b}
\end{figure}

\begin{figure}
  \centering
  \includegraphics[width=0.99\textwidth]{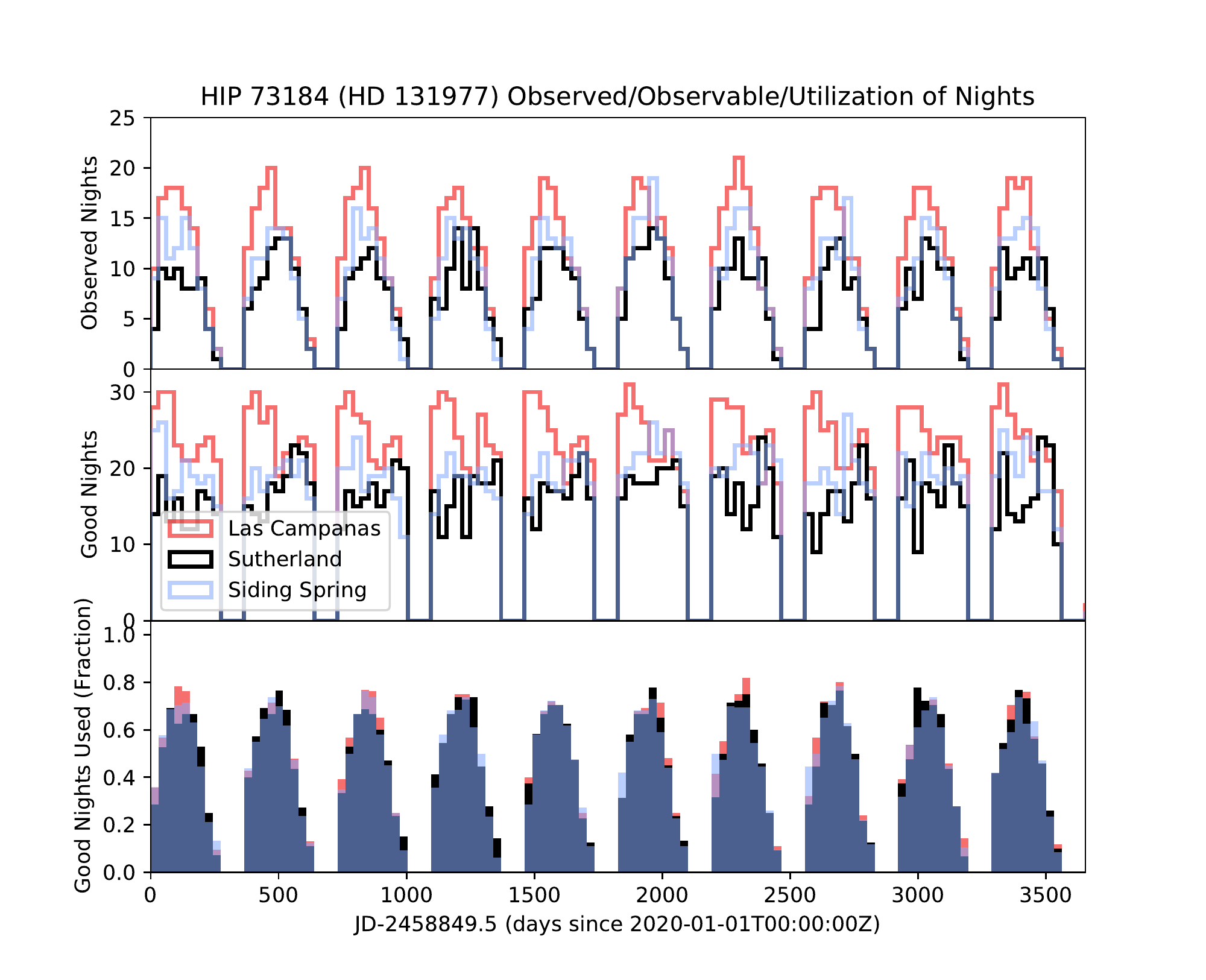}
  \caption{Simulated observation cadence and nightly availability for HIP 73184 (HD 131977) for Architecture I observed by three telescopes (one each at Las Campanas, Sutherland, and Siding Spring). Observations here are binned into 30.4375 day ``months''; actual observations were never more than 1/night. }
  \label{fig:HIP_74184c}
\end{figure}

Radial velocity time-series can be generated from the observation times of each star, although that was not done here for all architectures and telescope combinations. In Figures \ref{fig:HIP_74184a}, \ref{fig:HIP_74184b}, and \ref{fig:HIP_74184c}, we show a representative cadence and RV time-series. The full data outputs are available upon request 

\newpage
\section{Results: Architecture Exposure Times and Cadences} \label{sec:Results}

In this Section, we present summaries of the simulated radial velocity surveys for each of the architectures in turn. For each architecture, we present PDFs and CDFs of the stellar exposure times and number of observations of each star. These PDFs and CDFs are presented both for each telescope individually, and for certain combinations of telescopes within a given architecture. The 2/4/6 telescope combinations use the same number of telescopes in each hemisphere (1, 2, or 3), and we select the ``best'' ones first, e.g. those with the highest number of observations per star due to the different weather statistics. The best two sites are Mauna Kea and Las Campanas, then Kitt Peak and Sutherland, then Calar Alto and Siding Spring based upon the assumed weather statistics in Figure \ref{fig:SiteWeather}. This appeared to be consistent across all simulations of all architectures. 

Second, we include PDFs and CDFs of the fraction of observable days during which each target star was observed. We bin by Julian date as opposed to the local ``night'', which varies for each site in the global network. Third, we compute and present theoretically optimal estimates of the achievable SNR for the detection of a $K$ = 0.1 m/s planet, and the corresponding theoretical minimum detectable velocity semi-amplitude (for an SNR = 10) of a planet as per Equation \ref{eqn:SNR}. These estimates follow the idealized circular orbit and uniform cadence as described in ${\S}$\ref{sec:Introduction}, with the photon noise and instrument noise added in quadrature for the $\sigma$ term. Future injection and recovery tests, combined with stellar activity models, could explore the impact our realistic cadences have on survey sensitivity. Finally, for all of the above, we compute summary statistics for the distribution of these values within an architecture (e.g. median, quartiles, 5$^{th}$ and 95$^{th}$ percentiles), which we use for the overall comparisons of the performance across all architectures in ${\S}$\ref{sec:ArchComp}. 

\subsection{Per-Architecture Comments}
For architecture IIa, we only simulate the 6-m telescopes at Mauna Kea and Las Campanas. For the 4-m telescopes (Kitt Peak, Calar Alto, Sutherland, And Siding Spring), we re-use the simulations from architecture IIb; hence the identical PDFs and CDFs.

For architecture VIIIa, not all combinations of large and small telescopes are shown. In particular, the 2 and 4 small telescope cases (with and without large telescopes) are not shown to reduce clutter.

The higher RV precision requirements of the 10-m telescopes result in fairly similar exposure times compared to the 3.5-m apertures with lower RV precision requirements. Consequently, the 10-m apertures have a similar number of observations per star per usable night, despite the larger apertures. However, given the smaller time allocation available on the 10-m apertures, this results in an overall lower number of observations per star for those telescopes.

Architecture VIIIb is broadly similar to architecture VIIIa, though the small telescopes are six 2.4 m telescopes (comparable to architecture I), instead of four 3.5 m telescopes (comparable to architecture IIa). For legibility, in the plots that follow, only the small (2.4-m) telescopes are shown in the exposure time plots. The simulations of the 10-m telescopes from architecture VIIIa are reused here. Again as in architecture VIIIa, not all large/small telescope combinations are shown. Additionally, the detection calculations consider the differing instrument sensitivities using equation \ref{eqn:SNR2}.

\FloatBarrier
\newpage
\subsection{Example Architecture Results}
For clarity, only the results from Architecture I are shown here. The results from all other architectures are in Appendix \ref{sec:appendix}.
\FloatBarrier

\begin{figure}
\noindent \includegraphics[width=0.49\textwidth]{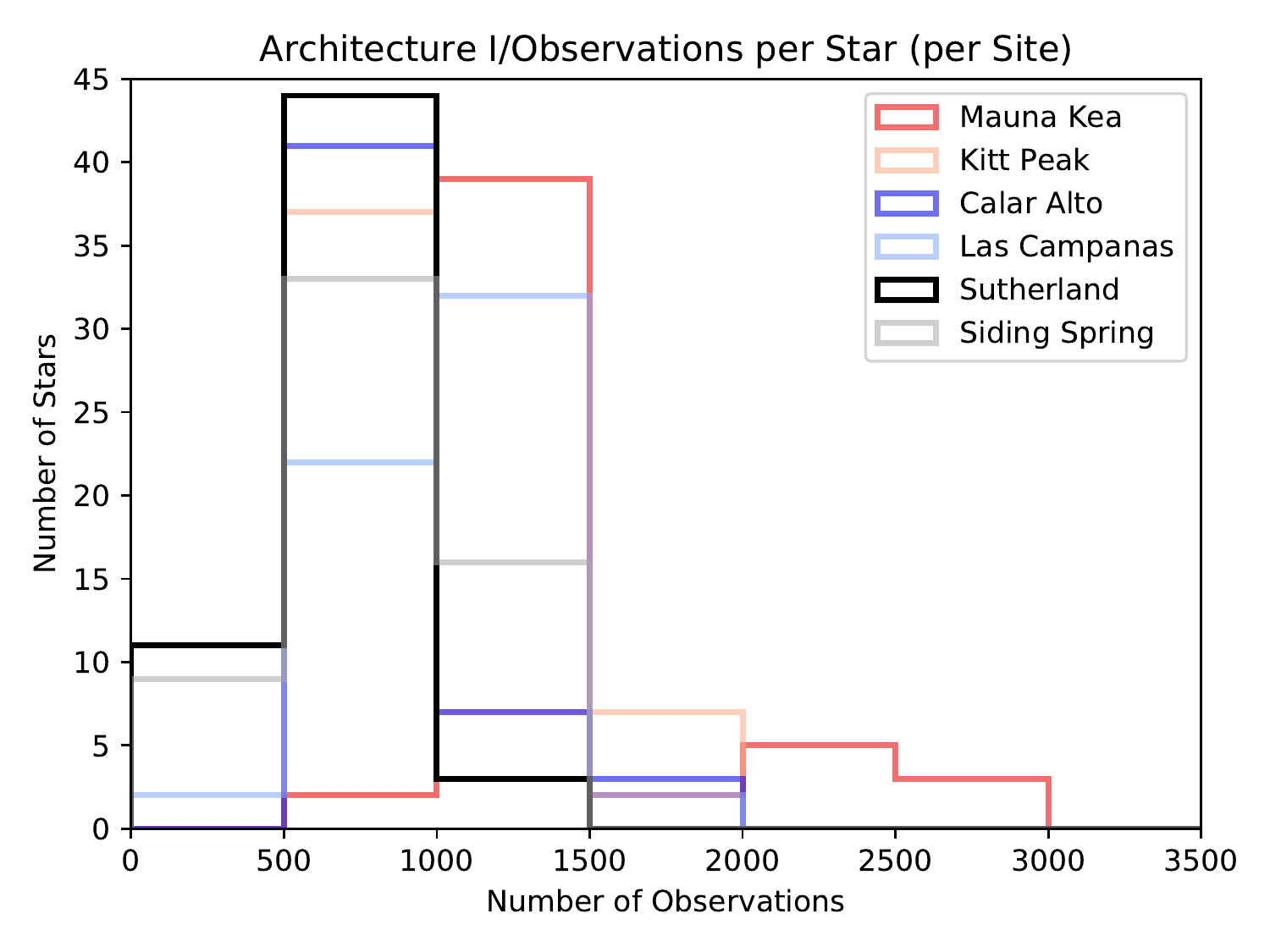}
\includegraphics[width=0.49\textwidth]{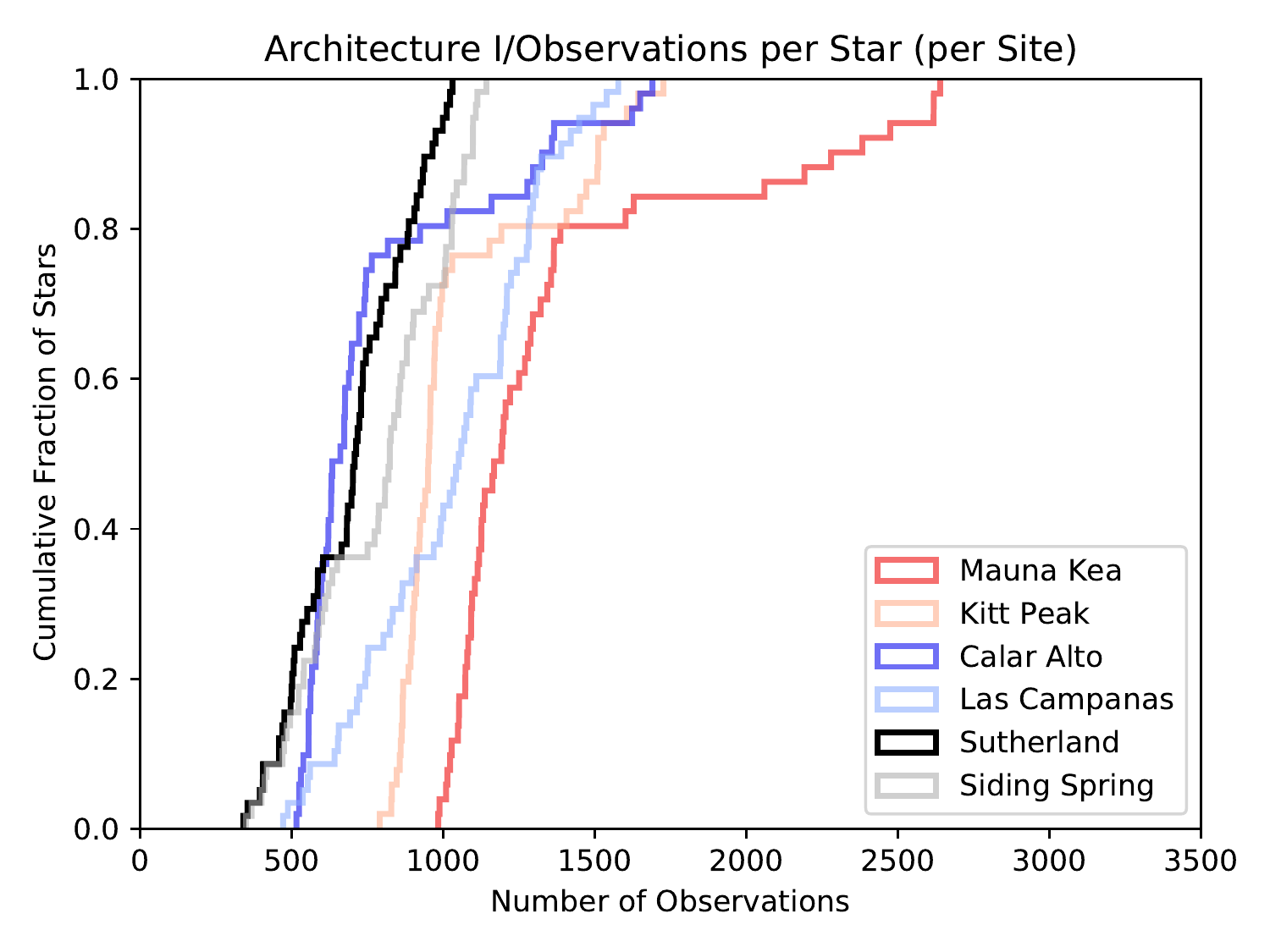}

\noindent \includegraphics[width=0.49\textwidth]{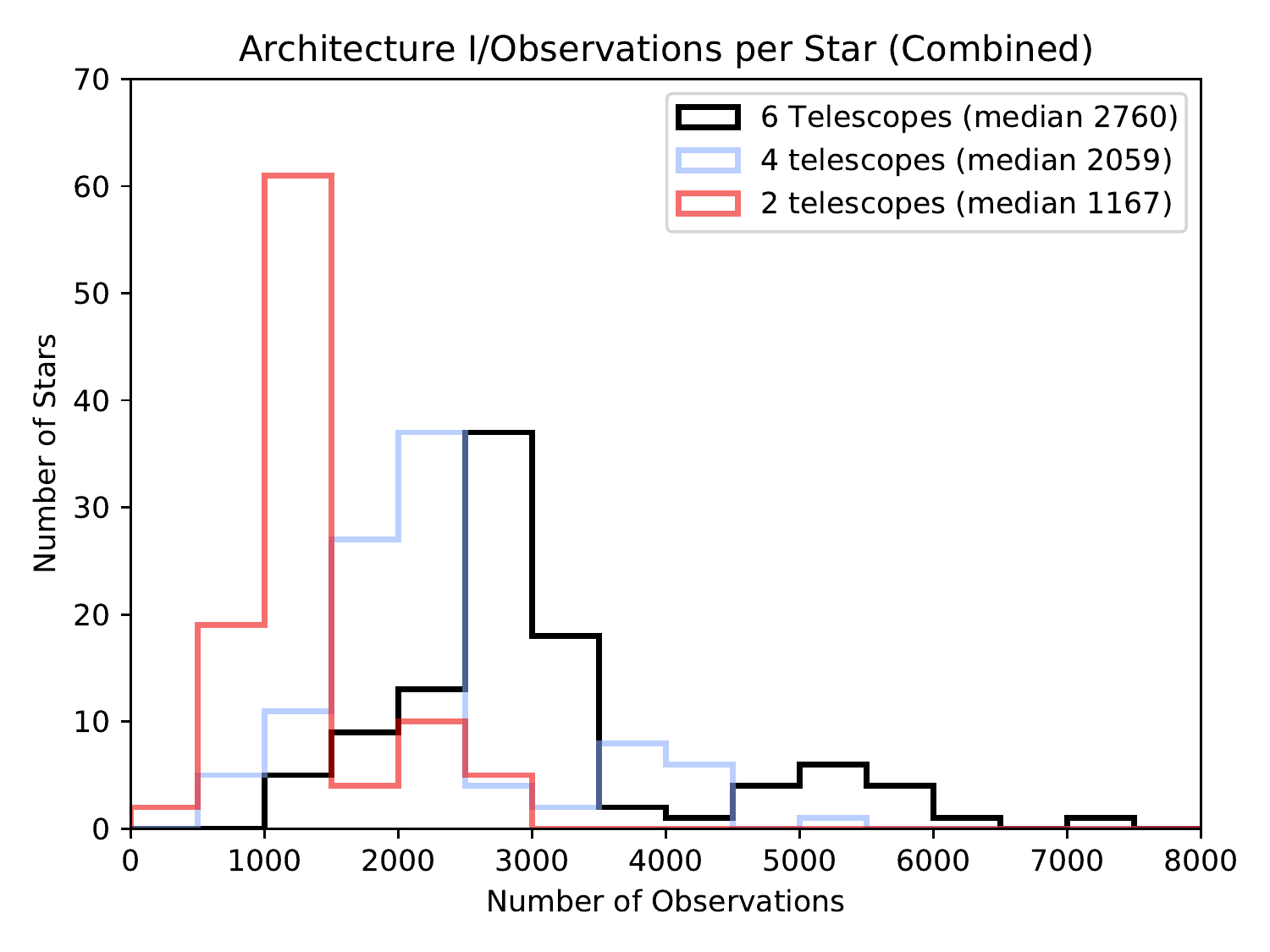}
\includegraphics[width=0.49\textwidth]{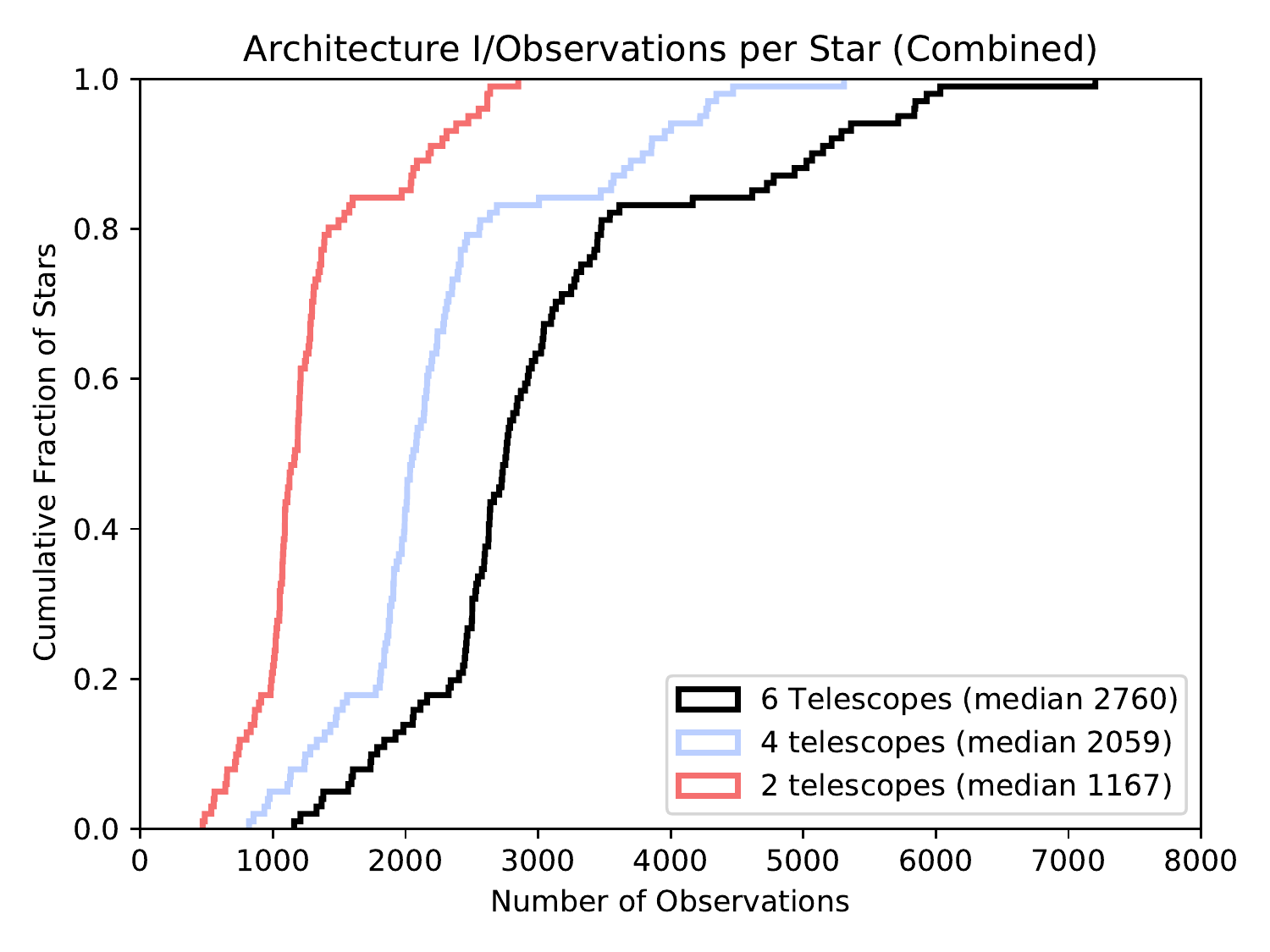}
\caption{Histograms of how often a given star was observed. Individual sites are shown (top), as well as groupings of different sites (bottom), with the 6 telescope grouping being the full architecture.}
\label{fig:ArchIobs}
\end{figure}

\begin{figure}
\noindent \includegraphics[width=0.49\textwidth]{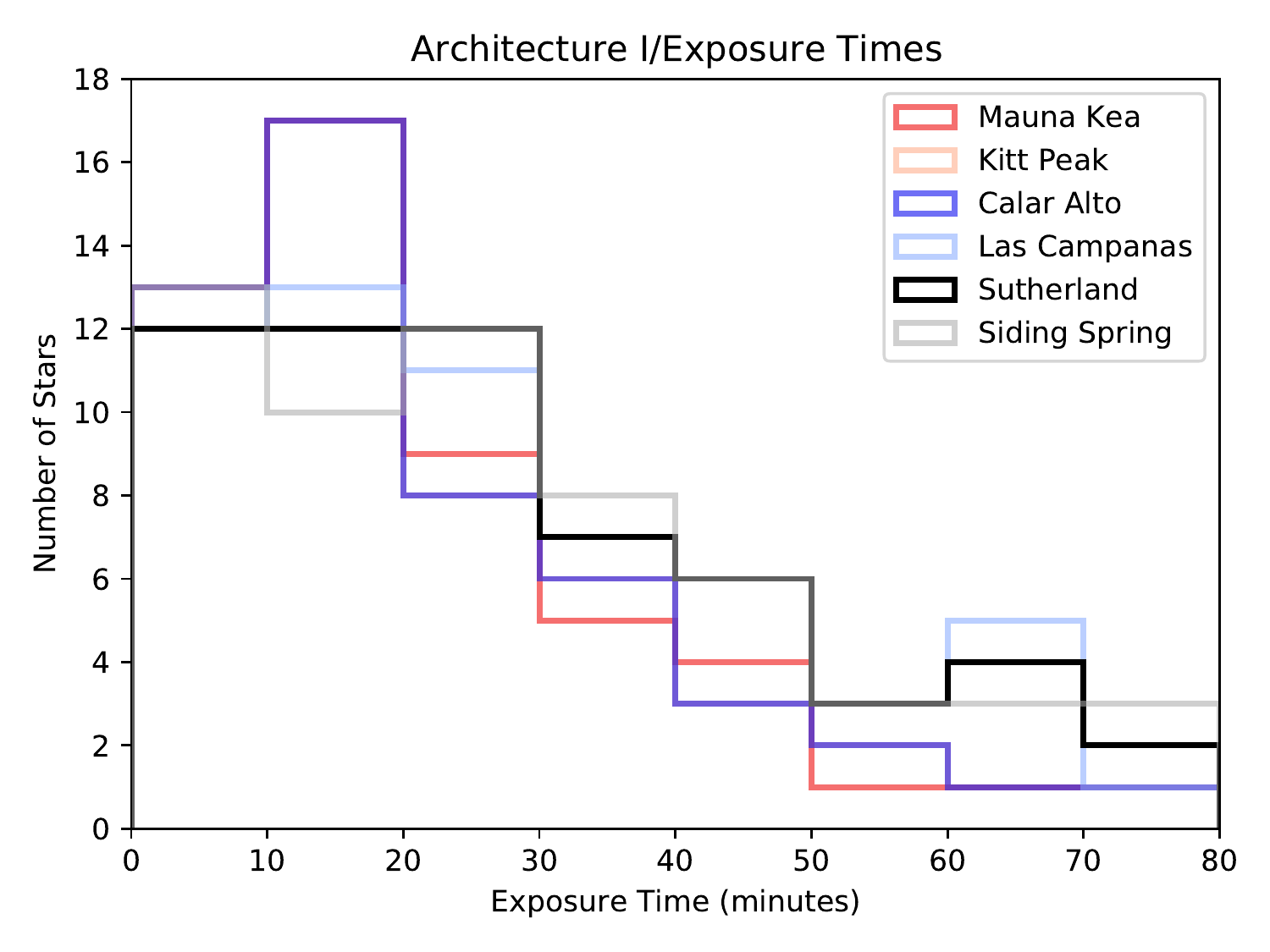}
\includegraphics[width=0.49\textwidth]{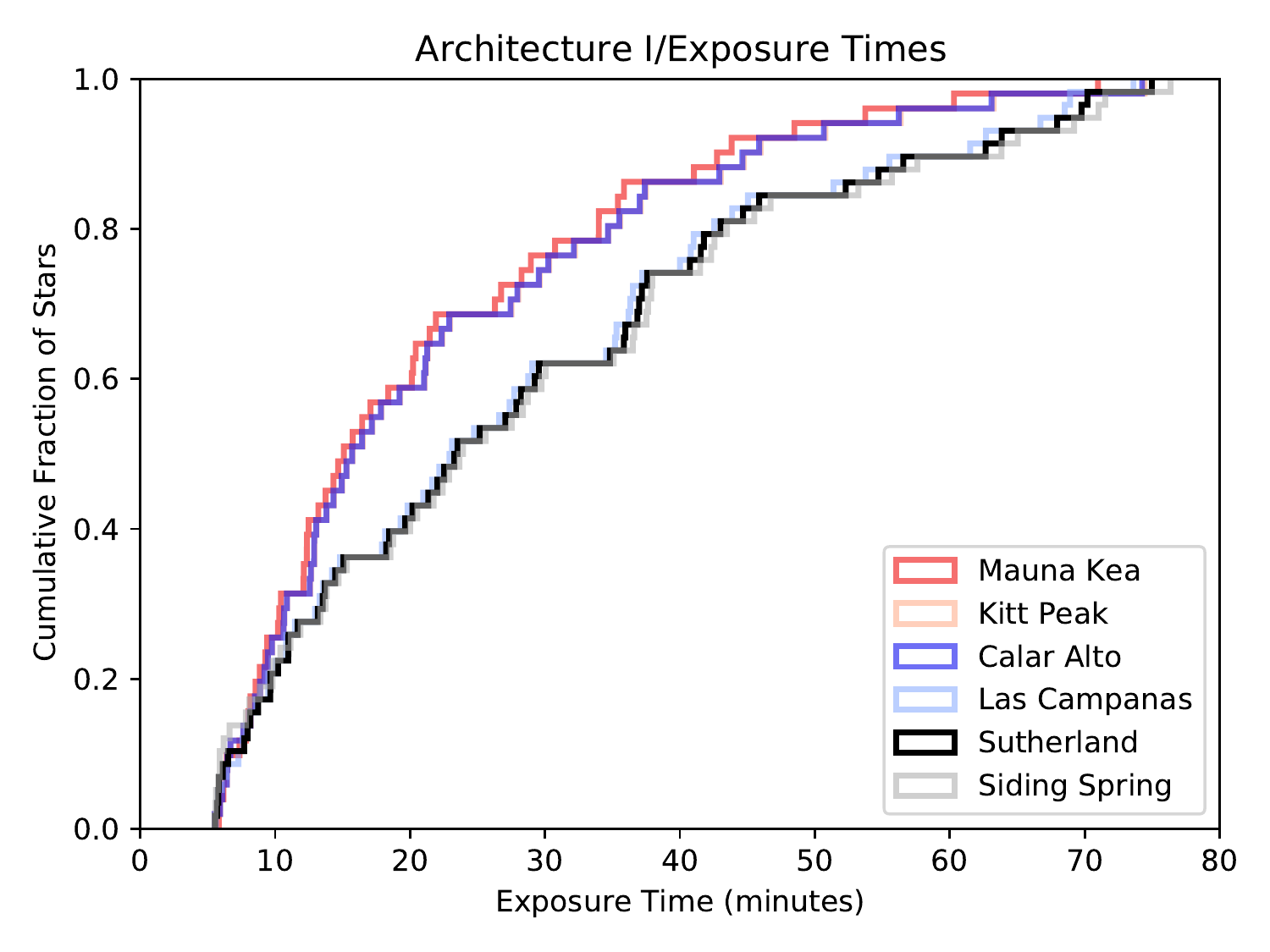}

\noindent \includegraphics[width=0.49\textwidth]{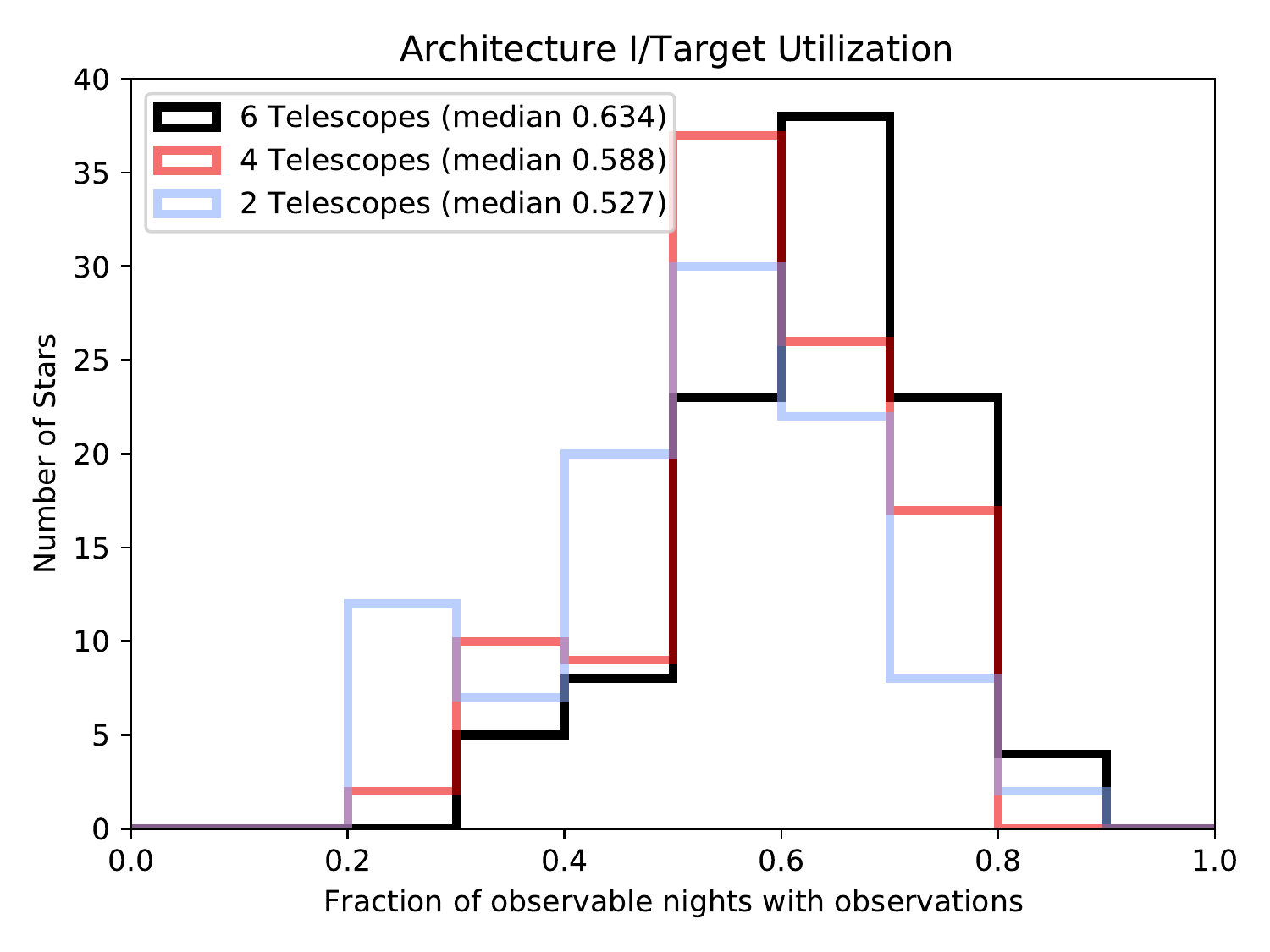}
\includegraphics[width=0.49\textwidth]{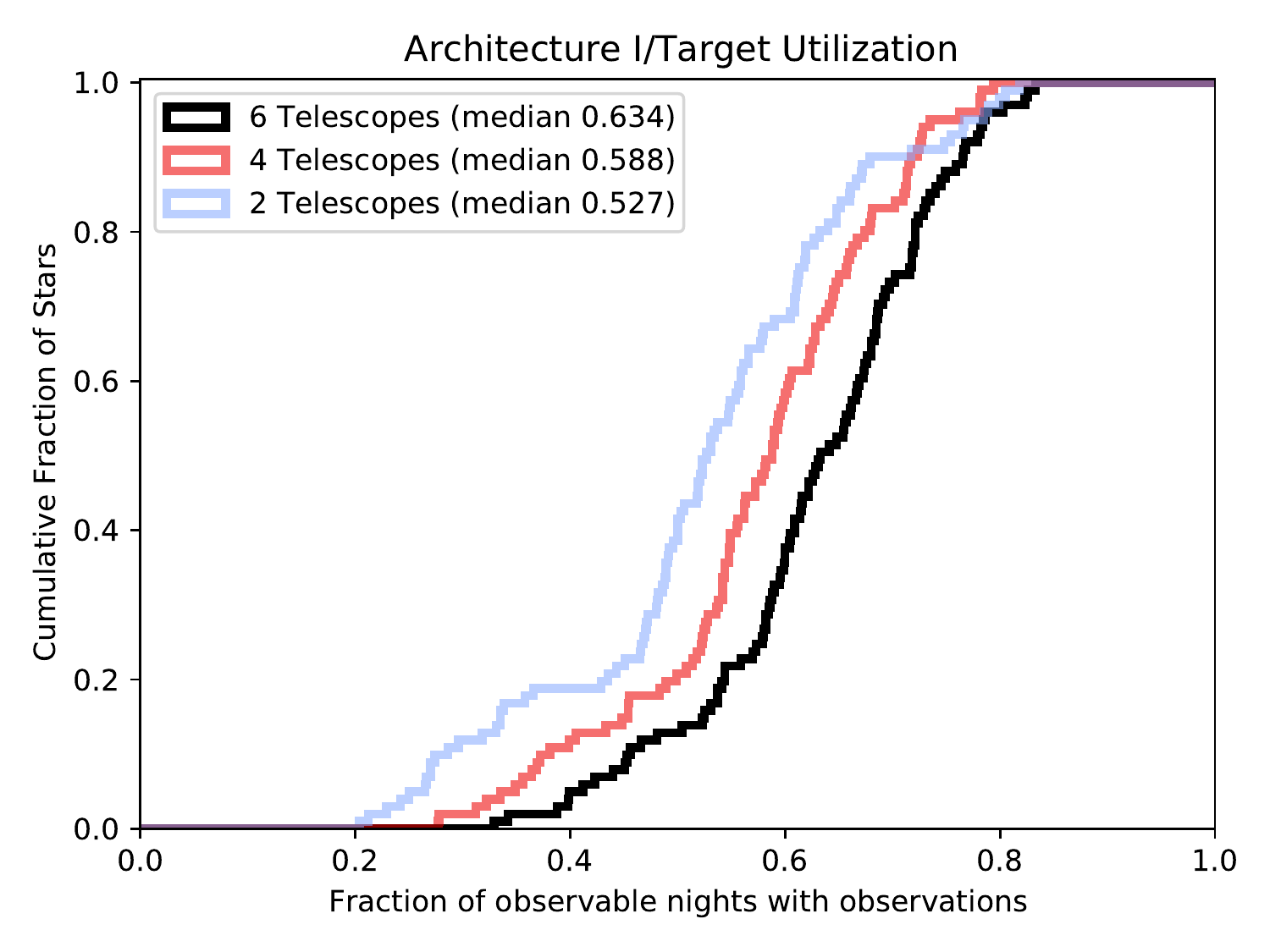}
\caption{Top: Histograms of exposure times per star calculated individually at each site. Bottom: how ``efficiently'' each star was observed. That is, for what fraction of nights where a star could be observed from at least one site it was observed at least once.}
\label{fig:ArchIexpfrac}
\end{figure}

\begin{figure}
\noindent \includegraphics[width=0.49\textwidth]{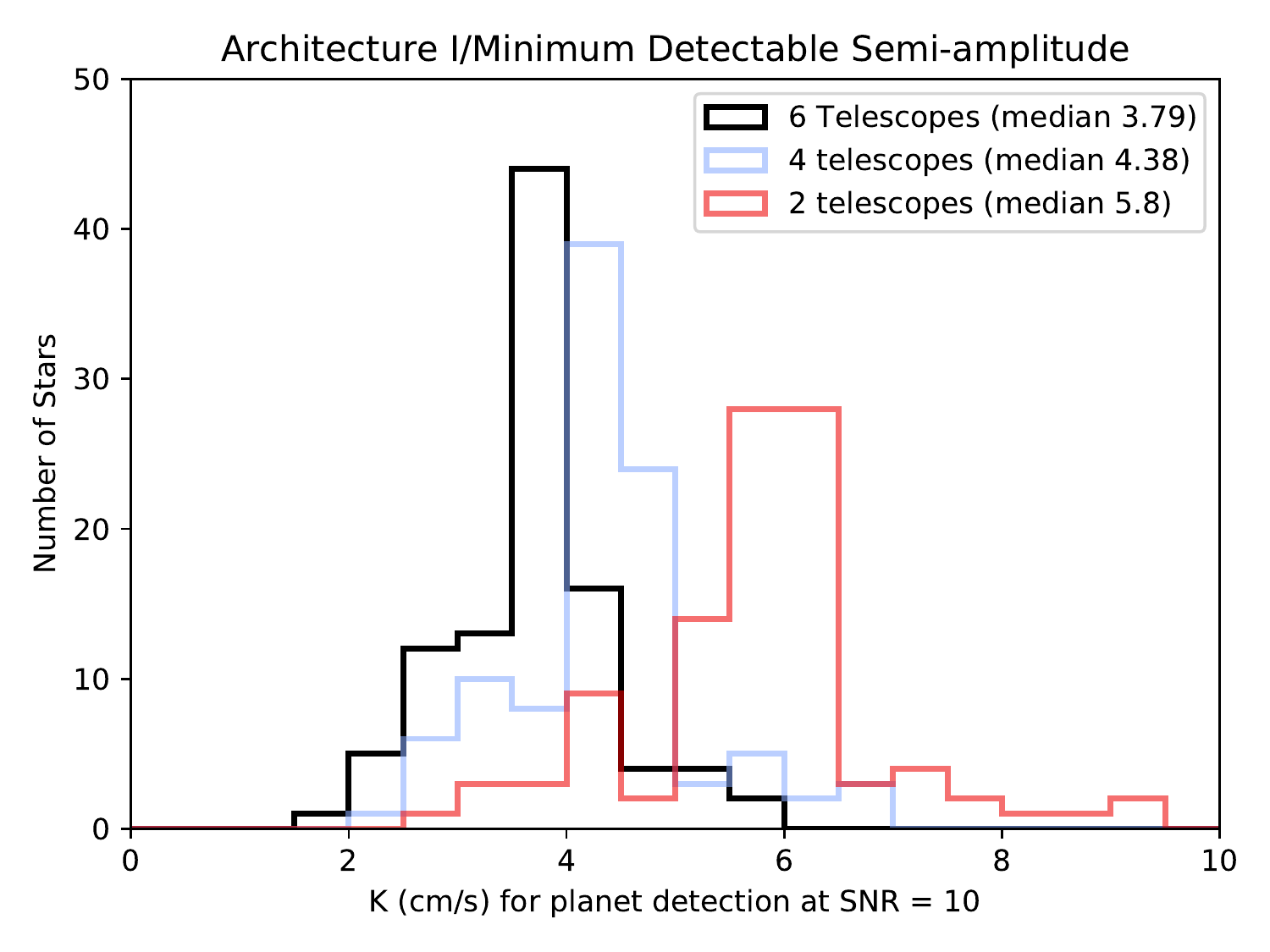}
\includegraphics[width=0.49\textwidth]{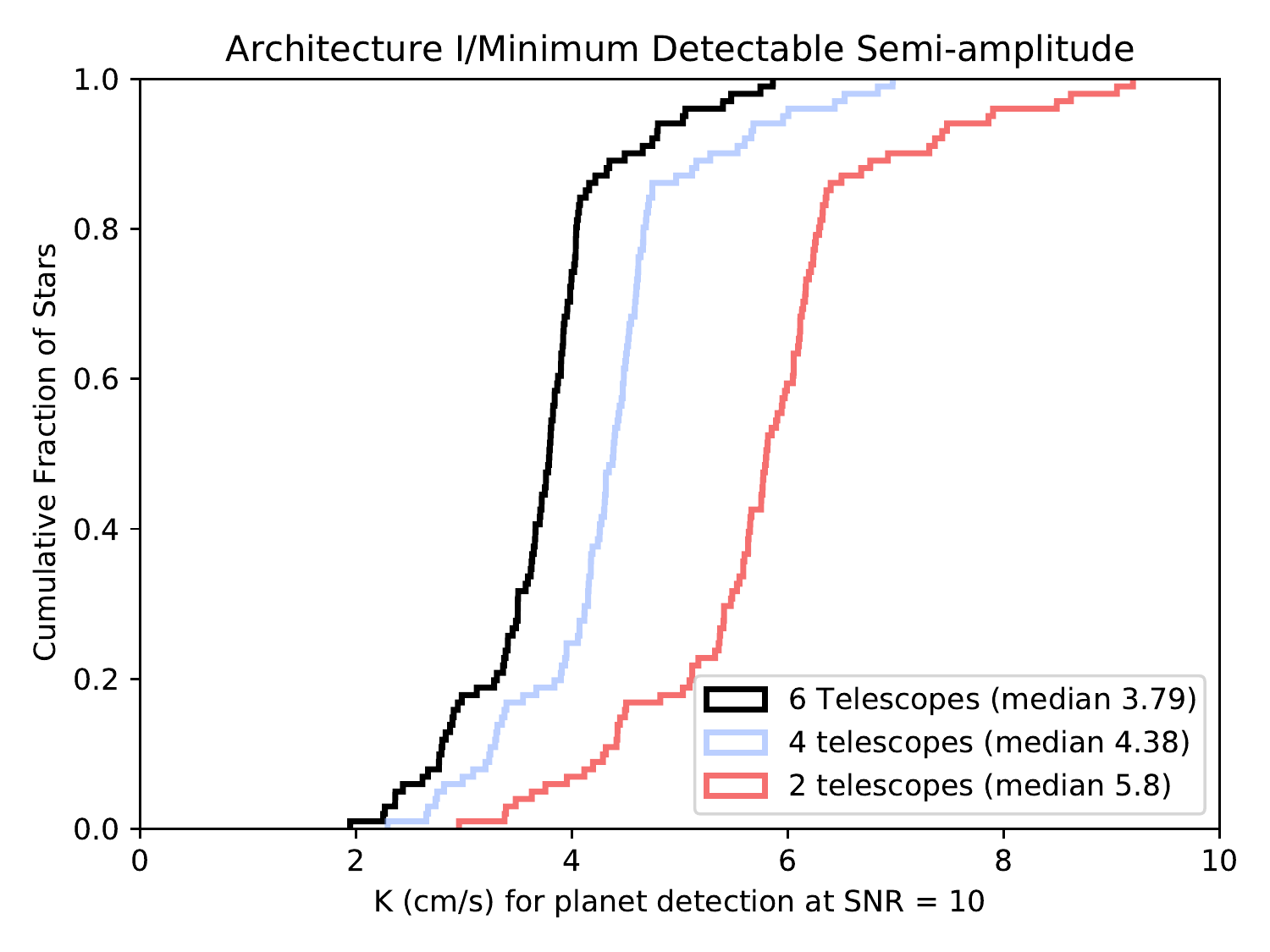}

\noindent \includegraphics[width=0.49\textwidth]{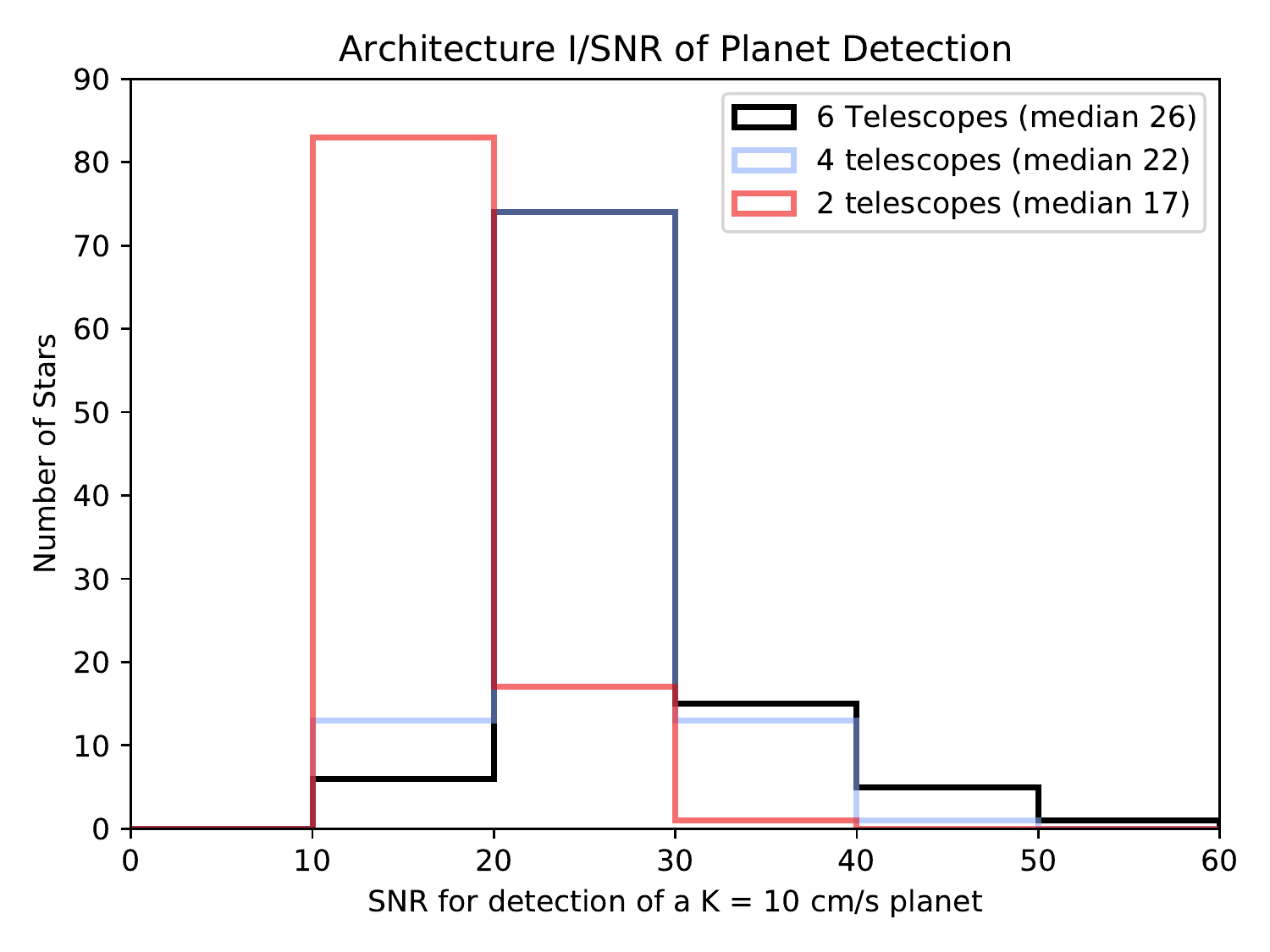}
\includegraphics[width=0.49\textwidth]{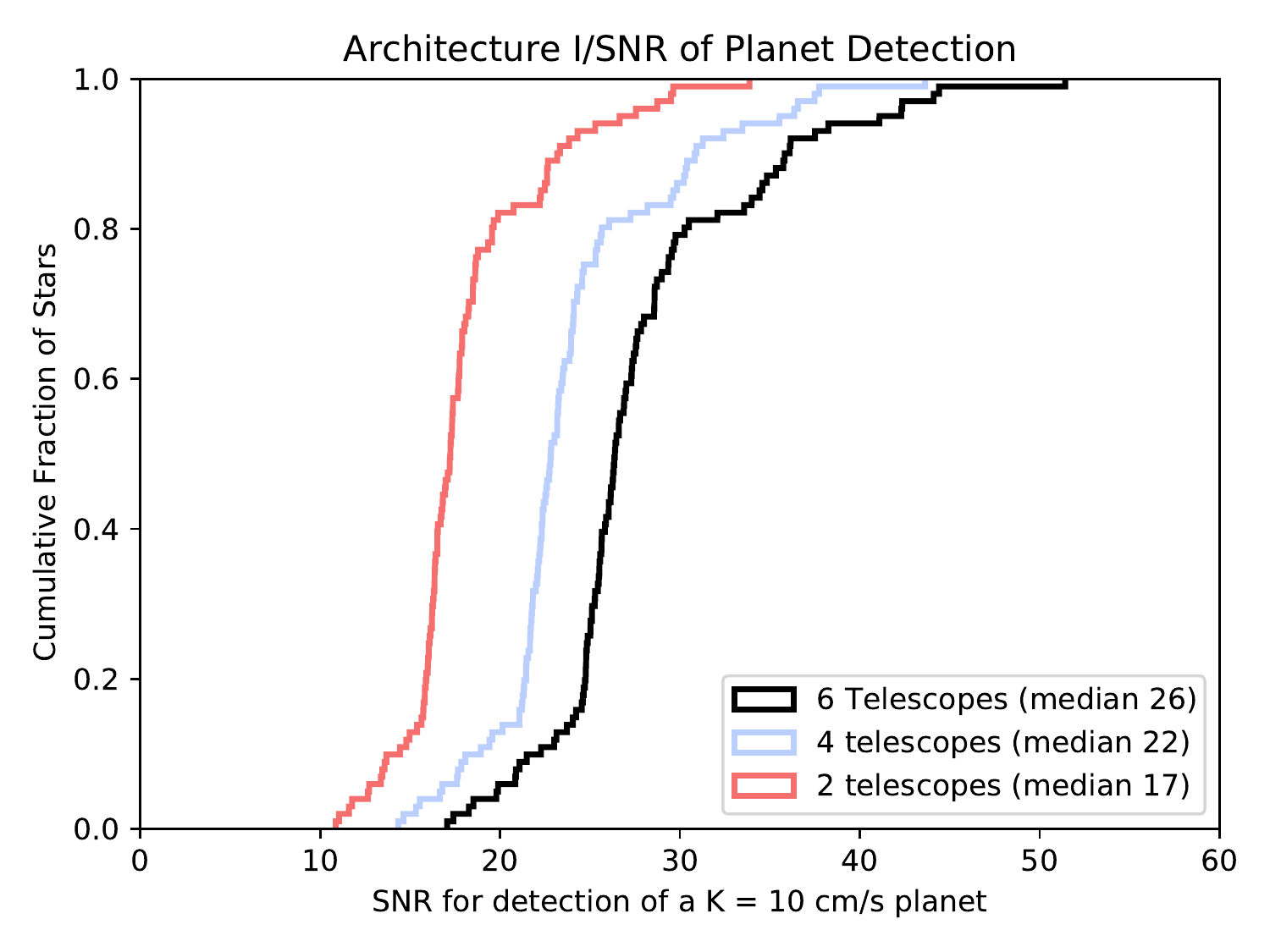}
\caption{Histograms of our sensitivity calculations from the observations. Top: smallest semi-amplitude $K$ for which a planet with a period much shorter than our 10 year survey could be found at an SNR of 10. Bottom: The SNR at which a planet with a period much shorter than our 10 year survey could be detected at if it had a semi-amplitude of 10 cm/s. In both cases, these values are calculated from equation \ref{eqn:SNR}.}
\label{fig:ArchIkSNR}
\end{figure}

\FloatBarrier
\section{Results: Architecture Comparison} \label{sec:ArchComp}

\begin{figure}[ht]
\includegraphics[width=0.49\textwidth]{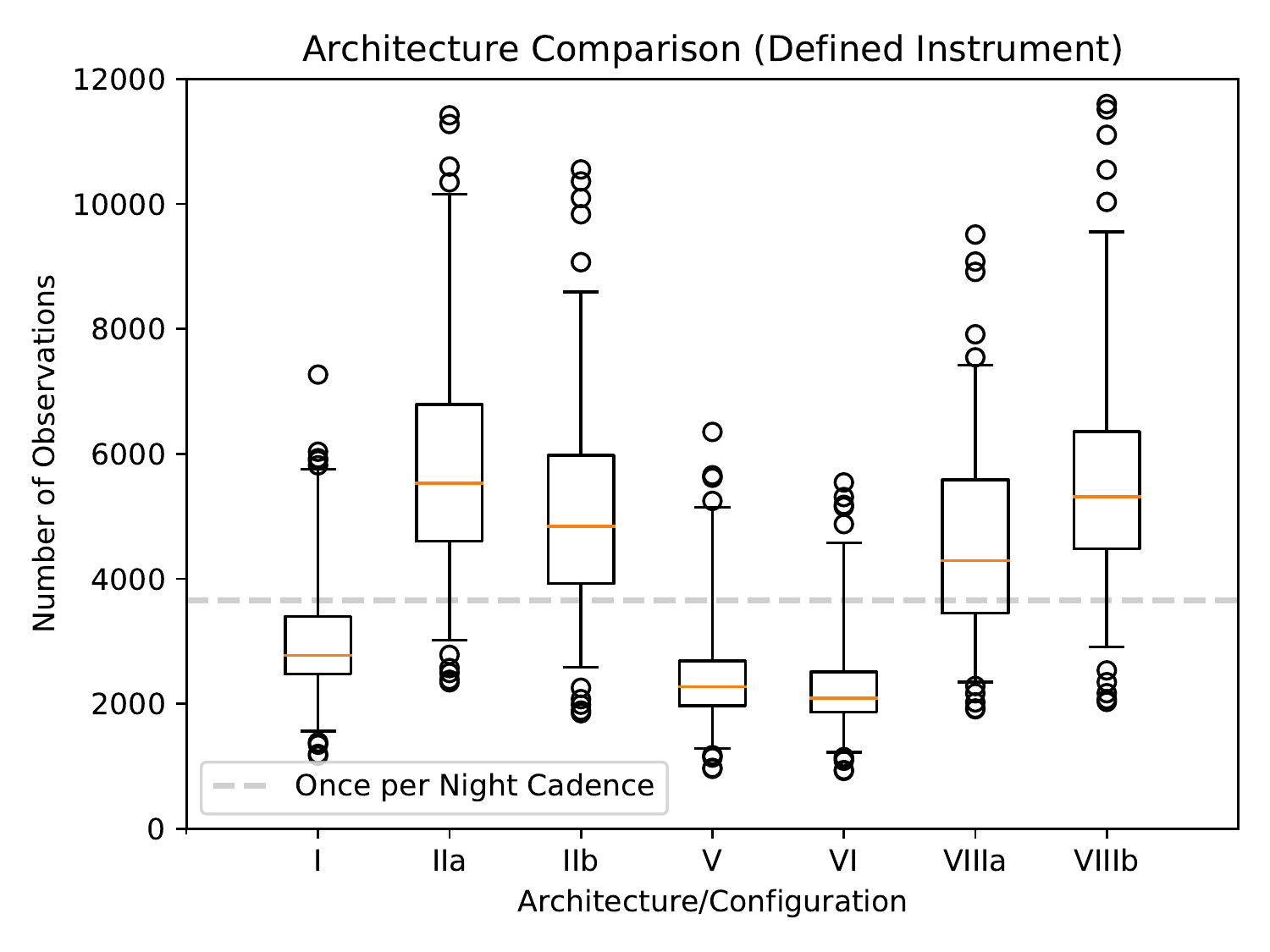}
\includegraphics[width=0.49\textwidth]{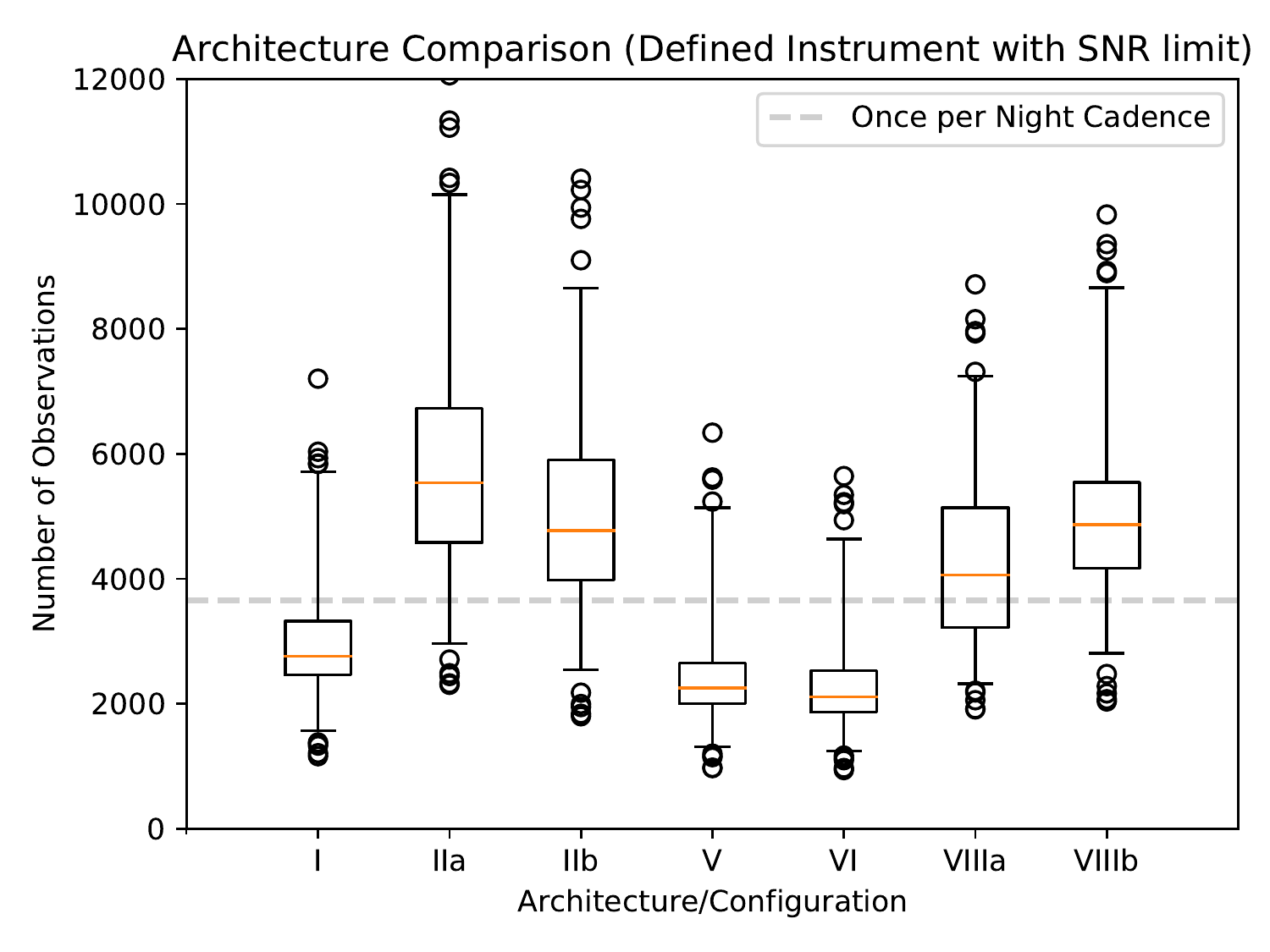}
\\
\includegraphics[width=0.49\textwidth]{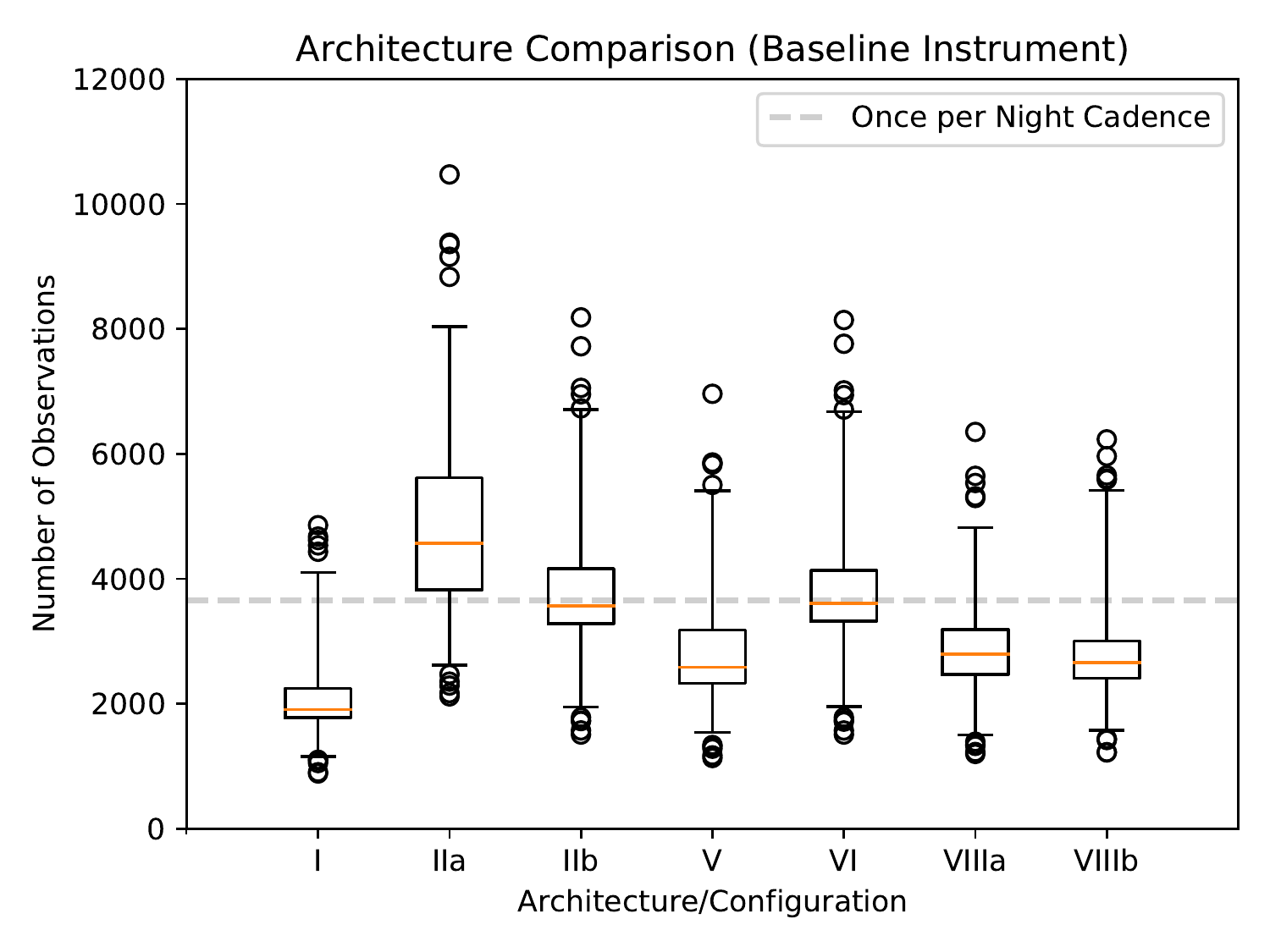}
\includegraphics[width=0.49\textwidth]{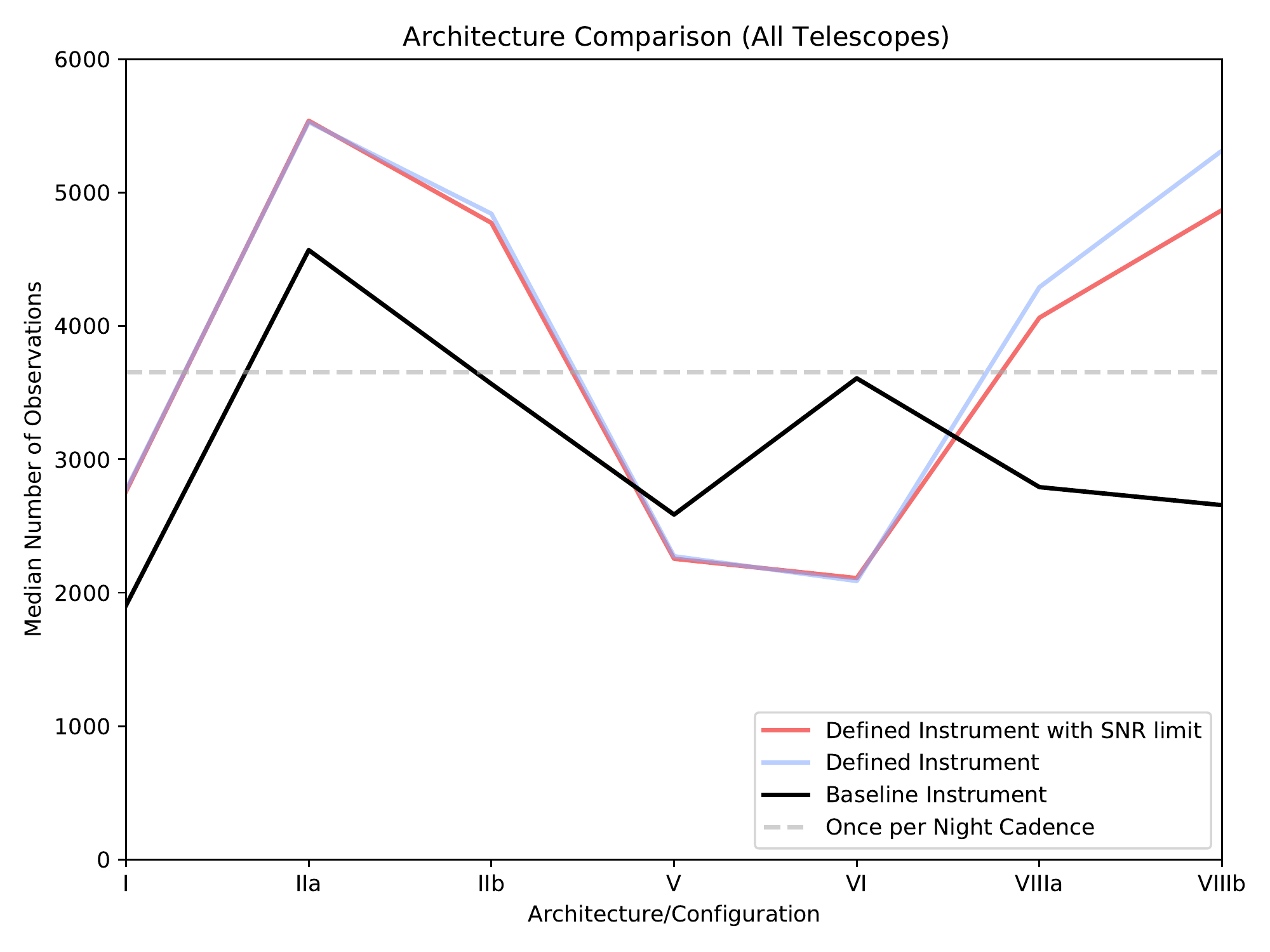}
\caption{Top Left, Top Right, Bottom left: Comparison of the distribution of the number of observations between each architecture presented in ${\S}$\ref{sec:ChampionDefinedArchitectures} for the three spectrograph assumptions respectively: the architecture-defined spectrograph with a RV precision photon noise requirement and no minimum SNR per spectral resolution element requirement (Top Left), the architecture-defined spectrograph with both a RV precision photon noise requirement and a minimum SNR per spectral resolution element requirement (Top Right), and the baseline instrument spectrograph with only a RV precision photon noise requirement and no minimum SNR per spectral resolution element requirement (Bottom Right). The boxes show the 25th/75th percentile, with the median marked in orange. The whiskers extend from the 5th to the 95th percentile, and the circles the stars below the 5th or above the 95th percentile. For all architectures, the number of observations is summed across all telescopes for the specified architecture. Bottom Right: Comparison of the median number of observations between each architecture presented in ${\S}$\ref{sec:ChampionDefinedArchitectures} for the three spectrograph and survey requirements combination assumptions considered: the architecture-defined spectrograph with both a RV precision photon noise requirement and a minimum SNR per spectral resolution element requirement (red line), the architecture-defined spectrograph with only a RV precision photon noise requirement and no minimum SNR per spectral resolution element requirement (blue line), and a baseline instrument spectrograph with only a RV precision photon noise requirement and no minimum SNR per spectral resolution element requirement (black line). The number of observations from an effective cadence of once per night per architecture (3653 epochs) is shown as a dashed horizontal line.}
\label{fig:Median_comparison}
\end{figure}

In this section we present our key findings from a comparison between architectures in our survey simulations. First, we present a summary of the number of observations per star for each architecture, summed across all telescopes within an architecture (Figure \ref{fig:Median_comparison}). The baseline case with the baseline spectrometer can be considered as an (aperture $\times$ time allocation) figure of merit, while the others leverage particular defined spectrograph specifications and SNR and RV precision optimizations as  listed in table \ref{table:ChampionDefinedArchitectures}.

\begin{figure}[ht]
\includegraphics[width=0.49\textwidth]{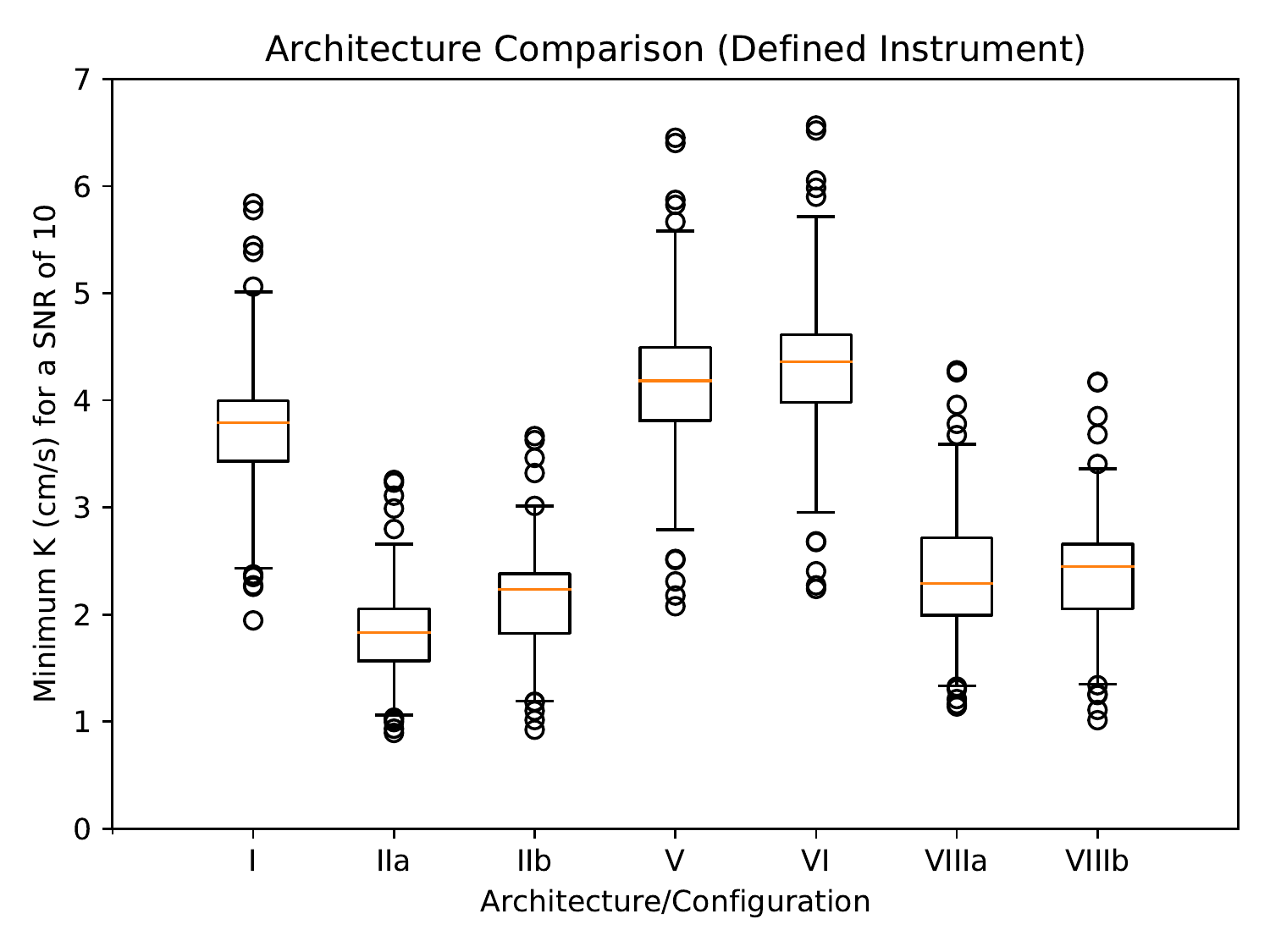}
\includegraphics[width=0.49\textwidth]{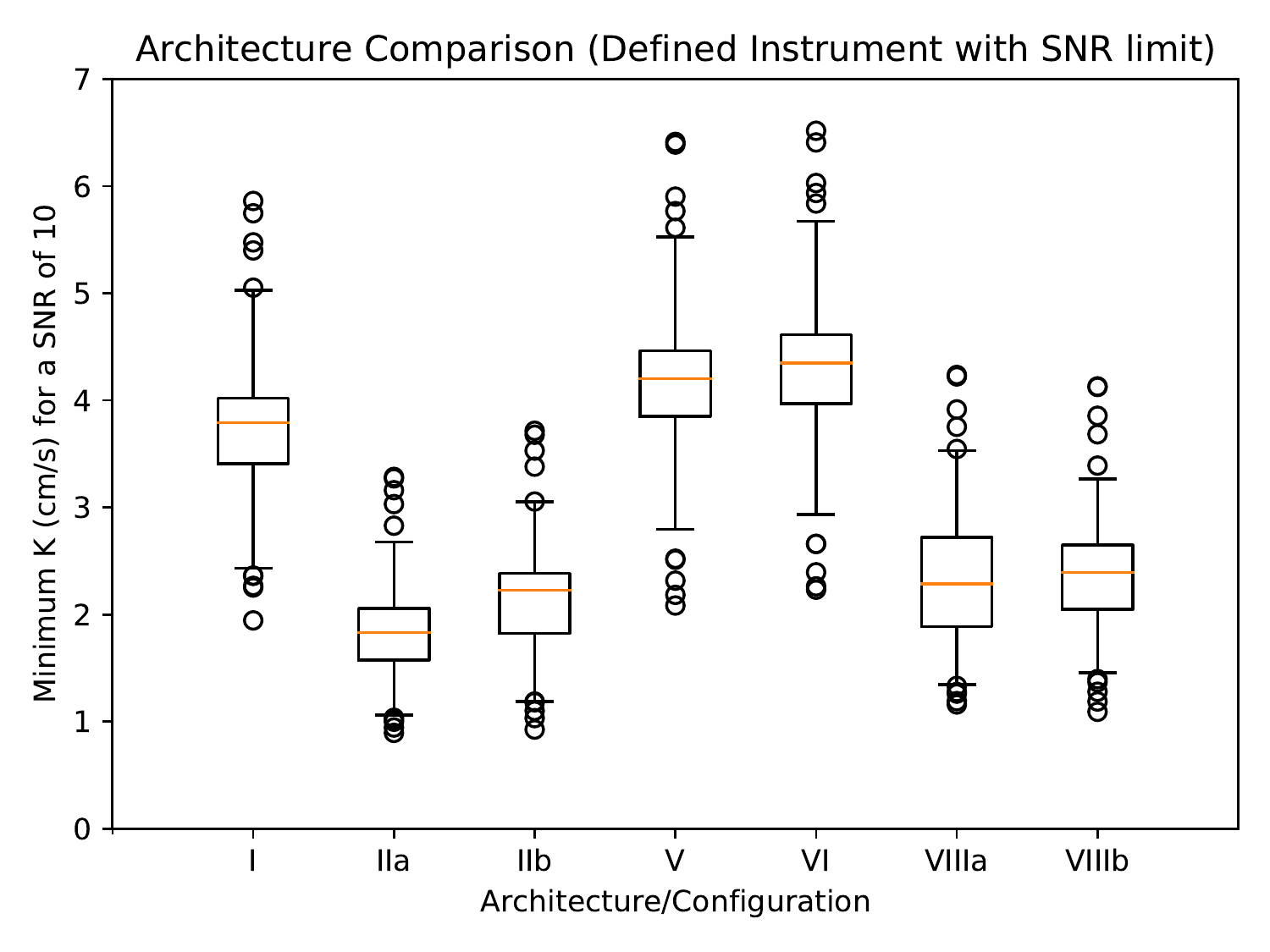}
\includegraphics[width=0.49\textwidth]{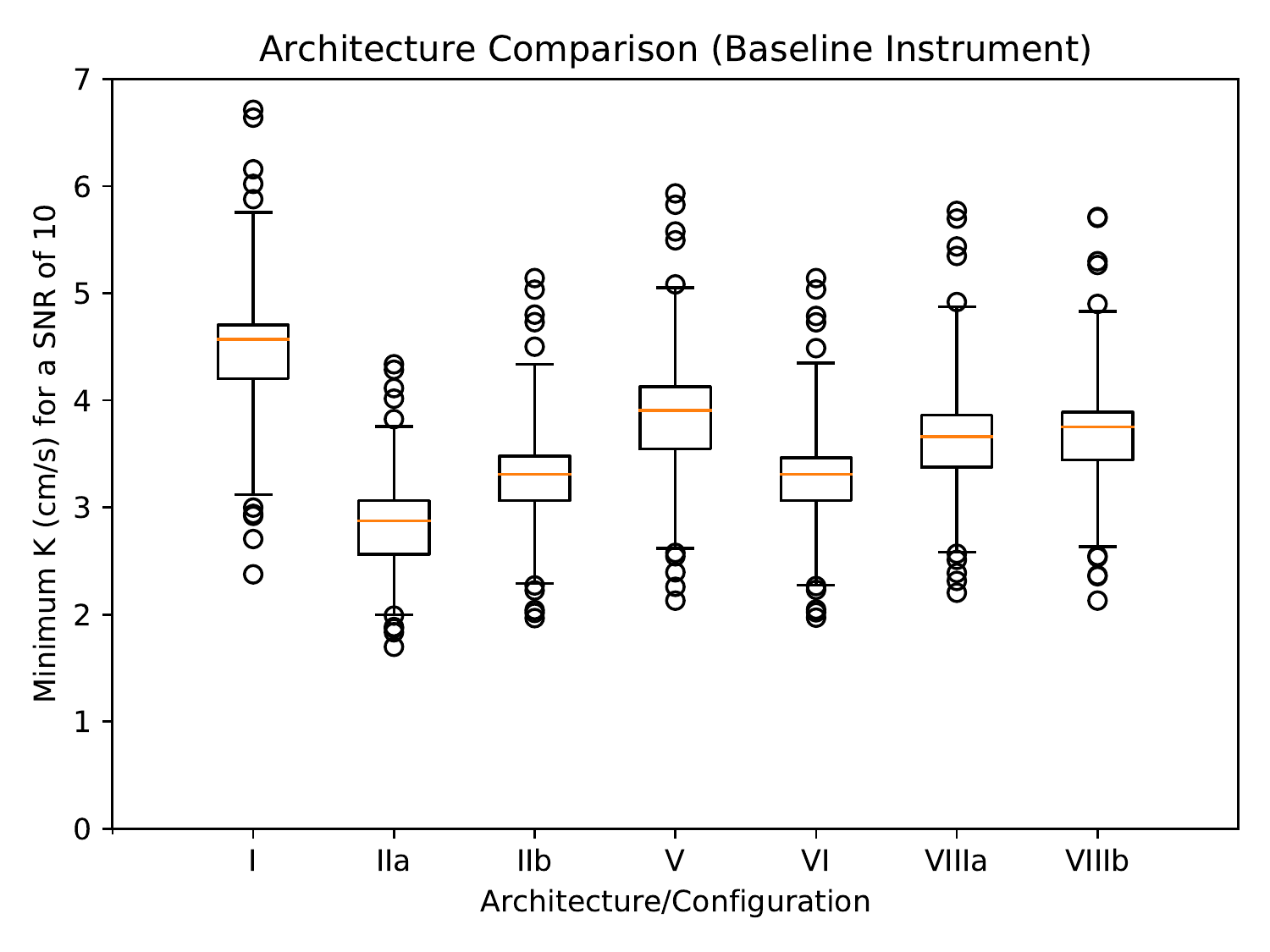}
\includegraphics[width=0.49\textwidth]{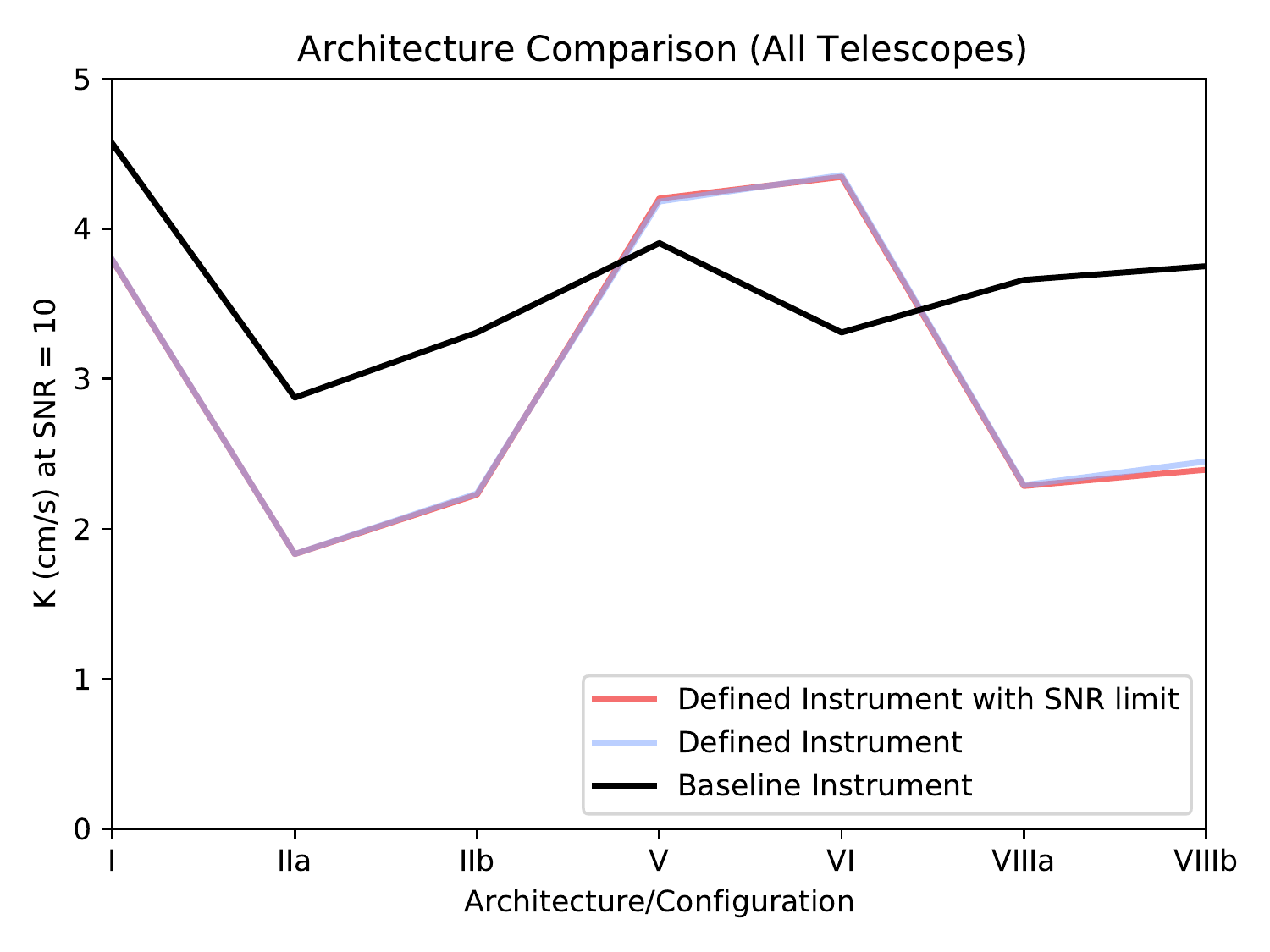}
\caption{Comparison across architecture simulations for minimum reflex velocity semi-amplitude sensitivity at a nominal SNR of 10, as inferred from equations \ref{eqn:SNR} and \ref{eqn:SNR2} for the number of observations per star and architecture single measurement precision. The box plots show the distribution of target stars under the different instrument configurations, while the line graph shows just the medians for all configurations, in the same layout as in Figure \ref{fig:Median_comparison}. Lower values are better, as they indicate sensitivity to smaller and/or more distant planets.}
\label{fig:KSNR}
\end{figure}

Second, we present a cross-architecture summary of the fraction of observable days during which each target star was observed; we use the same binning procedure as in section 3. This figure of merit represents the effective achieved cadences of each architecture given the size of the target list. Third, we present a cross-architecture summary of the maximum (photon noise) SNR for the detection of a $K$ = 0.1 m/s planet, and minimum (photon noise) detectable semi-amplitude for an SNR = 10 planet, per equations \ref{eqn:SNR} and \ref{eqn:SNR2}. This simple photon noise metric we have adopted represents a theoretical optimistic limit to what can be achieved in a real survey impacted by instrumental systematics, stellar activity, and other RV noise terms, but offers the most straightforward and direct means for assessing how the number of observations per star achieved for a given architecture maps to exoplanet sensitivity. The impacts of correlated noise on RV survey detection efficiency is the subject of follow-on work. Finally, in Figure \ref{fig:KSNR} we show the median detectable semi-amplitude $K$ and median SNR for a $K=0.1$ m/s for each architecture. These are directly anti-correlated and correlated with the achieved number of observations per star for a given architecture in Figure \ref{fig:Median_comparison}, as might be expected.

From these figures of merit, we can reach several conclusions. First, we find that the minimum SNR per resolution element requirements did not significantly increase exposure times relative to the required photon noise single measurement precision. In other words, they were approximately equivalent requirements. Second, whether or not an architecture achieves nightly cadence is determined largely by the collective telescope aperture, as might be expected. In the case of architecture VIIIb, the differences in specified surveys RV photon noise precision and time allocation of the 10-m telescopes also had significant effects. All architectures come close to or slightly exceed achieving an effectively nightly cadence for the specified target list, but that requires three telescope per hemisphere in order to do so; e.g. the typical cadence for a single telescope site within an architecture $\sim3$ days, given the assumptions, target list size, and this level of desired RV precision. Third, the differences in number of observations between sites in a given hemisphere correlates with number of clear nights, as might be expected. Future optimizations of a simulated RV survey could sculpt target lists and dynamically set priorities on a given telescope in a given hemisphere depending on the weather conditions at other sites.

\begin{figure}[ht]
\includegraphics[width=0.49\textwidth]{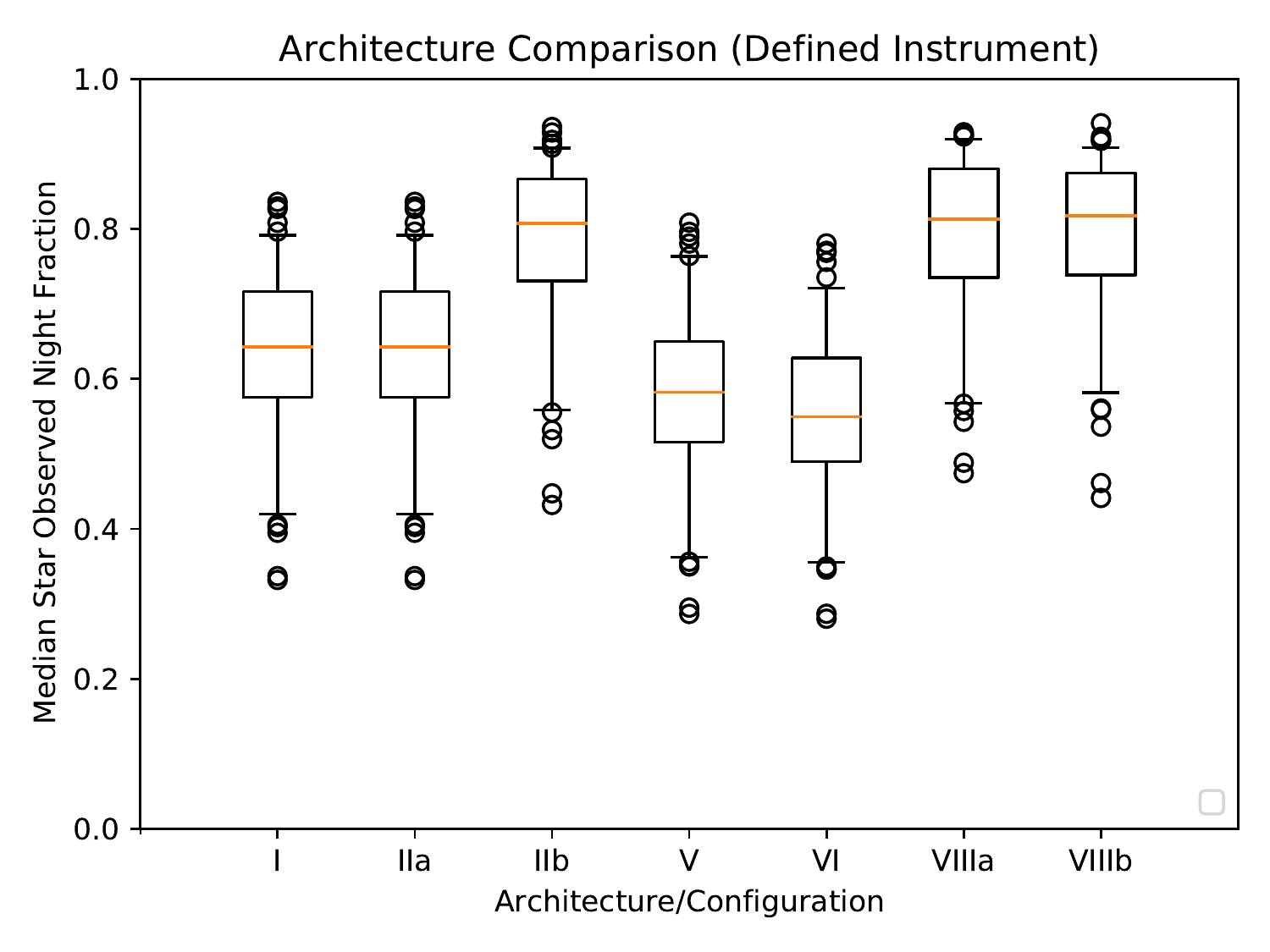}
\includegraphics[width=0.49\textwidth]{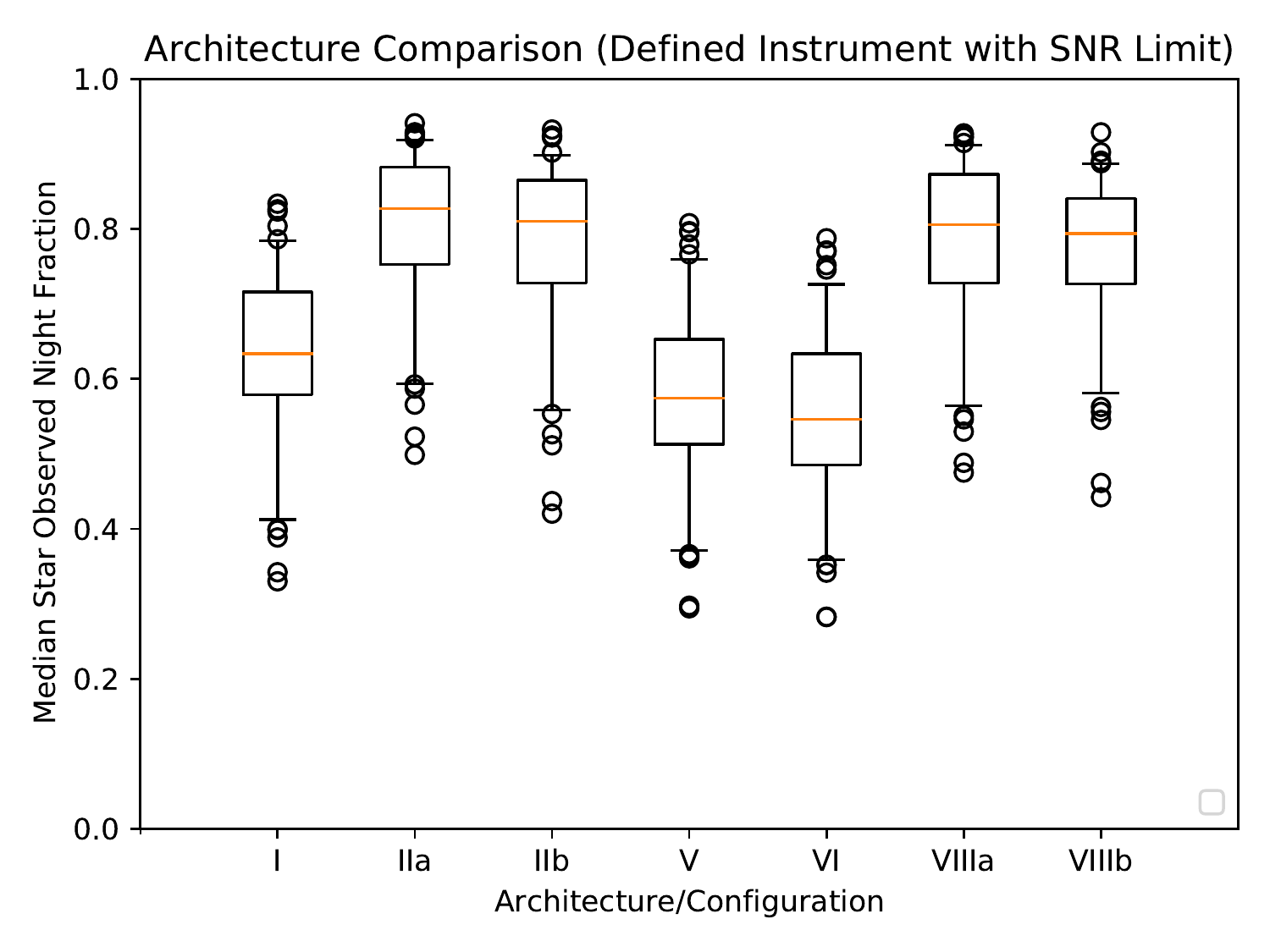}
\\
\includegraphics[width=0.49\textwidth]{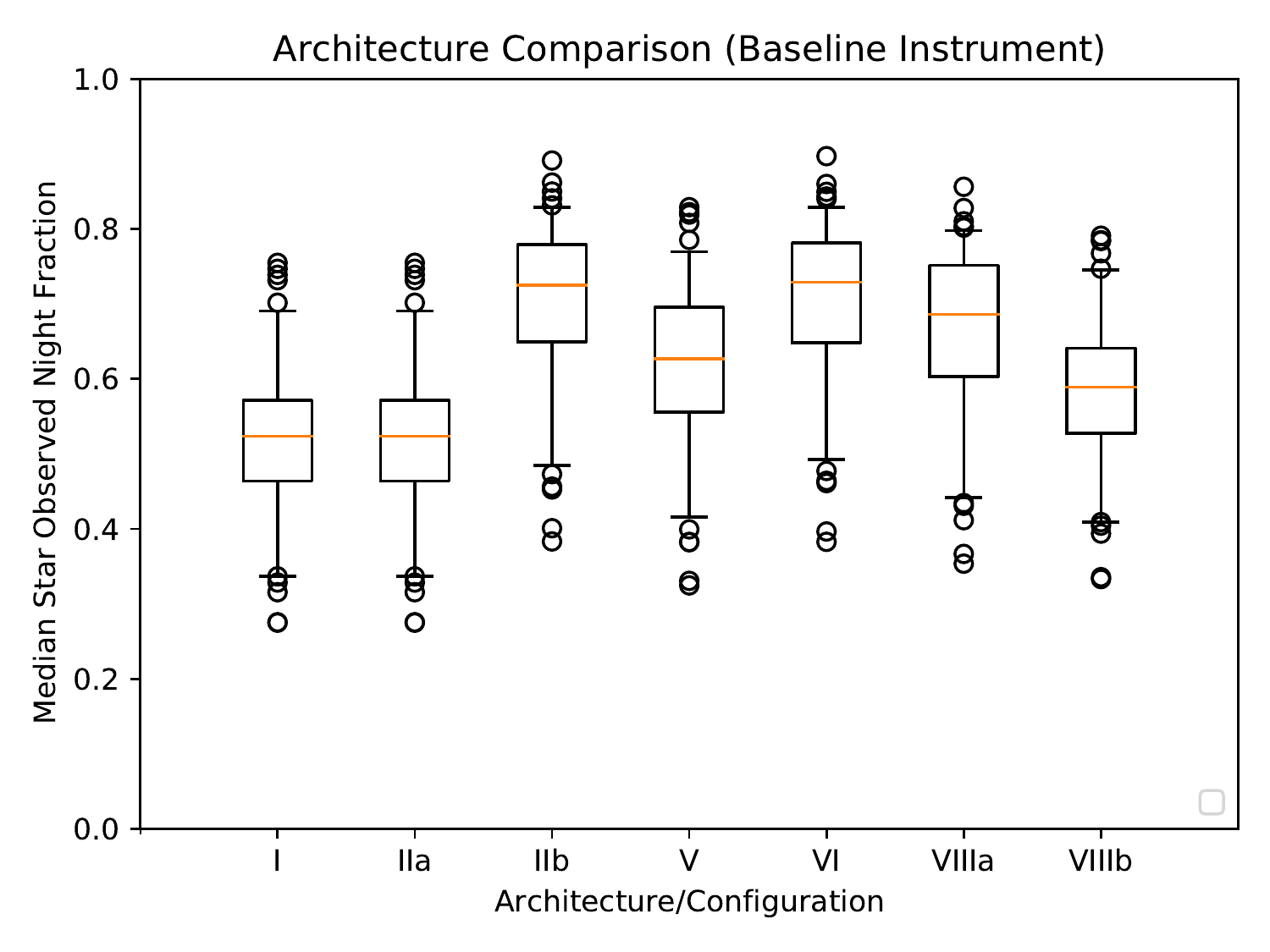}
\includegraphics[width=0.49\textwidth]{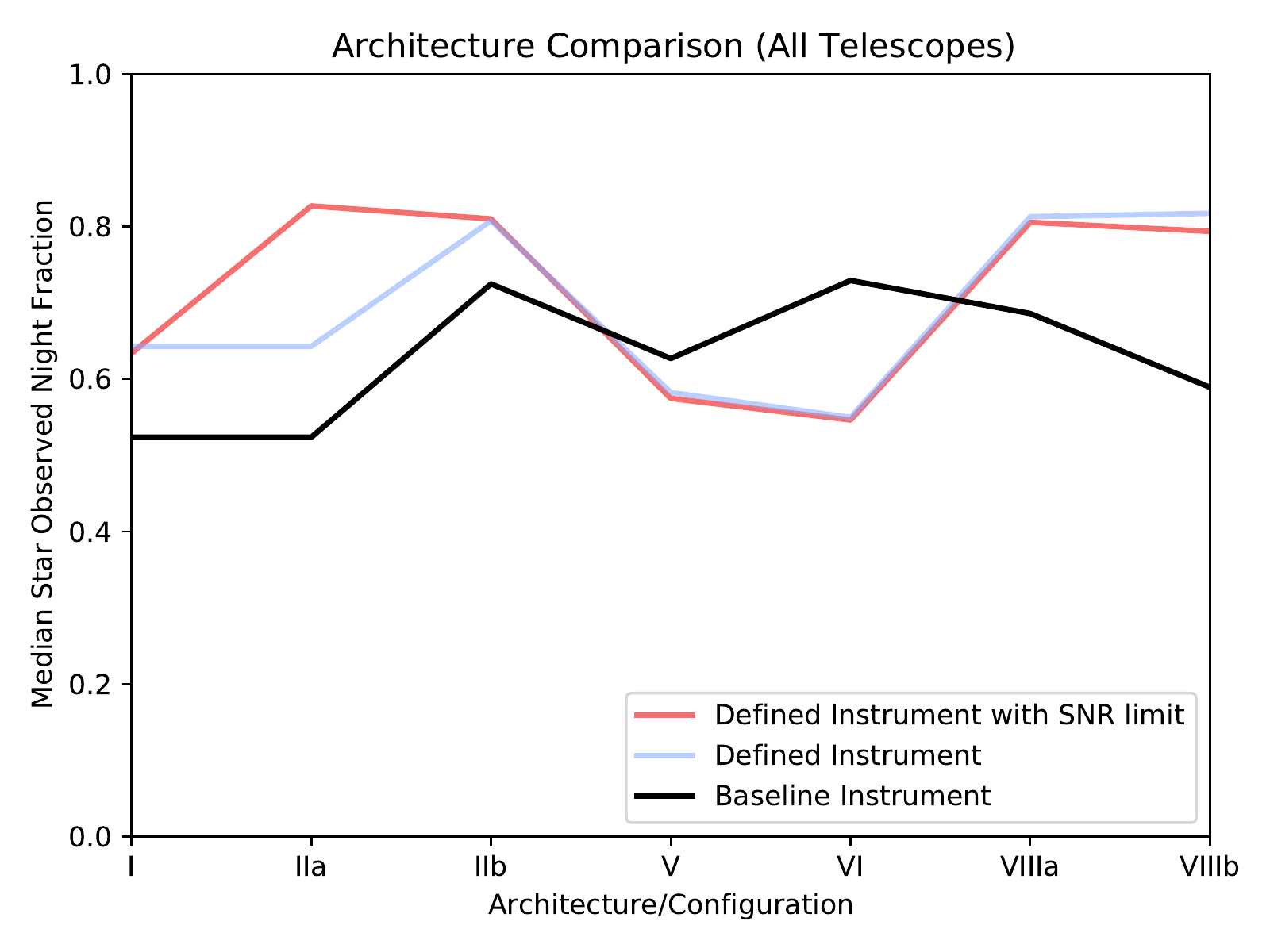}
\caption{Comparison across architecture simulations for the observable nights fraction: the number of nights in which a star is observed divided by the number of nights that star could be observed. A value of 1 means that for every night a star is observable by at least 1 telescope (e.g.:a telescope had time allocated, was not weathered out, and the star was above the pointing limits), at least 1 or more of the telescopes in the architecture does observe it in our simulations. If more than 1 telescope observes the same star on the same night, that star does not get double-counted, so the maximum possible fraction is 1.0. Given that each telescope site is modeled independently, we do not optimize this quantity (e.g., we do not prioritize a particular star at a given telescope higher if the other 2--5 telescopes are weathered out). The three box plots are for the distribution of observation fractions under different spectrograph instrument configurations as in Figure \ref{fig:Median_comparison}, while the line graph shows the medians for all. The color scheme is the same as Figure \ref{fig:Median_comparison} as well. The observable nights fraction is an approximate comparison of the survey efficiency for the given target list. No star in any architecture is observed for 100\% of possible nights.}
\label{fig:Frac_median_comp}
\end{figure}

Fourth, some stars are rarely observed for various reasons (right ascension, declination, and/or long exposure time), and they are least observed by telescopes with limited time allocation; e.g. the minimum outliers in the box-and-whiskers plots of Figures \ref{fig:Median_comparison} and \ref{fig:Frac_median_comp}. In this case, adding more telescopes increased the fraction of observable nights used. A more complex target prioritization scheme than the simple combination combination of hour angle and time since last observation metric used here will be necessary for some targets in order to achieve a more uniform number of observations per star. By comparison, architectures that were ``efficient'' with short exposure times were able to observe almost every available star every night. In that case, adding more sites increased the number of observations, but not the fraction of observable nights used.

Finally, an efficient RV survey will need to minimize target slew and acquisition time, particularly for larger aperture telescopes. We consider the following limiting survey case: if all of the exposure times are limited by p-mode oscillation time-scales and not photon noise or SNR exposure time requirements, then each star would have a five minute target dwell time (see ${\S}$\ref{sec:ExposureTimeCalculations}). Taking into consideration weather losses, slew and acquisition overhead, this limiting scenario would result in a survey with approximately $\sim$2500 observations on average per star per telescope. Thus, for a typical architecture with three telescopes per hemisphere, a p-mode oscillation time-scale limited exposure time survey would achieve $\sim$7500 observations per star. This number of observations is achieved for some of the most frequently observed targets in the architectures with $>$4-m apertures (IIa, IIb and VIIIa, VIIIb); for these telescopes and targets, the RV survey performance is limited by their minimum dwell times. However, we see that for the majority of the survey targets and architectures, even for the larger apertures, we are not limited by target dwell time, but rather the photon noise precision and SNR per resolution element exposure time requirements at this level of RV precision. This is distinctly different than the era of 1--3 m/s RV surveys, where the desired photon noise precision on large-aperture telescopes can be achieved within five minutes for most nearby, bright stars.

\begin{figure}[ht]
\includegraphics[width=0.49\textwidth]{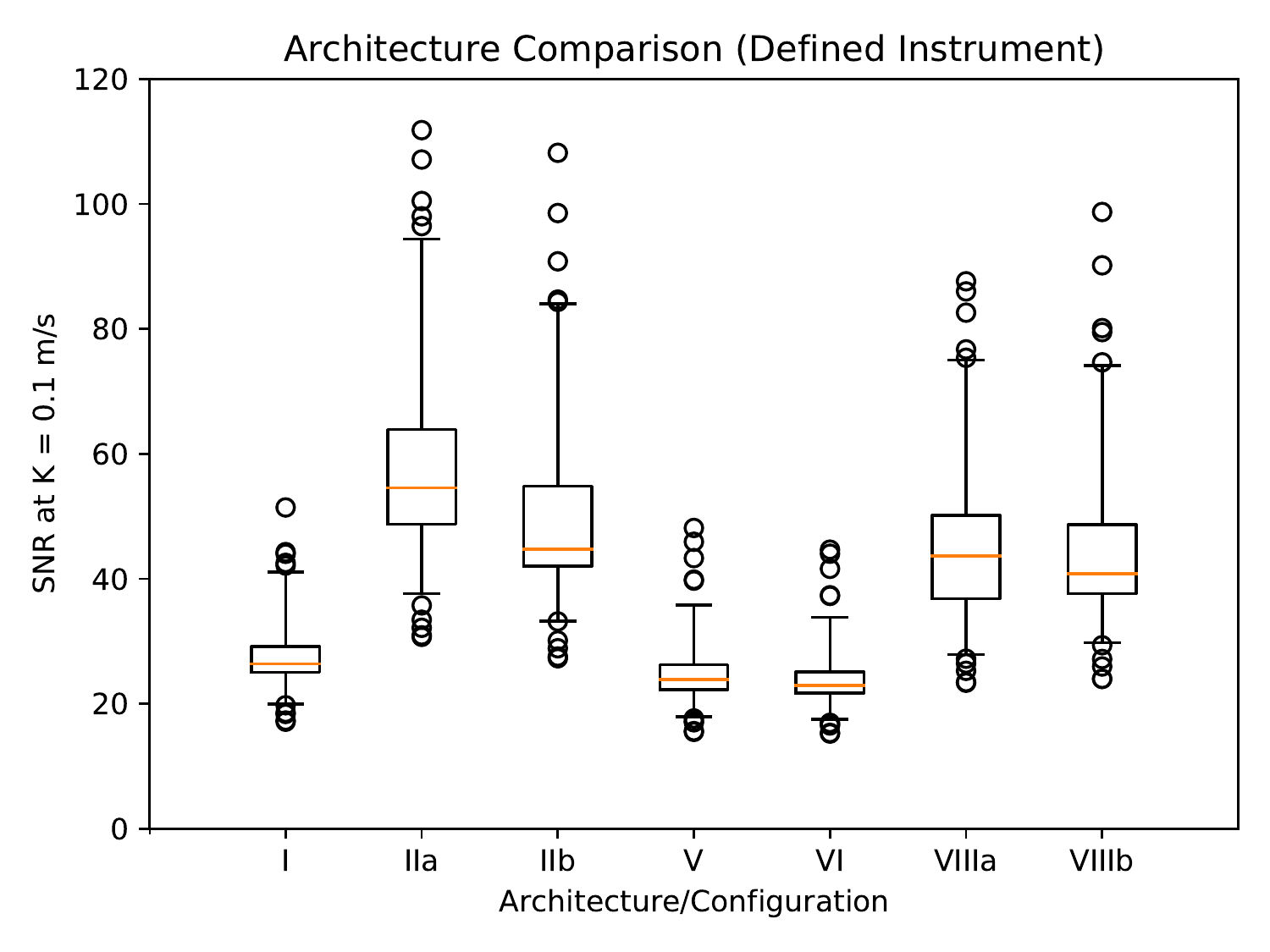}
\includegraphics[width=0.49\textwidth]{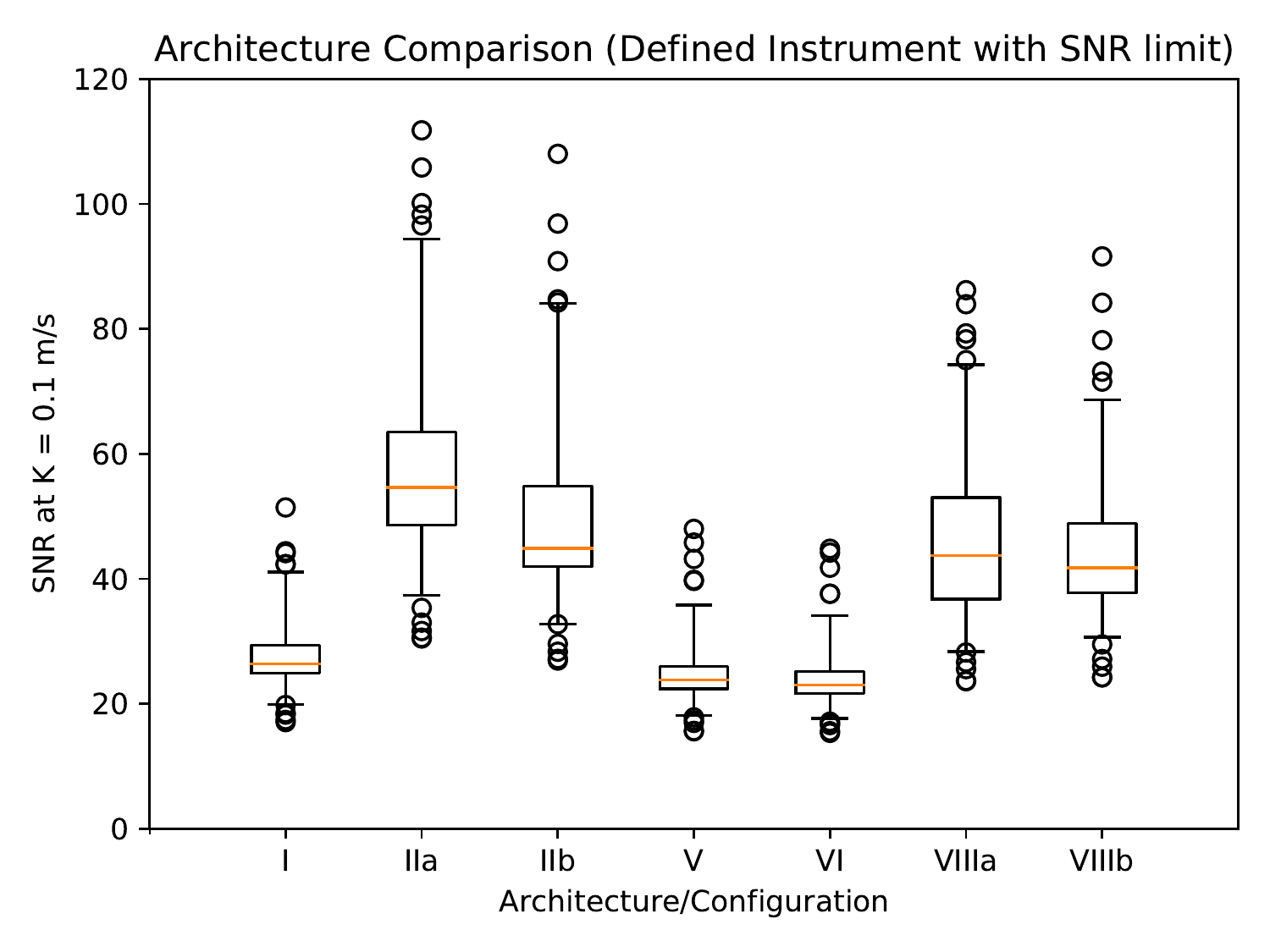}
\includegraphics[width=0.49\textwidth]{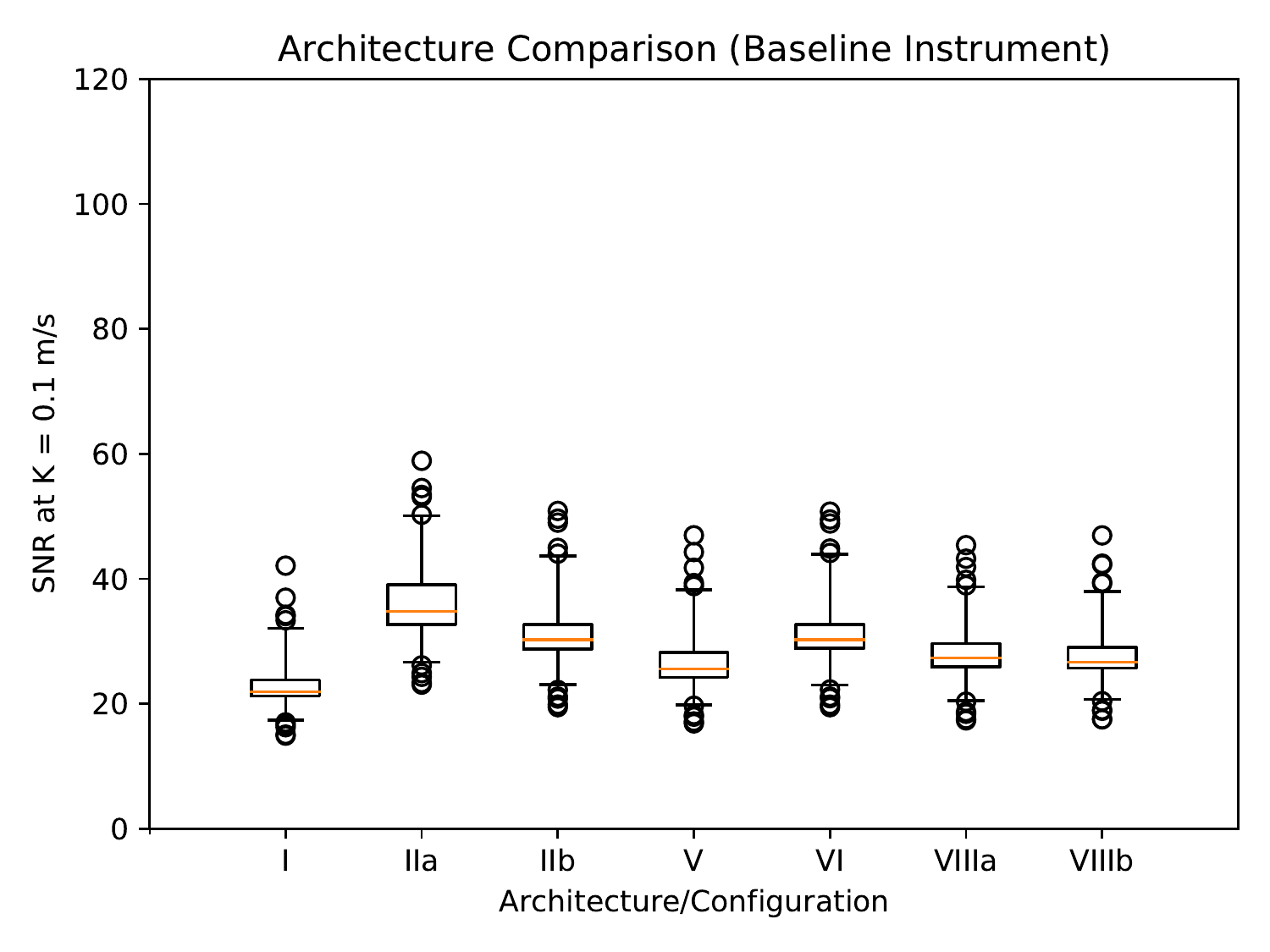}
\includegraphics[width=0.49\textwidth]{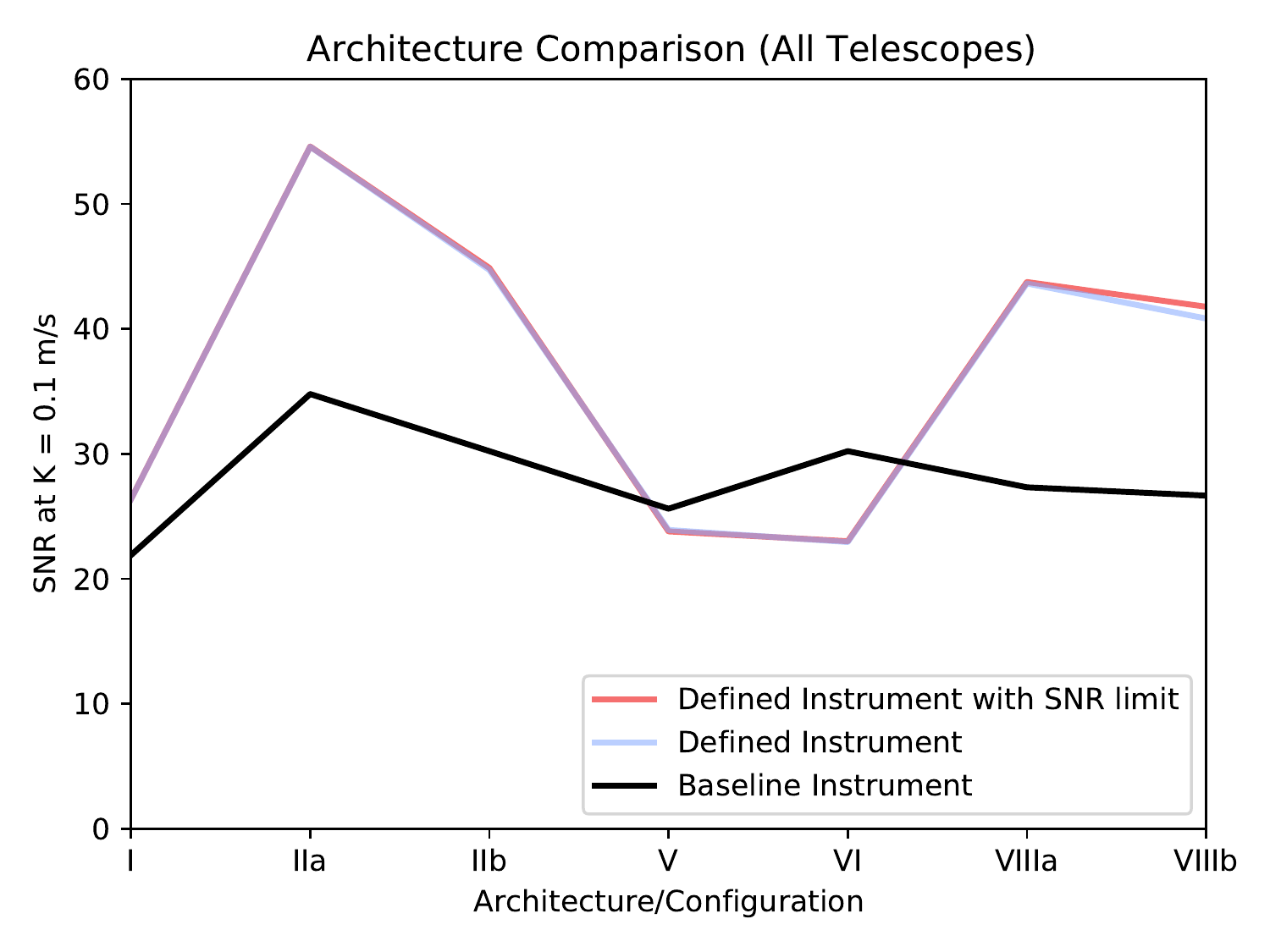}
\caption{Comparison across architecture simulations for the maximum signal to noise ratio (SNR) for a planet detection at a nominal reflex velocity semi-amplitude of 0.1 m/s, as inferred from equations \ref{eqn:SNR} and \ref{eqn:SNR2}. The box plots show the distribution of target stars under the different instrument configurations, while the line graph shows just the medians for all configurations, as in Figure \ref{fig:Median_comparison}. Higher values are better, as they indicate stronger detections of a planet at the nominal semi-amplitude.}
\label{fig:KSNR2}
\end{figure}
\FloatBarrier

\subsection{P-mode Oscillation timescale}
\label{sec:pmode}
In the above analysis, we did not take the p-mode oscillation time into account as a function of spectral type. To investigate the effect that varying p-mode oscillation times have on our results, we re-ran the simulations with the minimum exposure time set to 10 minutes. One would naively think that doubling the p-mode oscillation time would halve the number of observations per target. However, because of the 5-minute overhead time for slewing and target acquisition, a better naive assumption would be a decrease of 33\% when sufficient photons are available to reach the desired precision/SNR within the oscillation time.

Smaller telescope architectures see minimal impact on the number of observations per target, as they already have longer exposure times. For the larger architectures (6 and 10 m class), we observe a change of about 25\%. This is presumably because for some of our fainter targets we are already spending more time per target, and many of our stars having exposure times between 5 and 10 minutes.
Last, even with a 25\% change in the number of observations per target, this results in only a ~5-10\% effect in semi-amplitude, as this goes with the square root of the number of observations.
As a result, our results are relatively insensitive to choice of p-mode oscillation timescale to within a factor of about 10\%.

\begin{figure}[ht]
\includegraphics[width=0.49\textwidth]{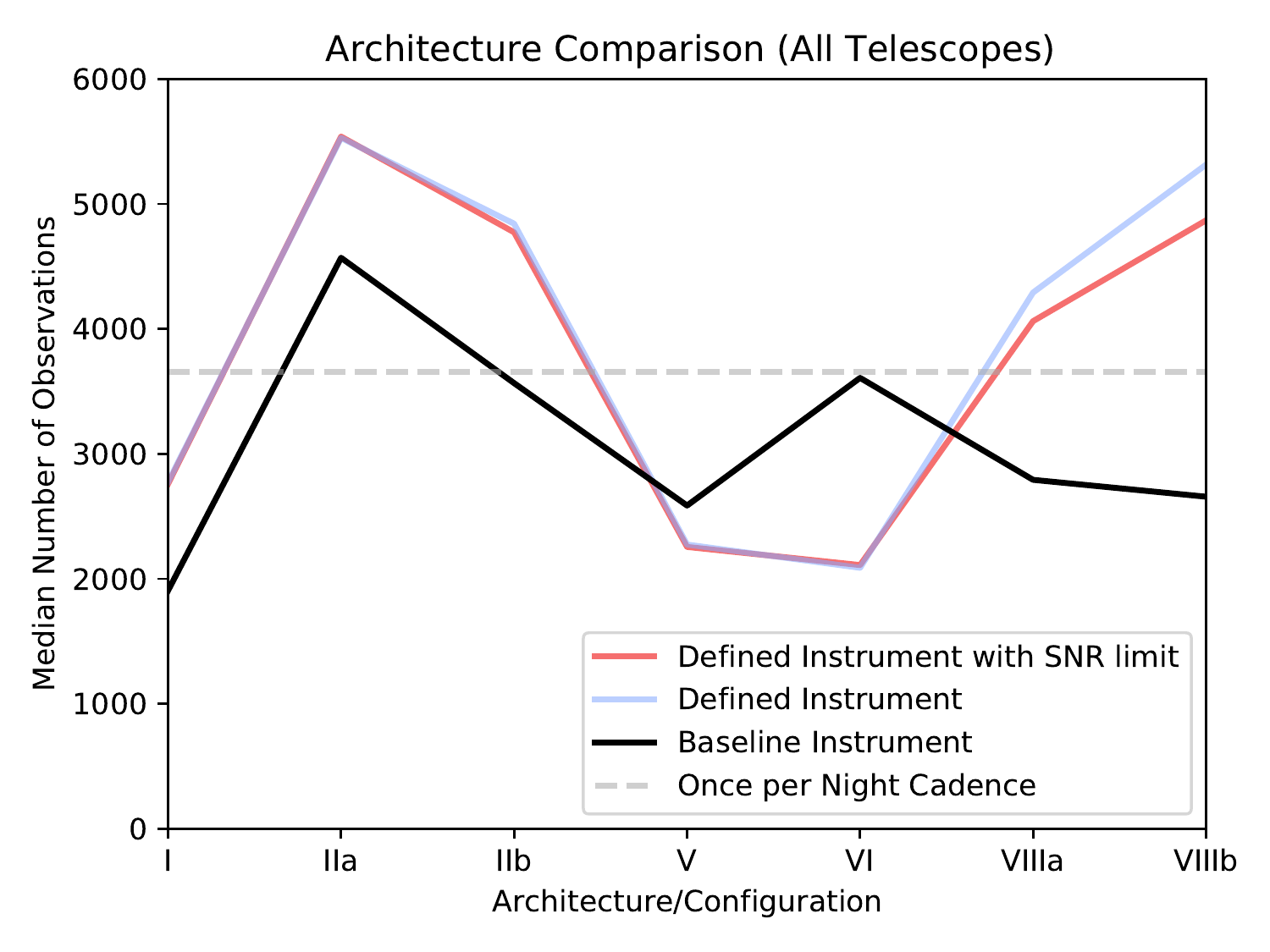}
\includegraphics[width=0.49\textwidth]{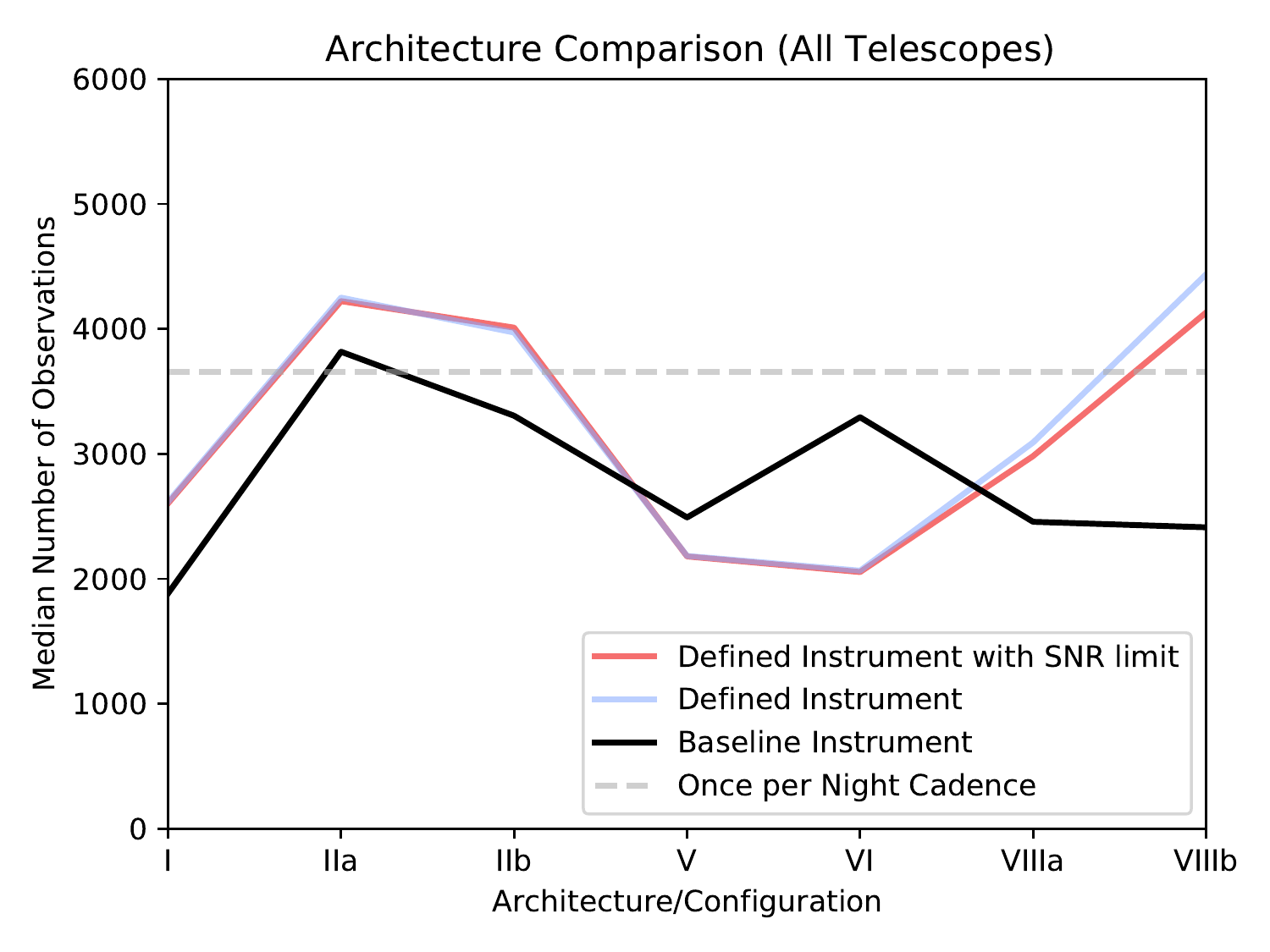}
\includegraphics[width=0.49\textwidth]{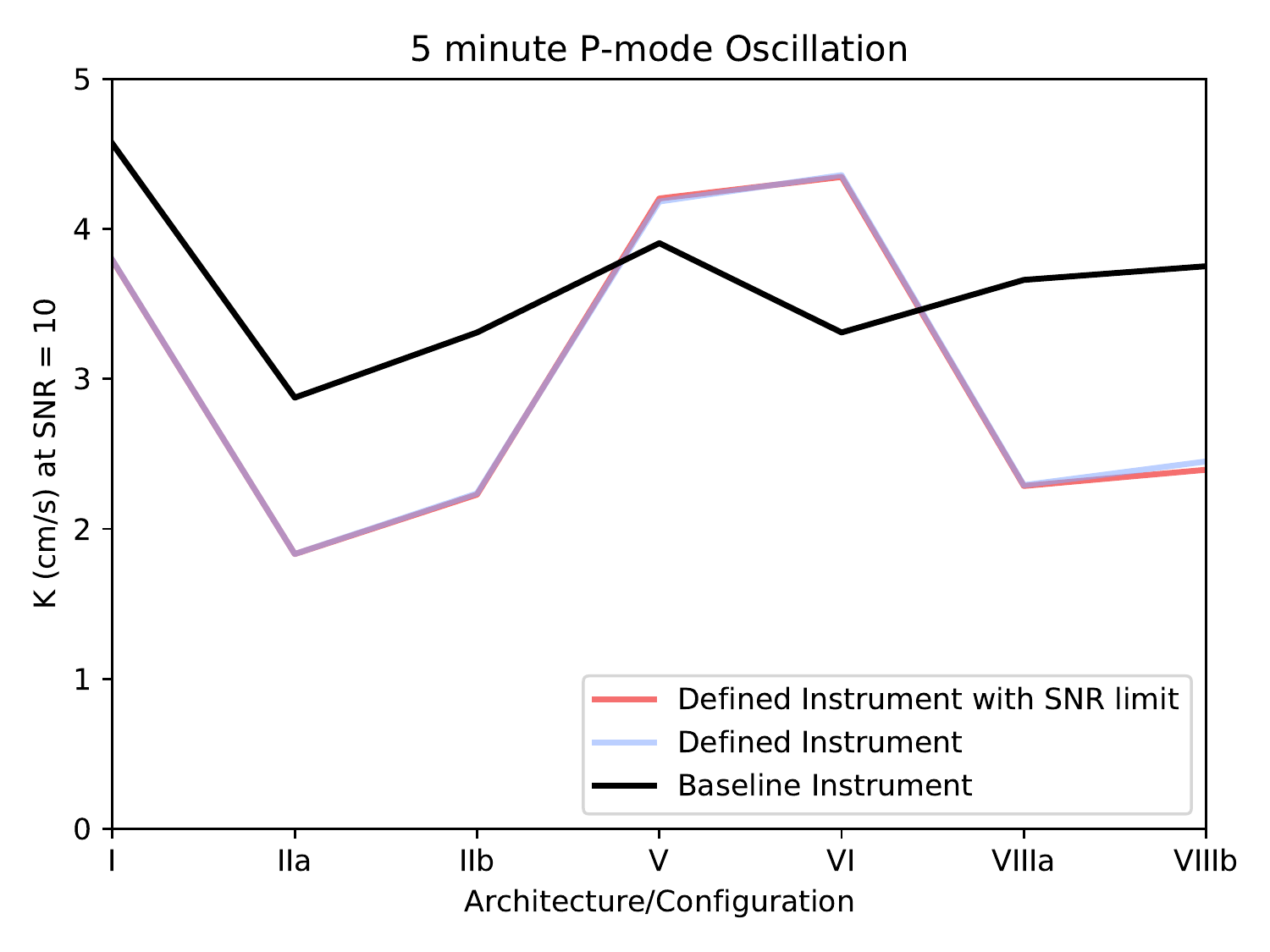}
\includegraphics[width=0.49\textwidth]{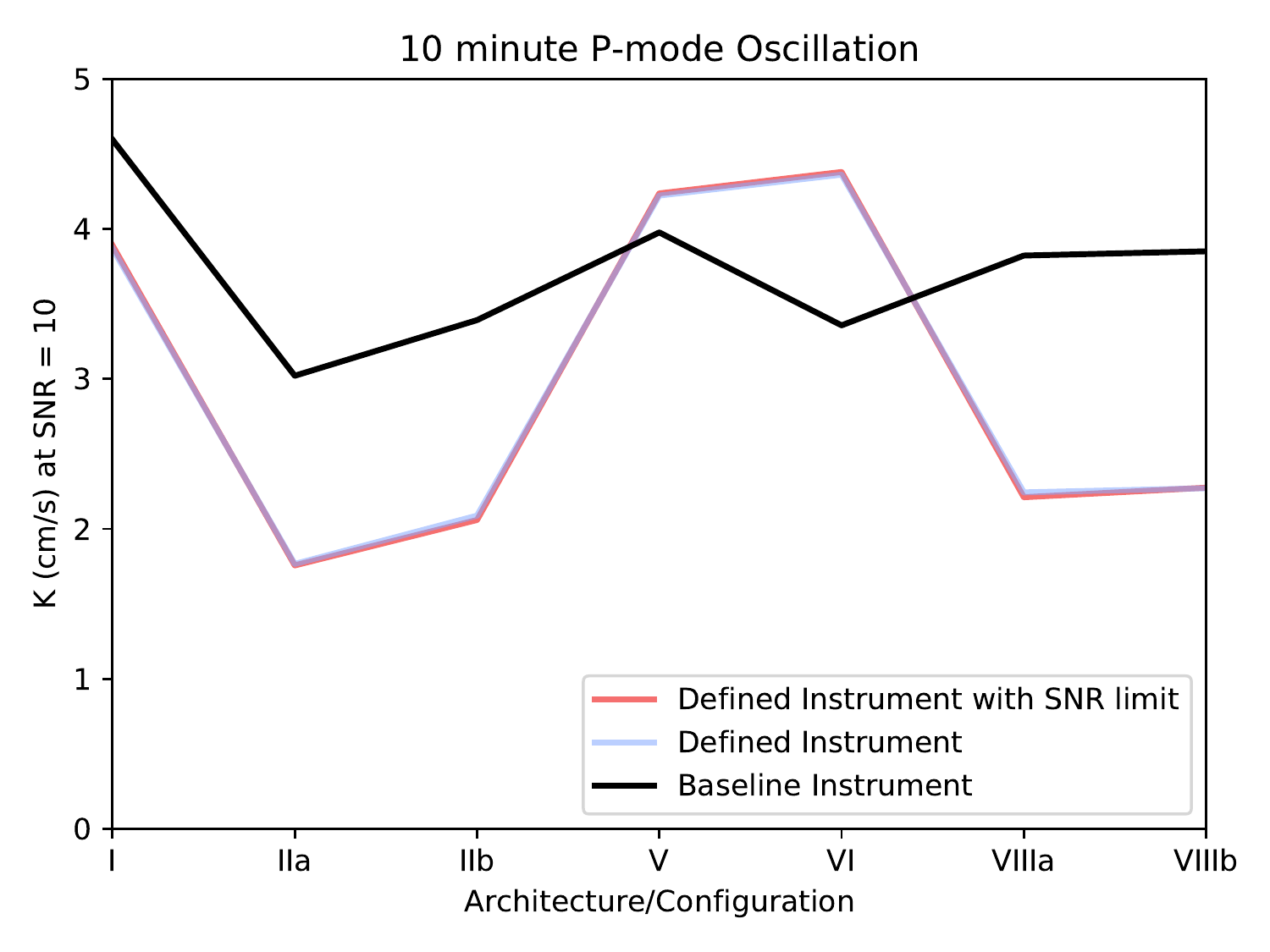}
\caption{Comparison across architecture simulations and P-mode oscillation assumptions for median number of observations (top), and minimum reflex velocity semi-amplitude sensitivity (bottom). The 5 minute (left) and 10 minute (right) graphs use the same vertical scale for ease of comparison.}
\end{figure}
\FloatBarrier

\section{Discussion} \label{sec:Discussion}

In this section, we discuss the relative performance of the architectures simulated herein, and we present an assessment in $\S$\ref{sec:Assumptions} of the impact of our survey simulation assumptions on our results. First, using the baseline instrument spectrograph for all architectures, architecture IIa has the largest median number of observations per star due to having the largest collecting area (on average). Architectures IIb and VI are close behind, with I, V, and VIII resulting in the fewest median number of observations per star. The sensitivity and SNR estimates follow from the number of observations per star; however the differences are more muted from these on account of the scaling with the square root of the number of observations. For the architecture-defined instruments, precision and wavelength coverage choices can have a large (about factor of 2) effect on number of observations per star. Including the instrument variations increase the spread in the number of observations per star and sensitivity estimates, with the architectures forming two distinct groups. Architectures IIa, IIb, VIIIa, and VIIIb are all in the high cadence/efficient/high sensitivity group, while architectures I, V, and VI are in the low sensitivity group for the reasons that vary between them -- telescope aperture, spectrograph efficiency, spectral grasp, etc. As might be expected, larger apertures can compensate for lower throughput and/or spectral grasp, and vice-versa.

Second, efficient (close to nightly cadence) observations are improved with additional telescopes at additional sites, up to a point. The peak achieved median survey cadence appears to be a bit above 80\% across the architectures considered, with all architecture achieving much lower cadences for some outlier stars. Presumably, a more sophisticated dispatch scheduler that coordinates the observing sequences across telescopes within an architecture could achieve a higher and more uniform cadence. This is an interesting topic for future study.

Third, all architectures considered were able to exceed the SNR and semi-amplitude sensitivities sufficient to find ``earth-mass planets in the habitable zone around sun-like stars'' for most of the targets in the survey in the optimistic photon noise limiting scenario. This is due to a combination of: some of the assumptions being optimistic (see section \ref{sec:Assumptions}), the relatively large number of telescopes compared to prior single-telescope RV surveys, most (and aside from architecture VIII, all) telescopes are assumed to be fully dedicated facilities, and the long survey duration of ten years. While several architectures failed to reach the equivalent of nightly cadence for most stars, even the down-scoped cases (less than 6 telescopes) had several hundred observations. Even though we did not consider the limiting impacts of stellar activity and correlated noise herein on the ability of these surveys to recovery Earth-mass analogs, all variants of all architectures achieved several thousand observations per star. As such, these architectures would gather sufficient data to enable the application of sophisticated stellar activity modeling and have sufficient excess sensitivity that the detection goals remain plausible.

\subsection{Survey simulation assumptions and their effects} \label{sec:Assumptions}

A number of simplifying assumptions were made for the survey simulations. In this sub-section, we categorize their effects as rendering these simulations optimistic, pessimistic, or neutral/ambiguous. We defer to a future work to quantify the impact of these assumptions.

Optimistic assumptions:
\begin{itemize}
\item Instrument design requirements can be achieved, including efficiency and instrumental noise as specified
\item There is no degradation in throughput over the decade-long survey due to, e.g., dust accumulation on the optics. \citep{2018SPIE10704E..01H}
\item Stellar activity can be mitigated completely, as no correlated noise models are included herein. This is investigated in \citep{Luhn2021}.
\item Tellurics can be perfectly corrected.
\item The atmospheric model does not consider line absorption. As such, it assumes that more light is available than in reality, as well as there being no noise from telluric absorption
\item Telescopes do not have significant downtime besides weather losses.
\item Weather losses continue at historical rates; climate change effects may be increasing clouds/haze at most sites. \citet{2020RAA....20...81C}
\item Weather conditions night-to-night are uncorrelated, which will significantly impact any injection and recovery tests which are not explored herein.
\item Time allocation is uniform; bright time gives additional limitations on when stars can be observed and introduce additional correlated sampling.
\item Our 10 degree moon separation is smaller than typical (up to 30 degrees during a full moon), though is directly taken from MINERVA.
\end{itemize}

Pessimistic assumptions:
\begin{itemize}
\item Observatories at different sites do not coordinate and optimize target lists and prioritization.
\item Telescope sizes, particularly for the architectures with the 4, 6, and 8 meter class observatories, are at the smaller end of each size class specified. For example, 3.5-m apertures are simulated for the architectures with 4-m class telescopes.
\item Pointing limits; many telescopes can go below the 2 airmass/30 degree above the horizon limit that we assumed, which can extend the observing season duration of targets and minimize annual gaps.
\item Site selection does not include any options in eastern Europe or Asia, and only existing sites were simulated.
\item Target Selection: While the target list is consistent between architectures and all stars on it are ``good'', no attempt to further optimize the list to increase cadence/sensitivity or number of stars at a fixed cadence/sensitivity was done.
\end{itemize}

Ambiguous Assumptions:
\begin{itemize}
\item Throughput in existing instruments is assumed to be constant as a function of wavelength. We assume a median throughput instead of peak throughput to account for this effect.
\item Constant spectrograph resolution. As with throughputs, we assume an intermediate value to reflect the overall system spectral resolution to account for this effect.
\item NEID is the assumed baseline specifications for future instruments, particular for the detectors and spectral grasp.
\item Site selection: a number of the chosen sites have alternatives far enough away to impact weather and declination effects, which were not optimized. Additionally, the large number and spread of sites mitigates the effects of any one site being good/bad.
\item Survey timing: A ten year survey with all dedicated instruments mitigates any effects from a specific site having an unusually good/bad year in terms of open-dome time.
\end{itemize}

\section{Conclusions} \label{sec:Conclusions}
We developed simulations of a set possible next-generation global-network RV surveys with realistic site parameters. We generated exposure times for a given star, from its temperature, radius, distance, metallicity, surface gravity, and rotational velocity, as well as the properties of the telescope and instrument measuring it. 
We gathered the parameters for a nominal target list of nearby, bright stars common to future direct imaging mission concepts, several existing and potential observatory sites, and a set of telescope/instrument combinations.
Through the results of these simulations, we estimated the distributions of observations per star, target observation frequency as a function of available nights, and approximate estimates of planet detection sensitivity. As expected, we find that the achieved cadences generally scaled with the collective telescope aperture. We find that most architectures considered herein can achieve the theoretical minimum (optimistic bound) number of observations required to detect at SNR=10 an Earth-mass analog producing a 9 cm/s stellar reflex velocity for all targets, with some margin to spare particularly for architectures II and VIII. Finally, we considered the major assumptions of the survey and how they could affect our results, which along with cost considerations can be explored in future work.

\acknowledgments
Much of the work presented herein resulted from the EPRV Working Group established by the NSF and NASA in 2018-2020. Part of this research was carried out at the Jet Propulsion Laboratory, California Institute of Technology, under a contract with the National Aeronautics and Space Administration (80NM0018D0004). PN and PP would like to acknowledge support from the JPL Exoplanet Exploration Program and NSF AAG 1716202 for this work. BSG was supported by the Thomas Jefferson Chair Endowment for Discovery and Space Exploration.

\software{Astropy \citep{astropy:2013, astropy:2018}, Matplotlib, Numpy} 

\newpage
\bibliography{references}

\FloatBarrier
\appendix
\section{Architecture Results} \label{sec:appendix}

In this Appendix, we present detailed architecture survey simulation results for the other architectures not presented in the main text.

\begin{figure}
\noindent \includegraphics[width=0.49\textwidth]{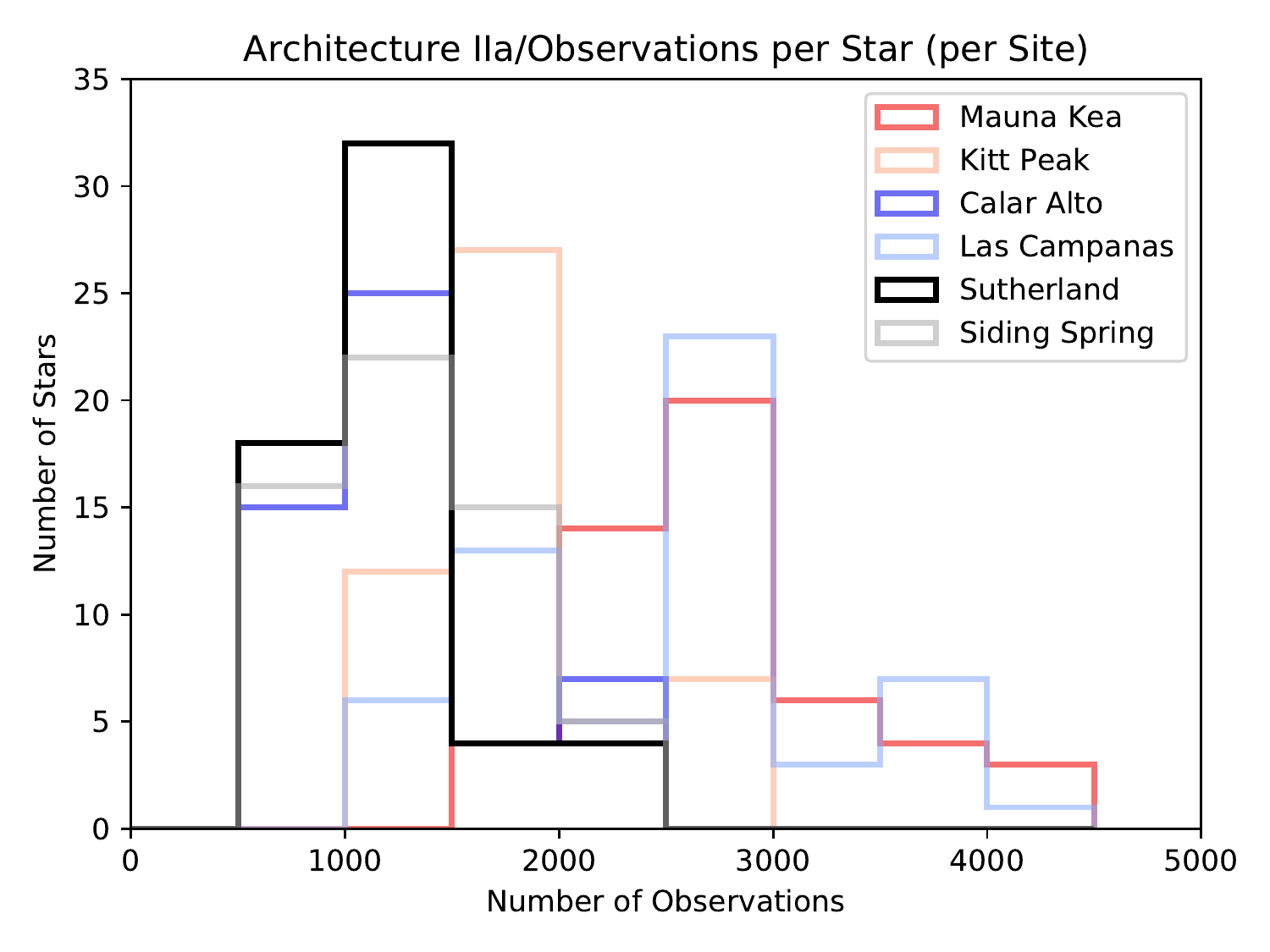}
\includegraphics[width=0.49\textwidth]{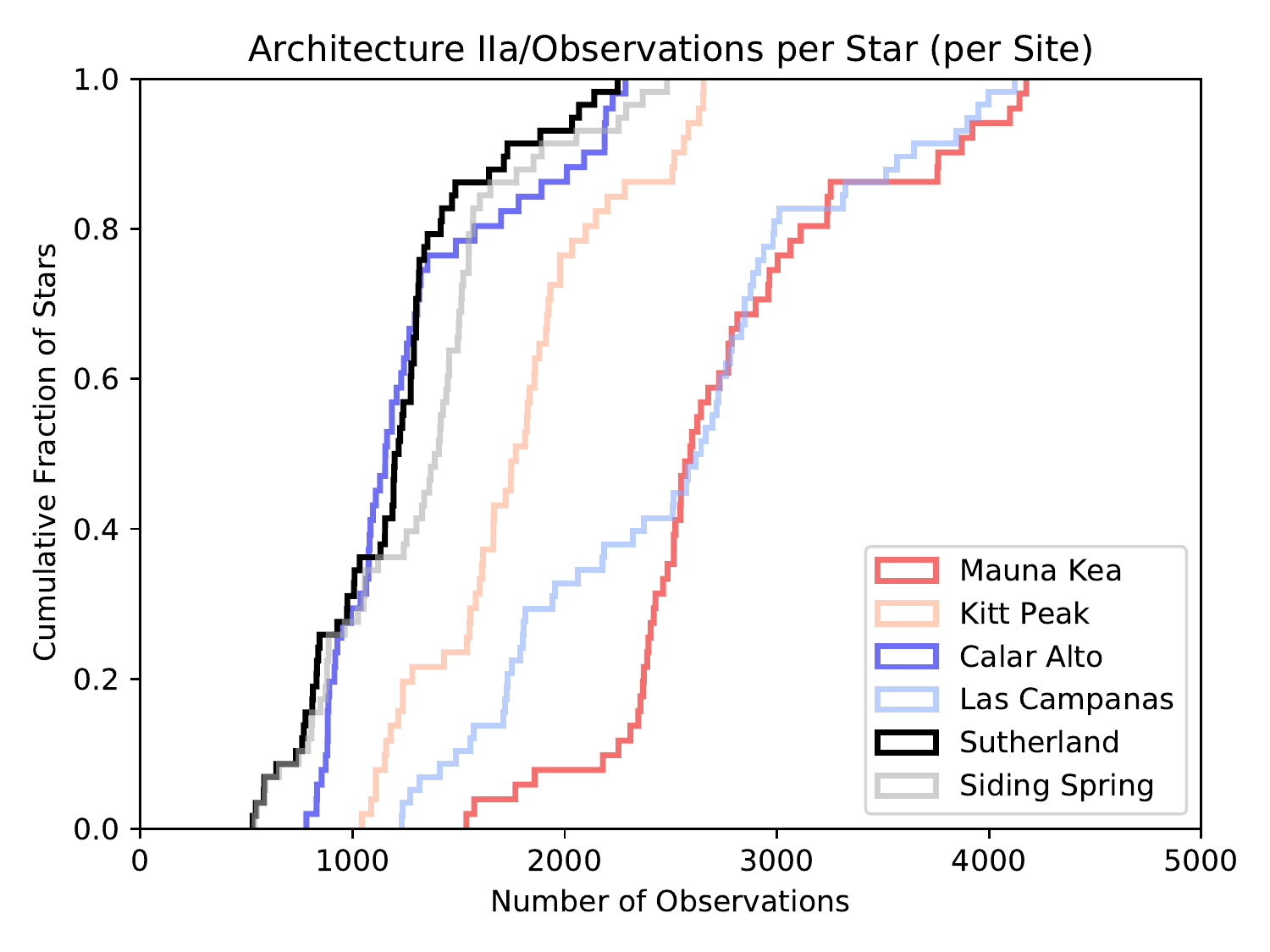}

\noindent \includegraphics[width=0.49\textwidth]{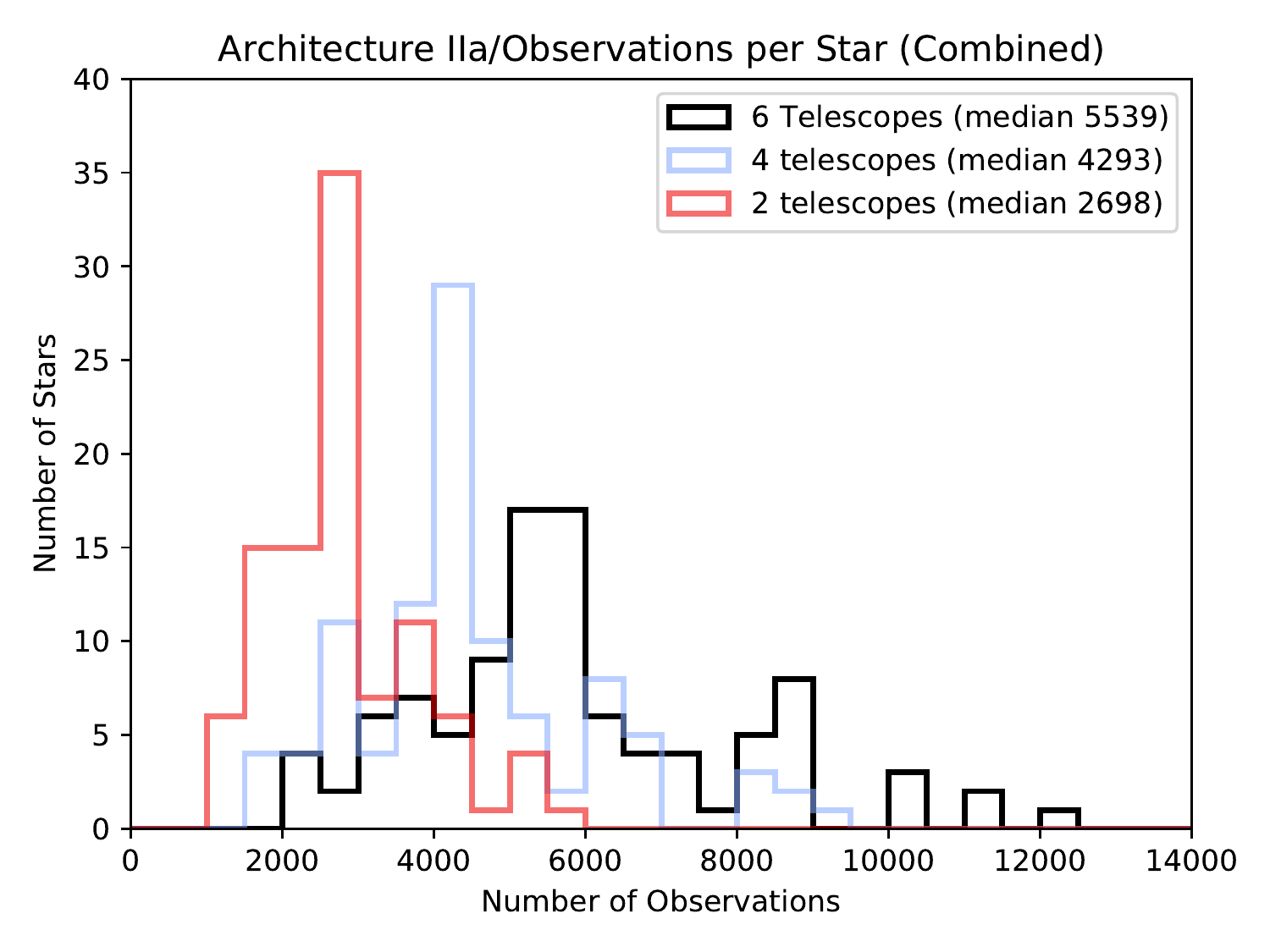}
\includegraphics[width=0.49\textwidth]{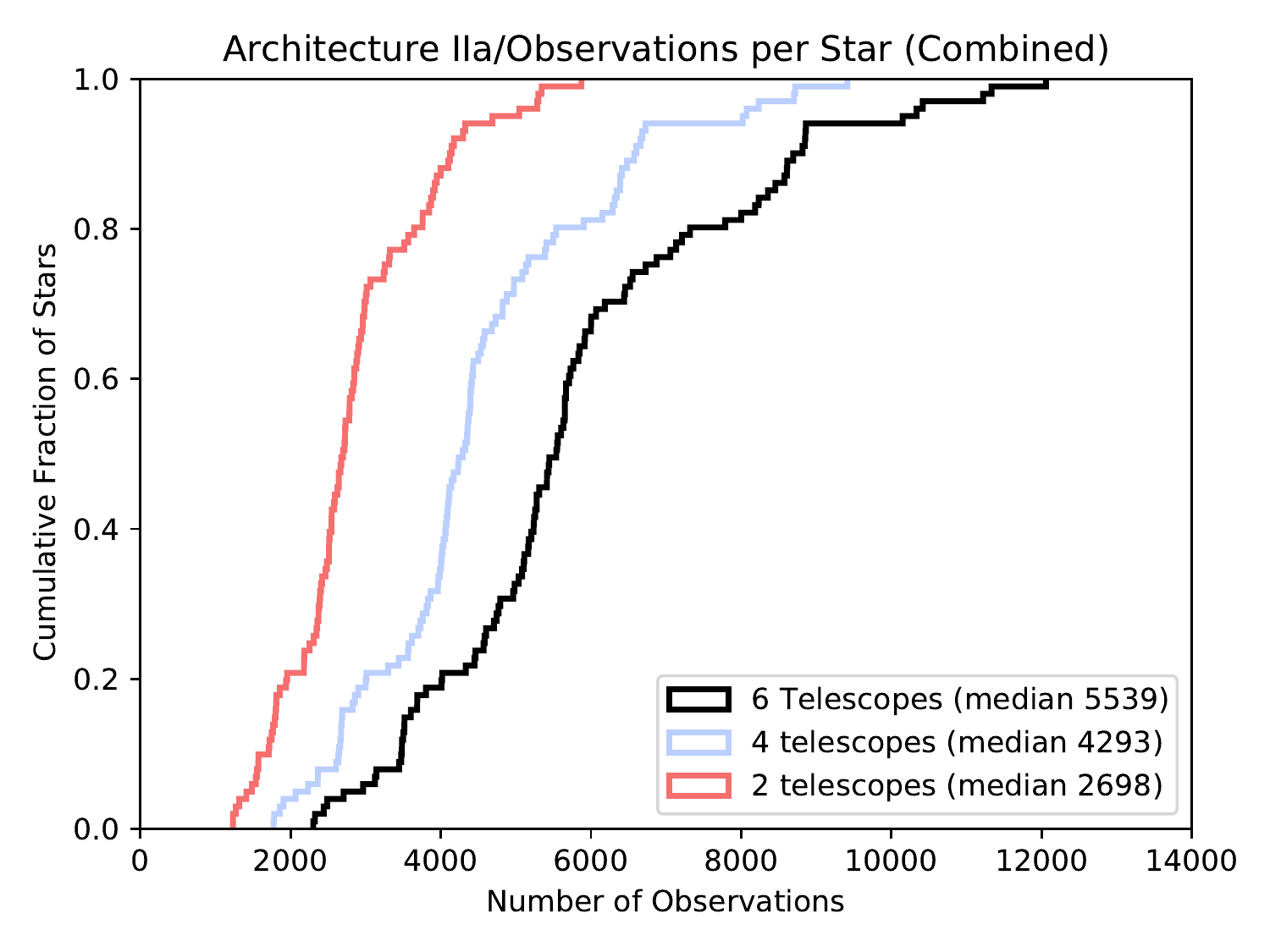}
\caption{Same as figure \ref{fig:ArchIobs}, but for architecture IIa.}
\end{figure}

\begin{figure}
\noindent \includegraphics[width=0.49\textwidth]{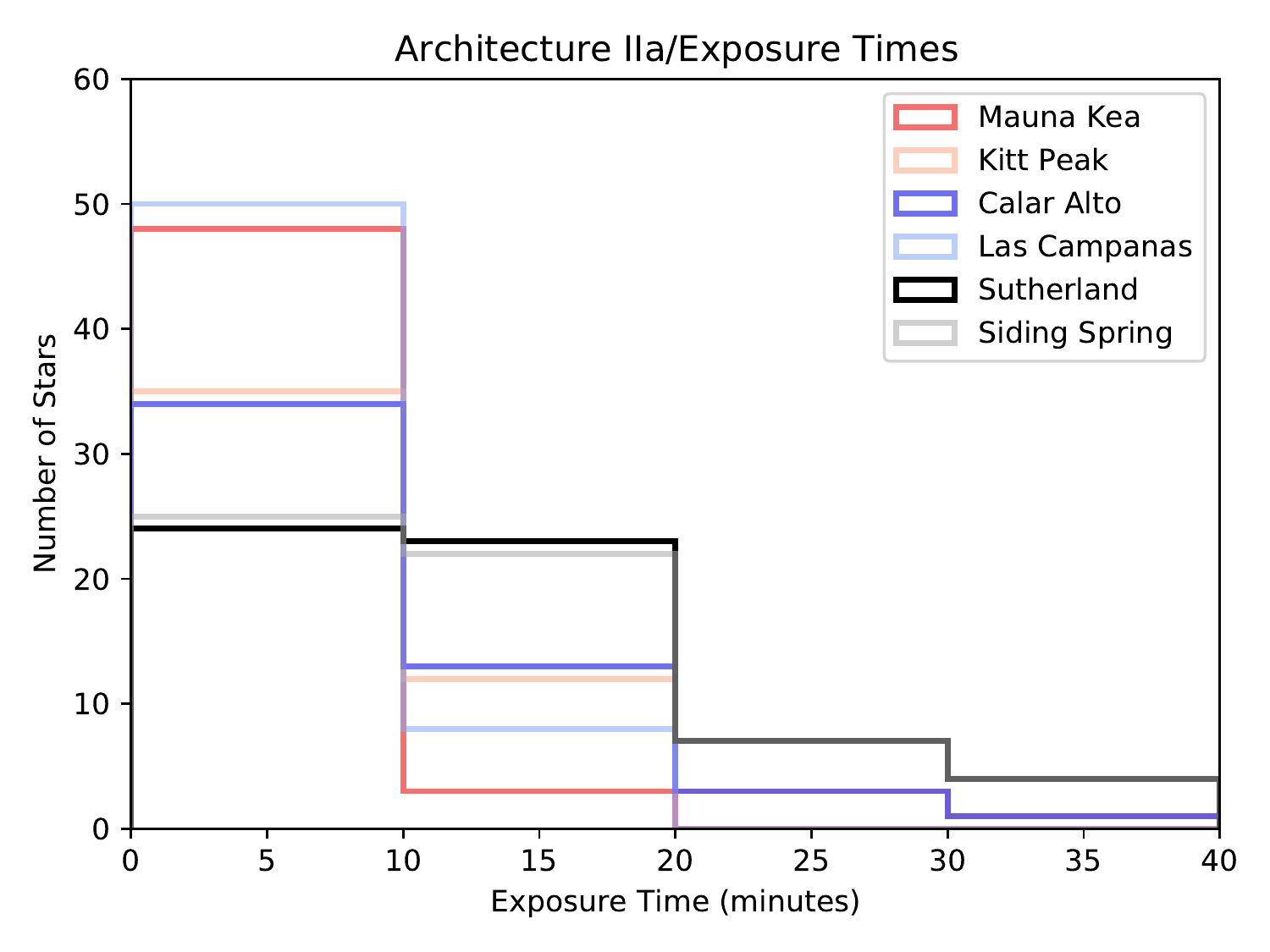}
\includegraphics[width=0.49\textwidth]{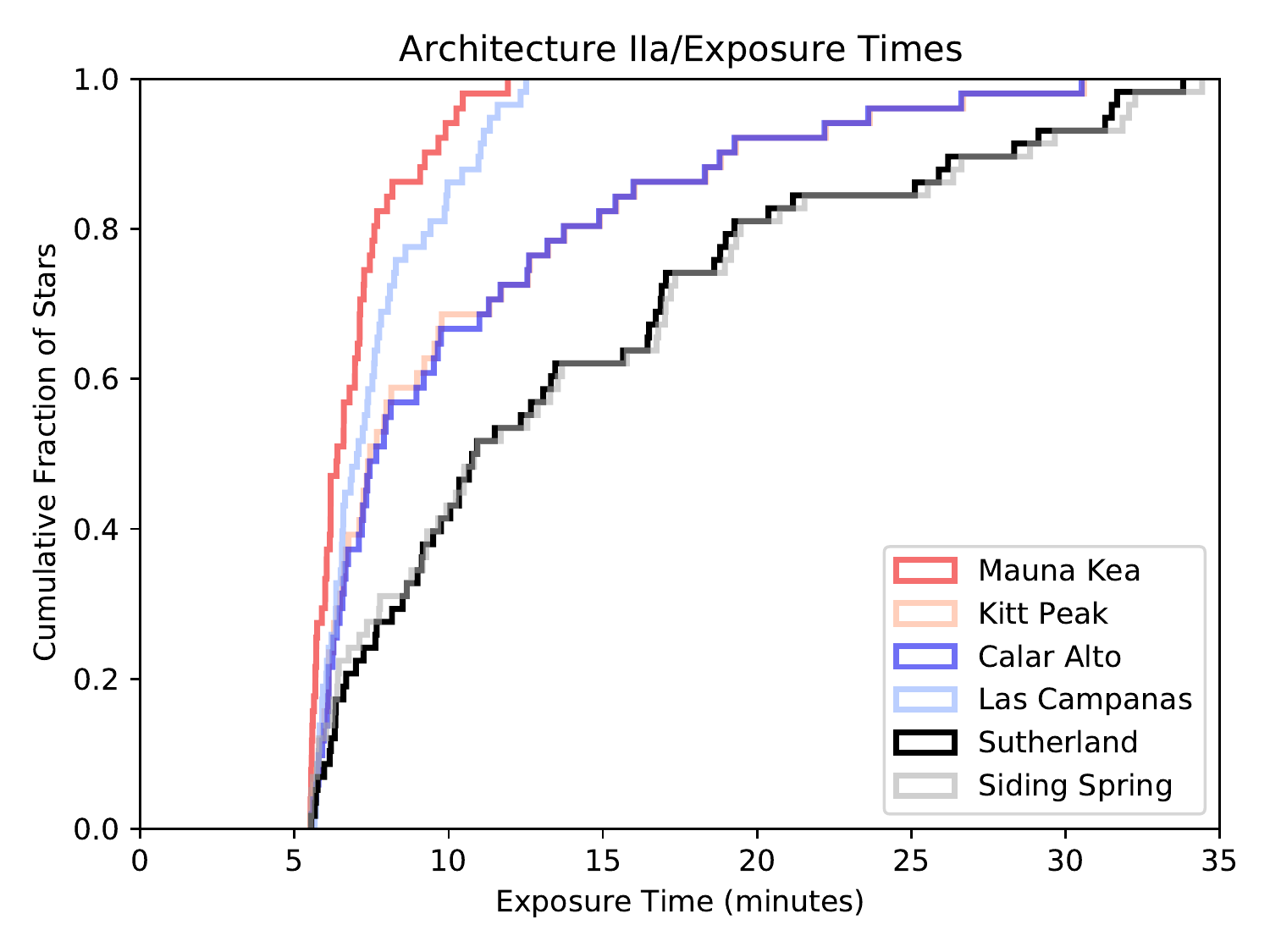}

\noindent \includegraphics[width=0.49\textwidth]{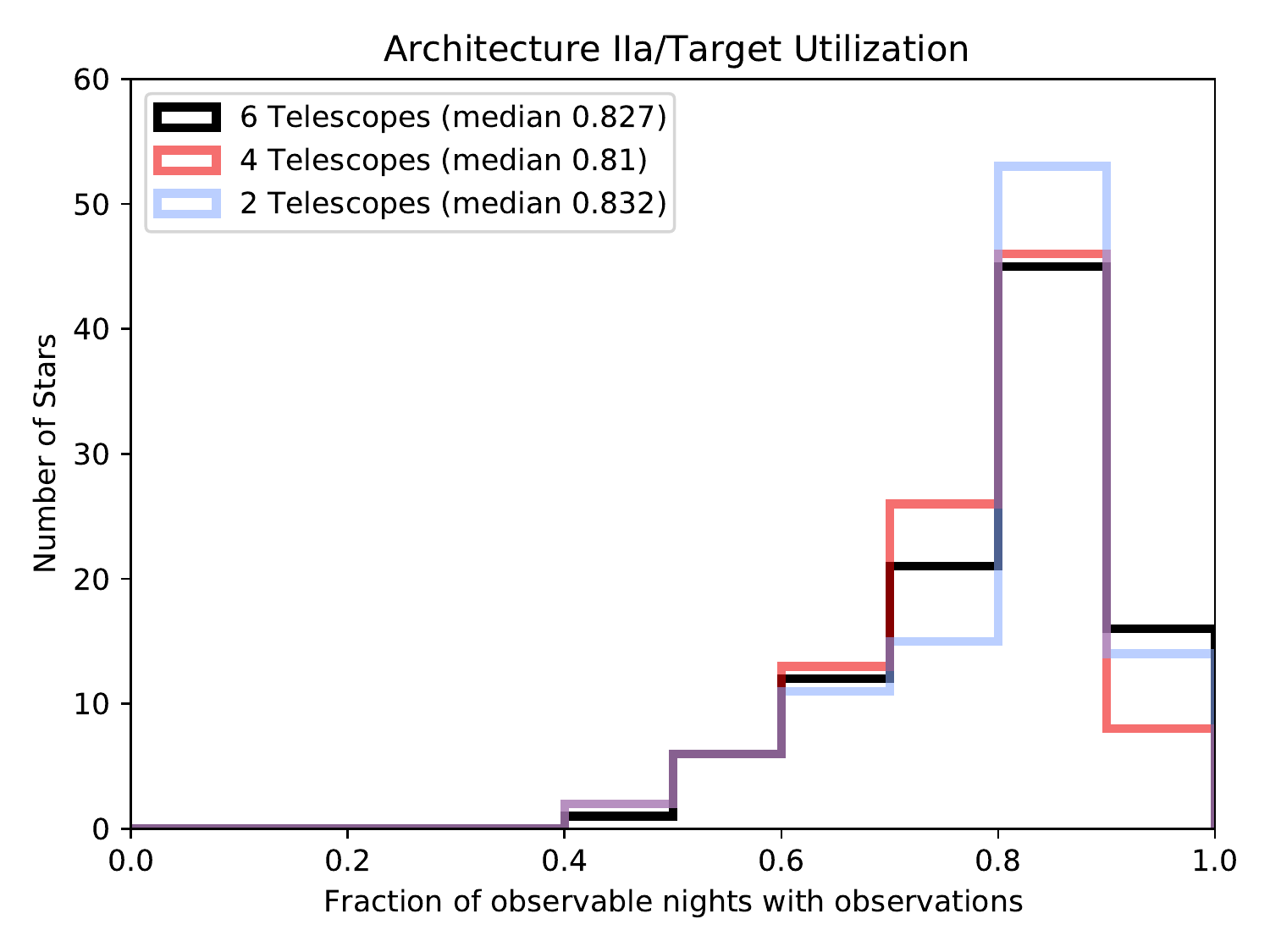}
\includegraphics[width=0.49\textwidth]{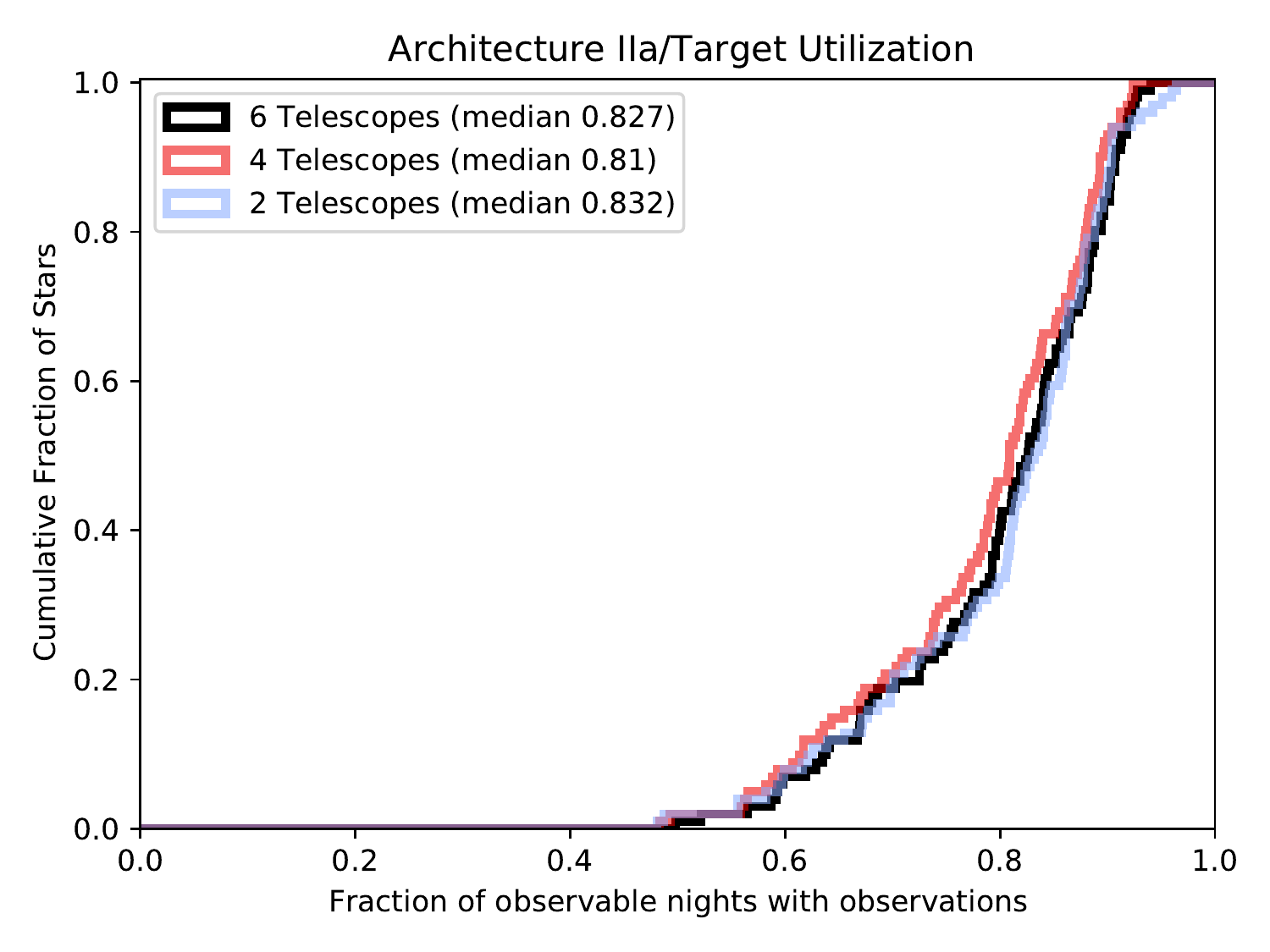}
\caption{Same as figure \ref{fig:ArchIexpfrac}, but for architecture IIa.}
\end{figure}

\begin{figure}
\noindent \includegraphics[width=0.49\textwidth]{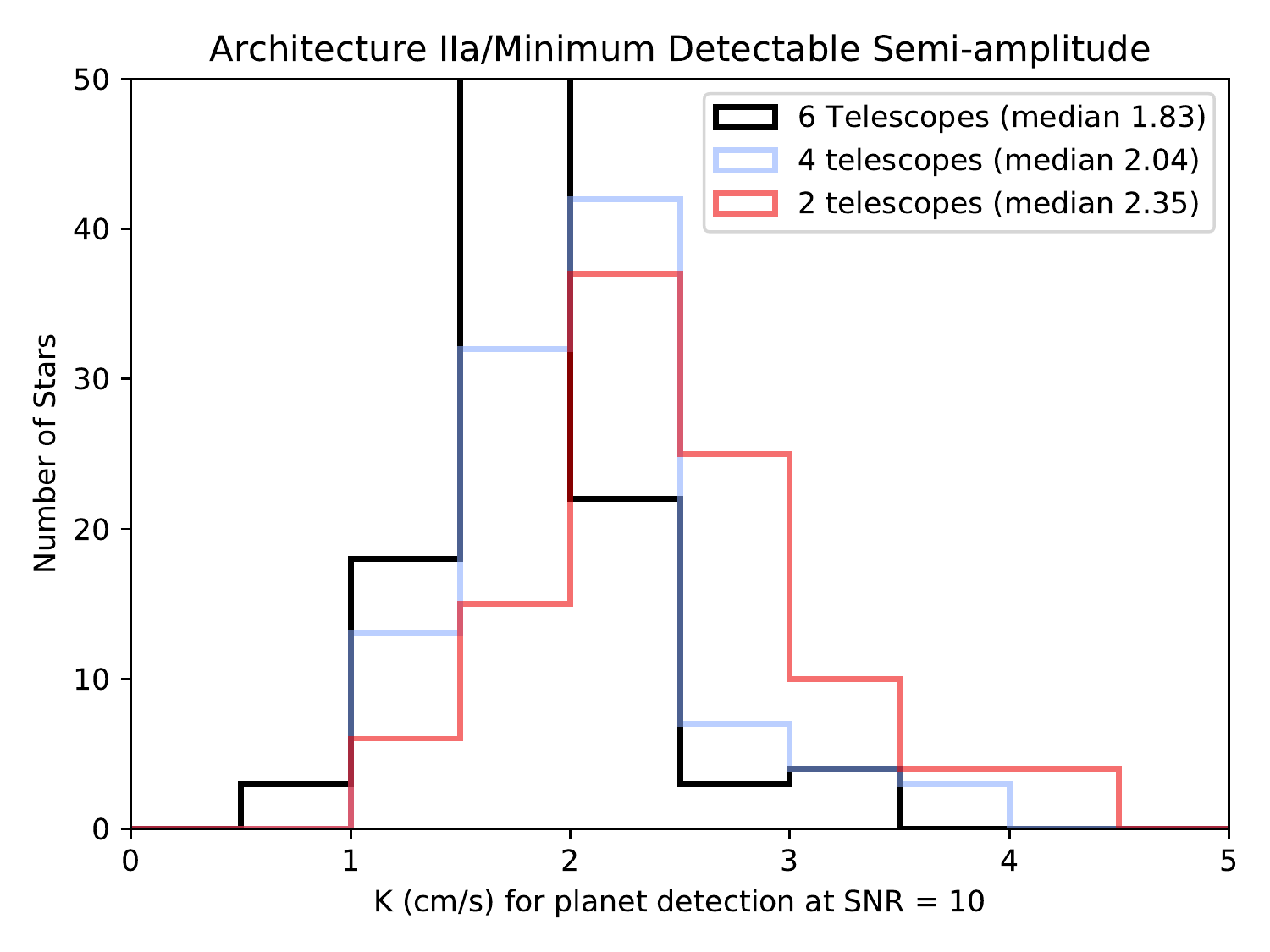}
\includegraphics[width=0.49\textwidth]{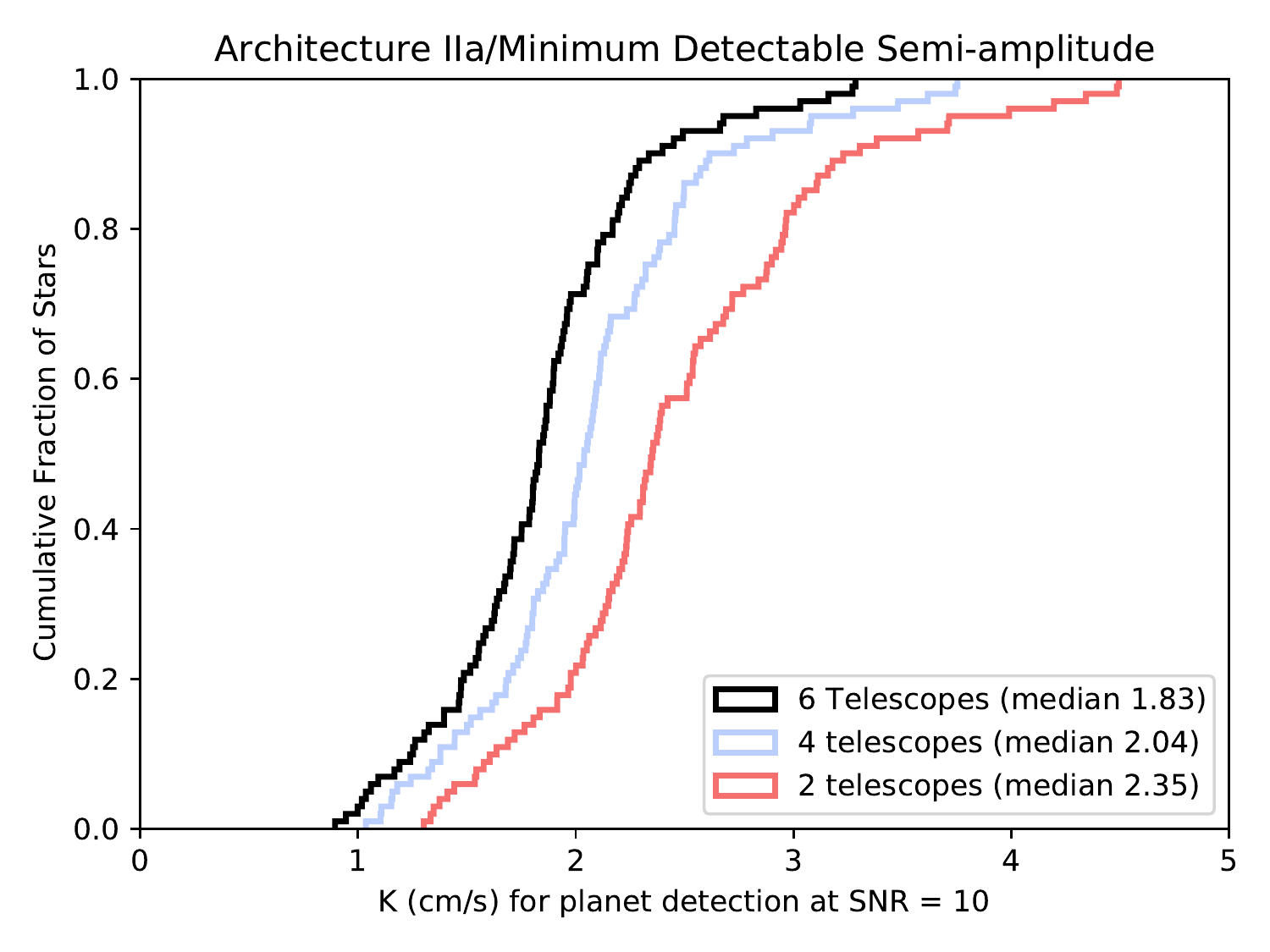}

\noindent \includegraphics[width=0.49\textwidth]{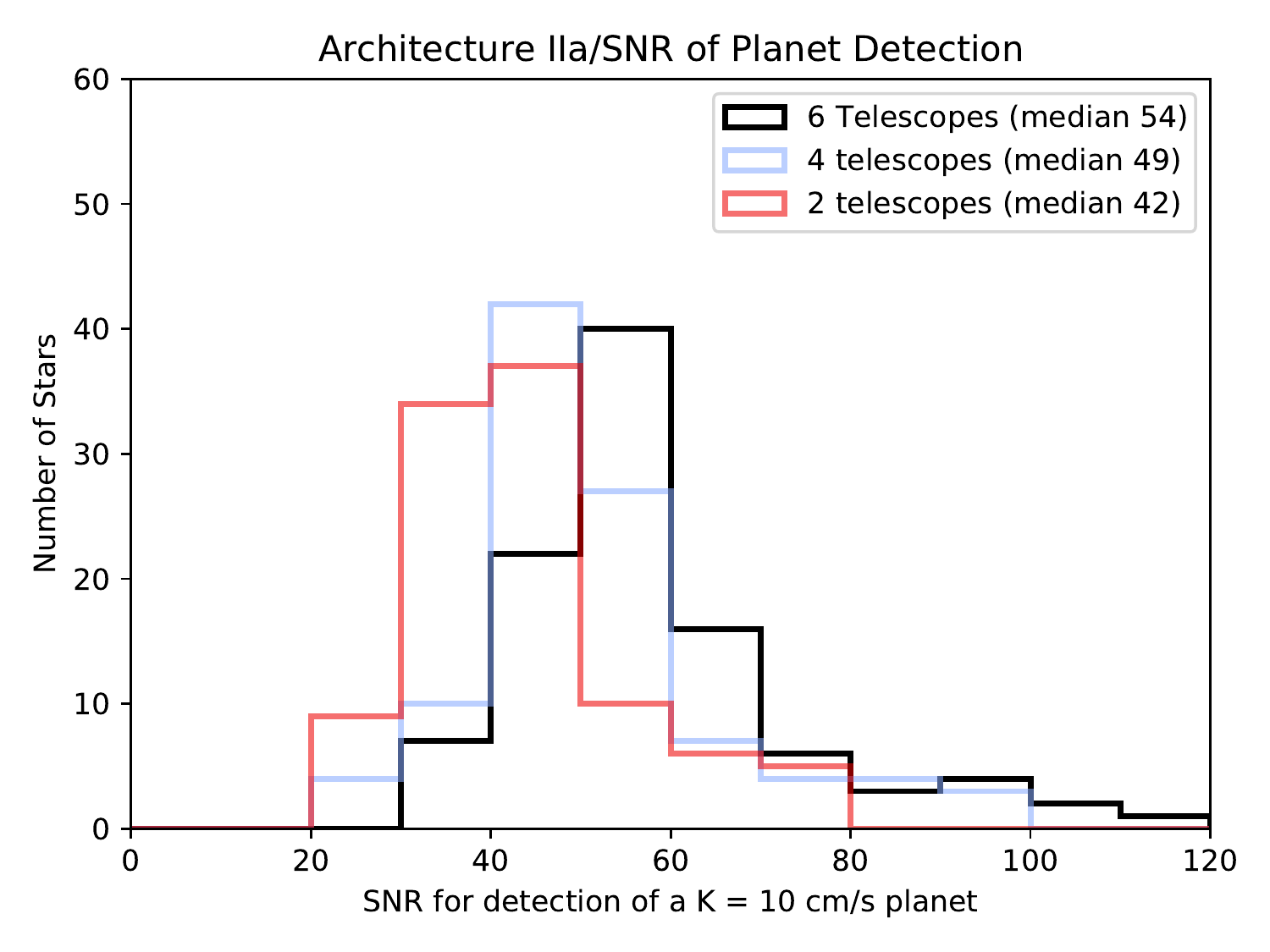}
\includegraphics[width=0.49\textwidth]{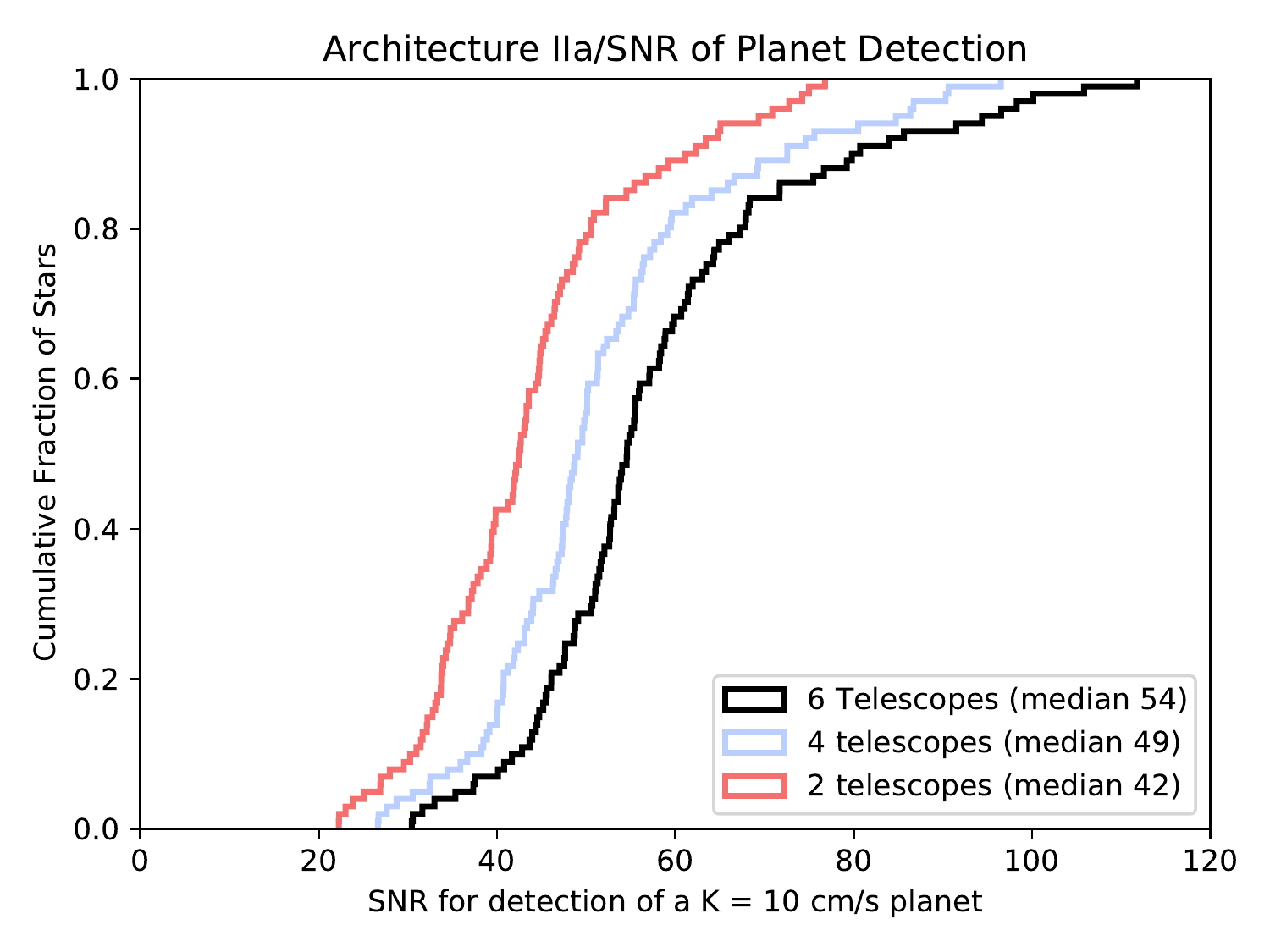}
\caption{Same as figure \ref{fig:ArchIkSNR}, but for architecture IIa.}
\end{figure}

\begin{figure}
\noindent \includegraphics[width=0.49\textwidth]{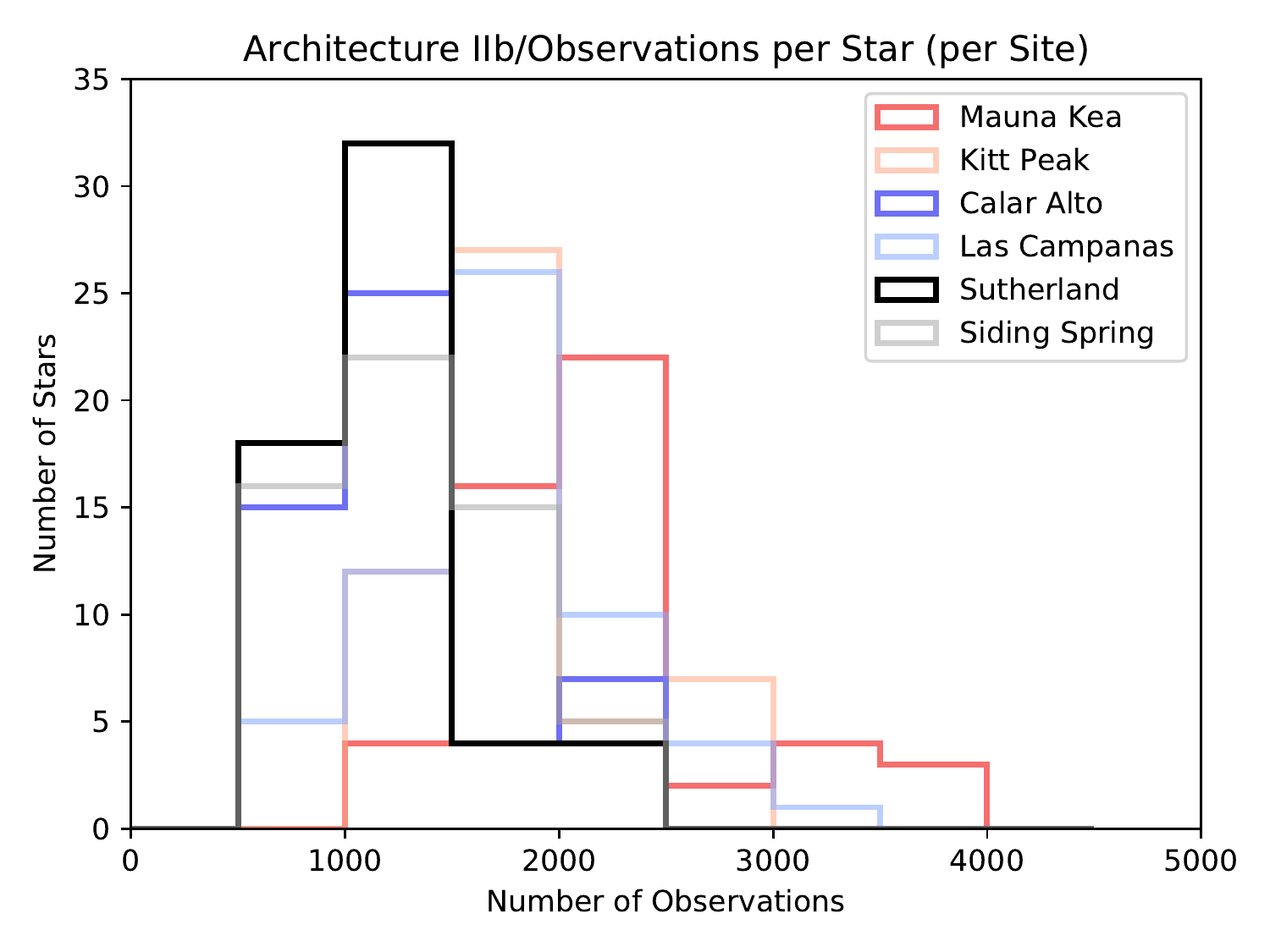}
\includegraphics[width=0.49\textwidth]{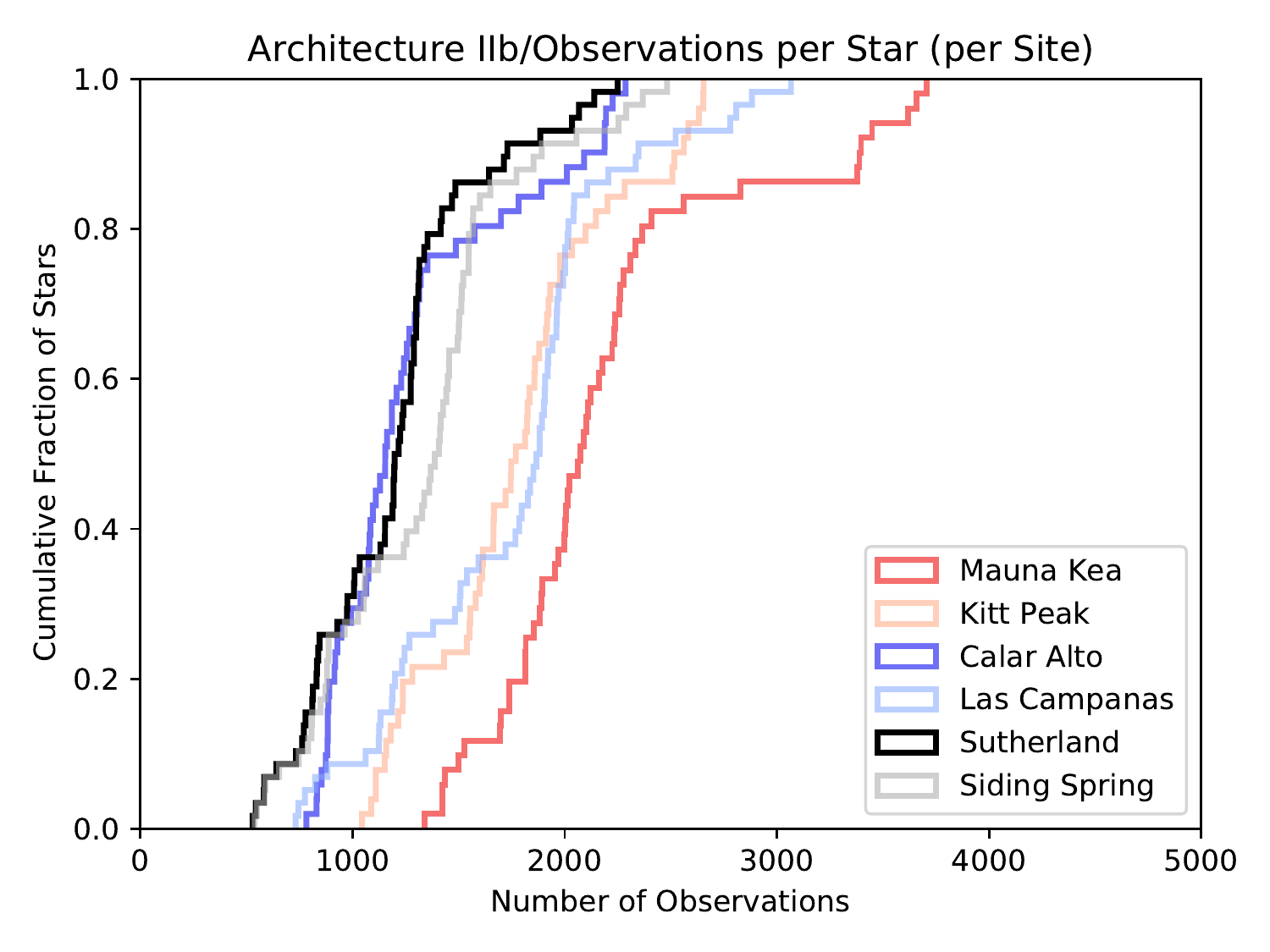}

\noindent \includegraphics[width=0.49\textwidth]{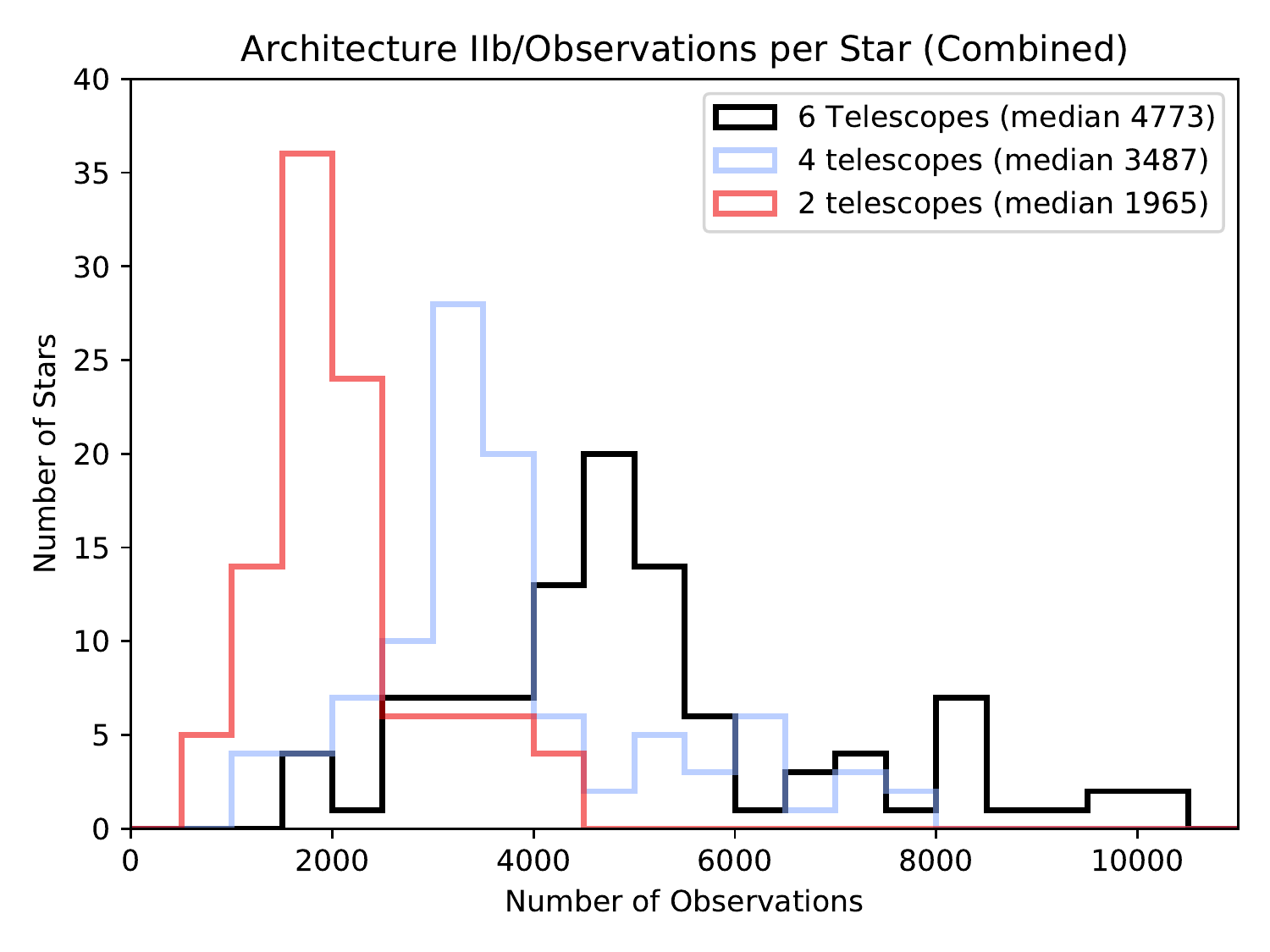}
\includegraphics[width=0.49\textwidth]{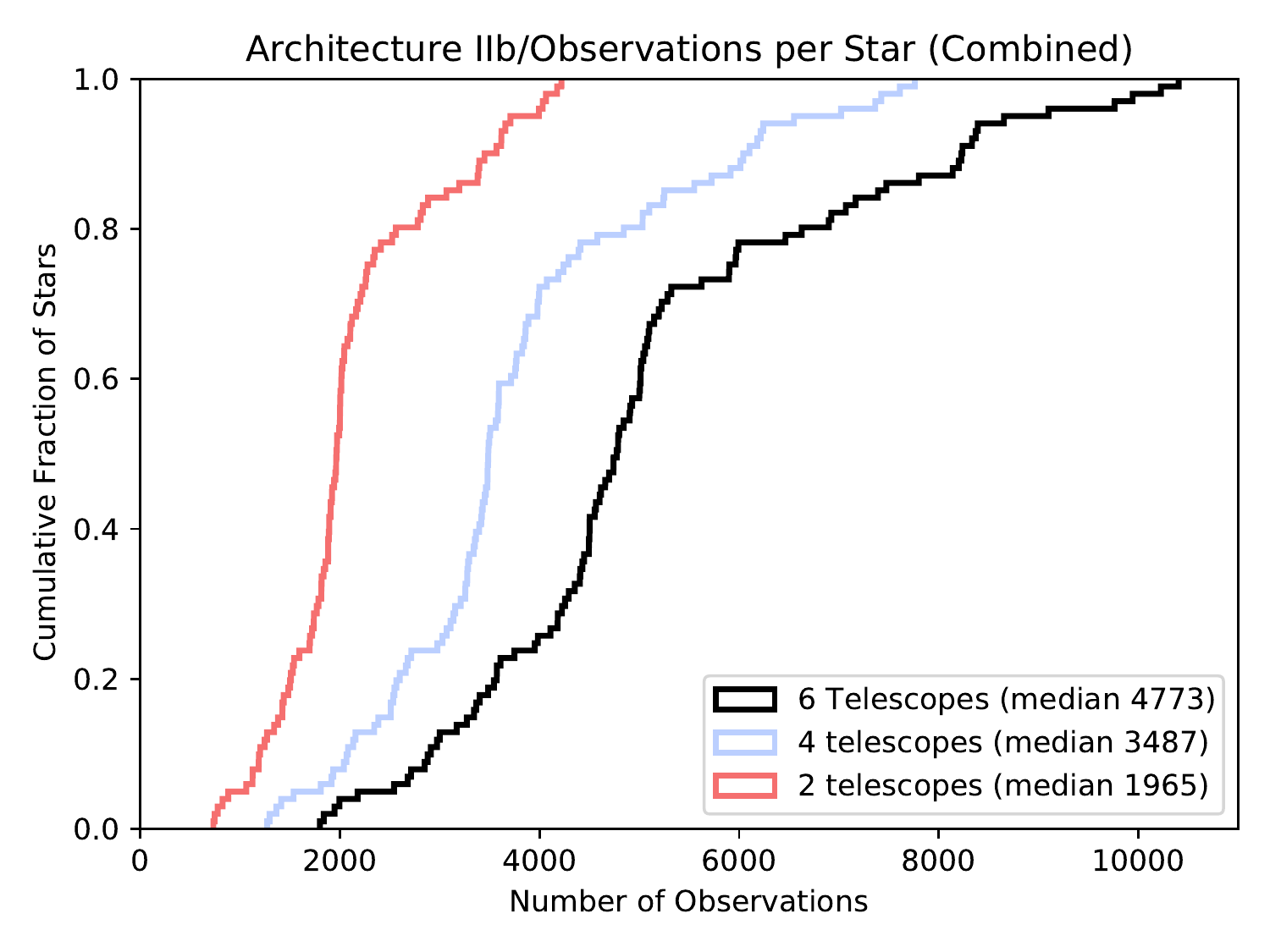}
\caption{Same as figure \ref{fig:ArchIobs}, but for architecture IIb.}
\end{figure}

\begin{figure}
\noindent \includegraphics[width=0.49\textwidth]{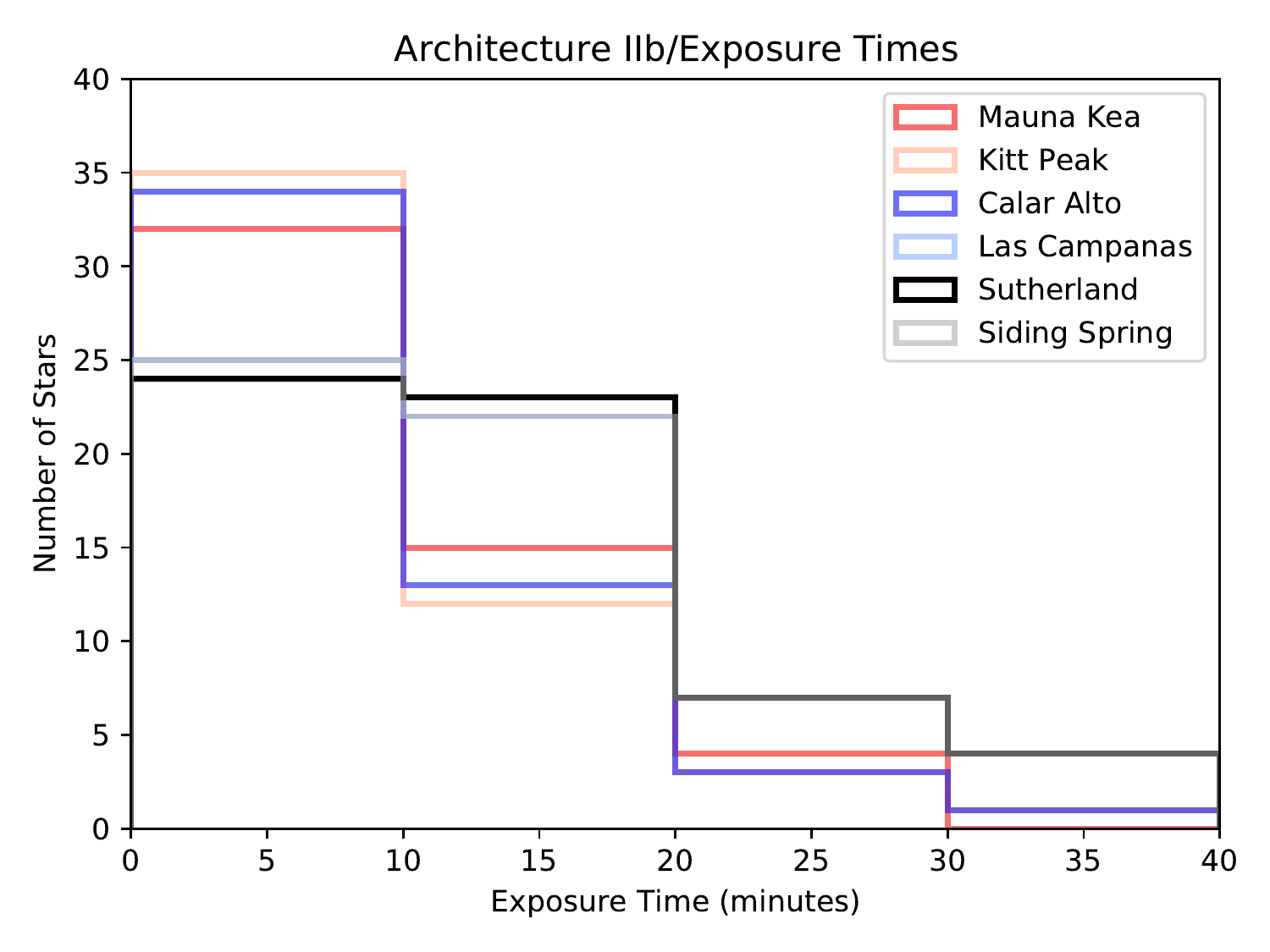}
\includegraphics[width=0.49\textwidth]{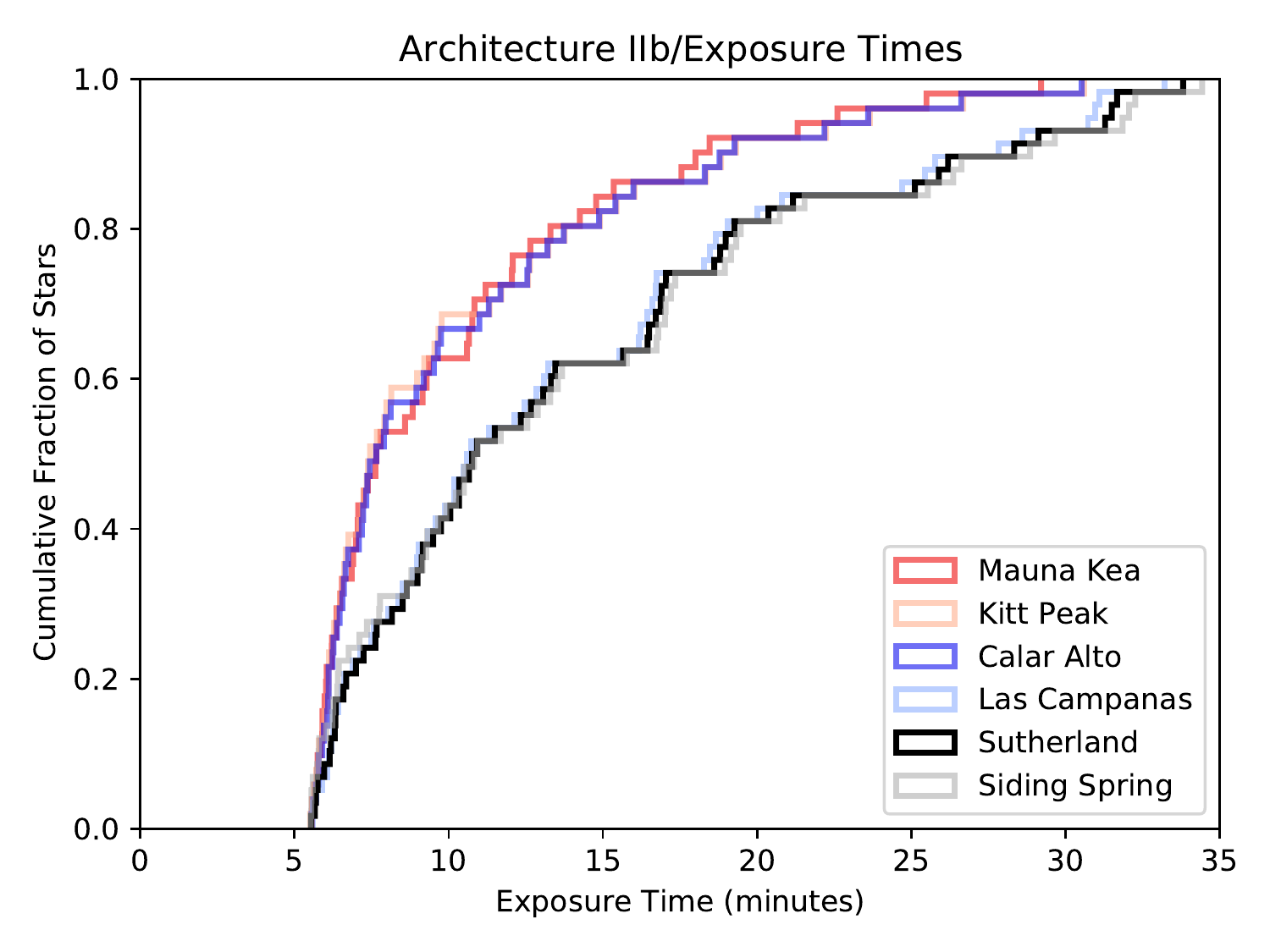}

\noindent \includegraphics[width=0.49\textwidth]{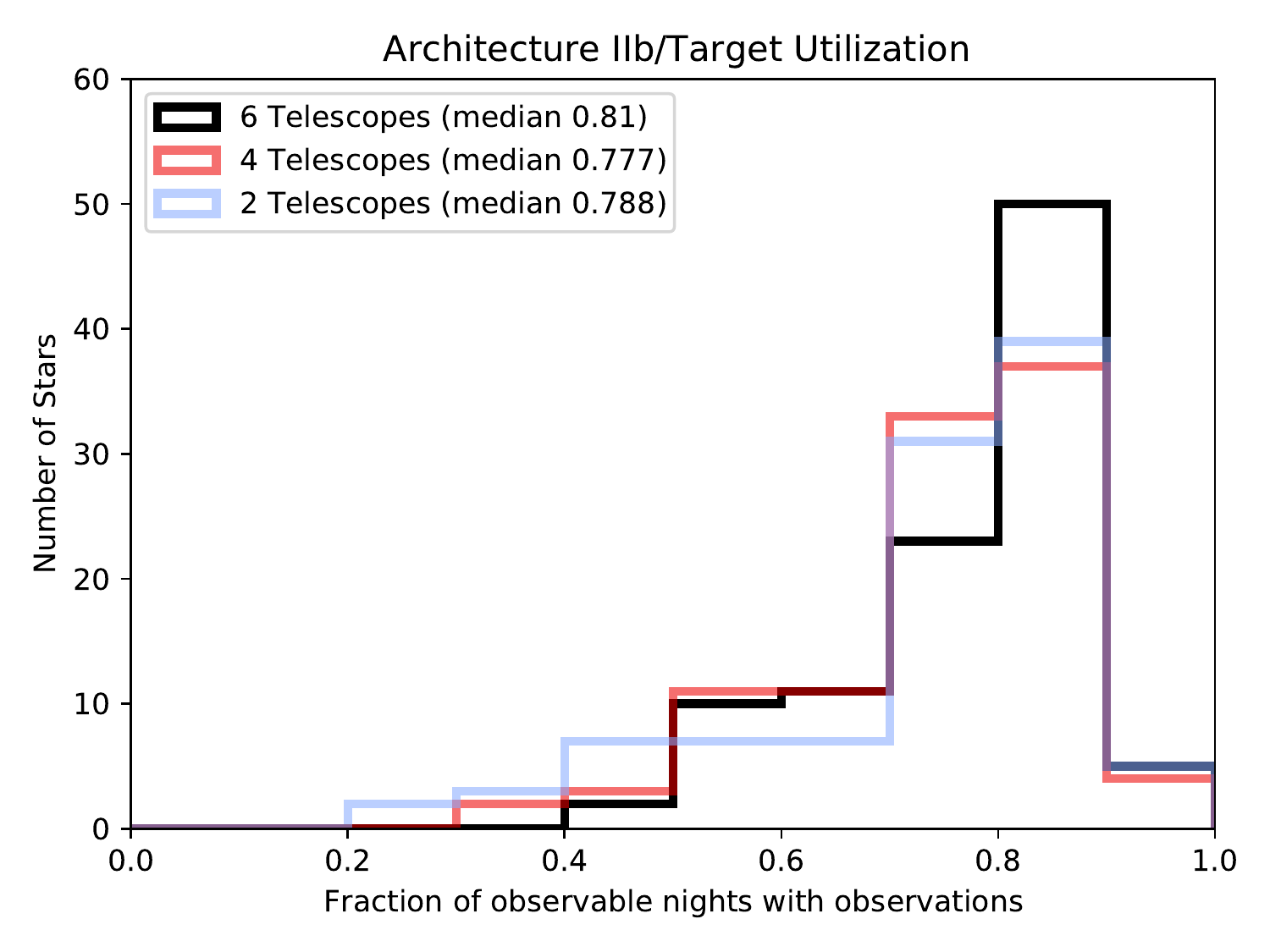}
\includegraphics[width=0.49\textwidth]{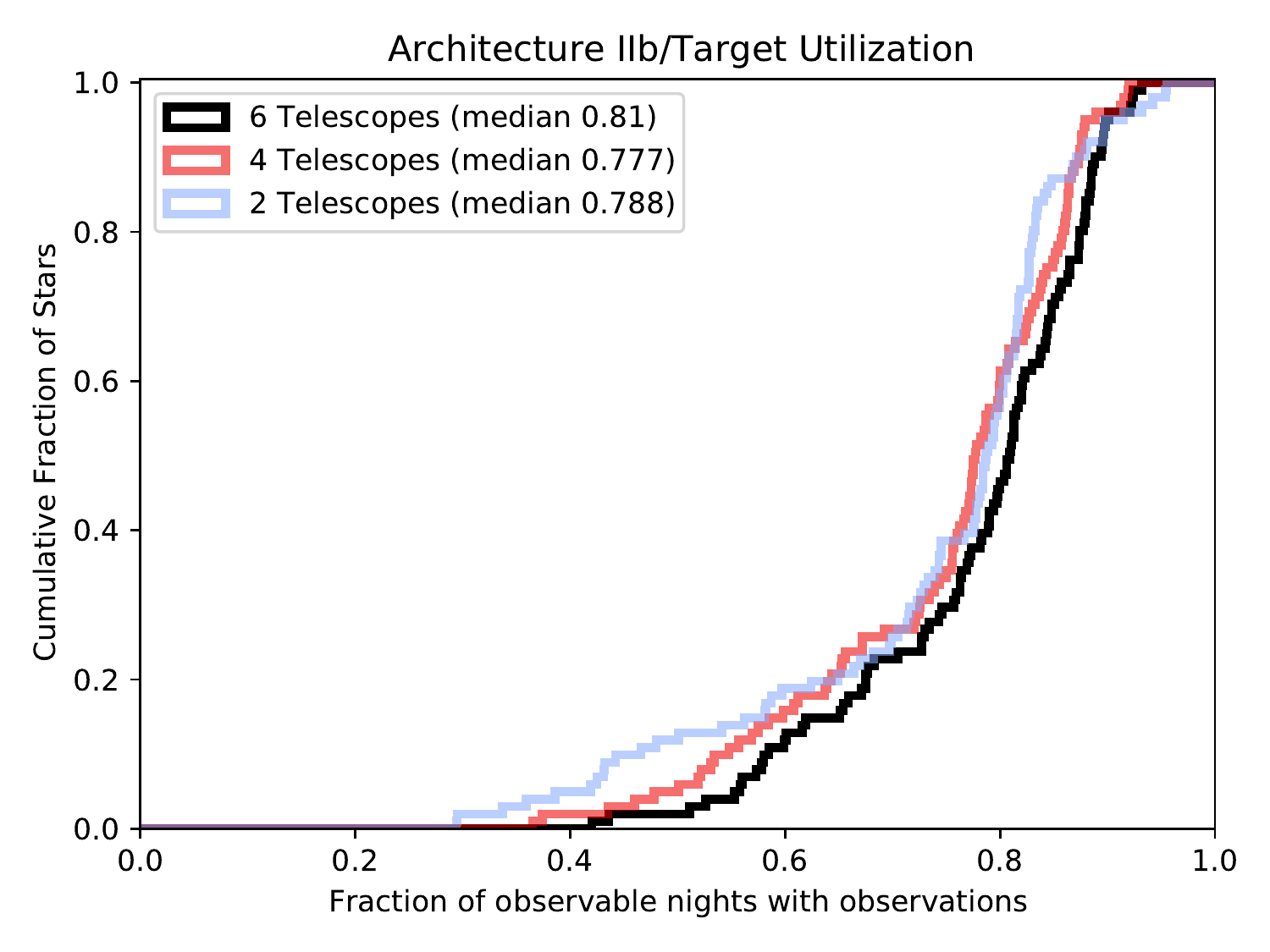}
\caption{Same as figure \ref{fig:ArchIexpfrac}, but for architecture IIb.}
\end{figure}

\begin{figure}
\noindent \includegraphics[width=0.49\textwidth]{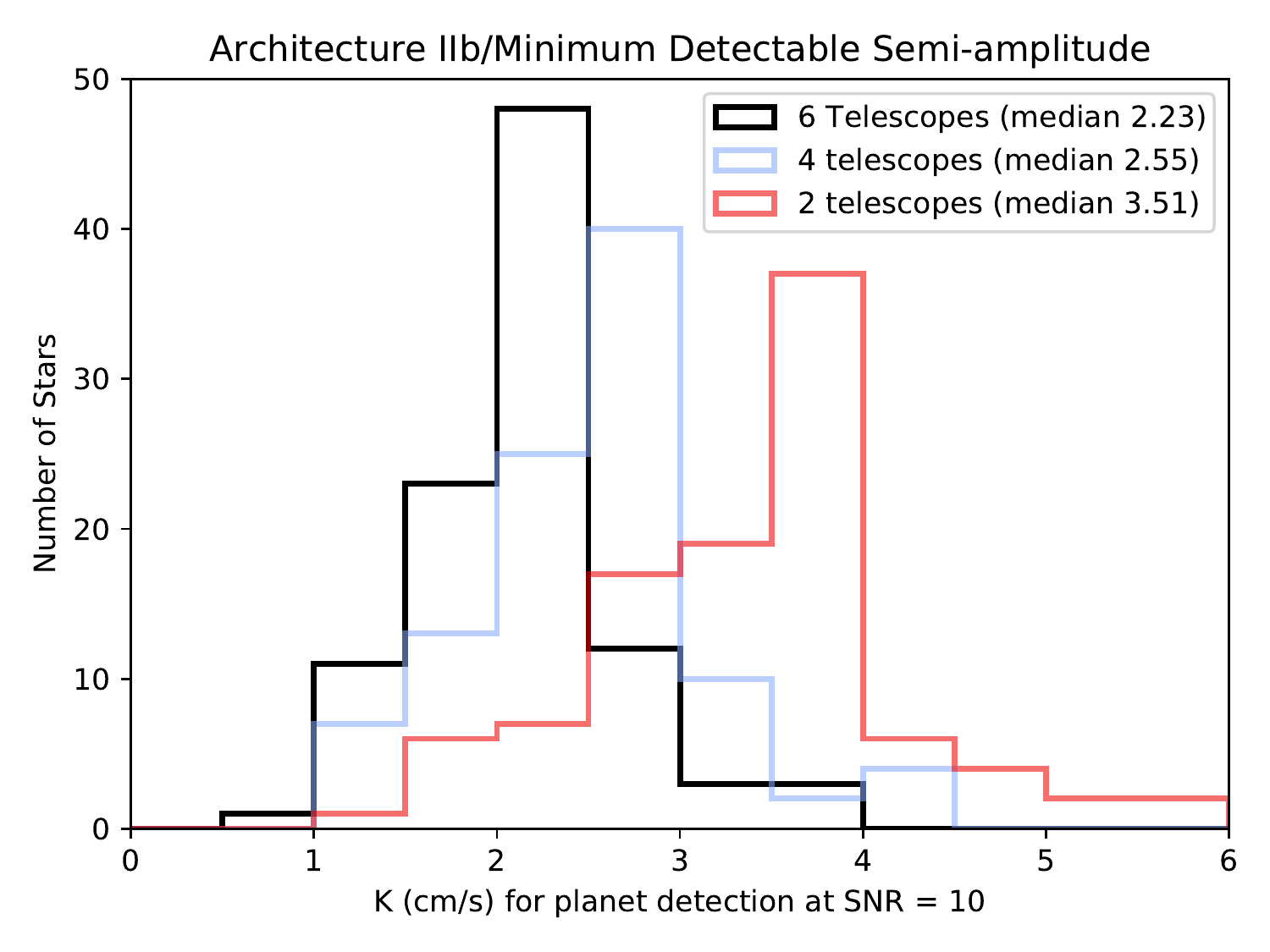}
\includegraphics[width=0.49\textwidth]{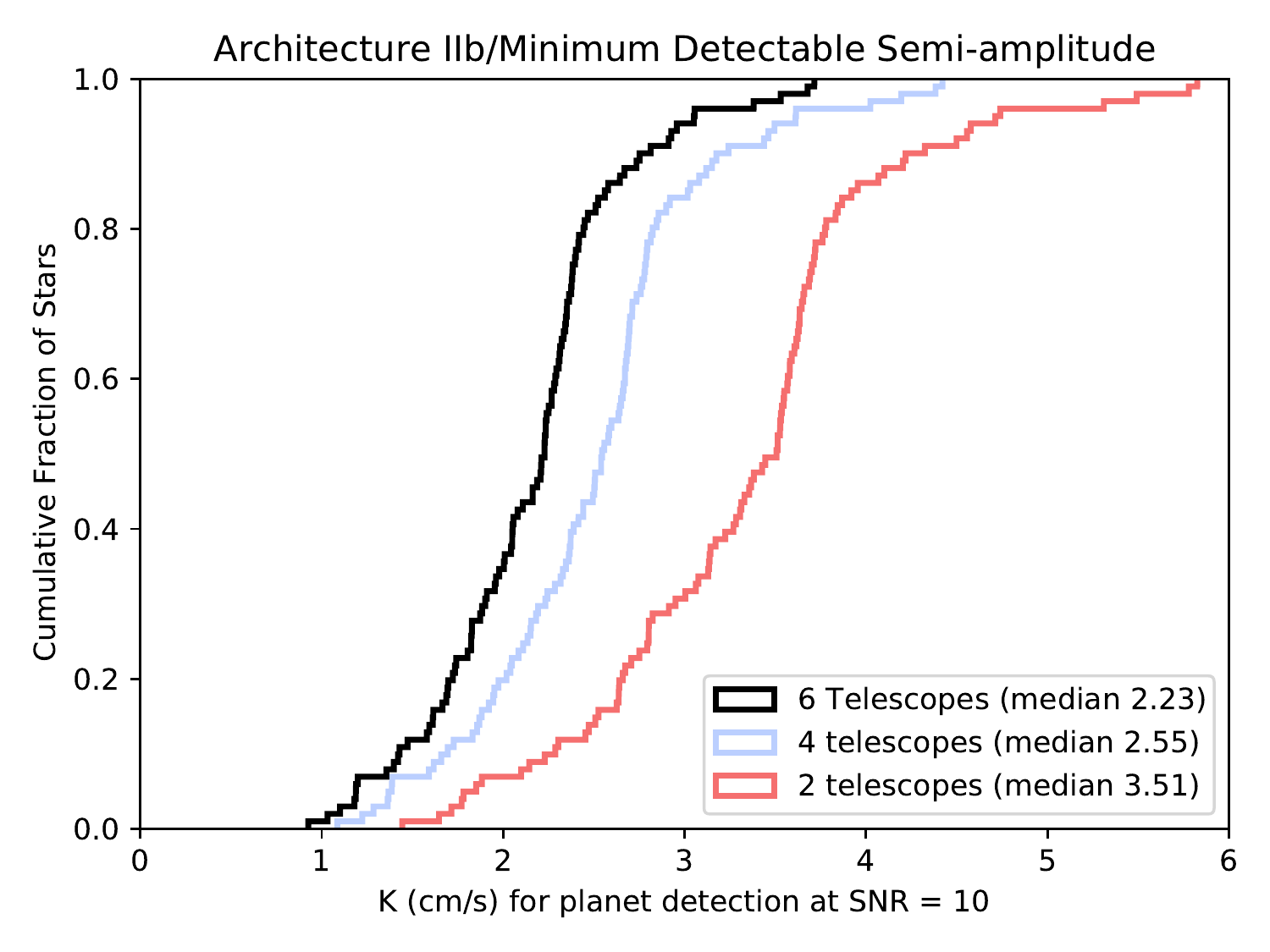}

\noindent \includegraphics[width=0.49\textwidth]{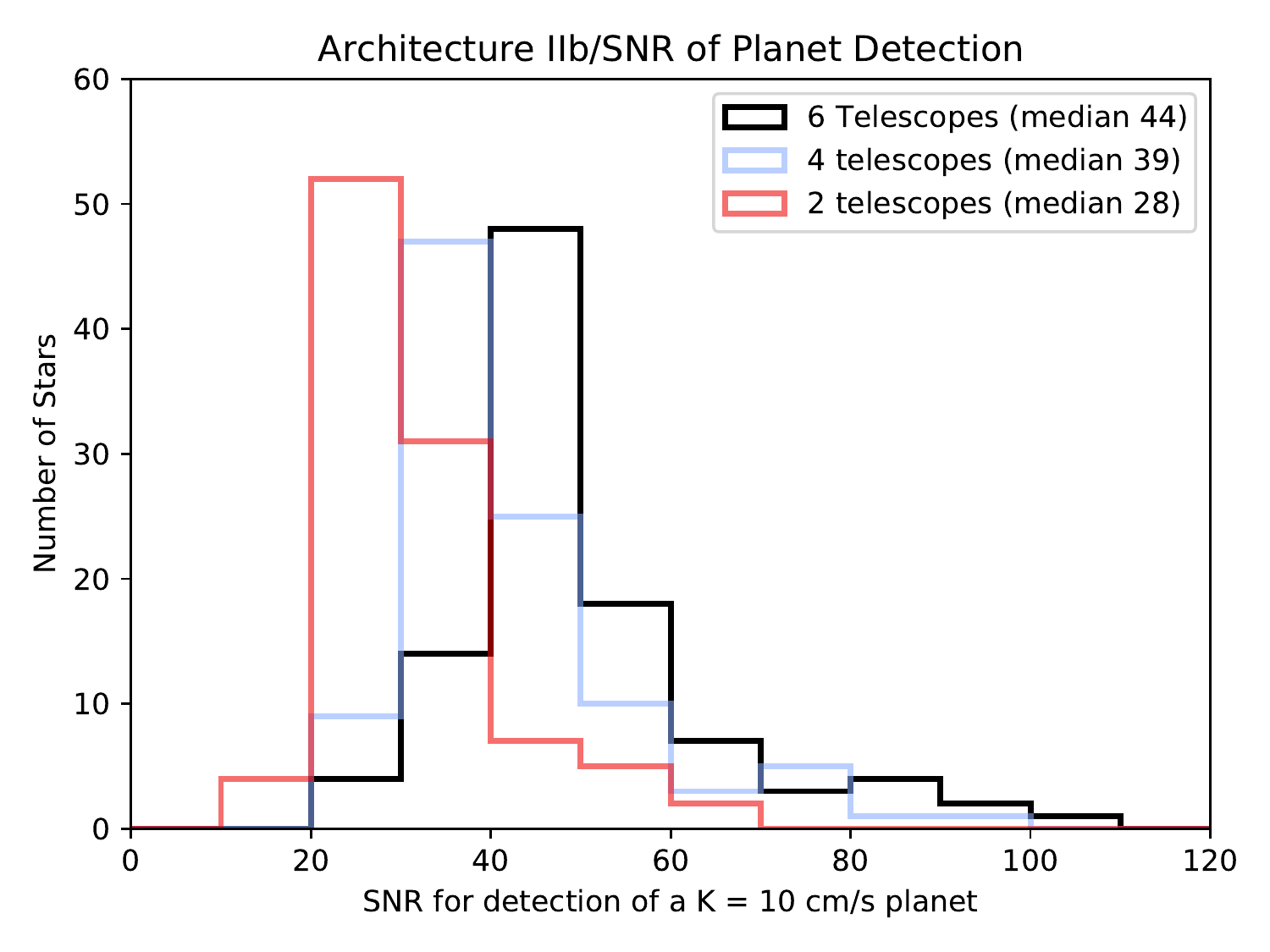}
\includegraphics[width=0.49\textwidth]{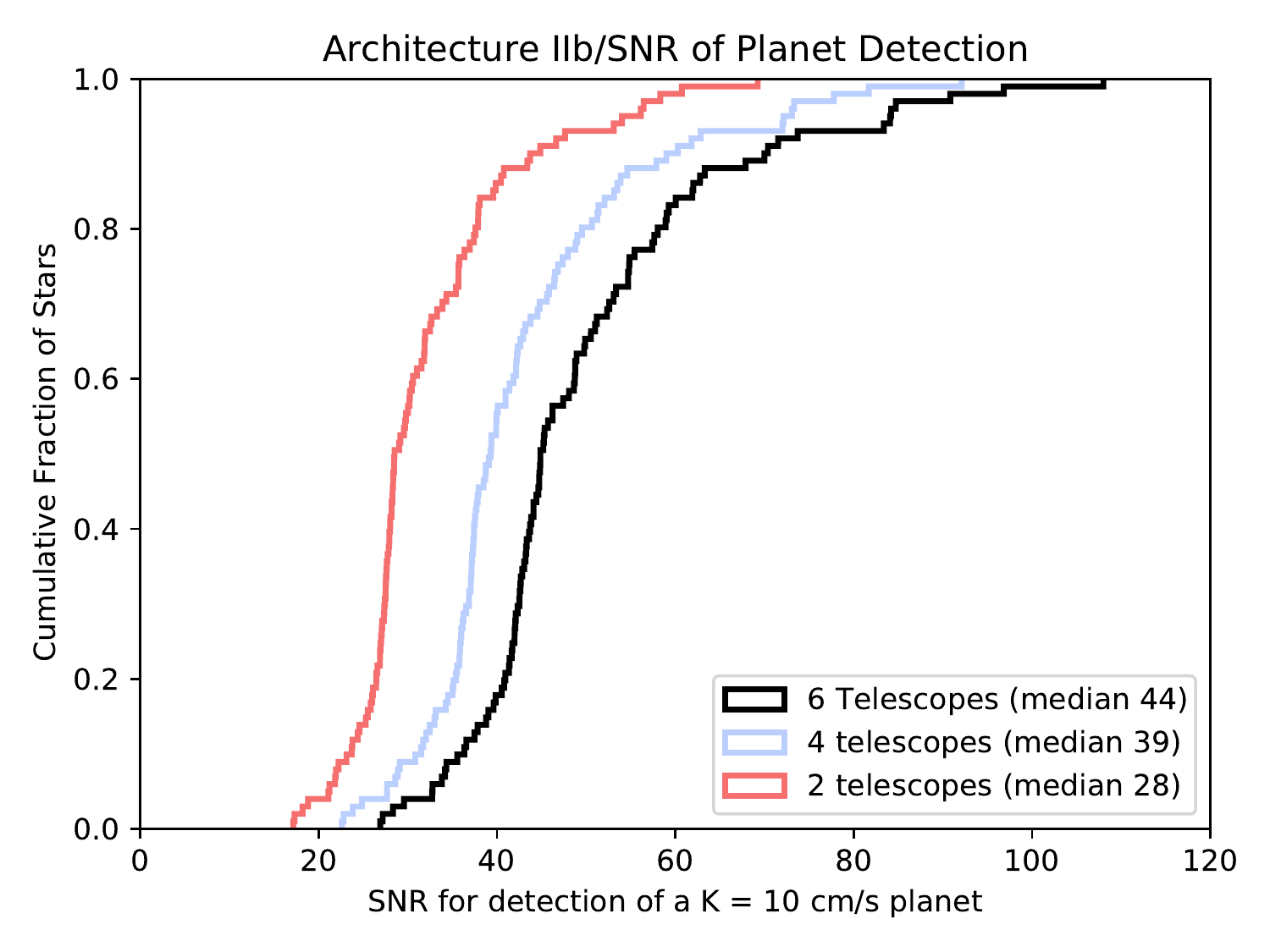}
\caption{Same as figure \ref{fig:ArchIkSNR}, but for architecture IIb.}
\end{figure}

\begin{figure}
\noindent \includegraphics[width=0.49\textwidth]{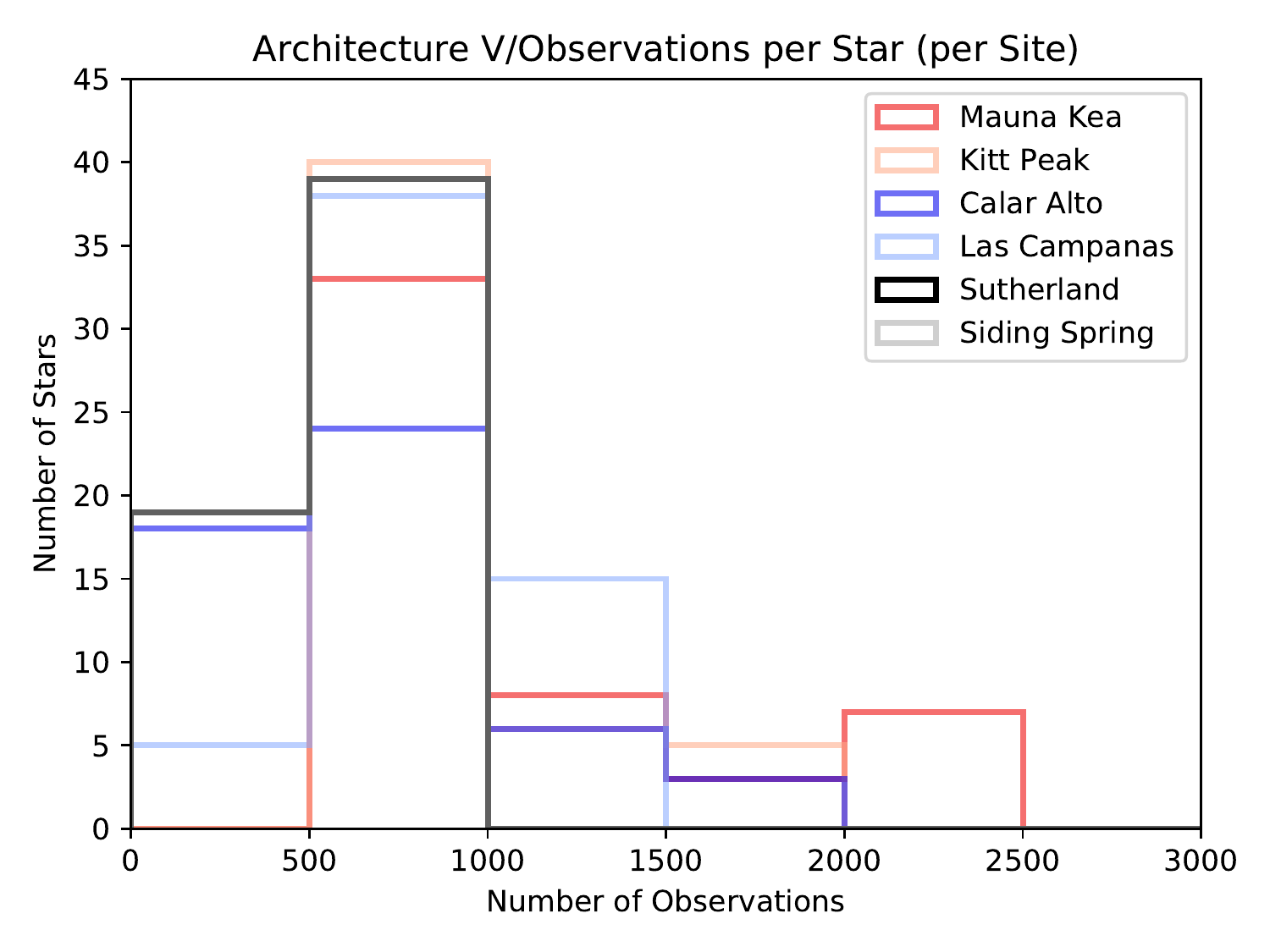}
\includegraphics[width=0.49\textwidth]{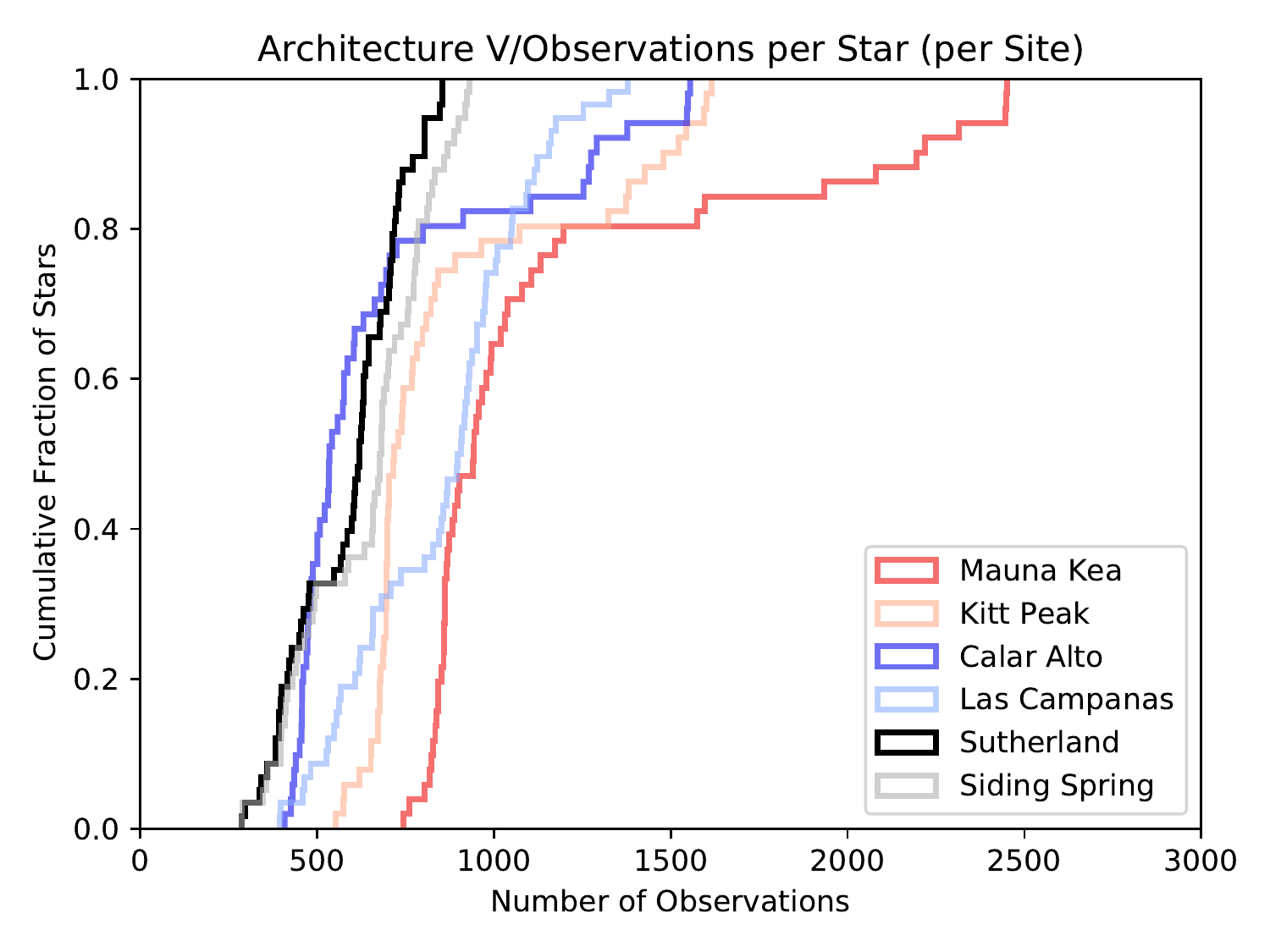}

\noindent \includegraphics[width=0.49\textwidth]{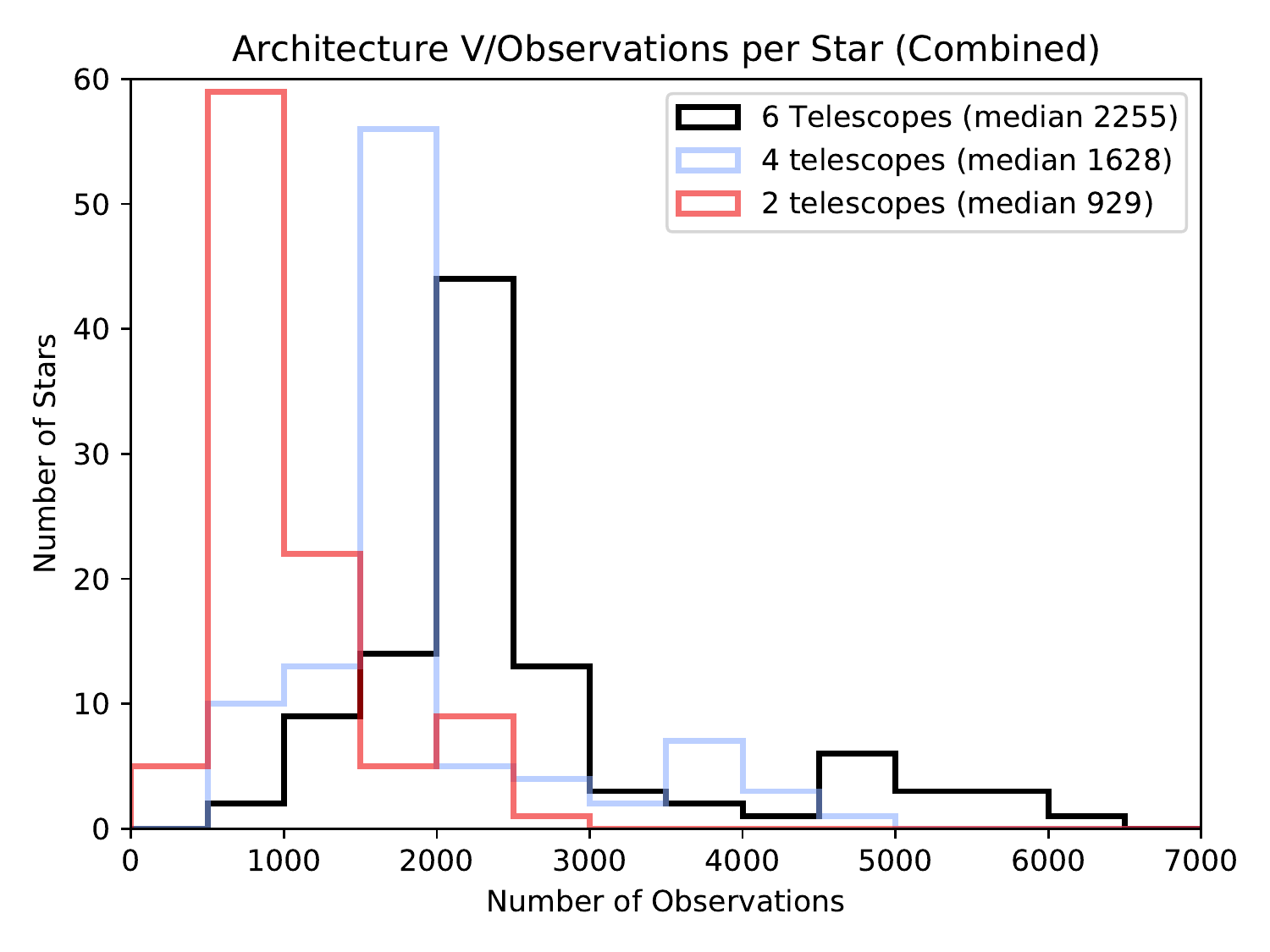}
\includegraphics[width=0.49\textwidth]{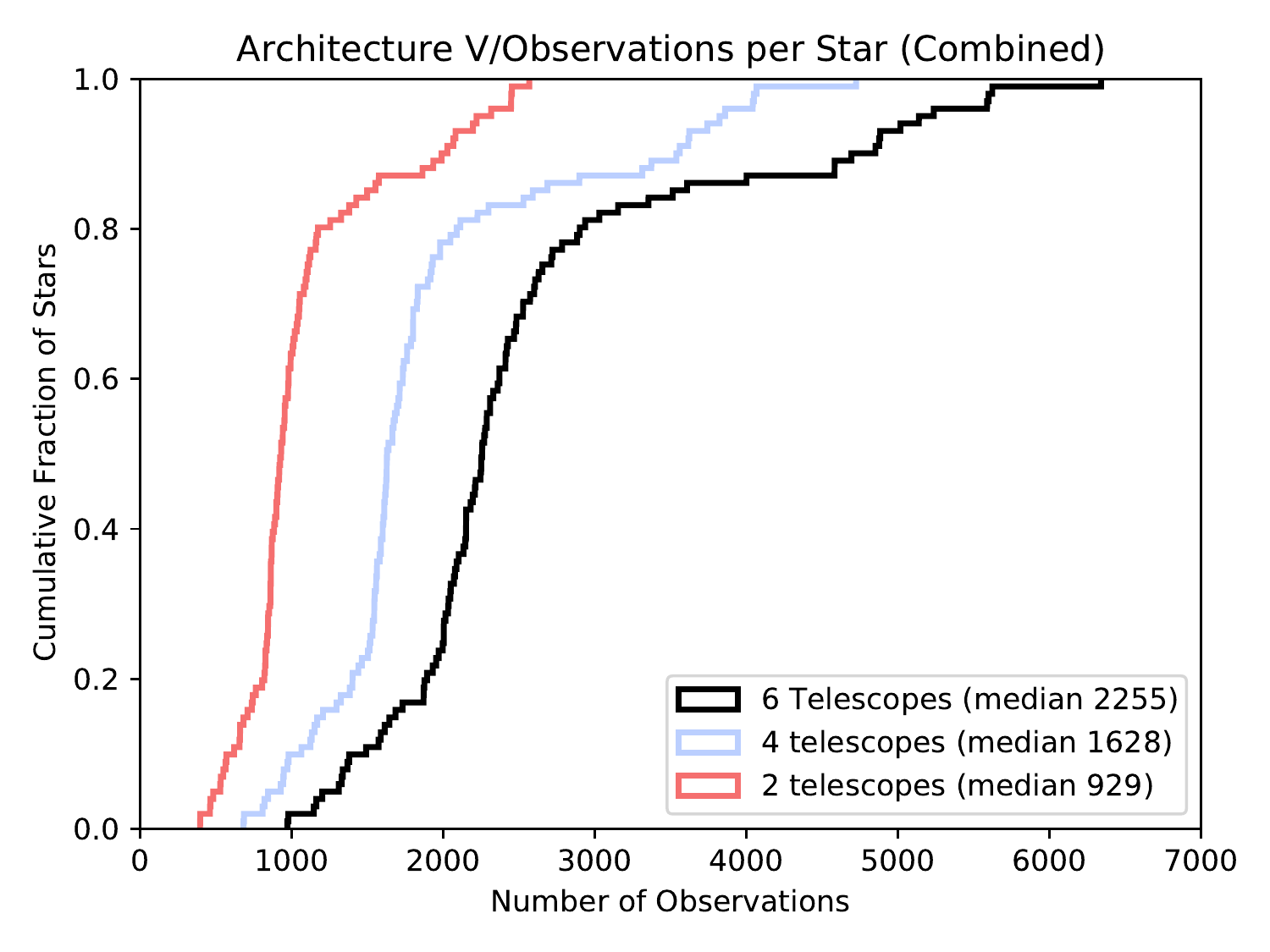}
\caption{Same as figure \ref{fig:ArchIobs}, but for architecture V.}
\end{figure}

\begin{figure}
\noindent \includegraphics[width=0.49\textwidth]{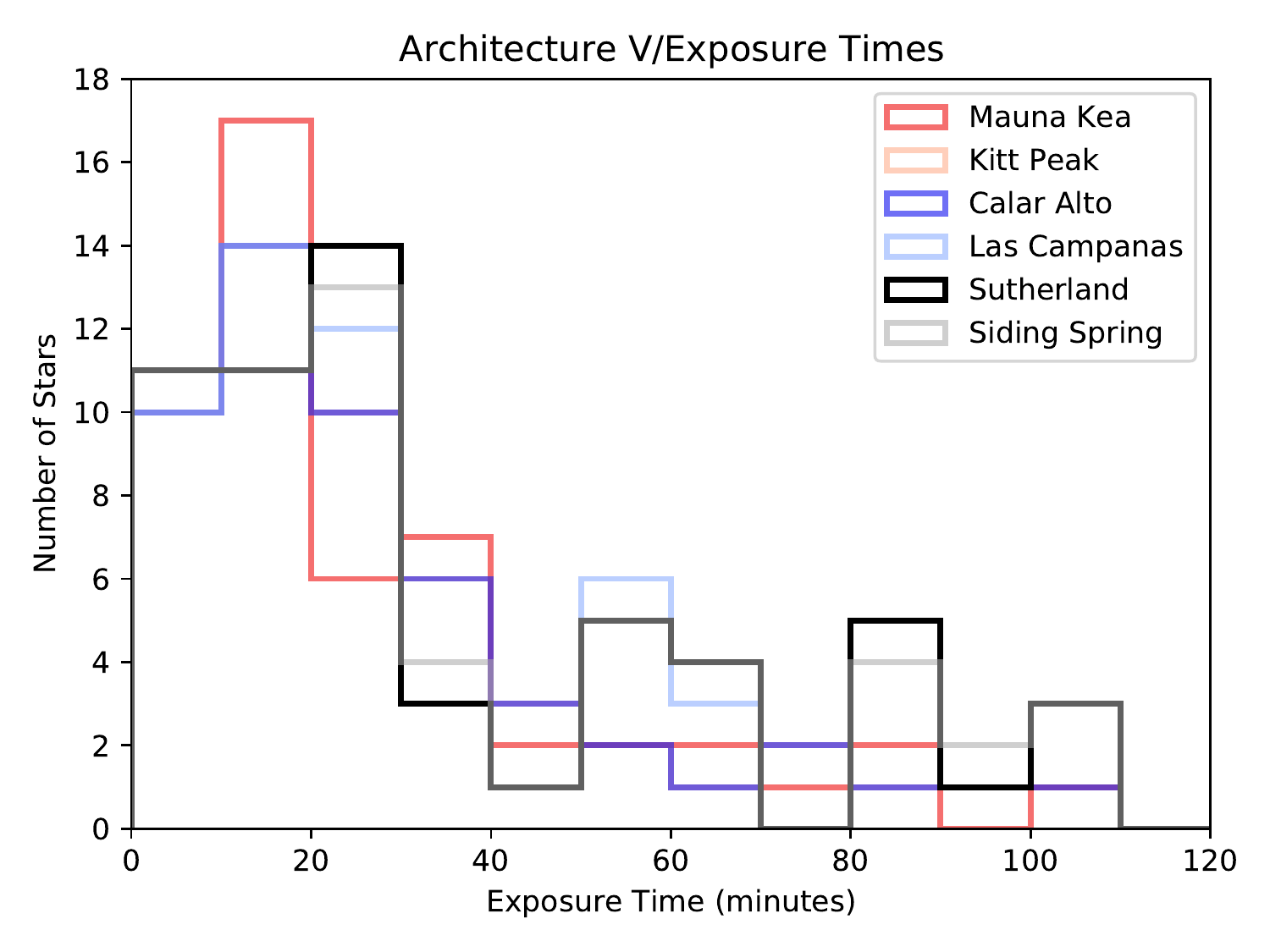}
\includegraphics[width=0.49\textwidth]{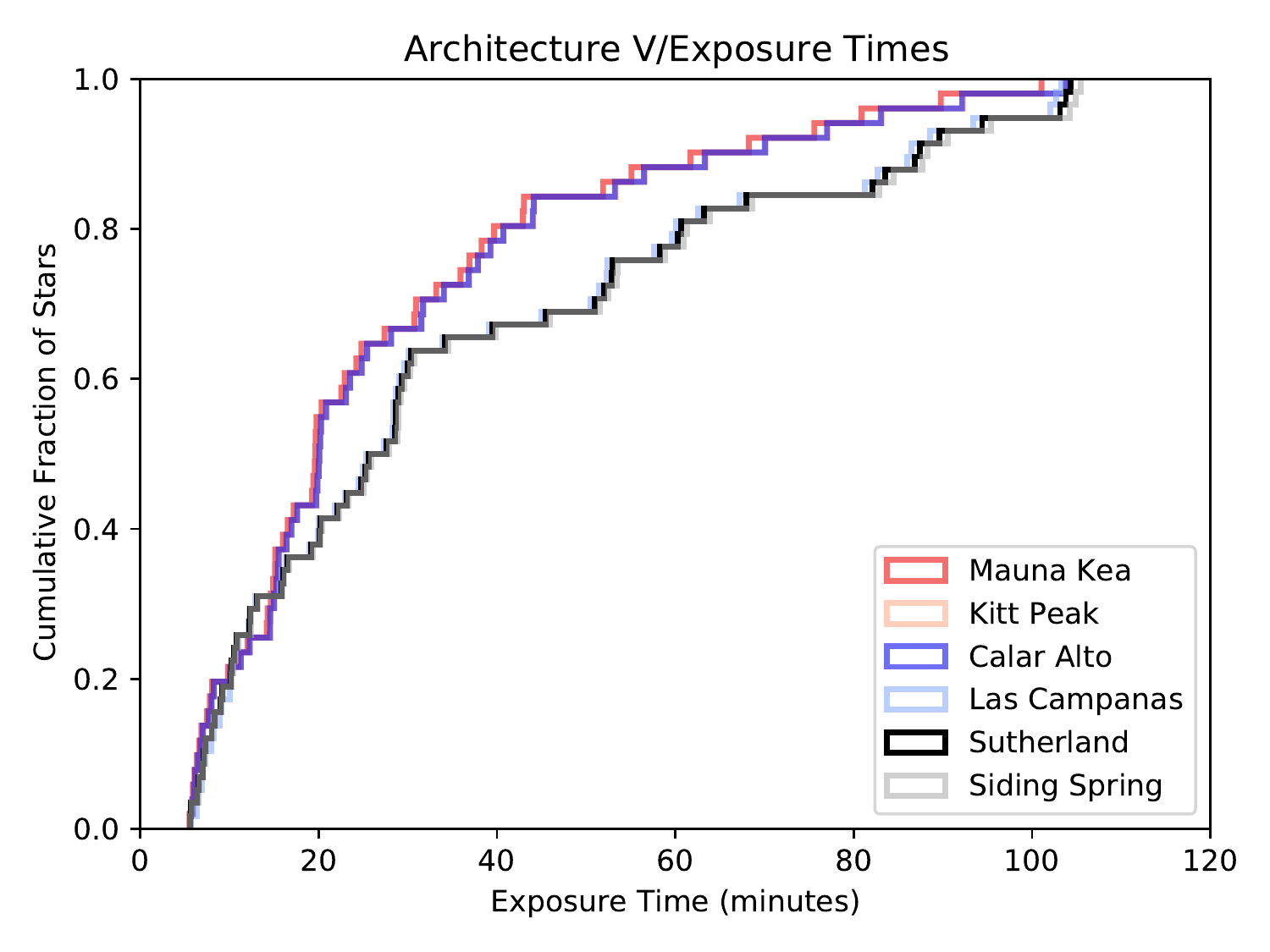}

\noindent \includegraphics[width=0.49\textwidth]{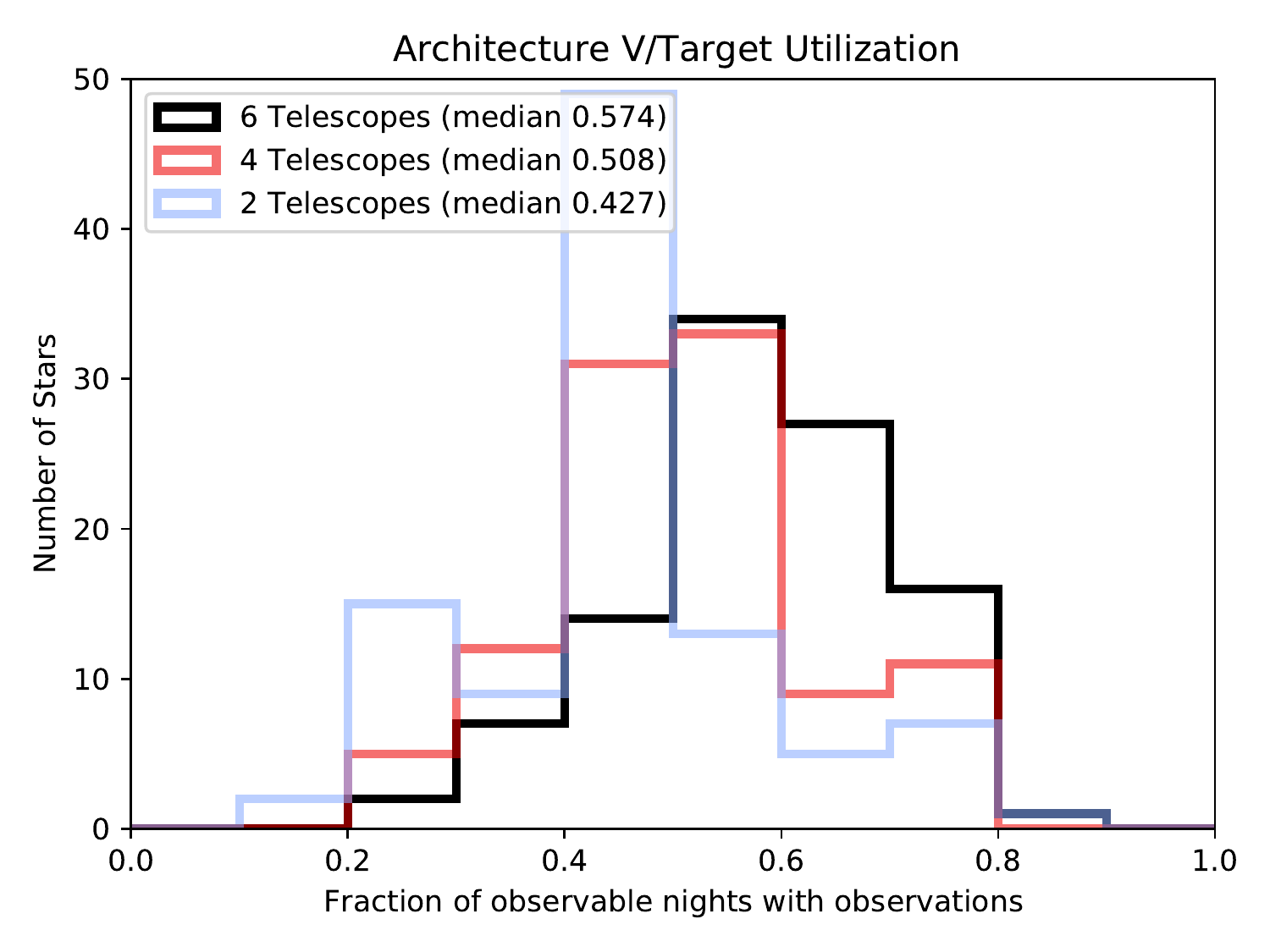}
\includegraphics[width=0.49\textwidth]{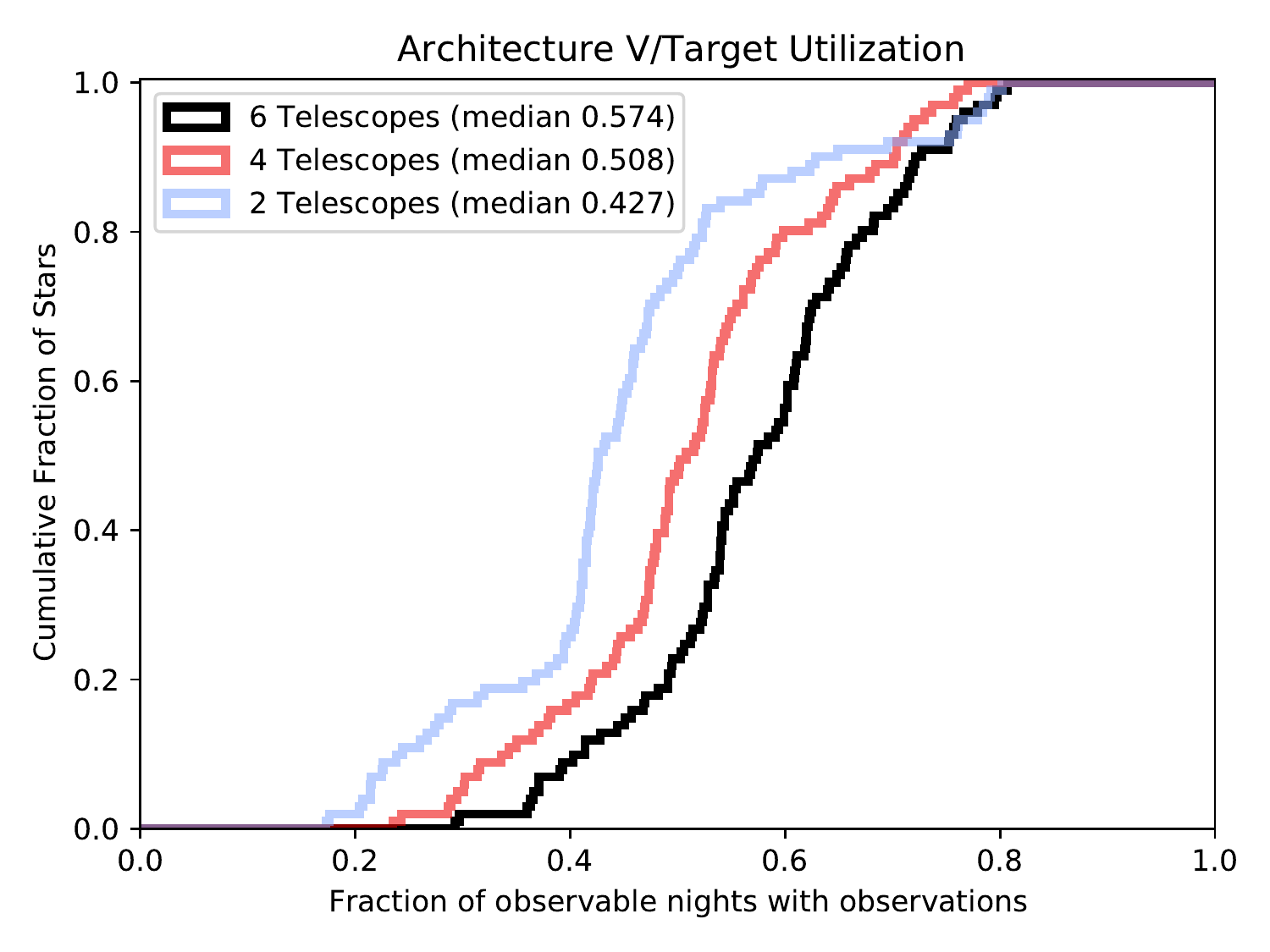}
\caption{Same as figure \ref{fig:ArchIexpfrac}, but for architecture V.}
\end{figure}

\begin{figure}
\noindent \includegraphics[width=0.49\textwidth]{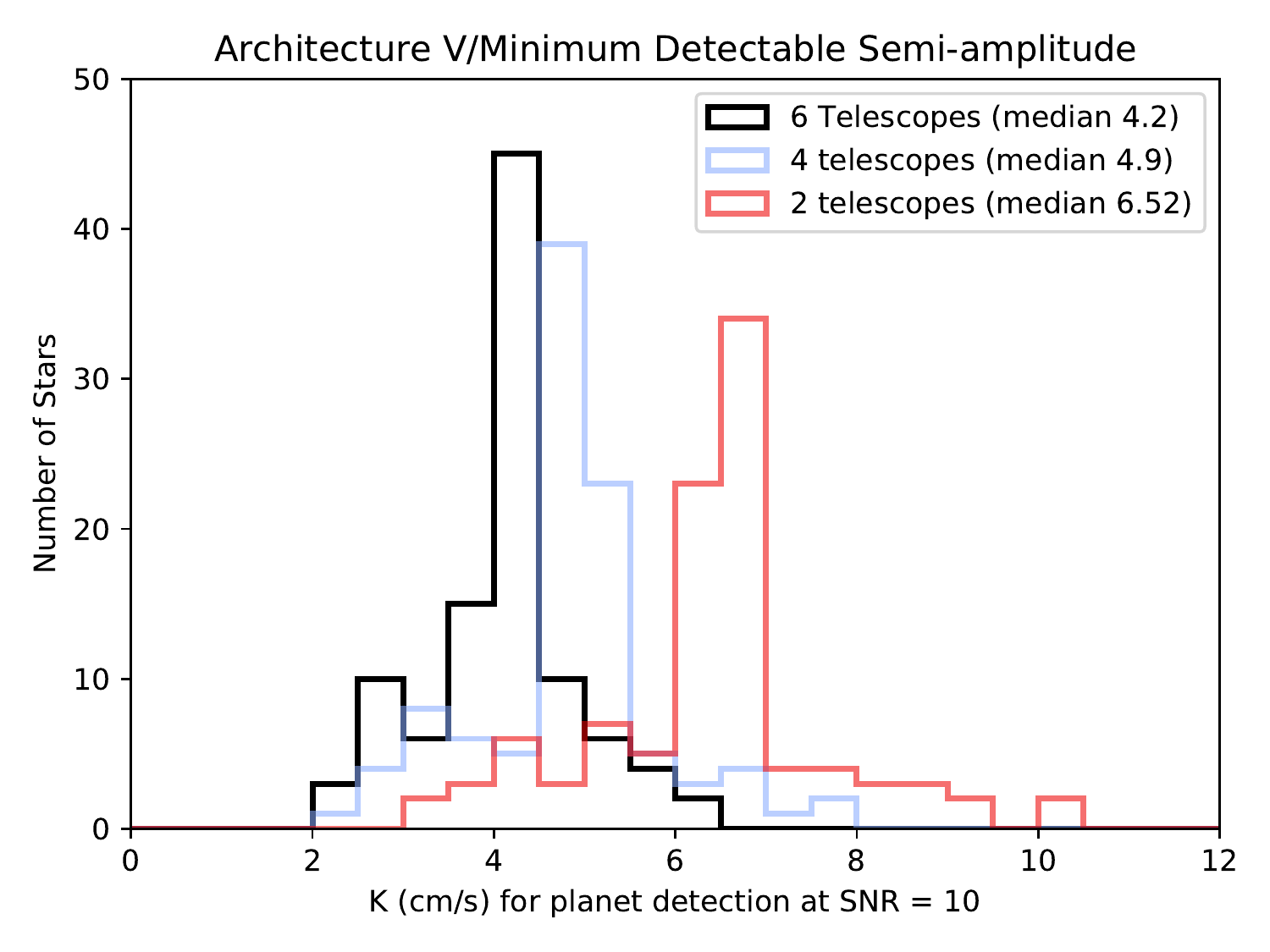}
\includegraphics[width=0.49\textwidth]{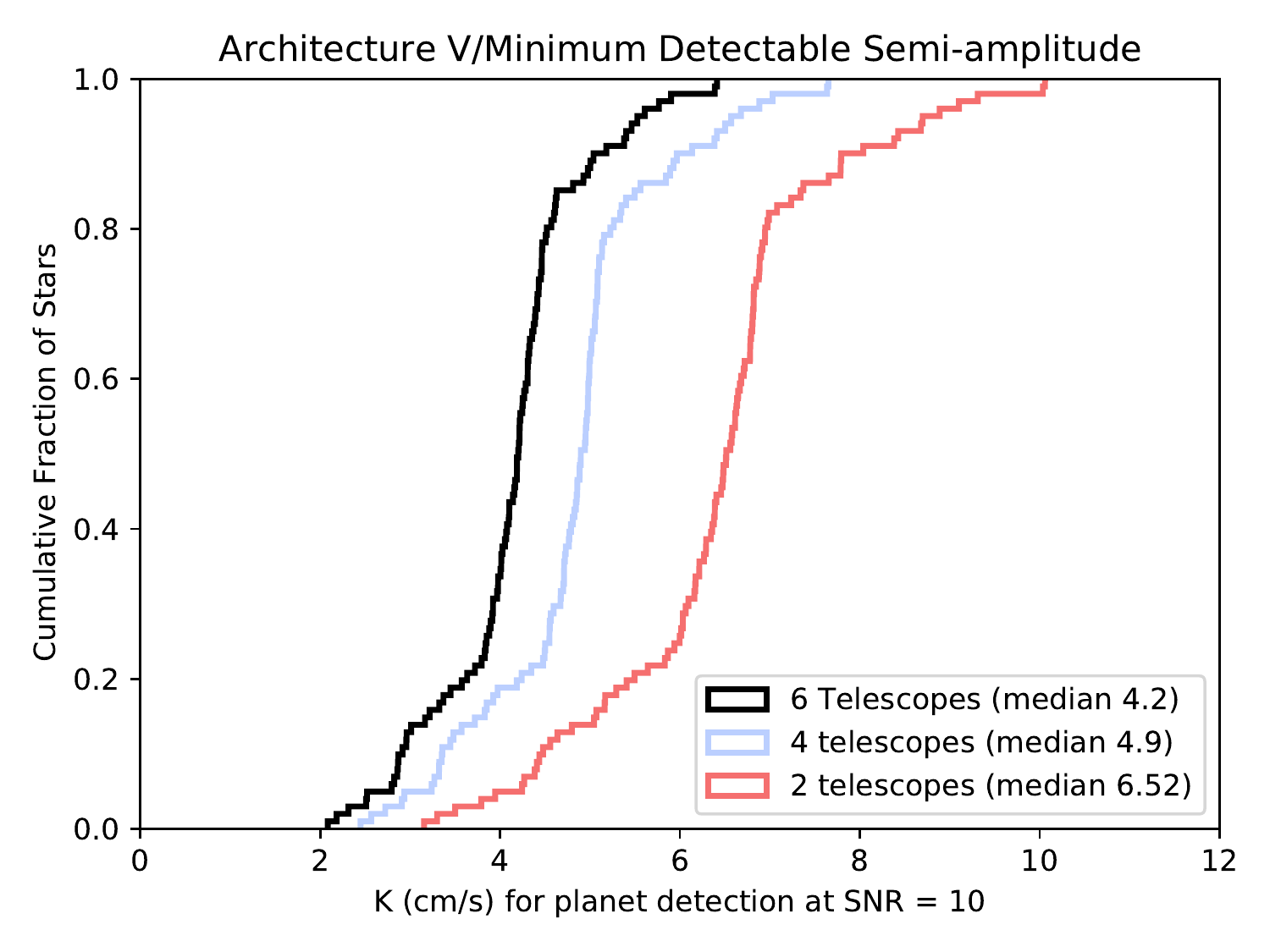}

\noindent \includegraphics[width=0.49\textwidth]{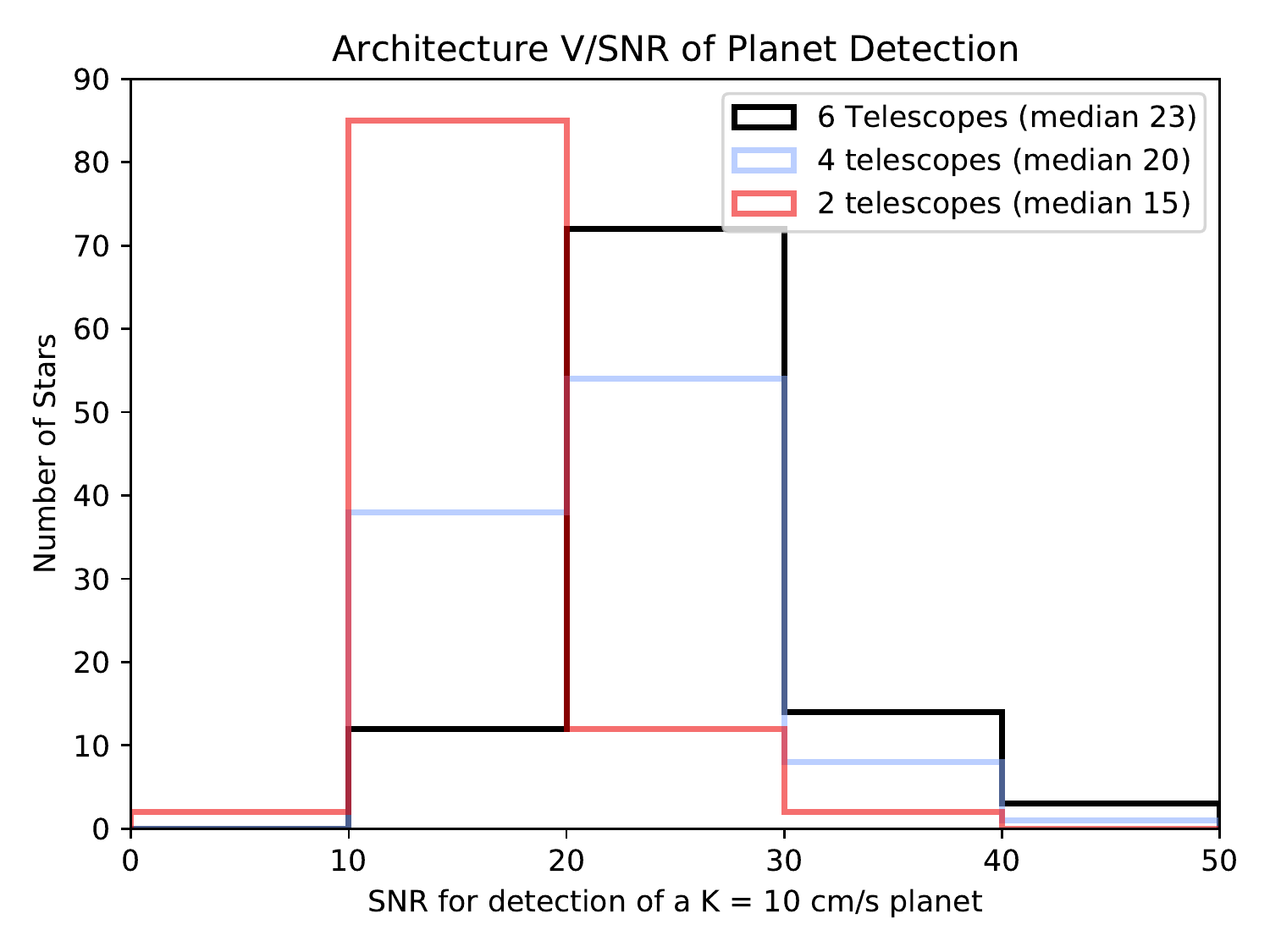}
\includegraphics[width=0.49\textwidth]{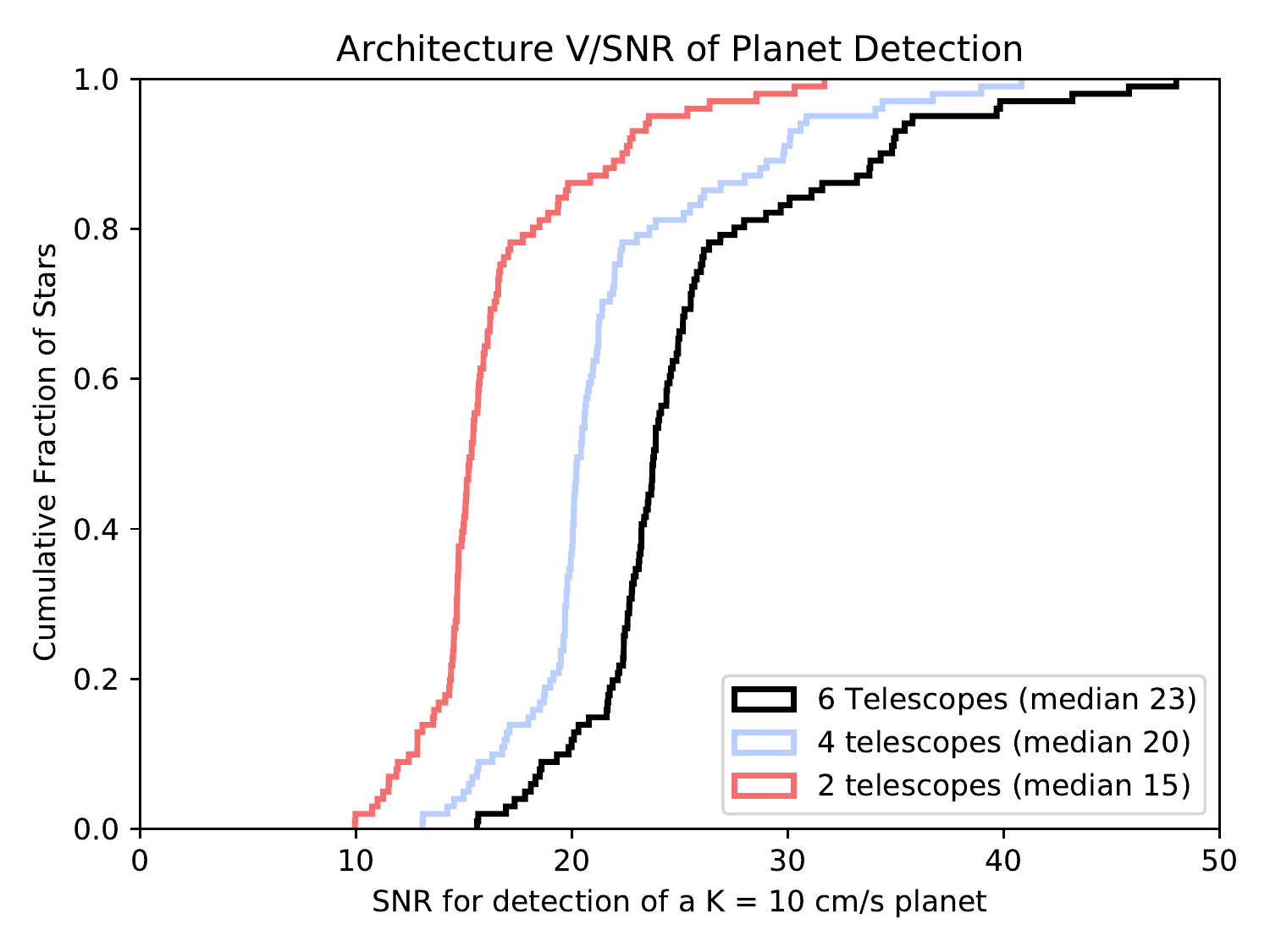}
\caption{Same as figure \ref{fig:ArchIkSNR}, but for architecture V.}
\end{figure}

\begin{figure}
\noindent \includegraphics[width=0.49\textwidth]{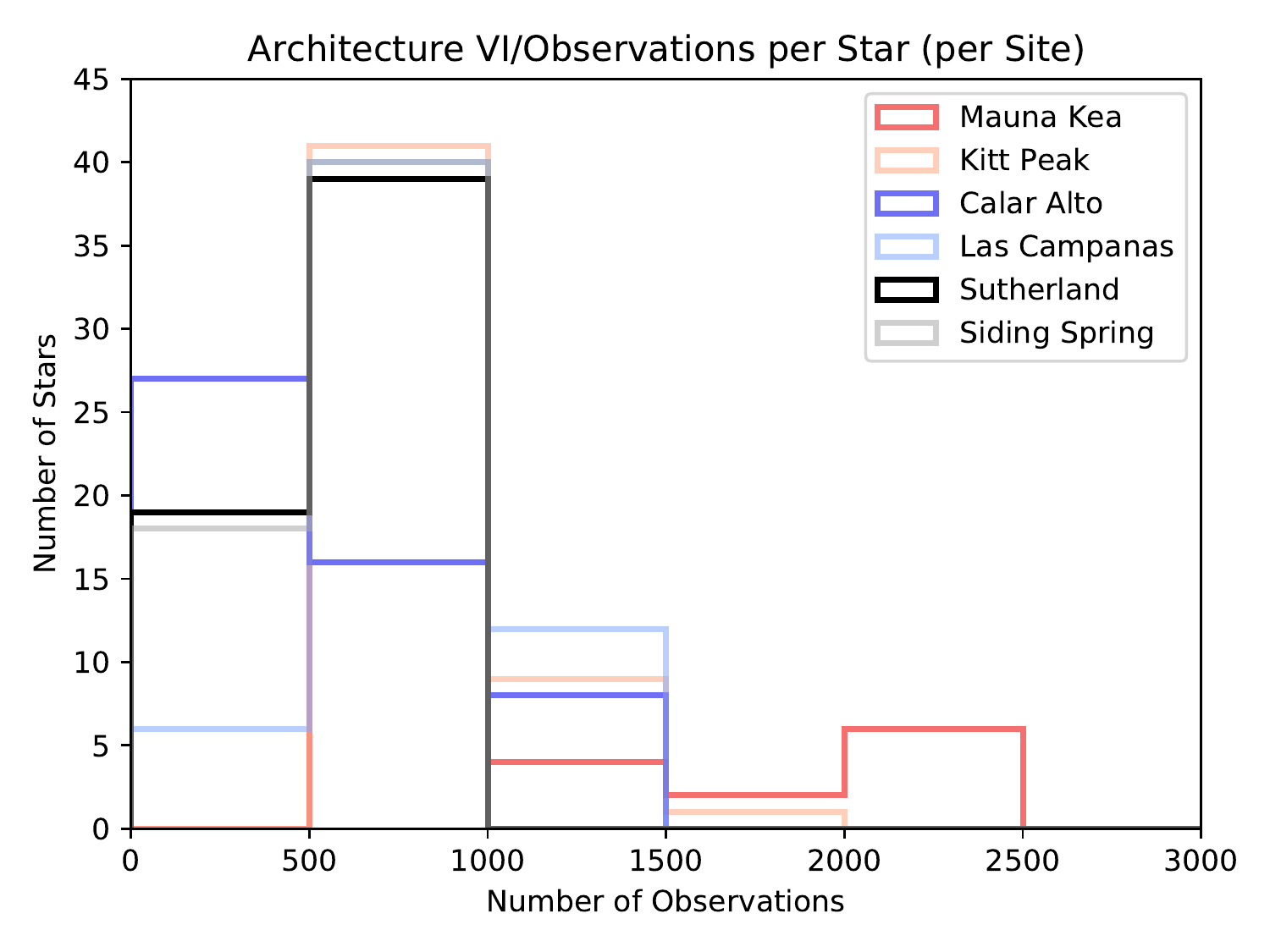}
\includegraphics[width=0.49\textwidth]{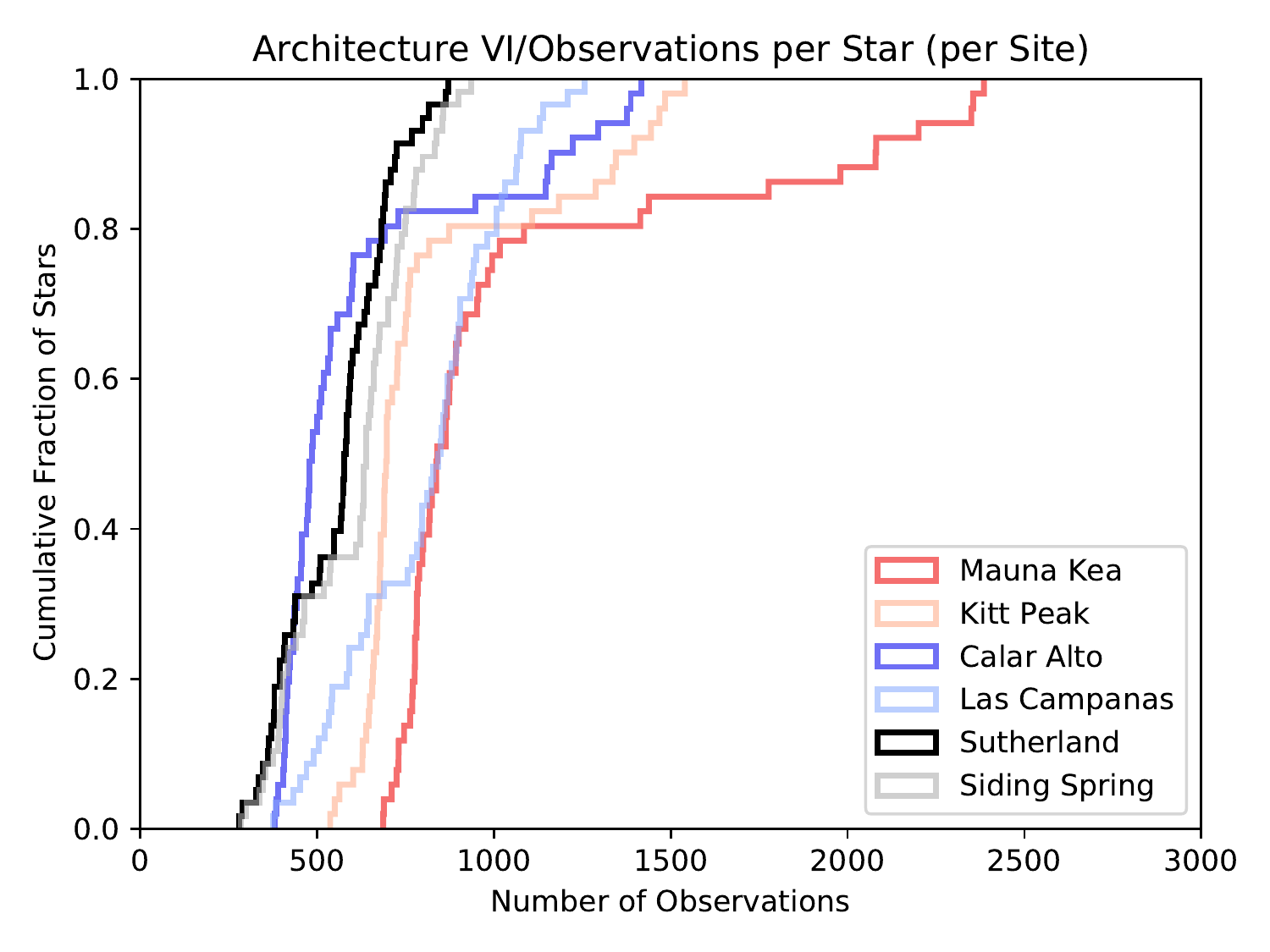}

\noindent \includegraphics[width=0.49\textwidth]{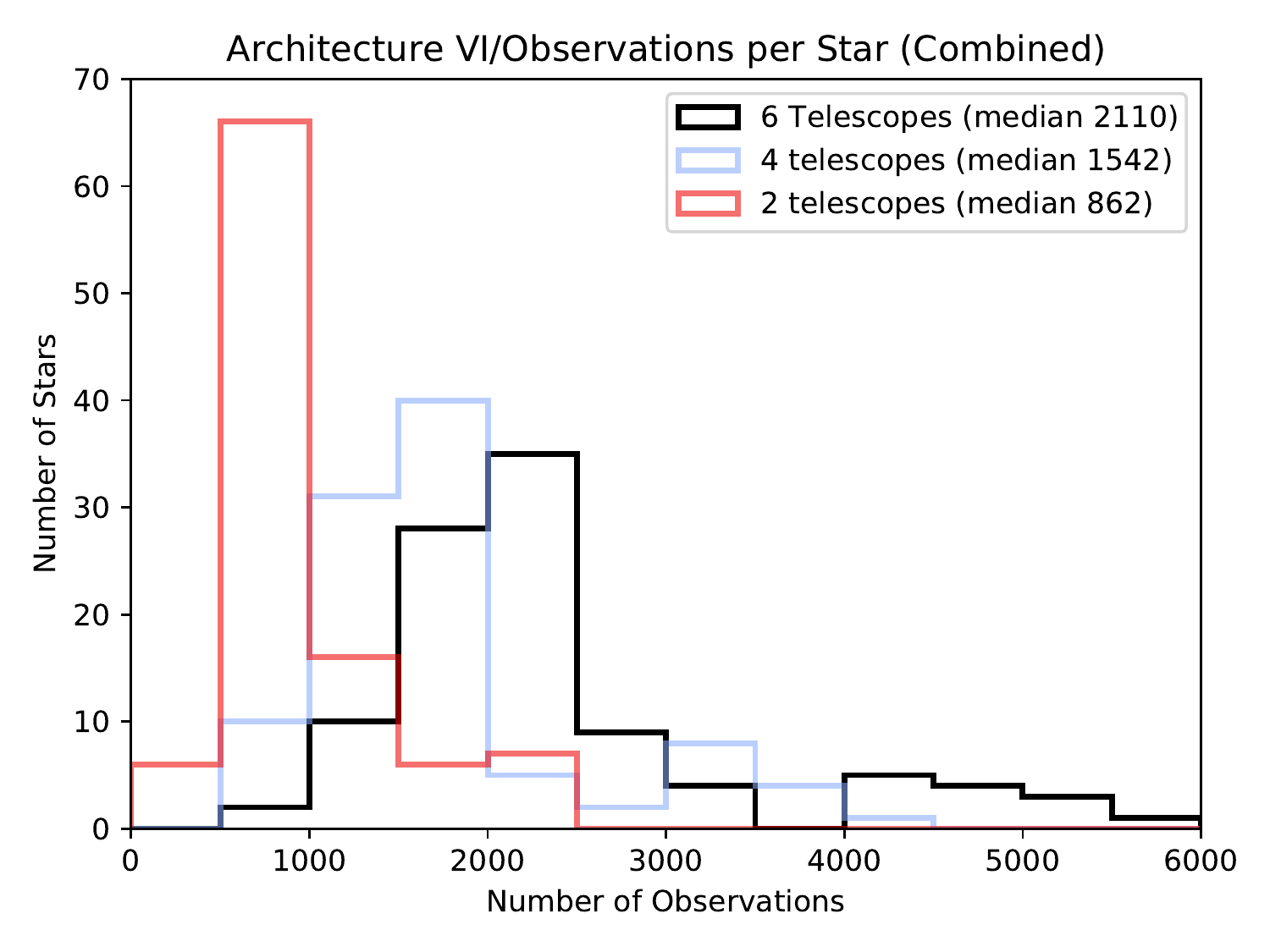}
\includegraphics[width=0.49\textwidth]{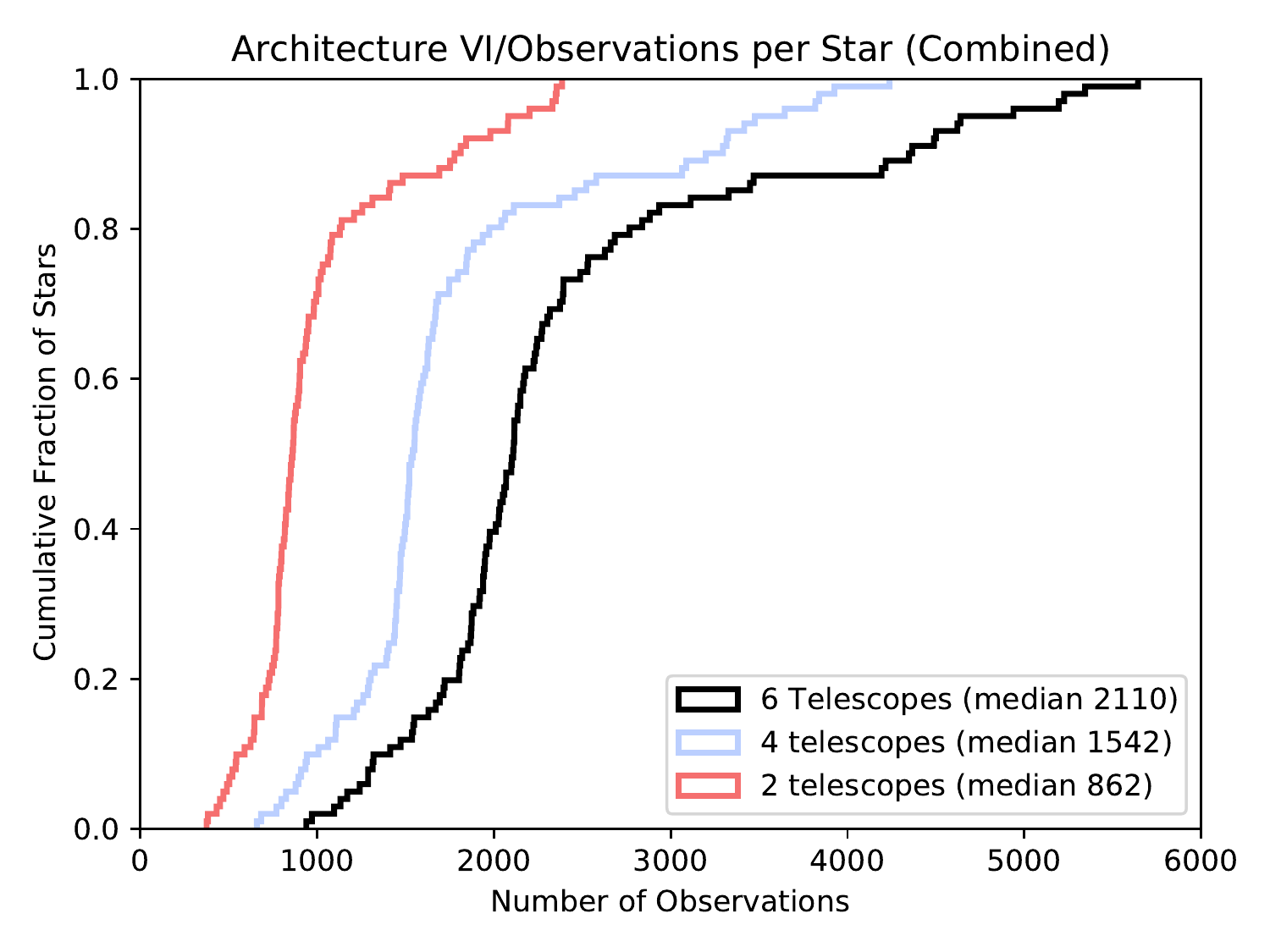}
\caption{Same as figure \ref{fig:ArchIobs}, but for architecture VI.}
\end{figure}

\begin{figure}
\noindent \includegraphics[width=0.49\textwidth]{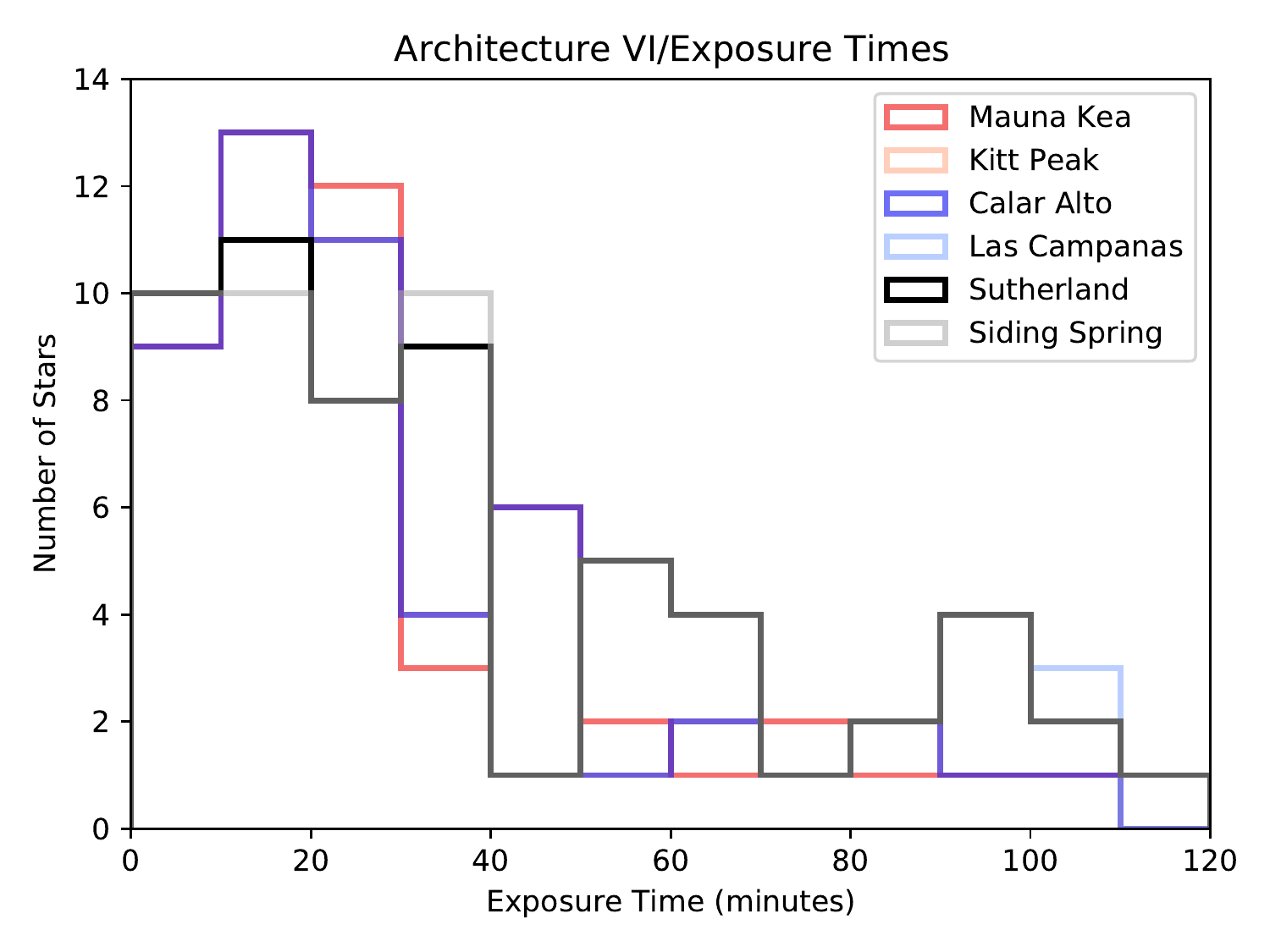}
\includegraphics[width=0.49\textwidth]{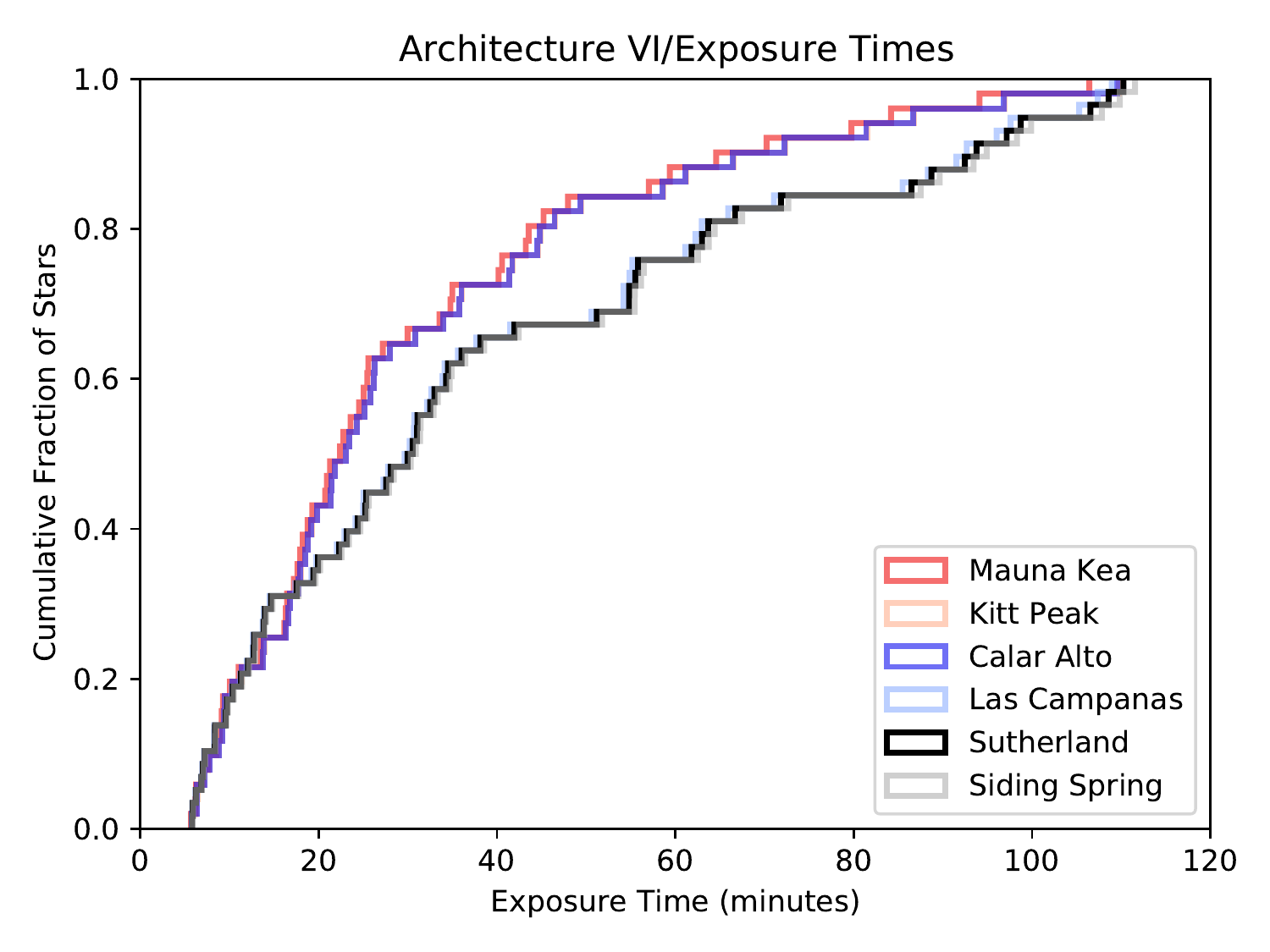}

\noindent \includegraphics[width=0.49\textwidth]{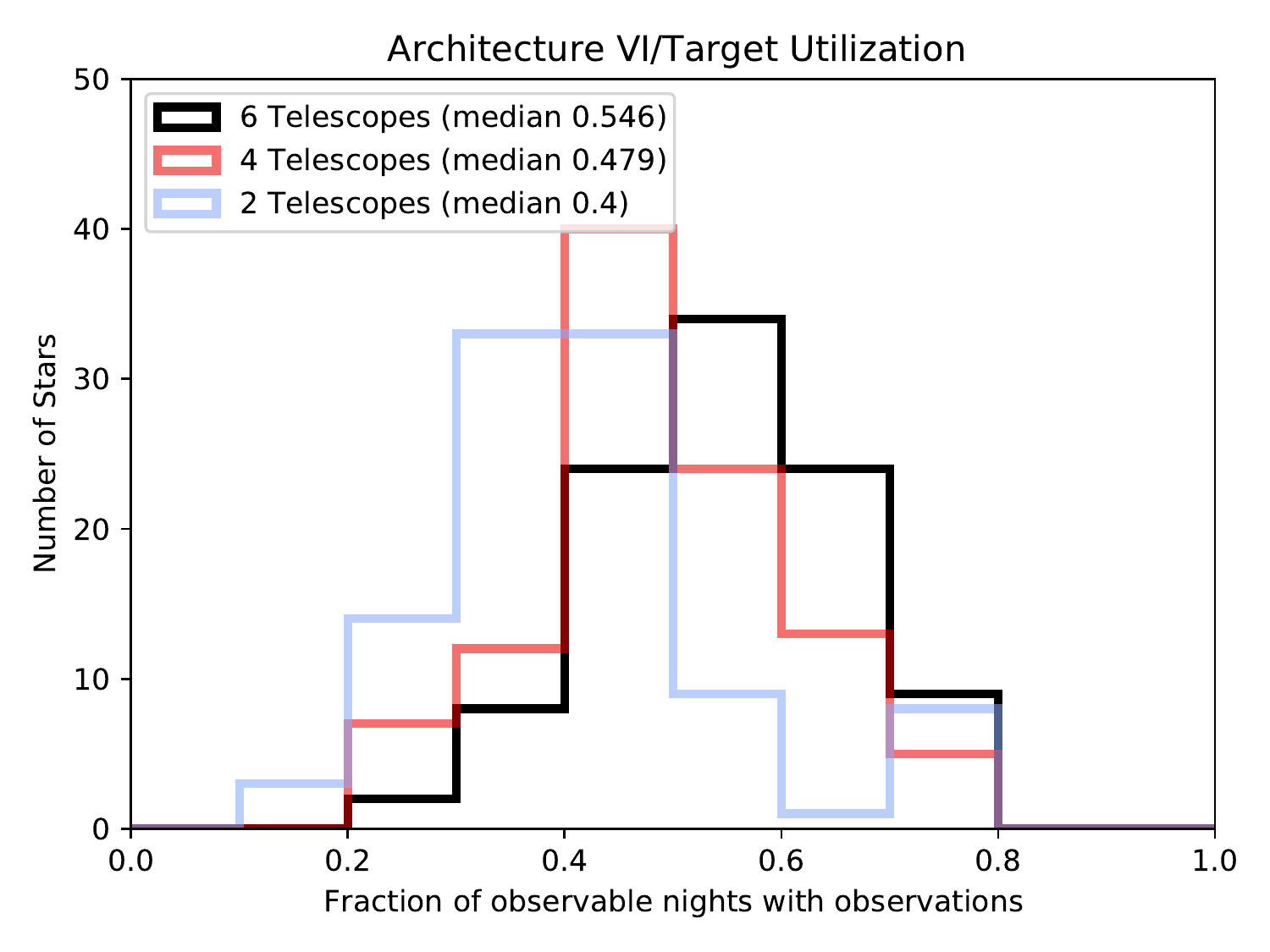}
\includegraphics[width=0.49\textwidth]{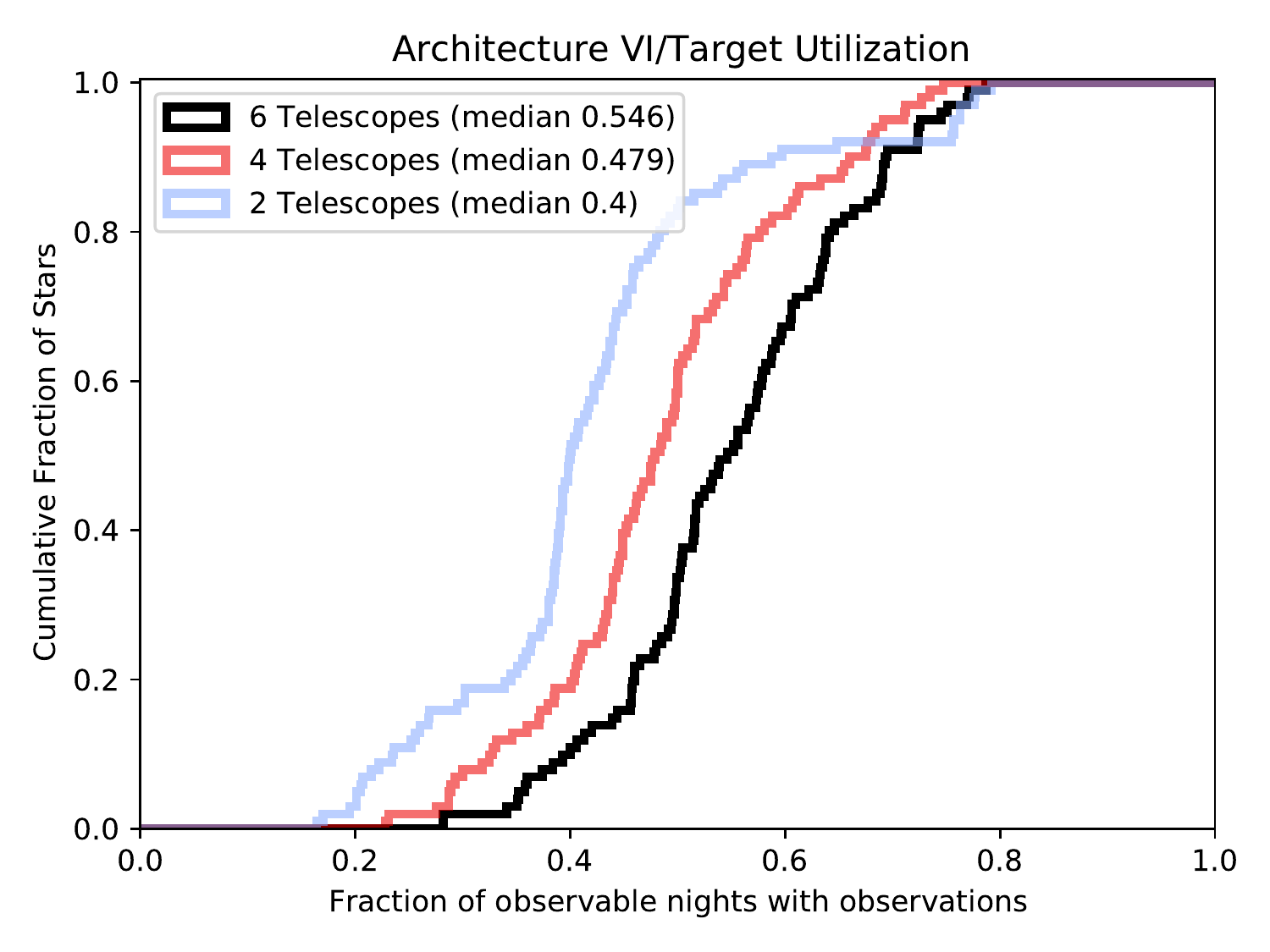}
\caption{Same as figure \ref{fig:ArchIexpfrac}, but for architecture VI.}
\end{figure}

\begin{figure}
\noindent \includegraphics[width=0.49\textwidth]{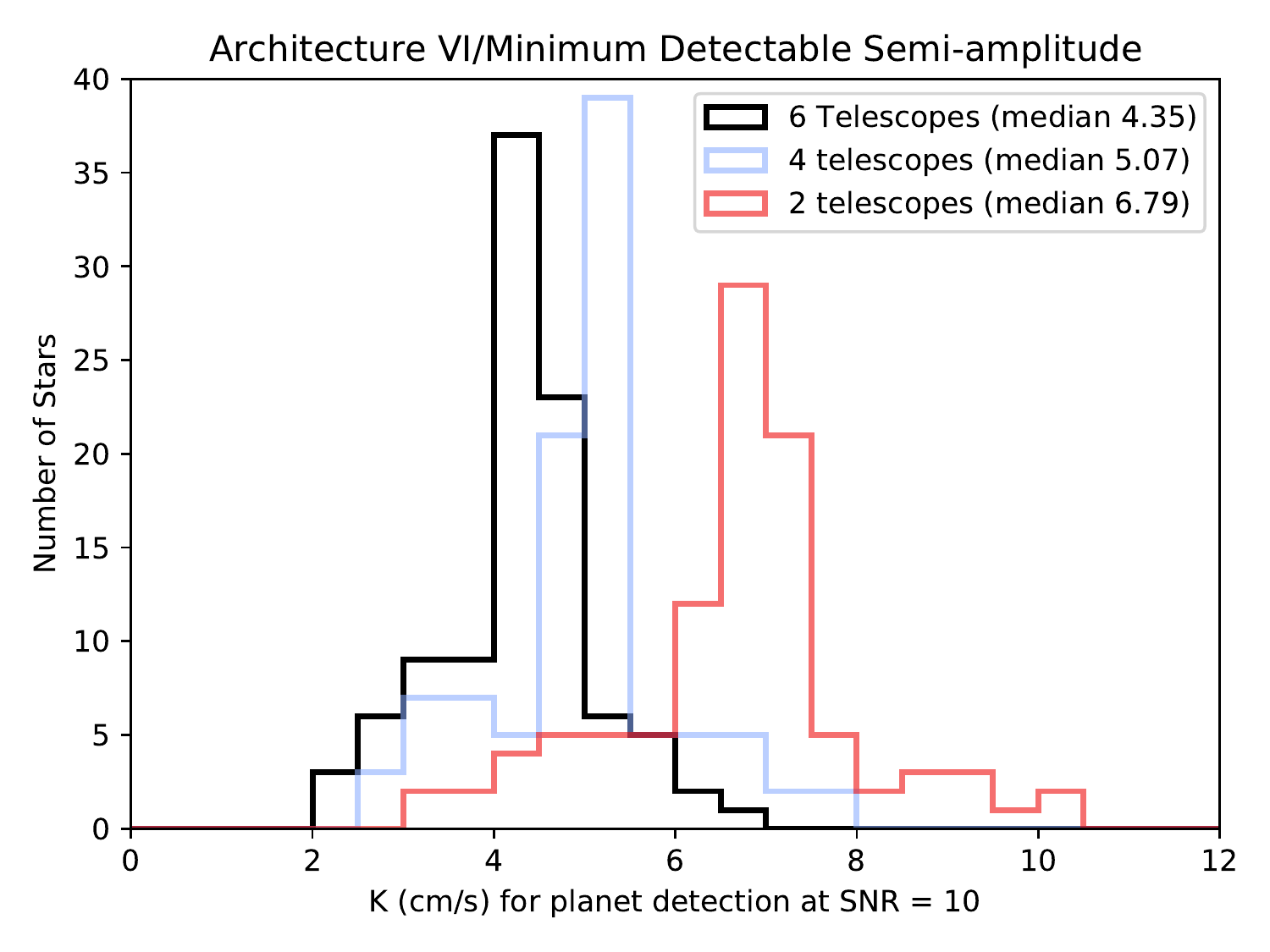}
\includegraphics[width=0.49\textwidth]{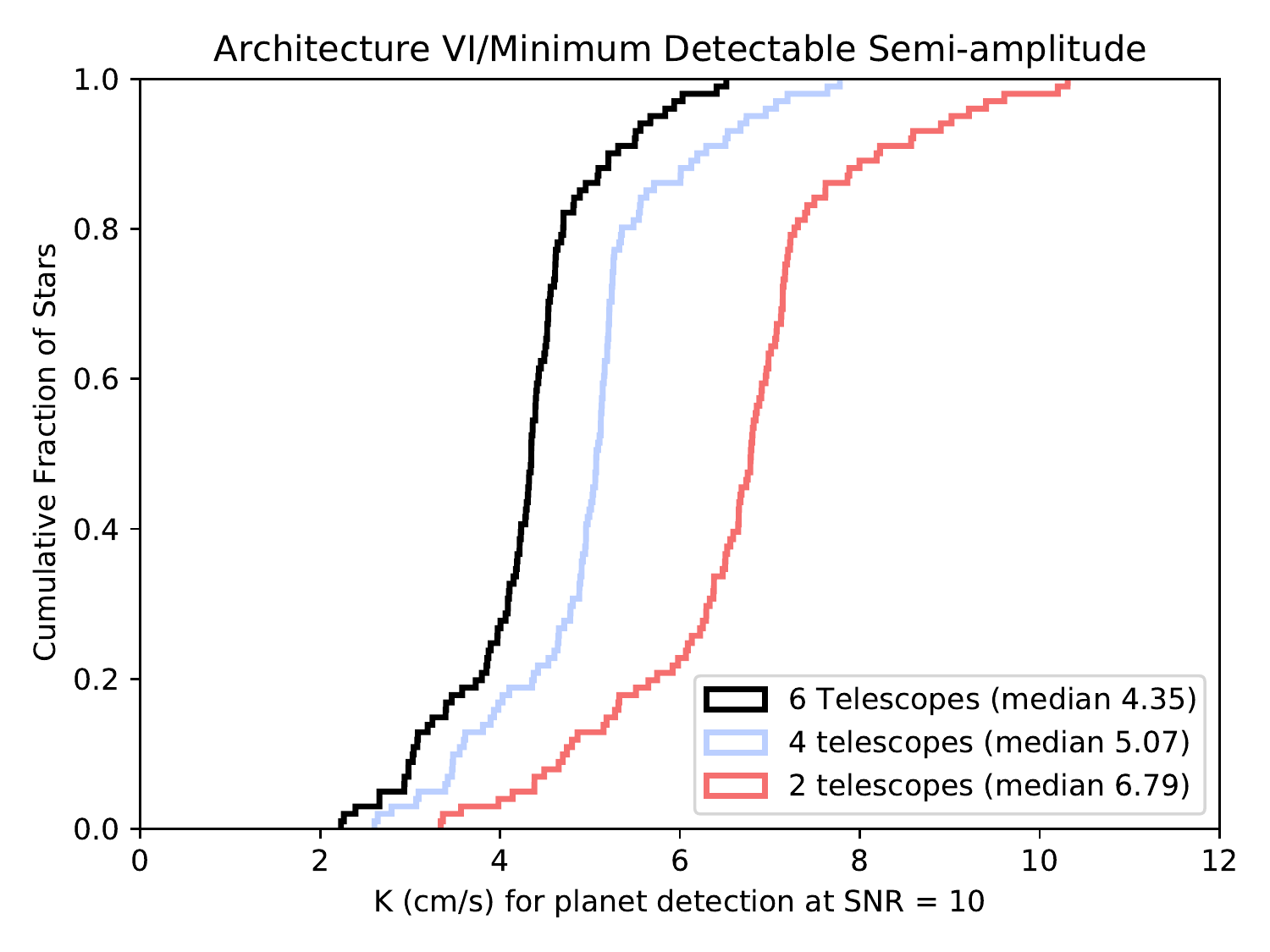}

\noindent \includegraphics[width=0.49\textwidth]{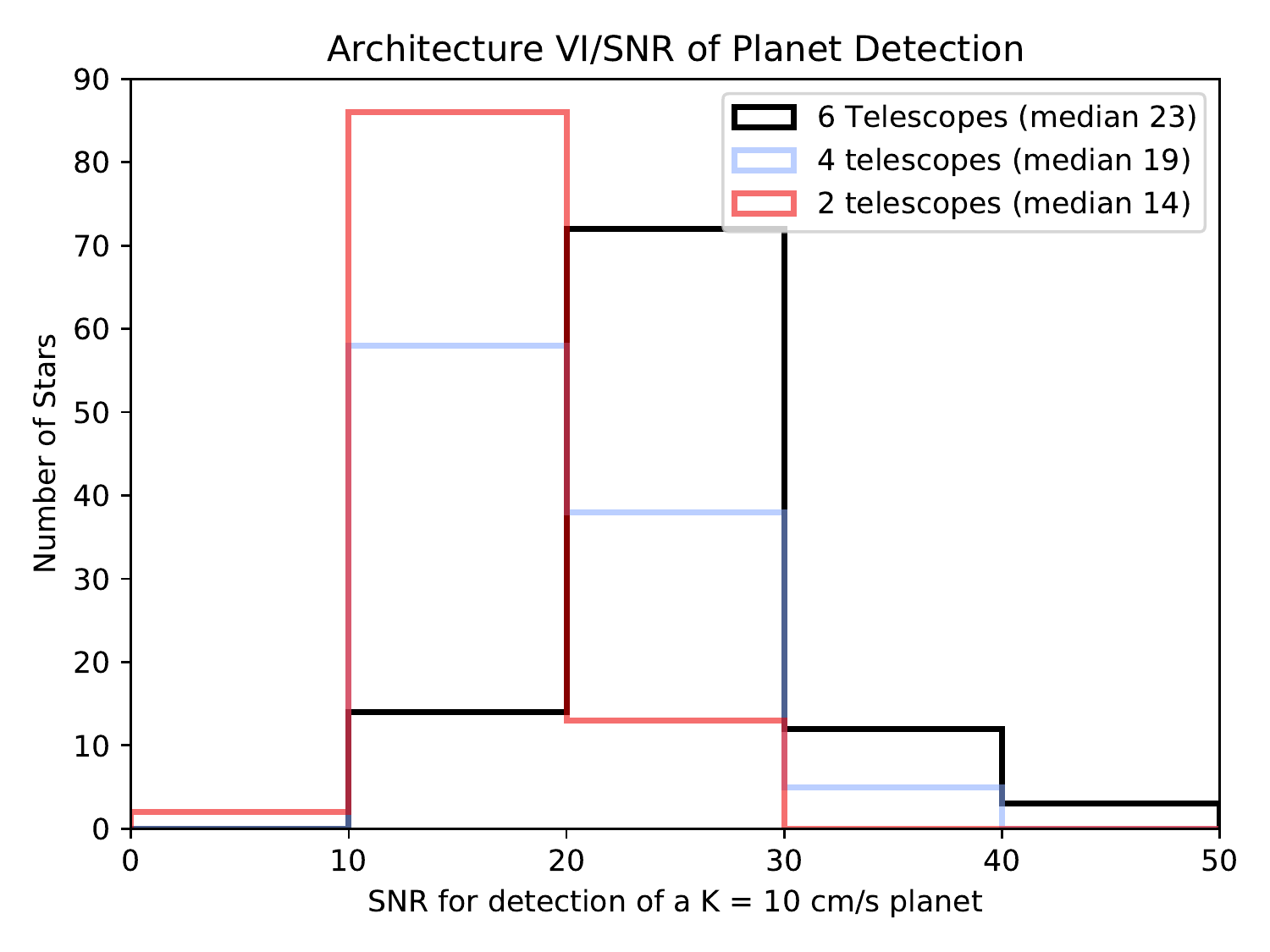}
\includegraphics[width=0.49\textwidth]{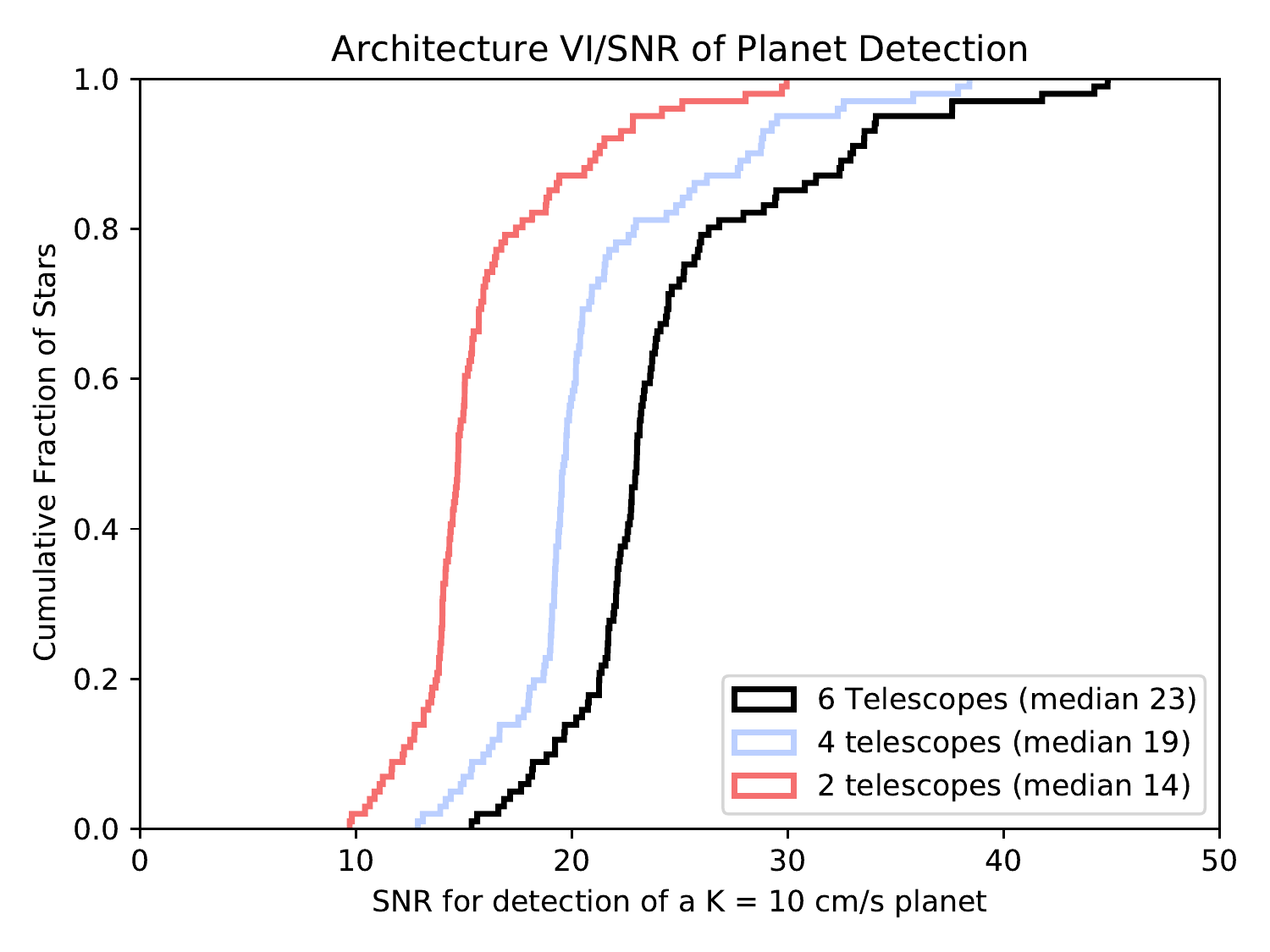}
\caption{Same as figure \ref{fig:ArchIkSNR}, but for architecture VI.}
\end{figure}

\begin{figure}
\noindent \includegraphics[width=0.49\textwidth]{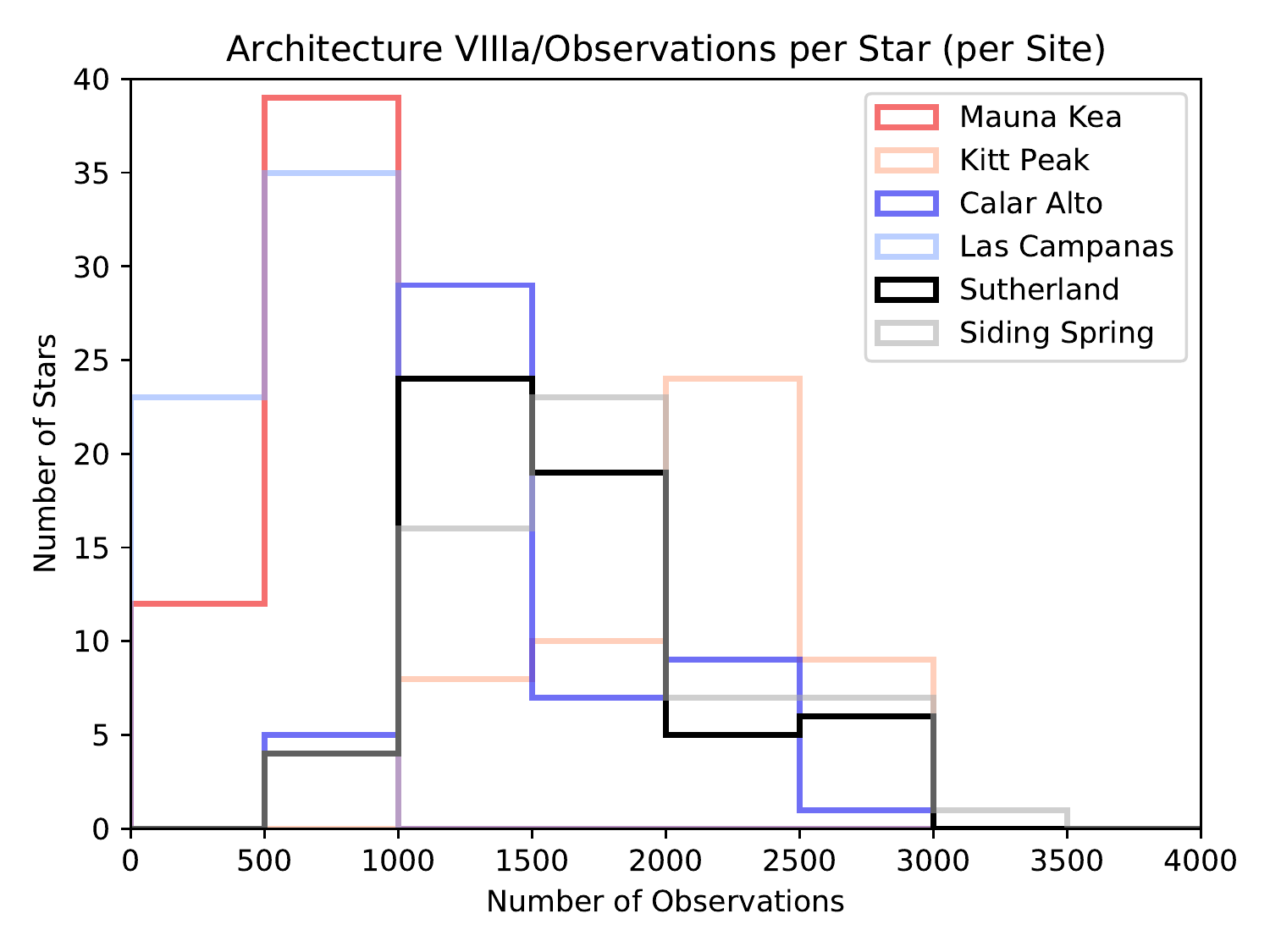}
\includegraphics[width=0.49\textwidth]{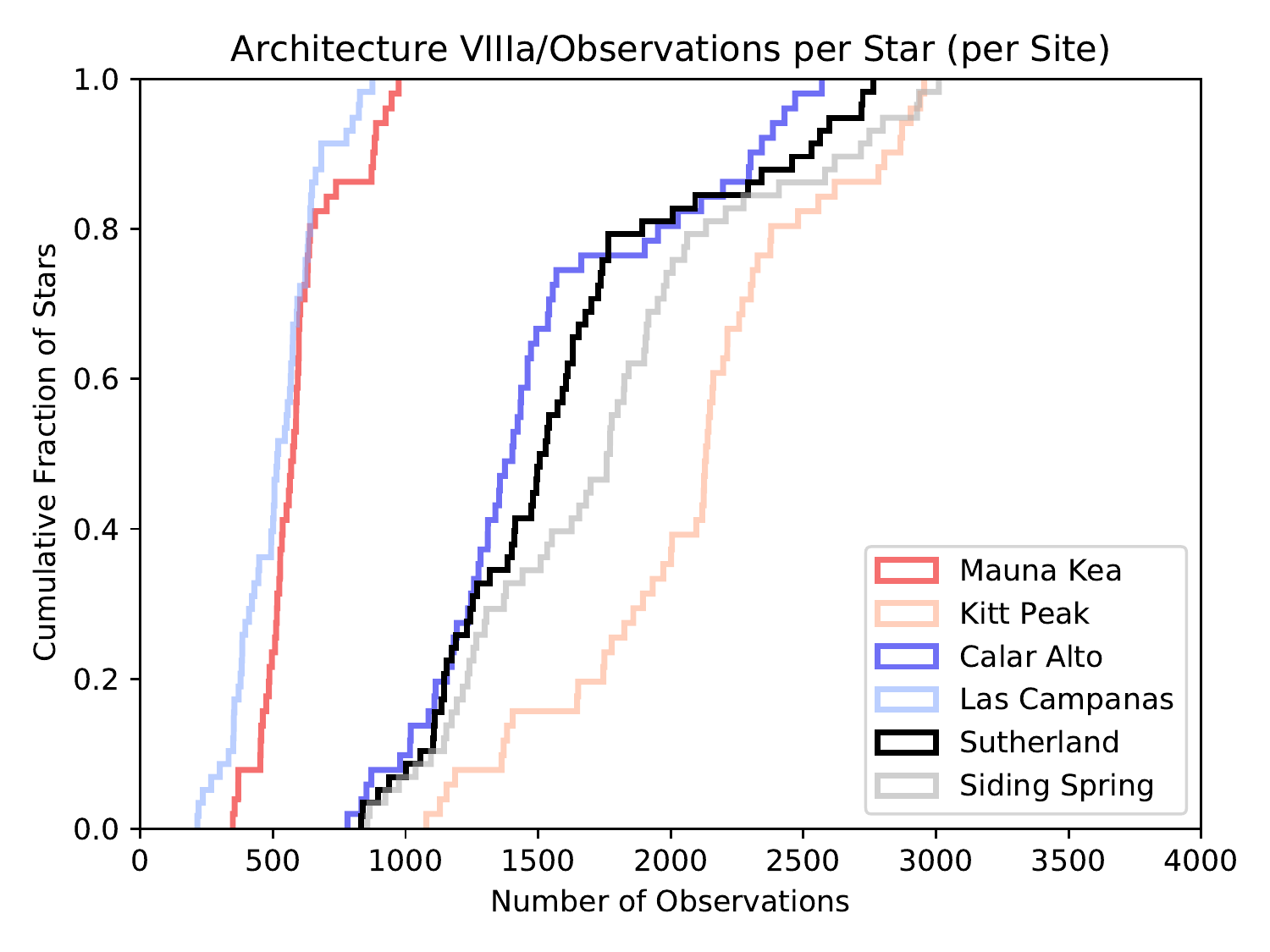}

\noindent \includegraphics[width=0.49\textwidth]{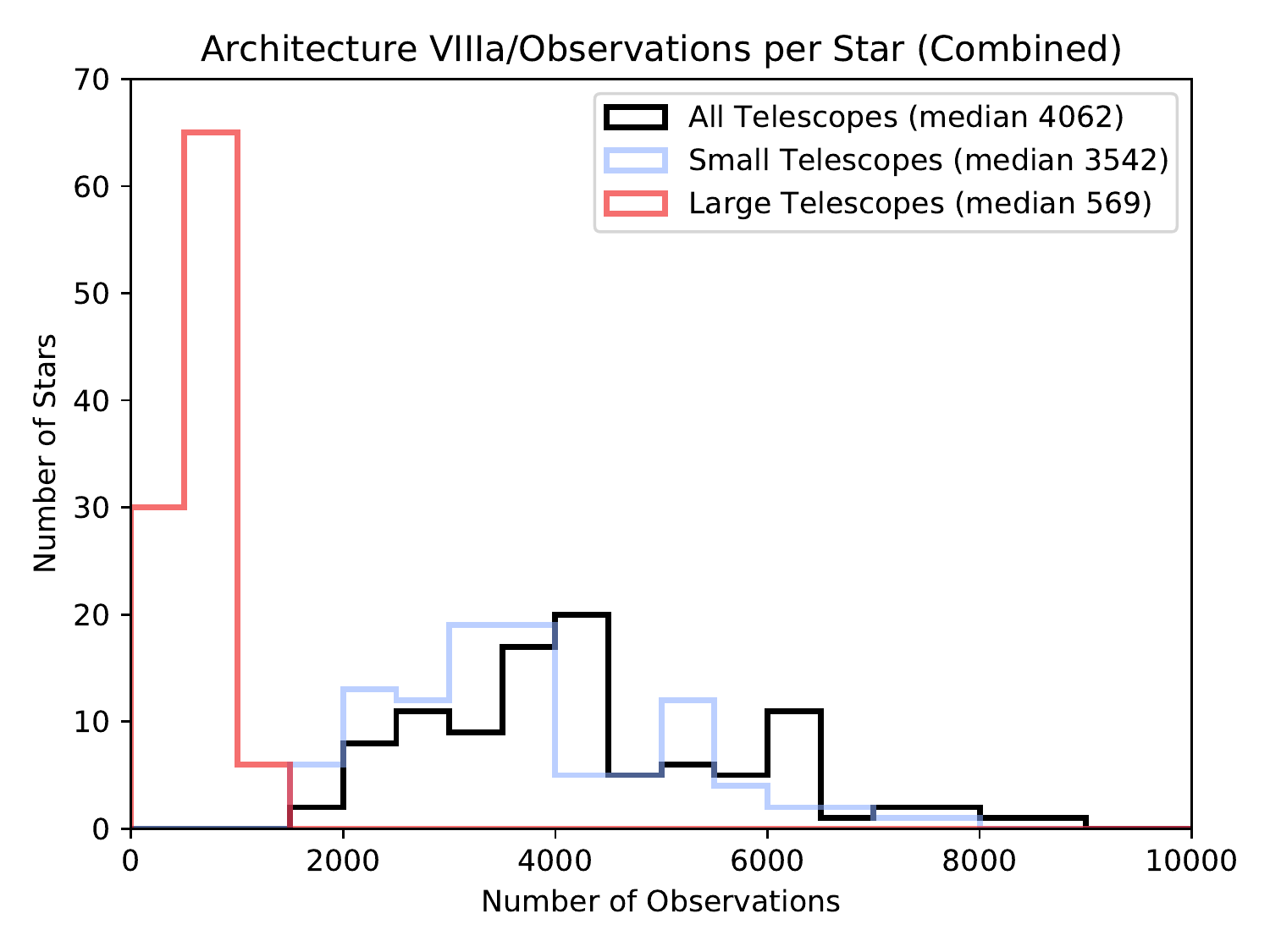}
\includegraphics[width=0.49\textwidth]{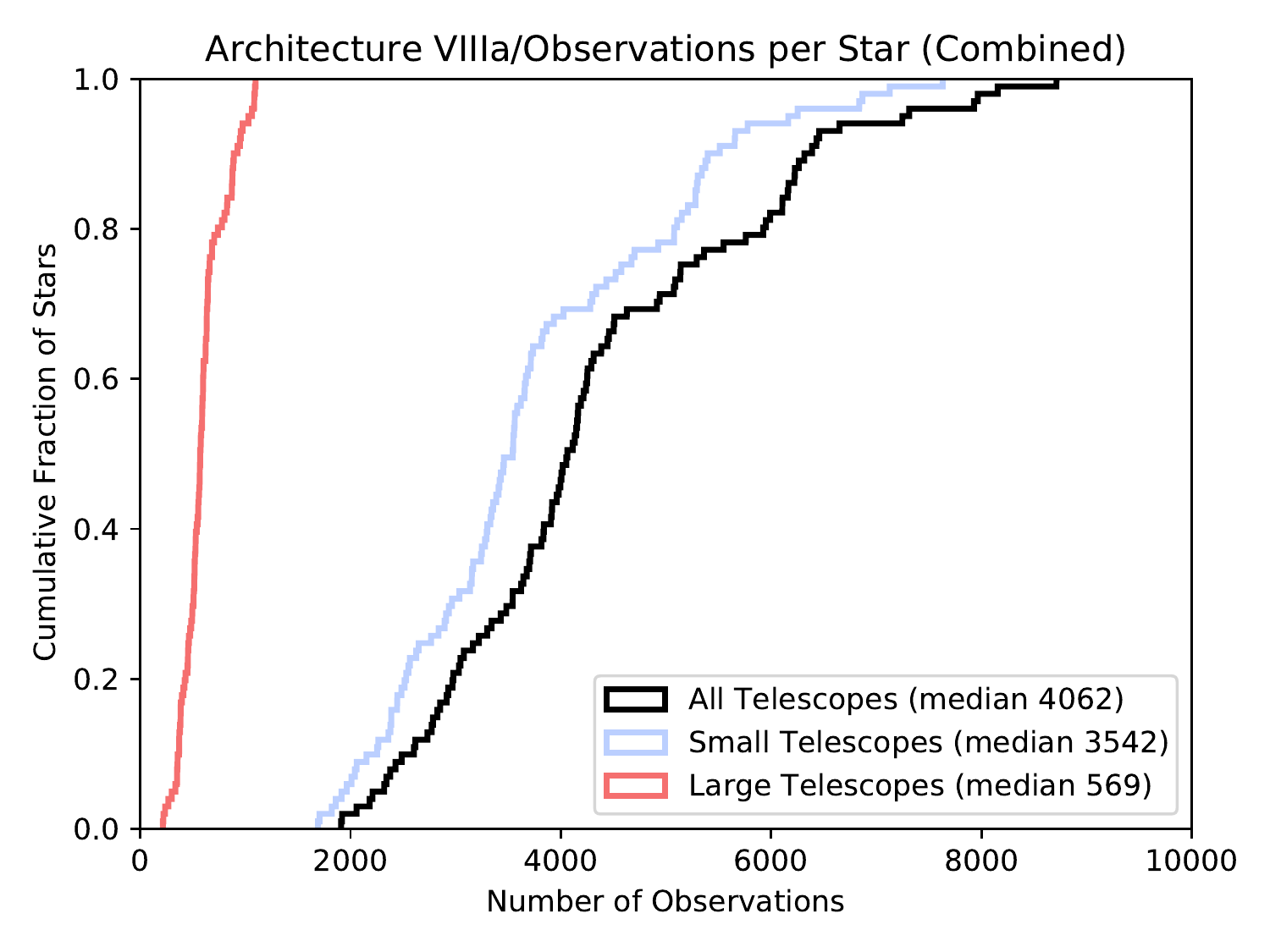}
\caption{Same as figure \ref{fig:ArchIobs}, but for architecture VIIIa.}
\end{figure}

\begin{figure}
\noindent \includegraphics[width=0.49\textwidth]{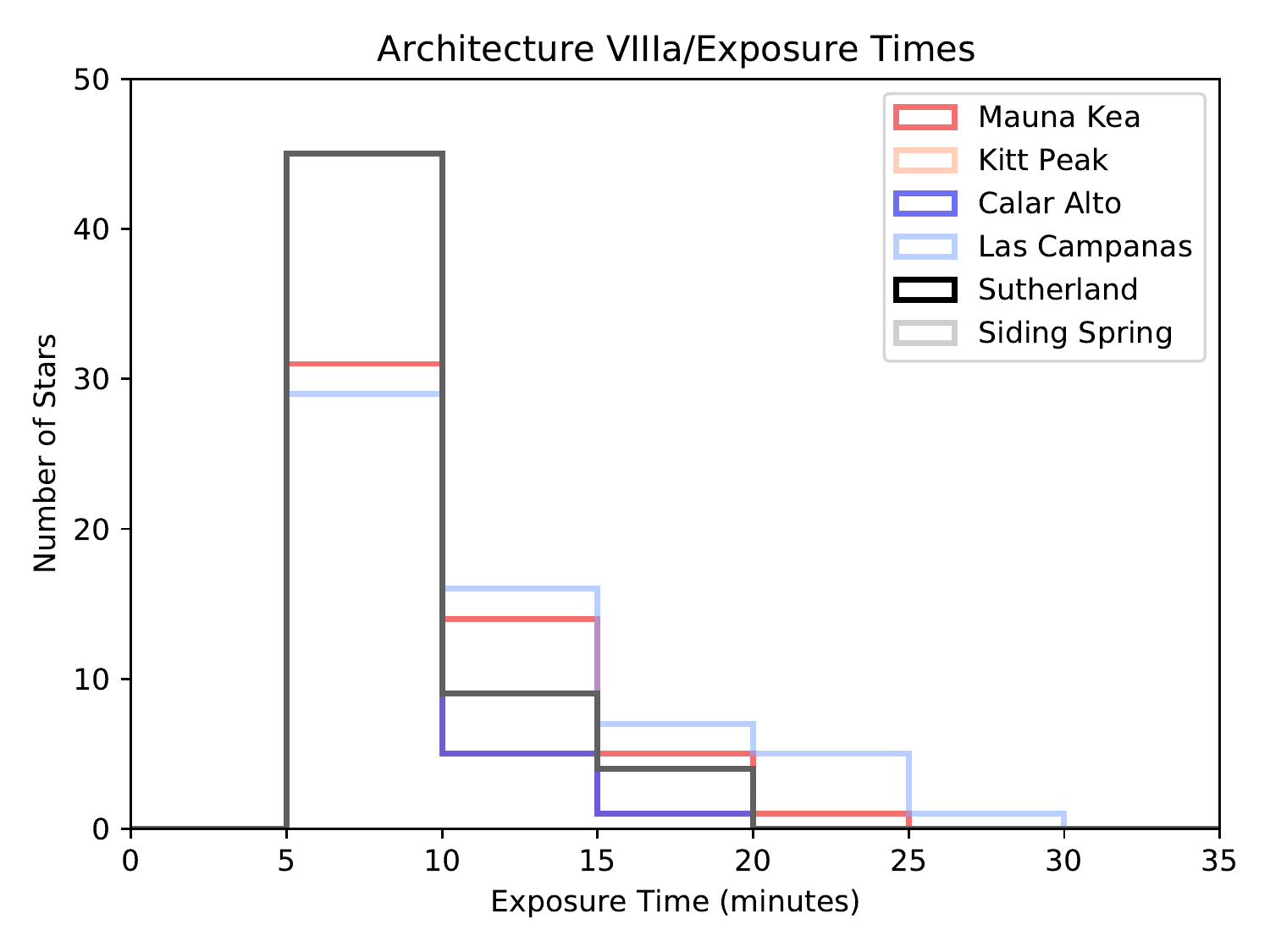}
\includegraphics[width=0.49\textwidth]{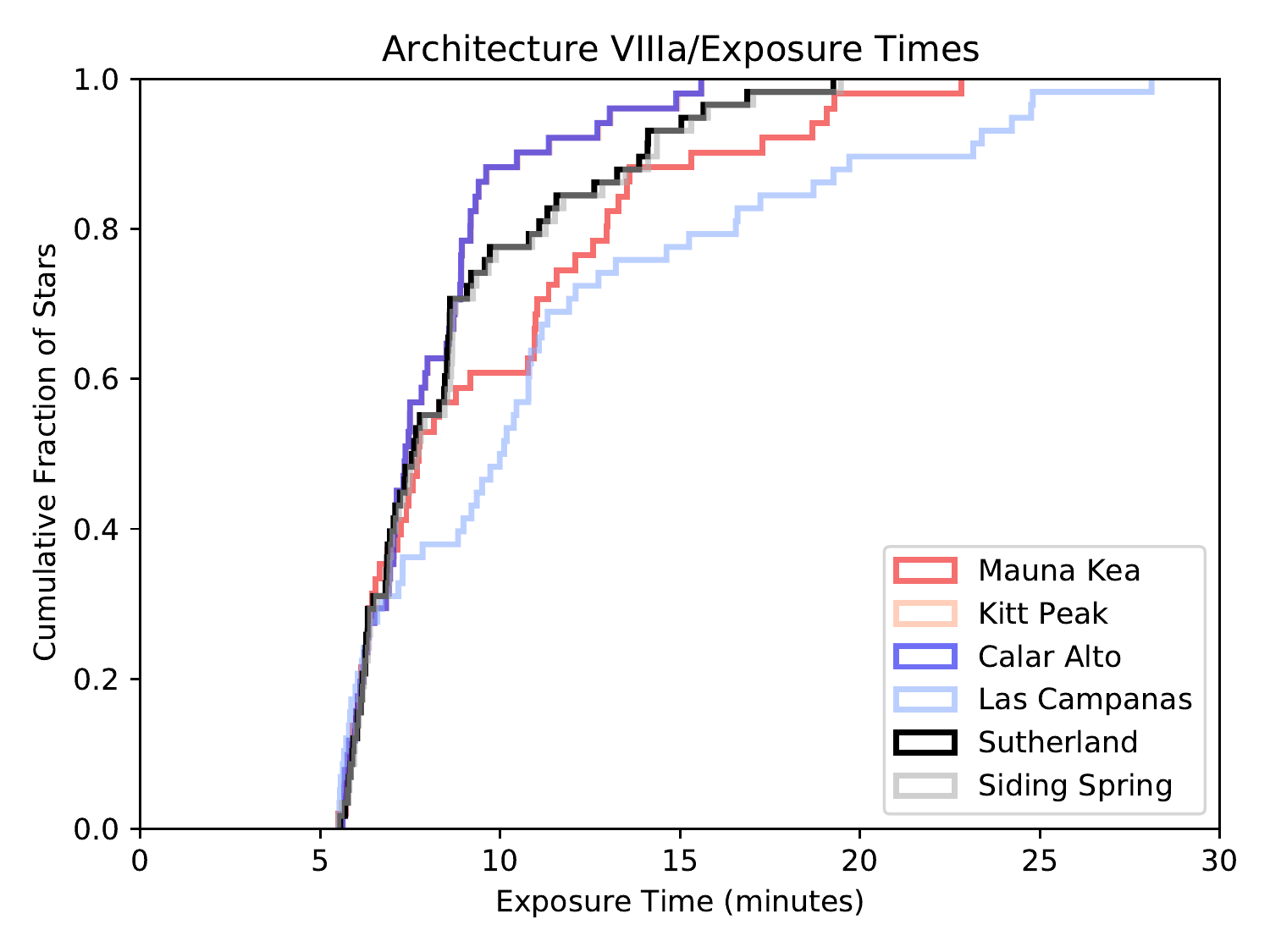}

\noindent \includegraphics[width=0.49\textwidth]{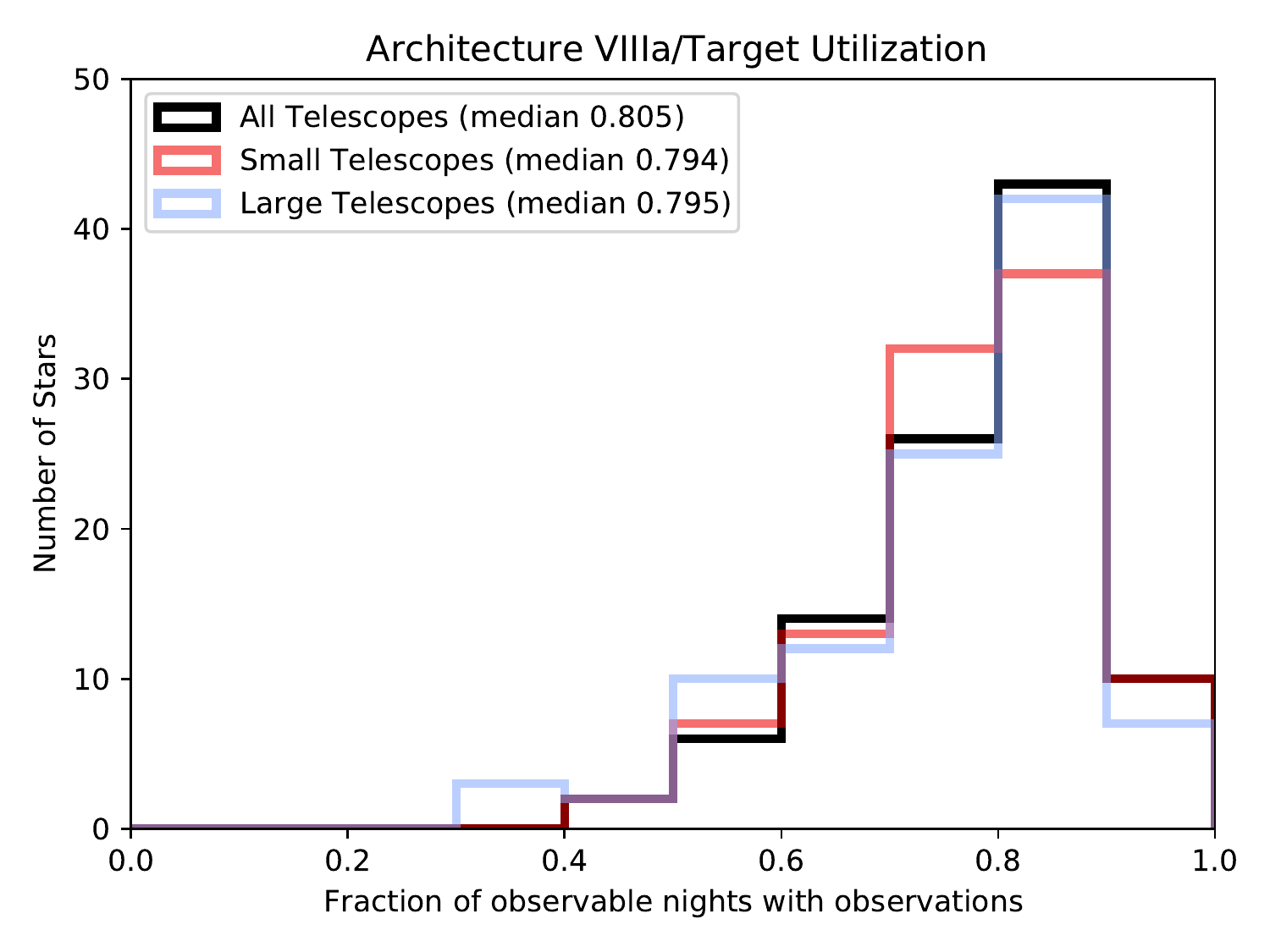}
\includegraphics[width=0.49\textwidth]{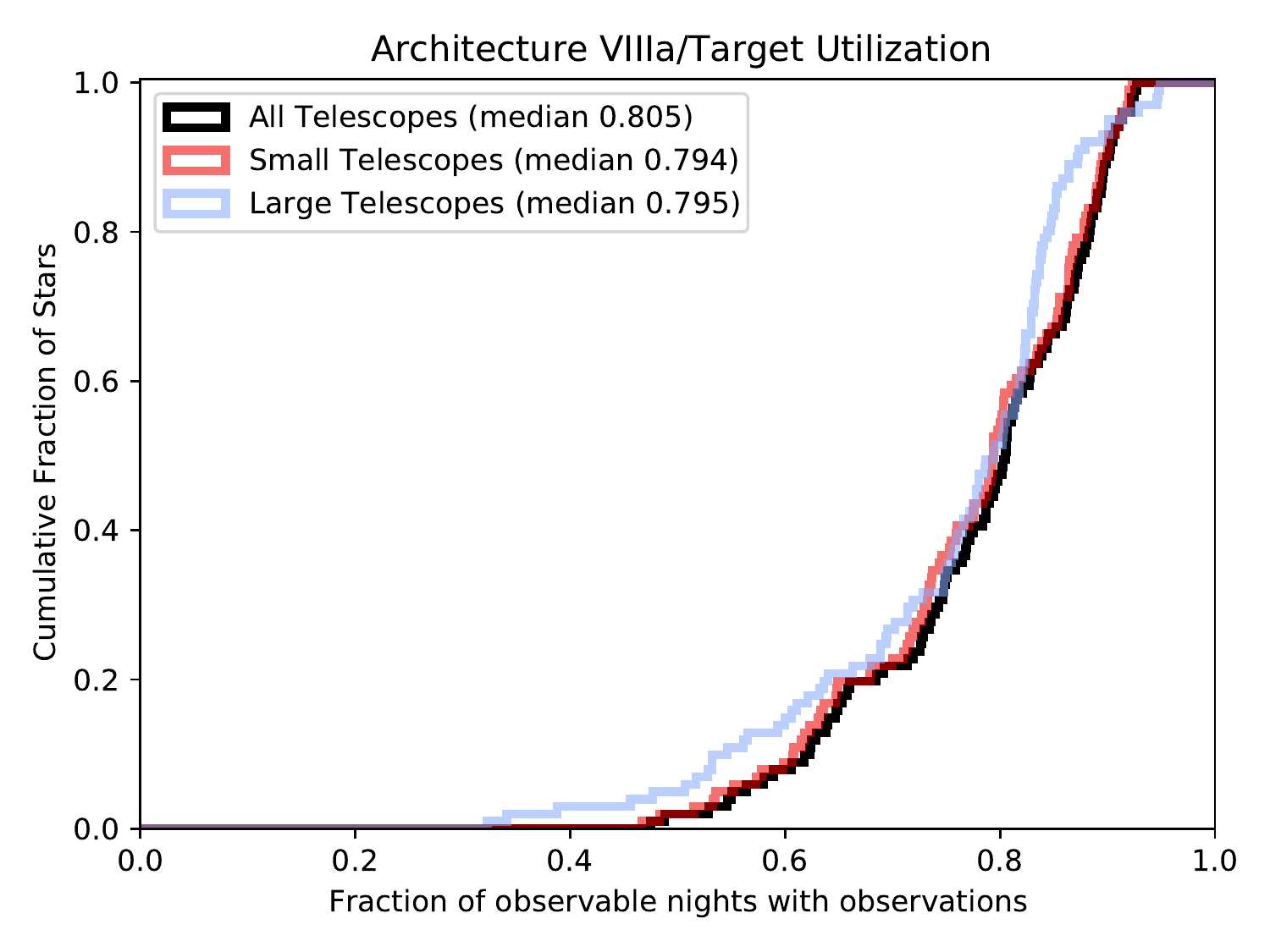}
\caption{Same as figure \ref{fig:ArchIexpfrac}, but for architecture VIIIa.}
\end{figure}

\begin{figure}
\noindent \includegraphics[width=0.49\textwidth]{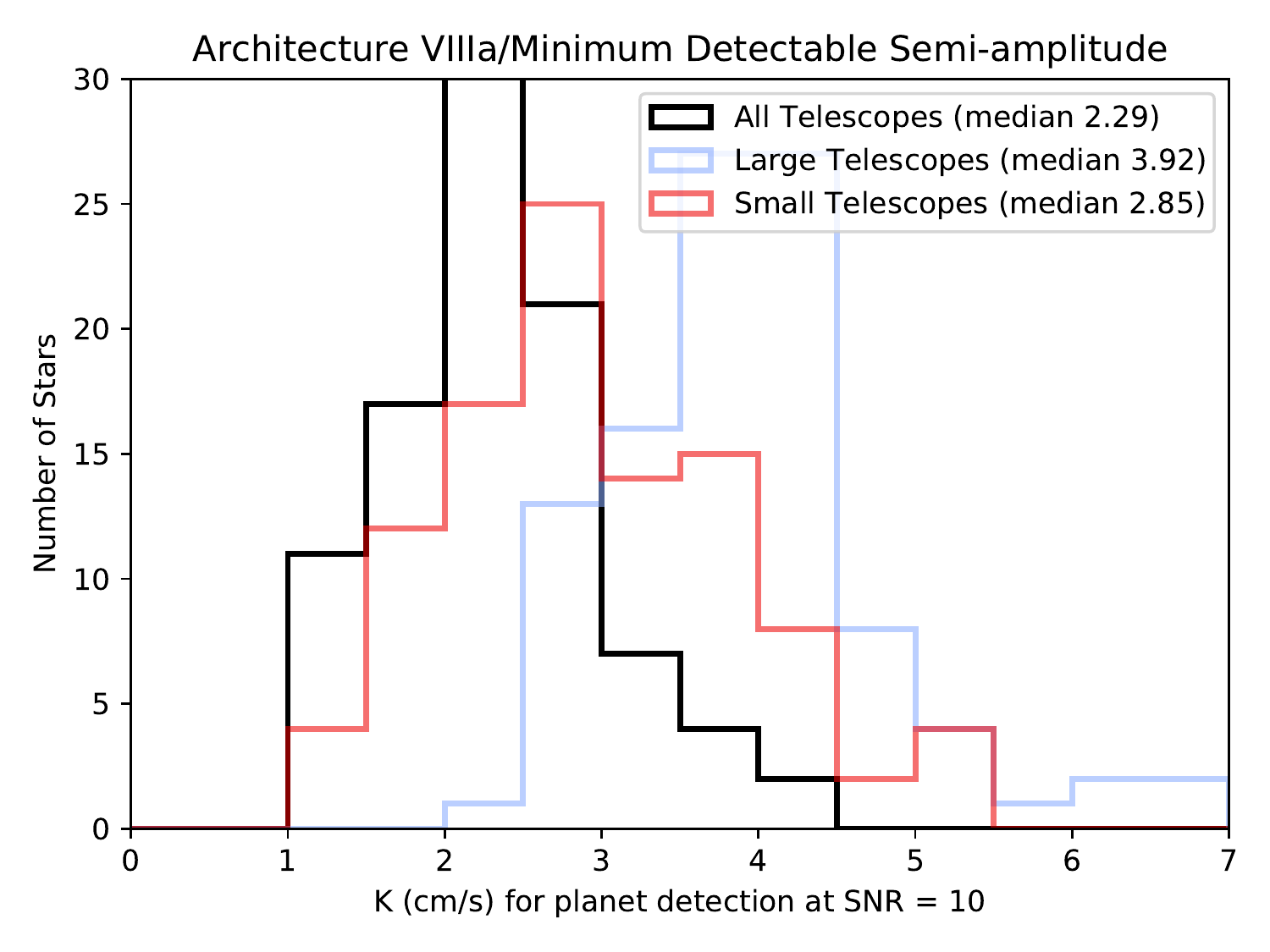}
\includegraphics[width=0.49\textwidth]{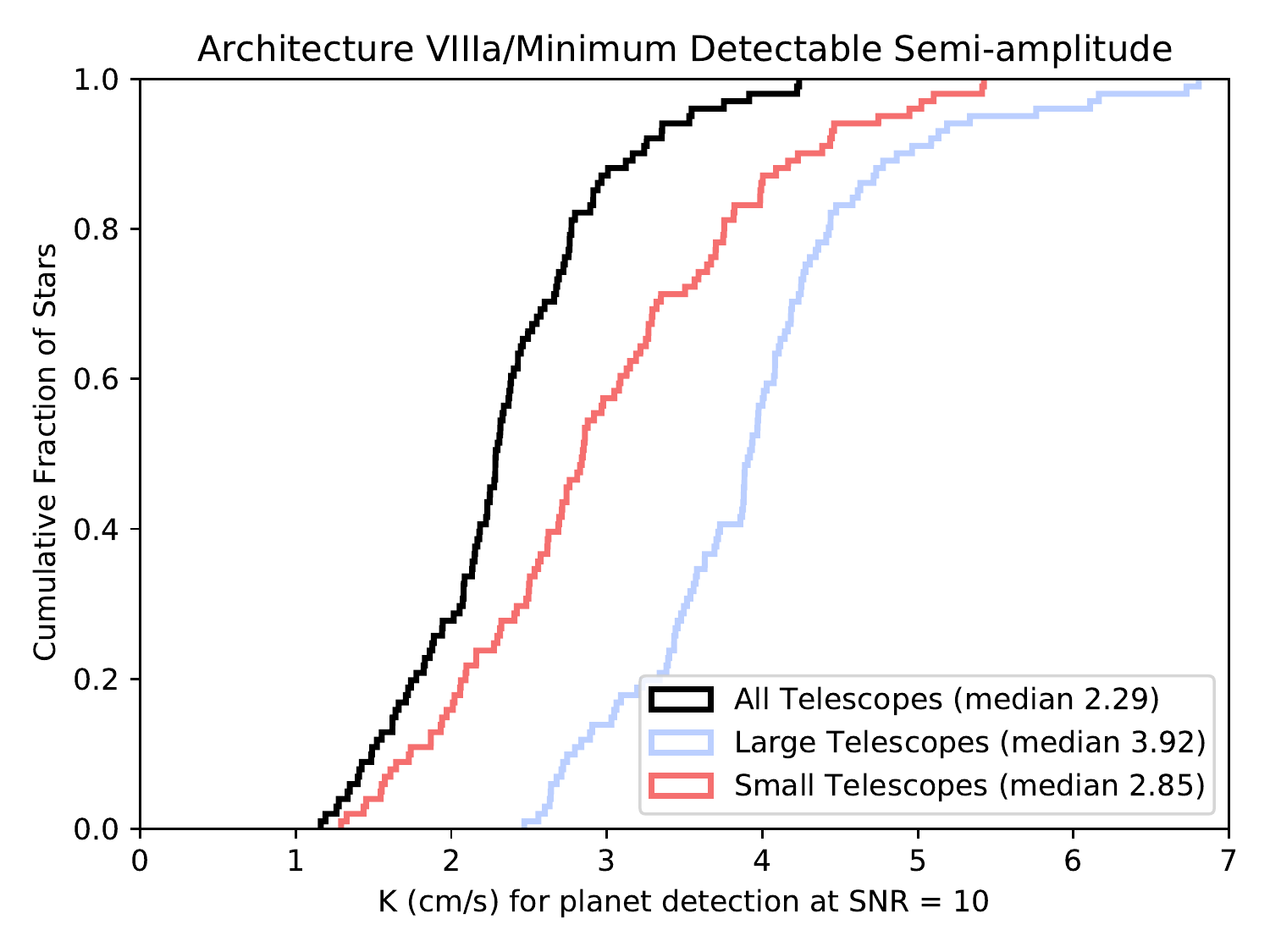}

\noindent \includegraphics[width=0.49\textwidth]{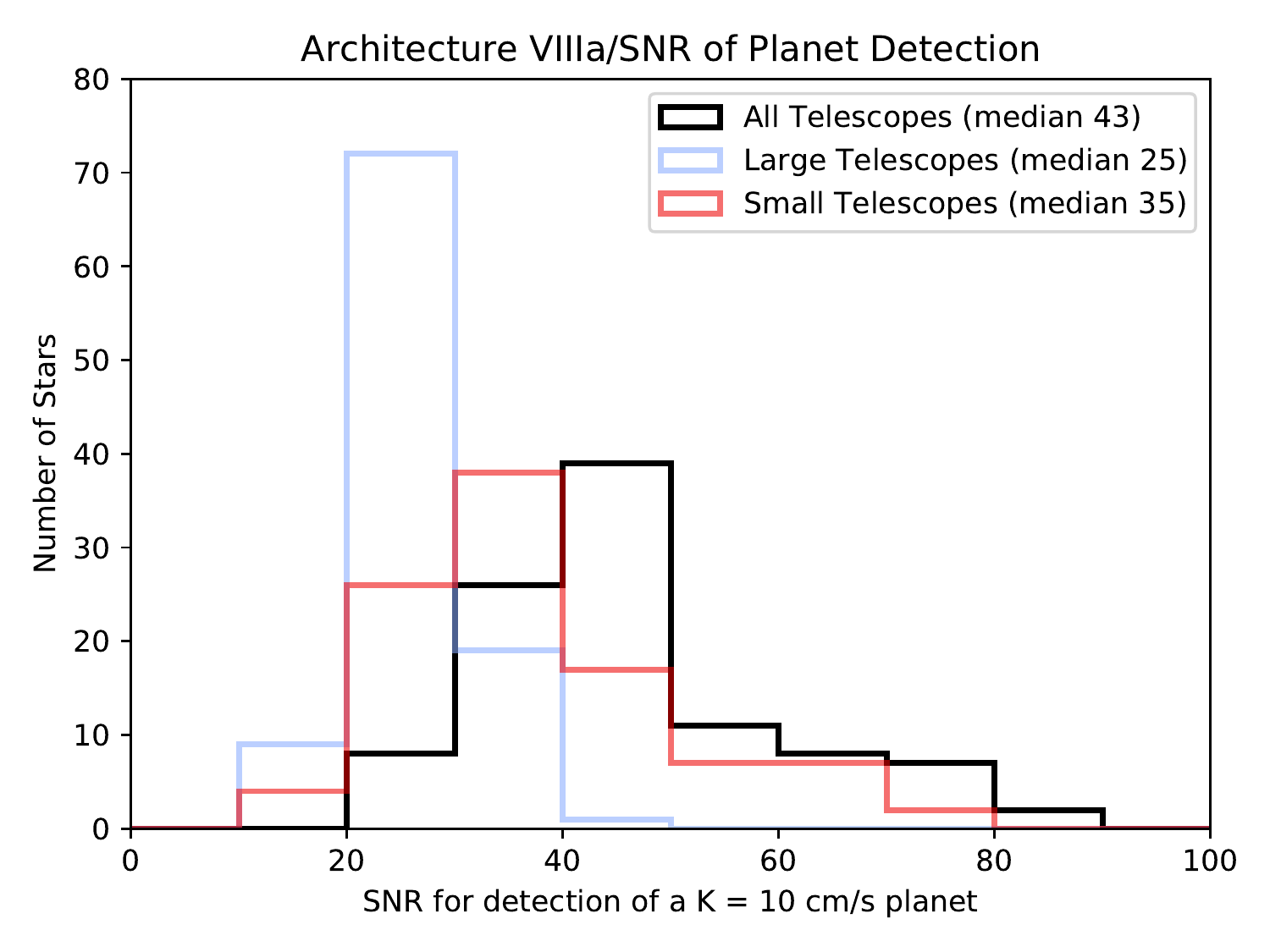}
\includegraphics[width=0.49\textwidth]{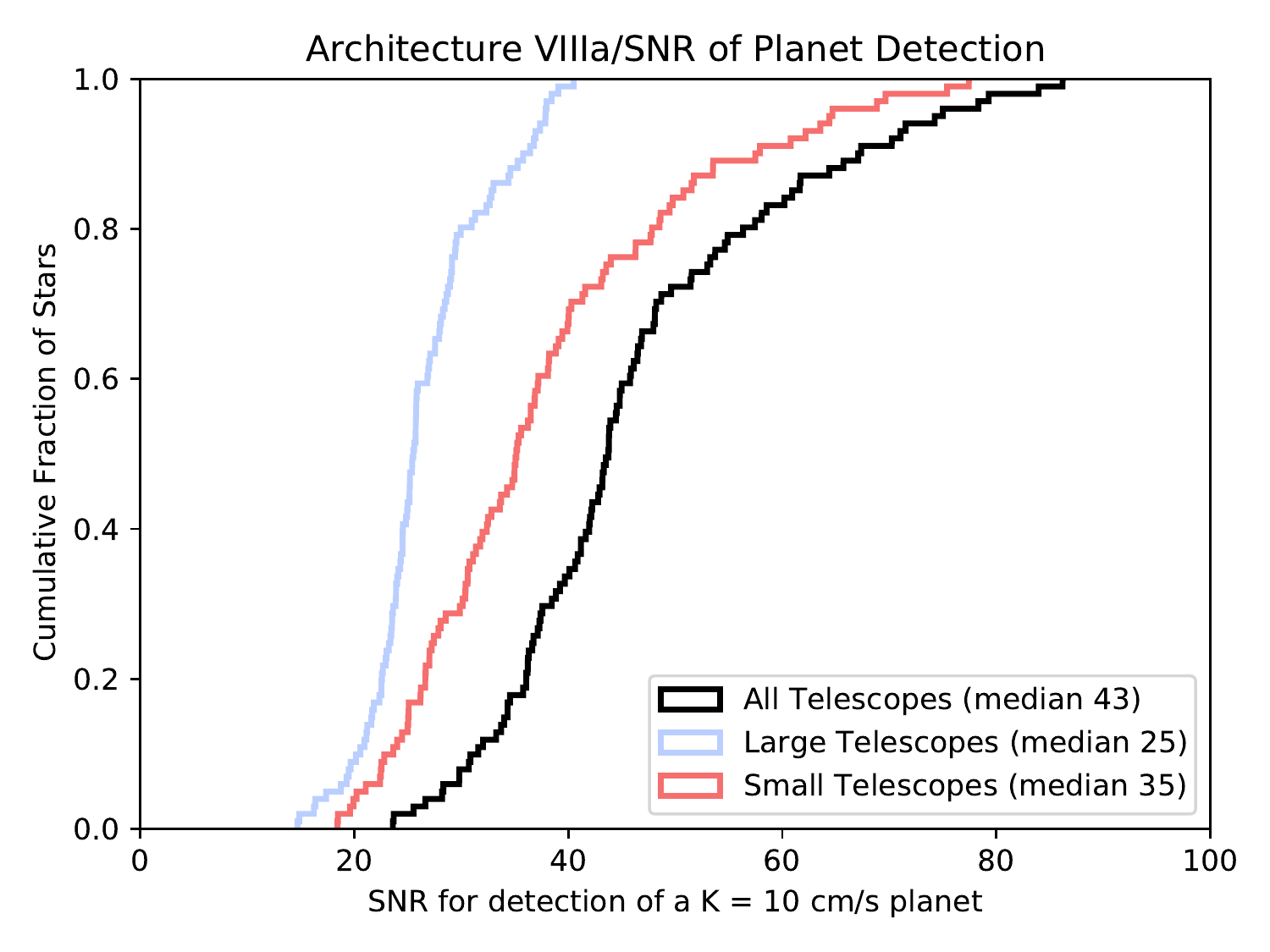}
\caption{Same as figure \ref{fig:ArchIkSNR}, but for architecture VIIIa. This includes using equation \ref{eqn:SNR2} for finding the sensitivity due to the varying precision of the measurements.}
\end{figure}

\begin{figure}
\noindent \includegraphics[width=0.49\textwidth]{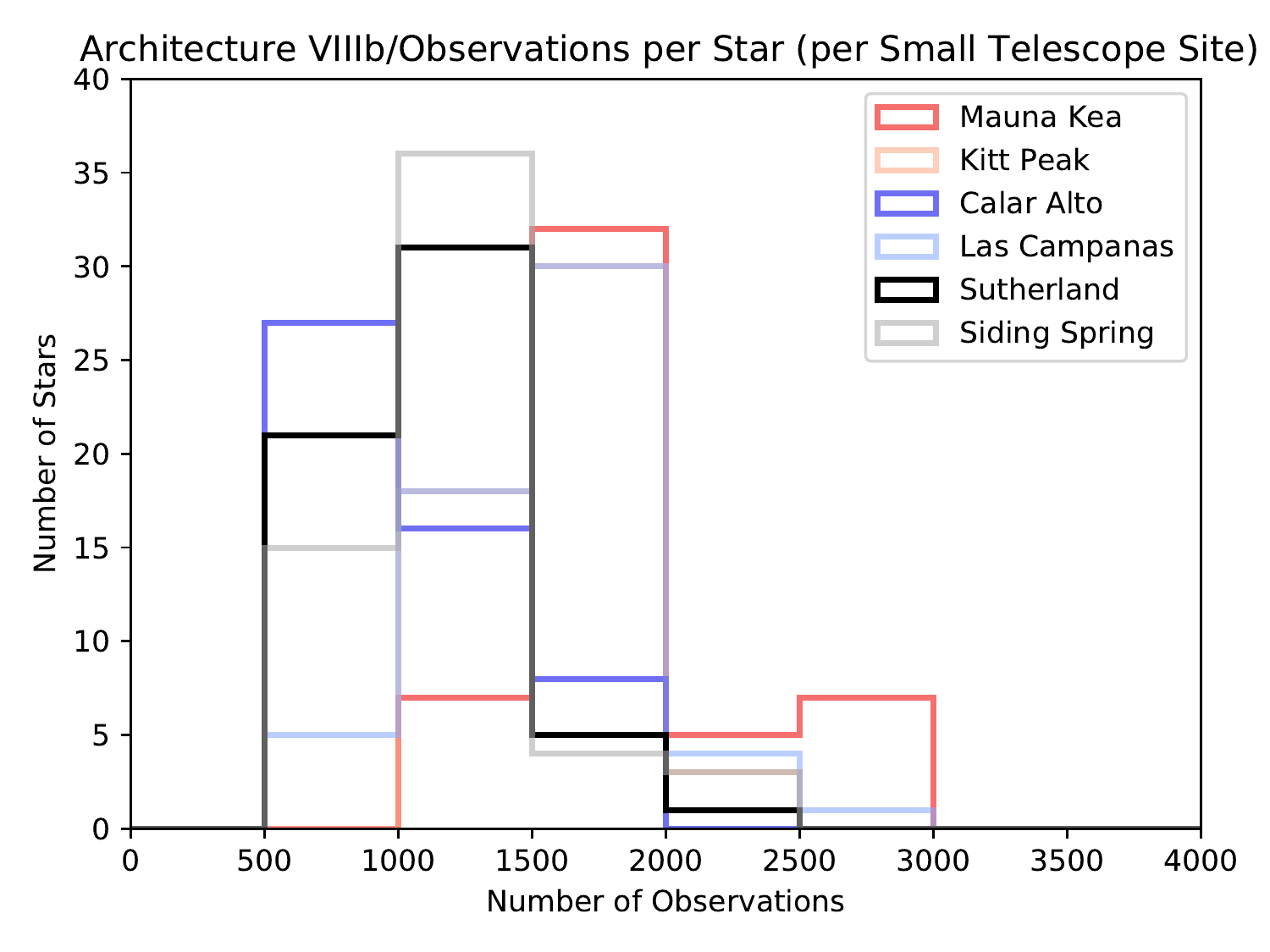}
\includegraphics[width=0.49\textwidth]{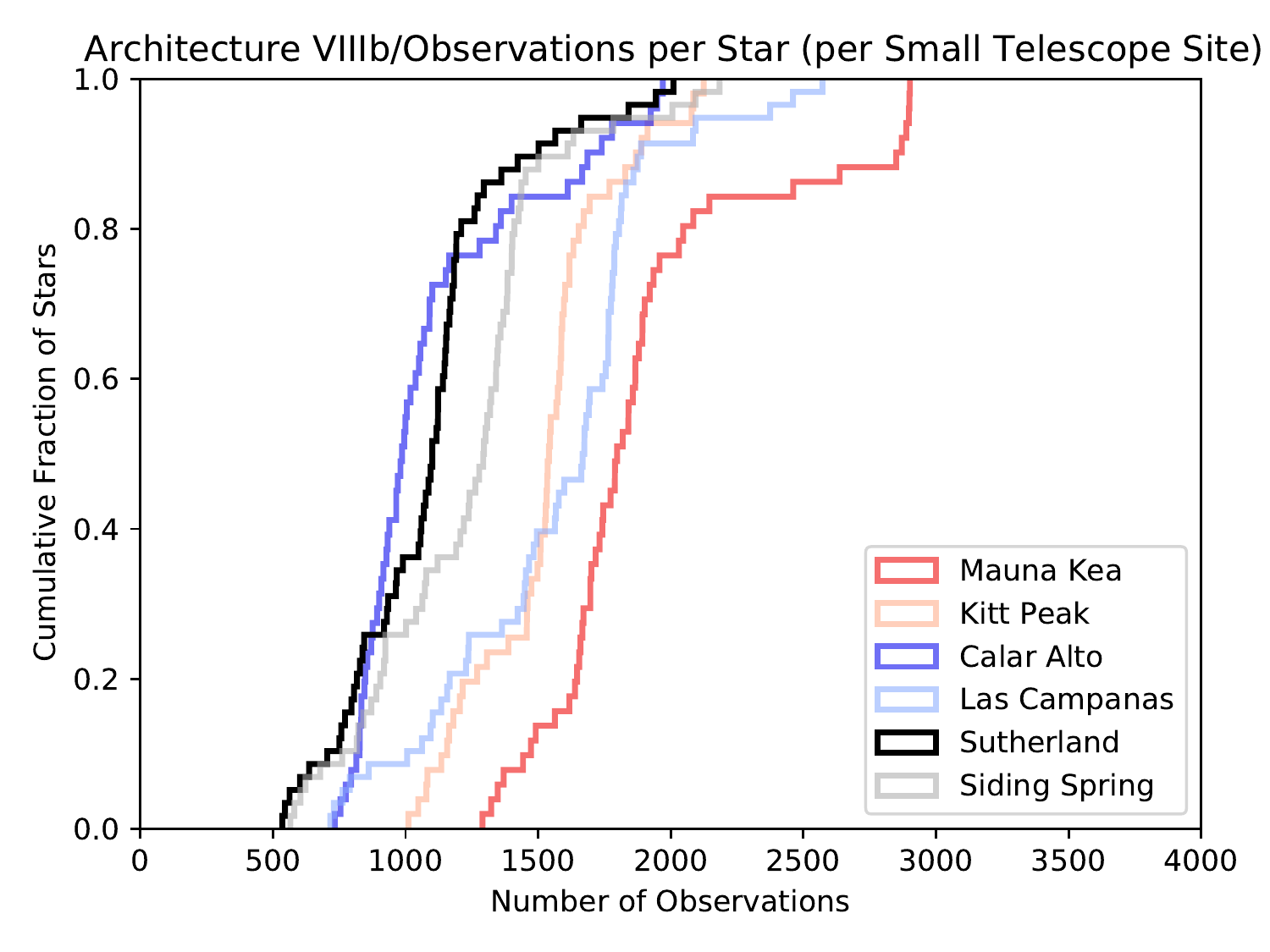}

\noindent \includegraphics[width=0.49\textwidth]{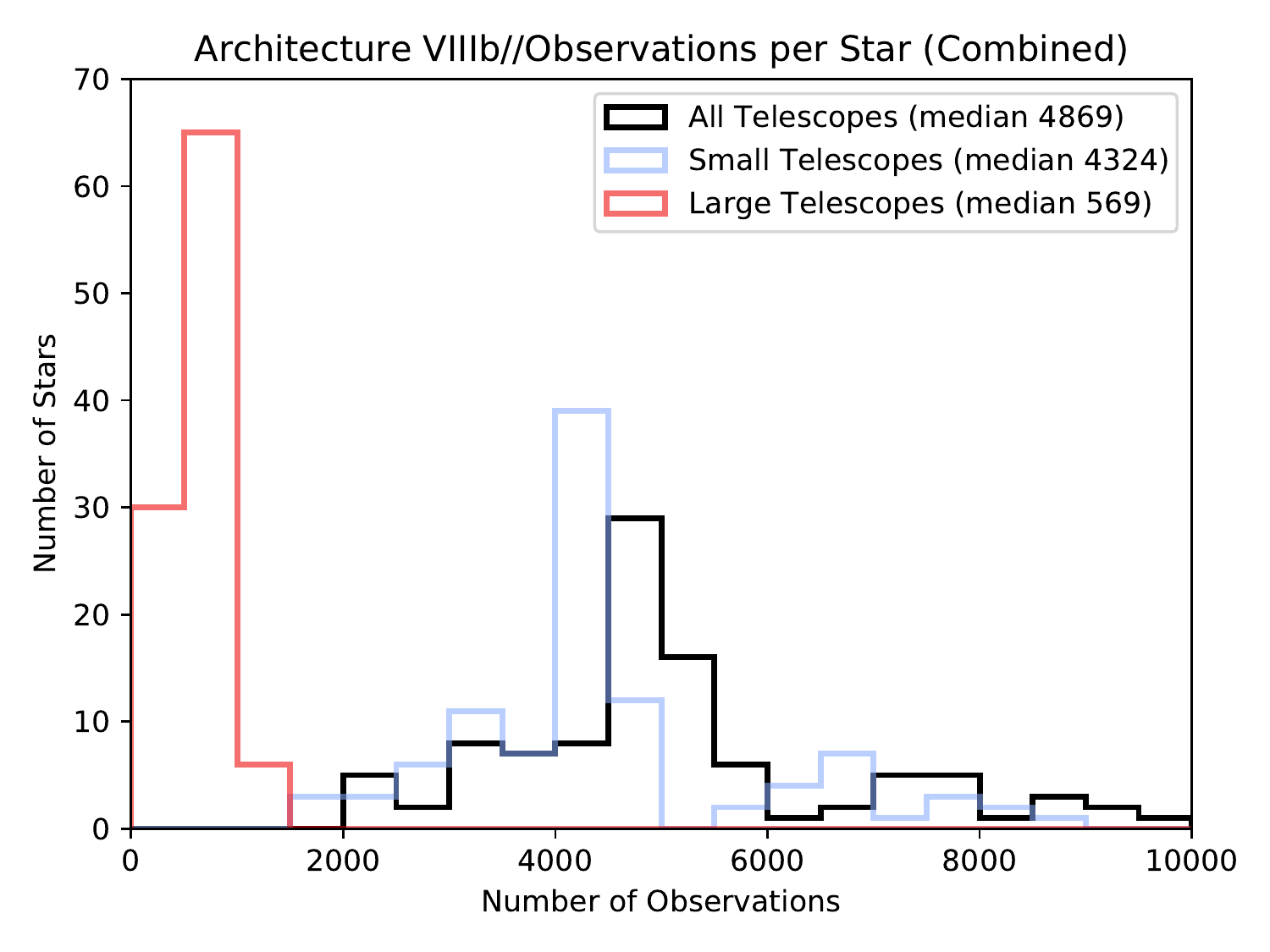}
\includegraphics[width=0.49\textwidth]{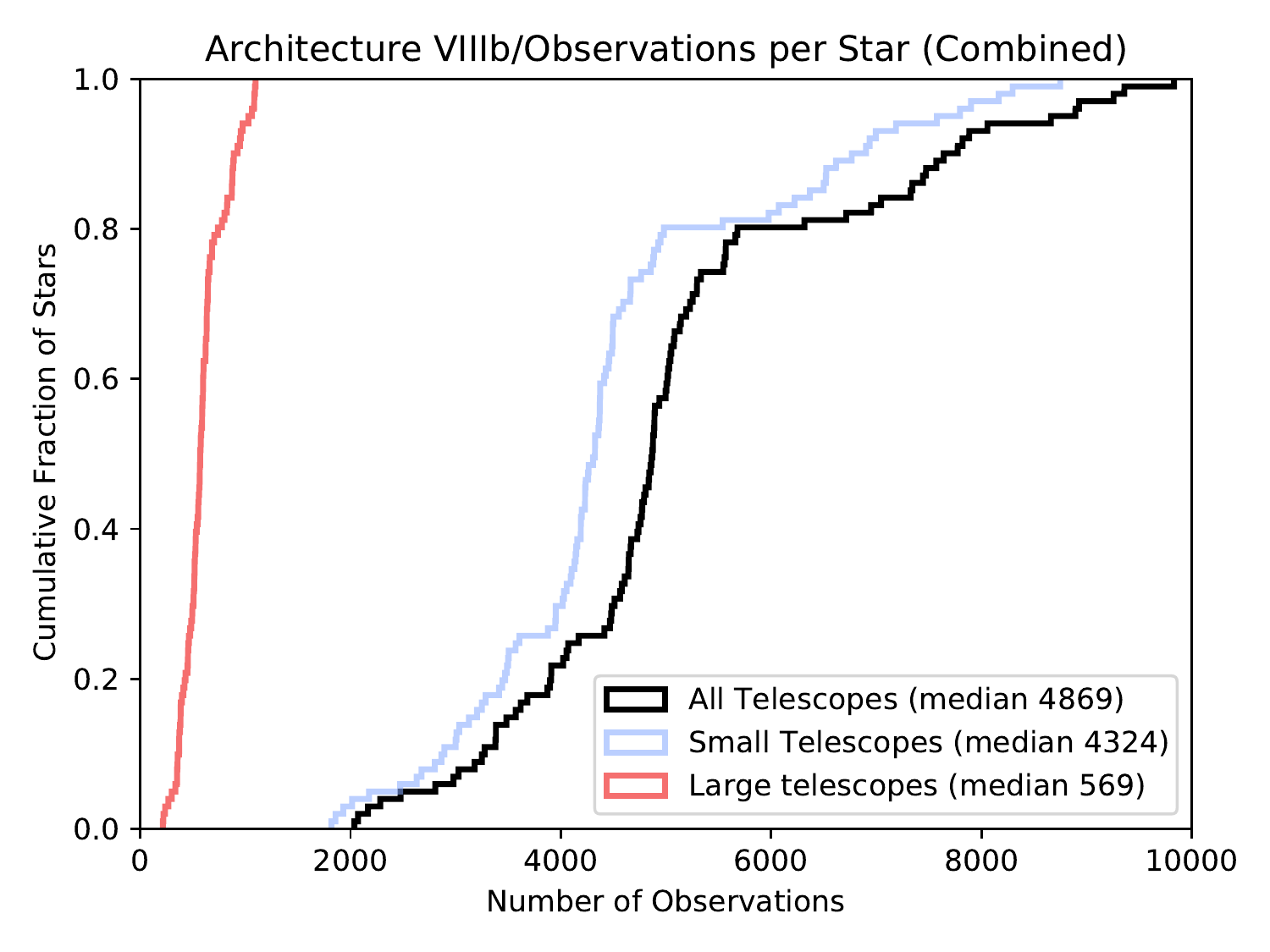}
\caption{Same as figure \ref{fig:ArchIobs}, but for architecture VIIIb.}
\end{figure}

\begin{figure}
\noindent \includegraphics[width=0.49\textwidth]{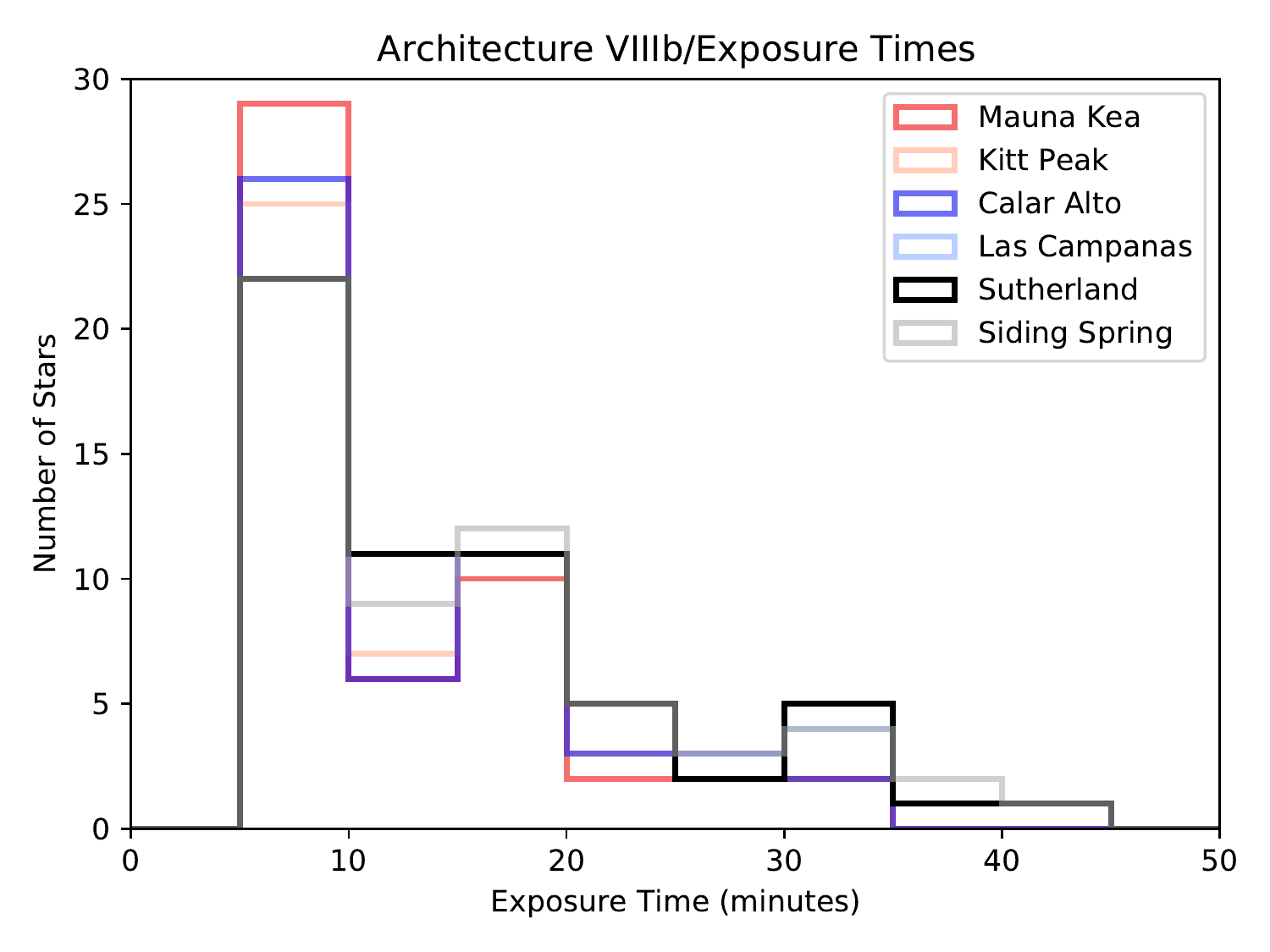}
\includegraphics[width=0.49\textwidth]{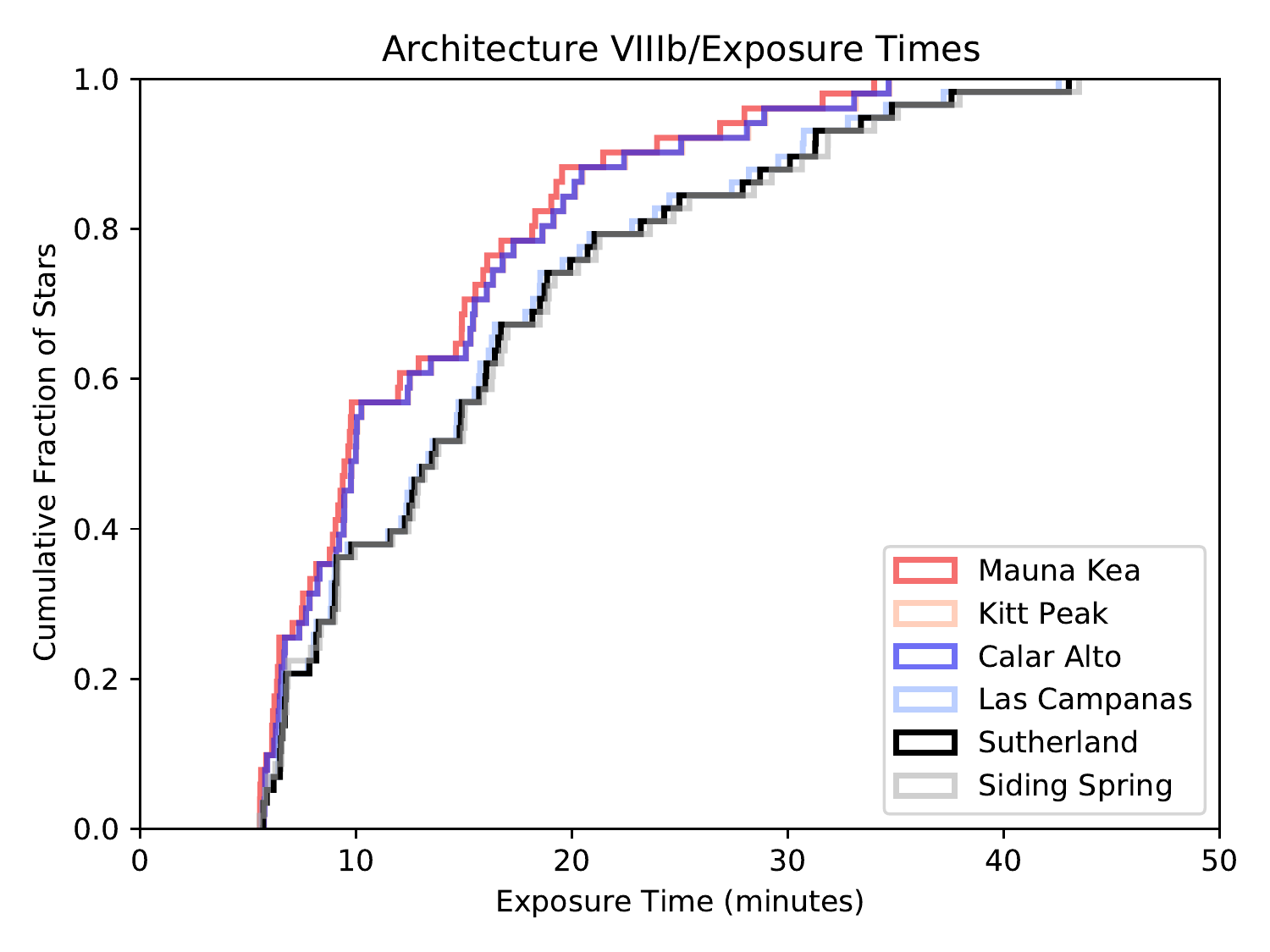}

\noindent \includegraphics[width=0.49\textwidth]{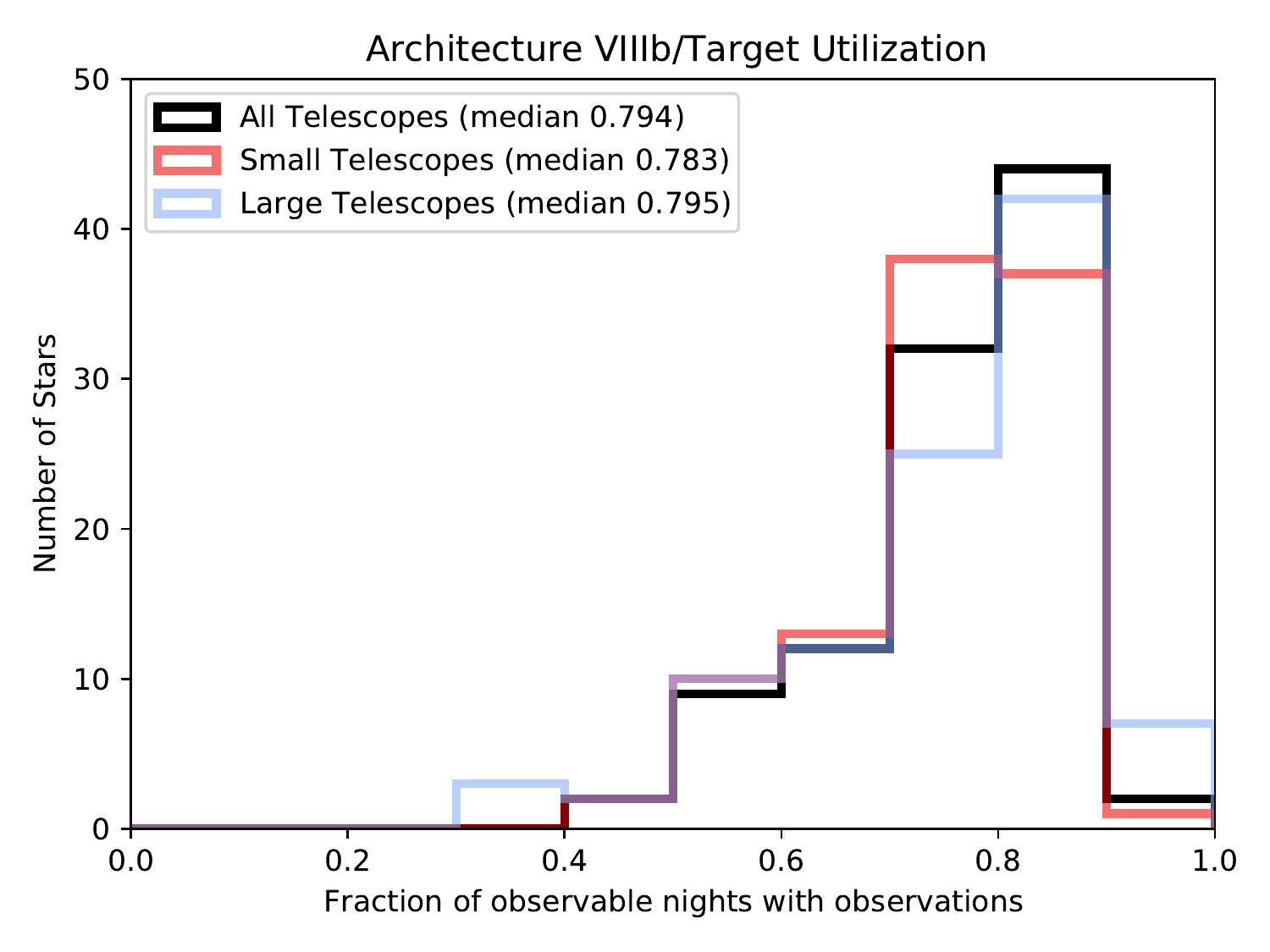}
\includegraphics[width=0.49\textwidth]{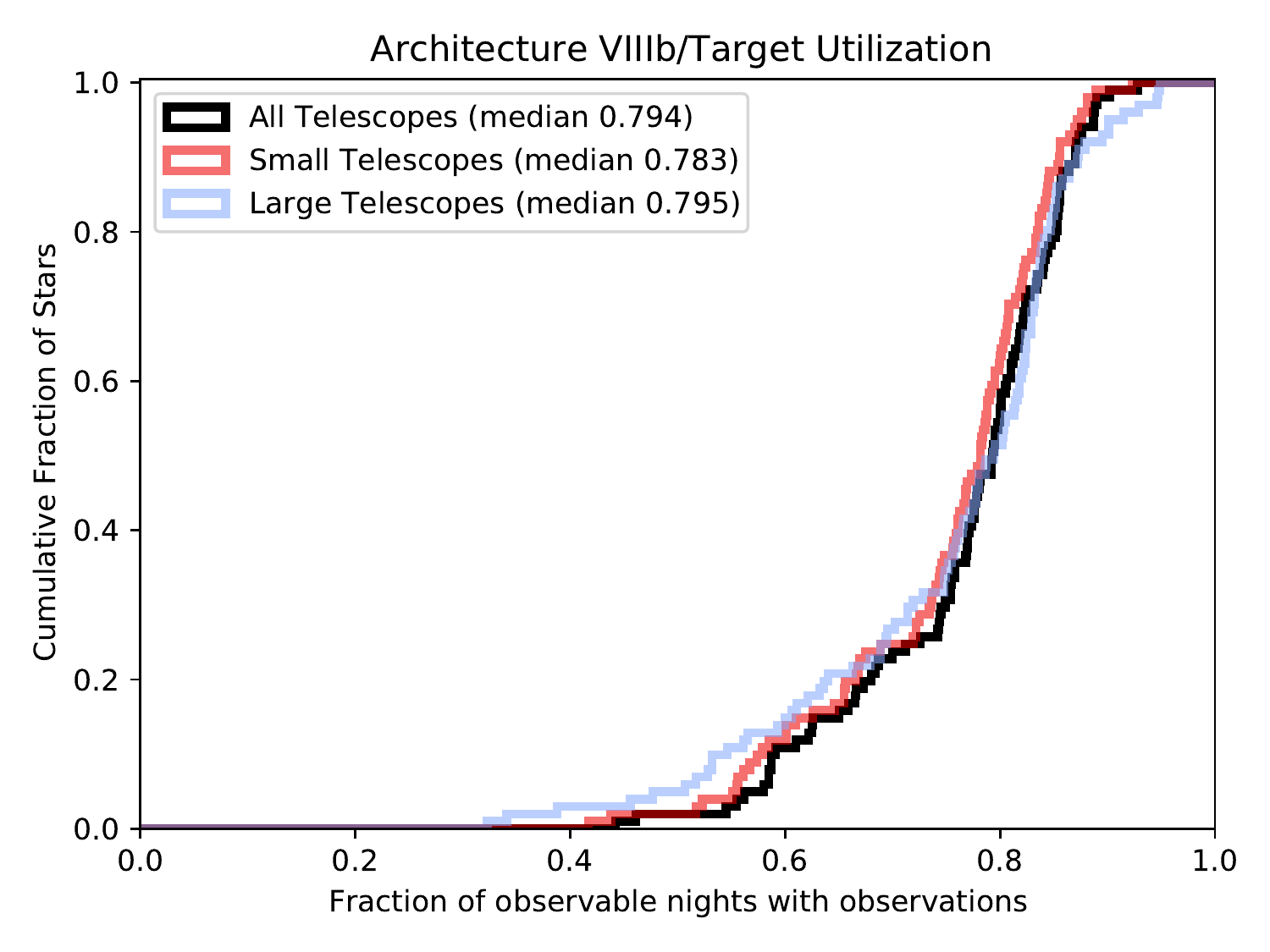}
\caption{Same as figure \ref{fig:ArchIexpfrac}, but for architecture VIIIb.}
\end{figure}

\begin{figure}
\noindent \includegraphics[width=0.49\textwidth]{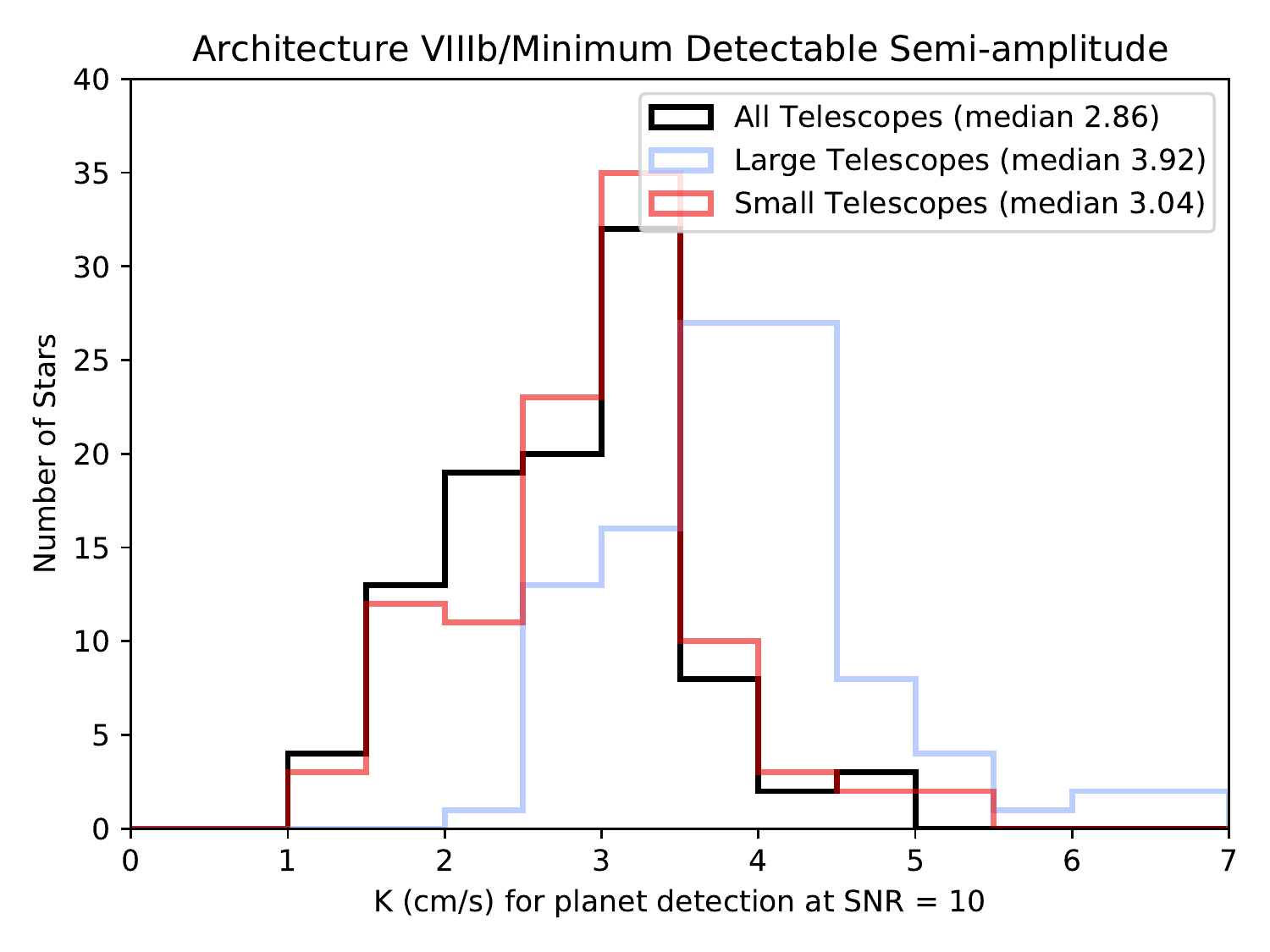}
\includegraphics[width=0.49\textwidth]{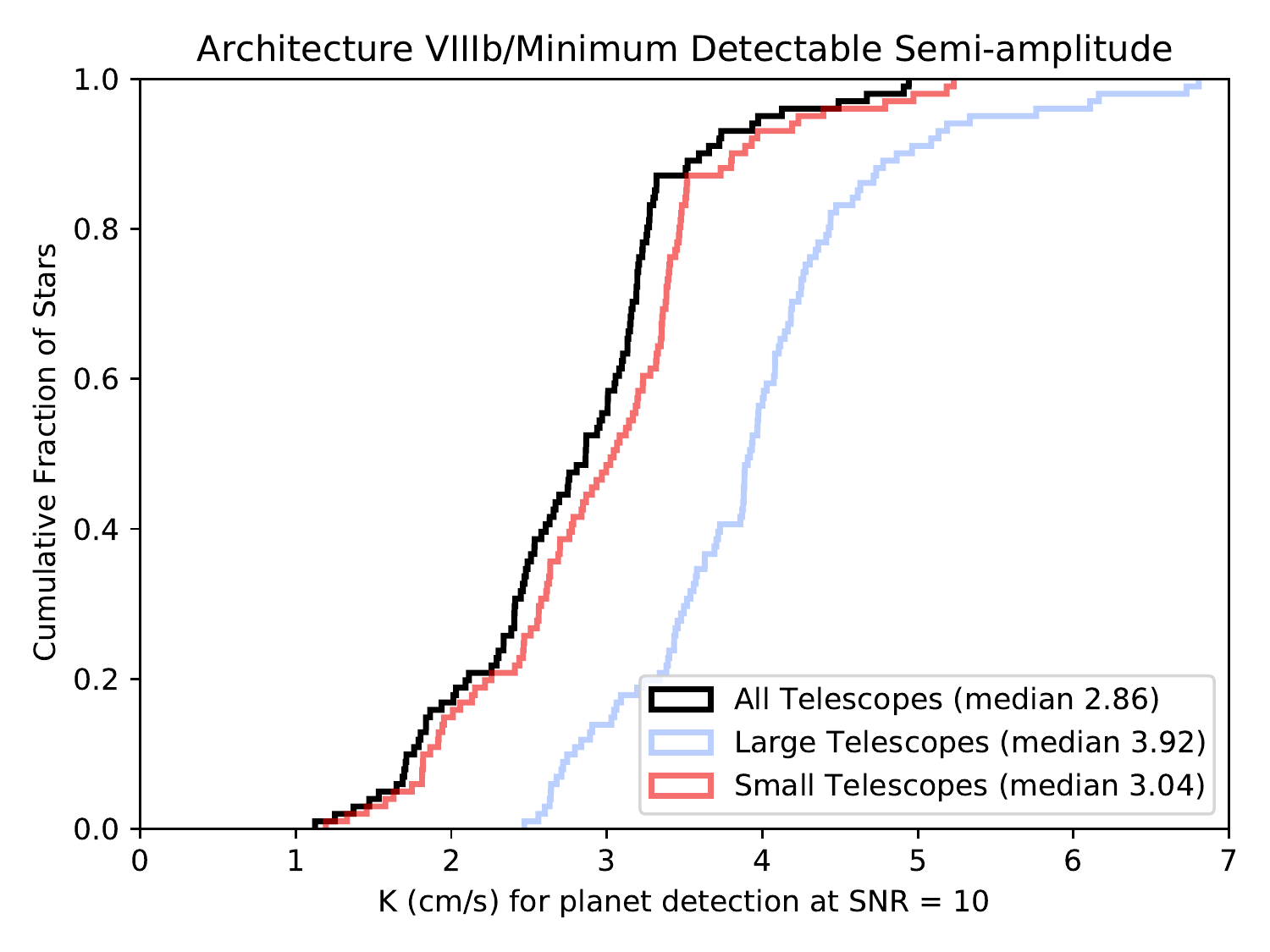}

\noindent \includegraphics[width=0.49\textwidth]{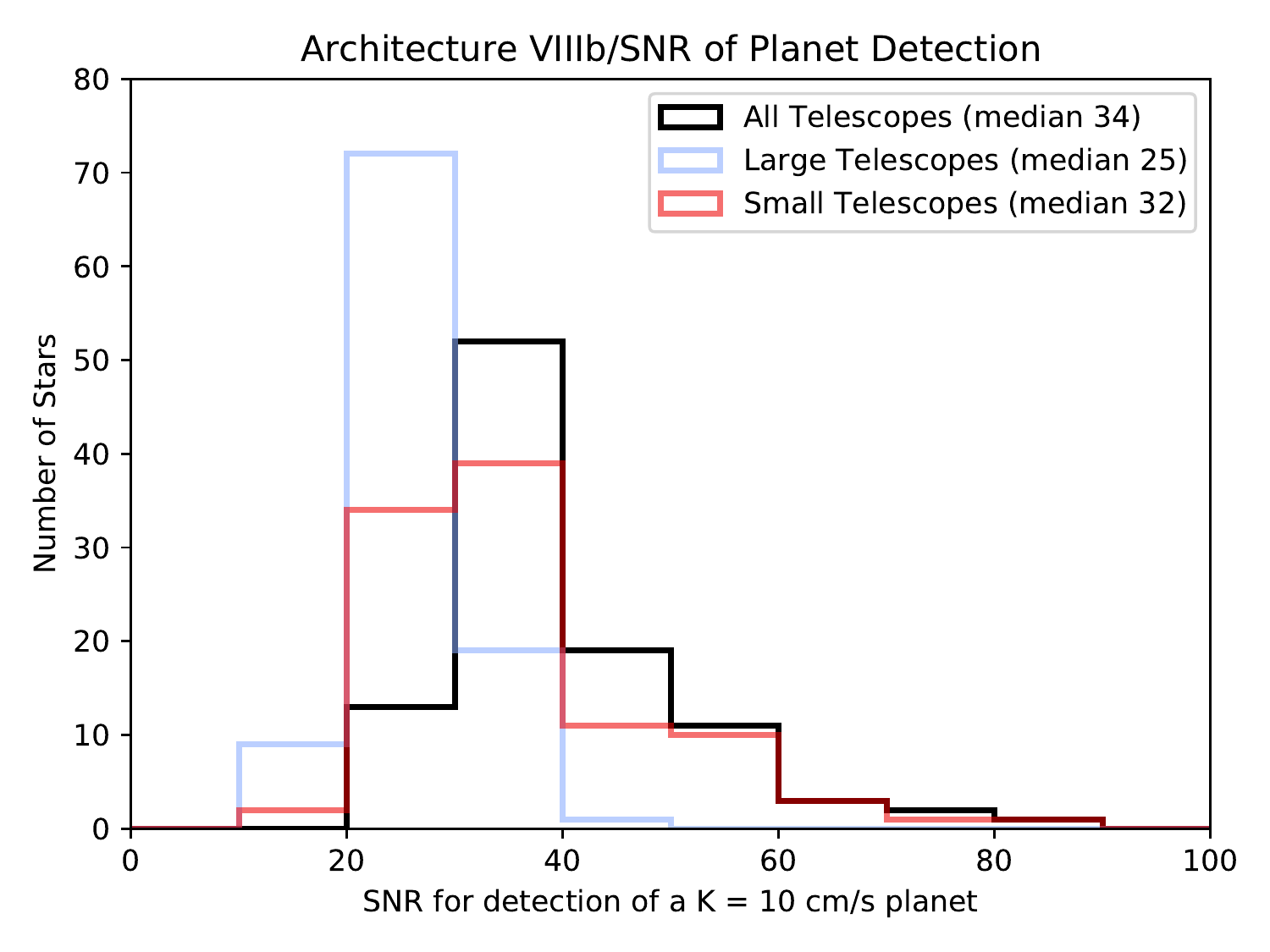}
\includegraphics[width=0.49\textwidth]{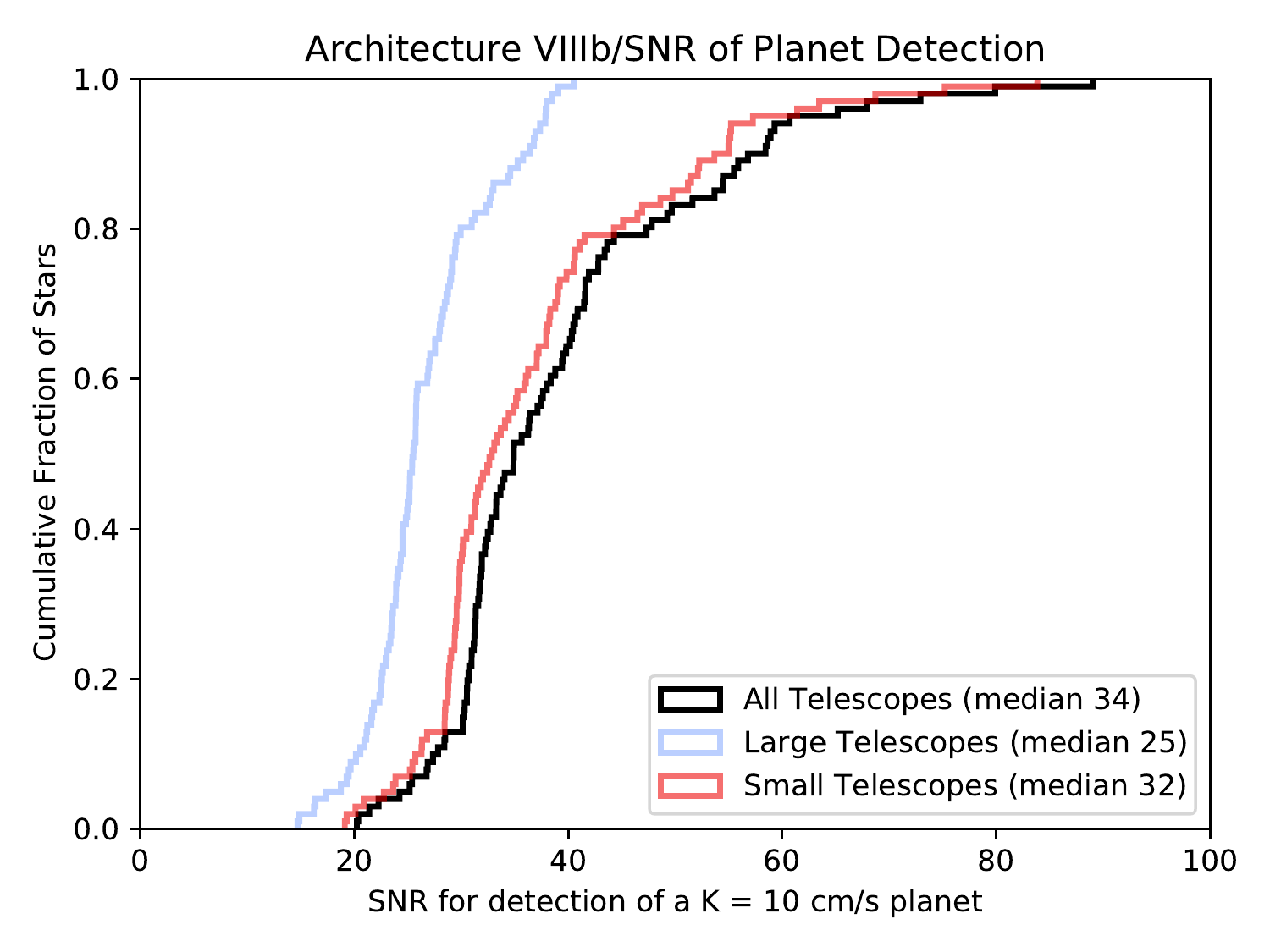}
\caption{Same as figure \ref{fig:ArchIkSNR}, but for architecture VIIIb. This includes using equation \ref{eqn:SNR2} for finding the sensitivity due to the varying precision of the measurements.}
\end{figure}

\end{document}